\documentclass[12pt]{article}
\usepackage{epsfig,amsfonts,amssymb}
\usepackage{hyperref}
\usepackage{cite}
\input epsf.sty
\def\ZZZ{{\hbox{ Z\kern-1.6mm Z}}}
\def\zzz{{\hbox{z\kern-1mm z}}}

\newcommand{\vt}{\vartheta}

\newcommand{\vtau} {\vec \tau}
\newcommand{\qq} {k}
\newcommand{\pp} {l}
\newcommand{\vj} {\vec J}
\newcommand{\vxi} {\vec \xi}
\newcommand{\vu} {\vec u}
\newcommand{\htau} {\vec \eta}
\newcommand{\vc}{\vec\chi}
\newcommand{\vpsi} {\vec \psi}
\def\RRR{{\hbox{ R\kern-2.4mm R}}}

\newcommand{\bF}{{\bar F}}

\newcommand{\eps}{\epsilon}

\newcommand{\vp}{\varphi}

\newcommand{\bx}{\bar x}
\newcommand{\bw}{\bar w}
\newcommand{\ws}{{\wt\sigma}}
\newcommand{\wrh}{{\wt\rho}}
\newcommand{\wv}{{\wt v}}

\newcommand{\VV}{{\cal V}}
\newcommand{\BB}{{\cal B}}
\newcommand{\DD}{{\cal D}}
\newcommand{\II}{{\cal I}}
\newcommand{\AAA}{{\cal A}}

\newcommand{\FF}{{\cal F}}
\newcommand{\JJ}{{\cal J}}
\newcommand{\HH}{{\cal H}}
\newcommand{\MM}{{\cal M}}
\newcommand{\CC}{{\cal C}}

\newcommand{\OO}{{\cal O}}

\newcommand{\EE}{{\cal E}}
\newcommand{\LL}{{\cal L}}

\newcommand{\wt}{\widetilde}
\newcommand{\wh}{\widehat}
\newcommand{\wc}{\wt}

\newcommand{\RR}{{\cal R}}
\newcommand{\NN}{{\cal N}}

\newcommand{\SSS}{{\cal S}}

\newcommand{\tI}{\wt\II}
\newcommand{\hI}{\wh\II}

\newcommand{\cp}{\wh\Phi}

\newcommand{\be}{\begin{equation}}
\newcommand{\ee}{\end{equation}}
\newcommand{\ben}{\begin{eqnarray}\displaystyle}
\newcommand{\een}{\end{eqnarray}}

\newcommand{\bea}[1]{\begin{eqnarray}\label{#1} }
\newcommand{\eea}{\end{eqnarray}}

\newcommand{\refb}[1]{(\ref{#1})}
\newcommand{\p}{\partial}
\newcommand{\sectiono}[1]{\section{#1}\setcounter{equation}{0}}
\newcommand{\subsectiono}[1]{\subsection{#1}\setcounter{equation}{0}}

\def\one{{\hbox{ 1\kern-.8mm l}}}
\def\zero{{\hbox{ 0\kern-1.5mm 0}}}

\newcommand{\eq}{\refb}

\begin{document}
{}~
{}~

~\hfill  arXiv:0708.1270

\vskip .6cm

{\baselineskip20pt
\begin{center}
{\Large \bf
Black Hole Entropy Function, Attractors and Precision
Counting of Microstates} 

\end{center} }

\vskip .6cm
\medskip

\vspace*{4.0ex}

\centerline{\large \rm
Ashoke Sen}

\vspace*{4.0ex}

\centerline{\large \it Harish-Chandra Research Institute}

\centerline{\large \it  Chhatnag Road, Jhusi,
Allahabad 211019, INDIA}

\vspace*{1.0ex}

\centerline{E-mail: sen@mri.ernet.in, ashokesen1999@gmail.com}

\vspace*{5.0ex}

\centerline{\bf Abstract} \bigskip

In these lecture notes we describe 
recent progress in our understanding of attractor
mechanism and entropy of extremal black holes
based on the entropy function formalism. We also describe
precise computation of the microscopic degeneracy of a class
of quarter BPS dyons in $\NN=4$ supersymmetric string theories,
and compare the statistical entropy of these dyons, expanded in
inverse powers of electric and magnetic charges, with a similar
expansion of the corresponding black hole entropy. This comparison
is extended to include the contribution to the entropy from
multi-centered black holes as well.

\vfill \eject

\baselineskip=18pt

\tableofcontents

\renewcommand{\theequation}{\thesection.\arabic{equation}}

\sectiono{Motivation} \label{s00}

It is well known that low energy limit  
of string theory gives rise to gravity
coupled to other fields.
As a result
these theories typically have black hole solutions.
Thus string theory gives a framework for studying classical
and quantum properties of black holes.

Classically black holes are solutions of Einstein's equations with
special properties.
They have a hypothetical surface -- known as the 
event horizon -- surrounding them
such that no object inside the event horizon can escape the black hole.
 In quantum theory however the black hole behaves as a black body
with finite temperature -- known as the Hawking temperature.
Consequently it emits Hawking radiation in accordance with the
laws of black body radiation, and in its interaction with matter
it behaves as a thermodynamic system characterized
by entropy and other thermodynamic quantities. 
In the low curvature approximation where we ignore terms in the
action with more than two derivatives,
this entropy, 
known as the
Bekenstein-Hawking entropy $S_{BH}$,
 is given by a simple expression
\be\label{s0.1}
S_{BH} = A / (4 G_N) \, ,
\ee
where
$A$ is the area of the event horizon
and
 $G_N$ is the Newton's constant.
One of the important questions is:
Can we understand  this entropy from 
statistical viewpoint \i.e.\ as logarithm of the number of
quantum states associated with the black hole?

Although we do not yet have a complete answer to this question,
for a special class of black holes in string theory, -- known as 
extremal black holes, -- this question has been answered in the
affirmative. 
These 
black holes have zero temperature and hence do not Hawking radiate
and are usually stable. 
 Often, but not always, extremal black holes are also invariant under
certain number of supersymmetry transformations.
In that case they are called BPS black holes. Due to the stability and
the supersymmetry properties one has some control over the dynamics
of the microscopic configuration (typically
involving D-branes, fundamental strings and other solitonic
objects) representing these black holes. This in turn allows us to calculate
the degeneracy of such states at weak coupling where gravitational
backreaction of the system can be ignored. Supersymmetry allows
us to continue the result to strong coupling where gravitational
backreaction becomes important and the system can be described
as a black hole.
In string theory one finds that 
for a wide class of extremal BPS black holes we have, in the limit where
the size of the black hole is large\cite{9601029},
\be \label{s0.2} 
S_{BH}(Q) = S_{stat}(Q)\, ,
\ee
where $S_{BH}(Q)$ denotes the Bekenstein-Hawking entropy of an
extremal black hole carrying a given set of charges labelled by $Q$, and
$S_{stat}(Q)$ is defined as
\be\label{s0.3}
S_{stat}(Q) = \ln d(Q)\, ,
\ee
where $d(Q)$ is the degeneracy of BPS states in the theory
carrying the same set of charges.
This clearly gives a good
understanding of this Bekenstein-Hawking
entropy from microscopic viewpoint.

The initial comparison between
$S_{BH}$ and $S_{stat}$
was carried out in the limit of large charges. Typically
in this limit the horizon size is large so that the
curvature and other field strengths at the horizon
are small and hence we can calculate the entropy via eq.\refb{s0.1}
without worrying about the higher derivative corrections to the
effective action of string theory. On the other hand the computation
of $S_{stat}(Q)$ also simplifies in this limit since the
dynamics of the corresponding microscopic system 
is  often described by a 1+1 dimensional conformal field theory
(CFT)
with the
spatial coordinate compactified on a circle. An extremal black hole
with large charges typically  corresponds to a state of this
CFT with large $L_0$ (or $\bar L_0$) eigenvalue and
zero $\bar L_0$ (or $L_0$) eigenvalue. The degeneracy
of such states can be computed via the Cardy formula in terms of the
left- and right-handed central charges $(c_L,c_R)$ 
of the conformal field theory and the $L_0$ 
(or $\bar L_0$) eigenvalue without knowing the details of the
conformal field theory:
\ben \label{e0.4}
S_{stat}(Q) &\simeq& 2\pi \sqrt{c_L L_0\over 6} \quad
\hbox{for $\bar L_0=0$} \nonumber \\
&\simeq& 2\pi \sqrt{c_R \bar L_0\over 6} \quad
\hbox{for $L_0=0$}\, .
\een
The pleasant surprise is that these two completely different
computations, -- one for $S_{BH}( Q)$ and the other for
$S_{stat}(Q)$, -- give the same answer.

Given this success, it is natural
to carry out this comparison to finer details.
When we move away from the large charge limit, the curvature and
other field strengths at the horizon are no longer negligible.
Thus we must take into account the  effect of higher derivative terms
in the effective action on the black hole entropy.
Typical example of such
higher derivative terms are terms involving 
square and higher powers of the
Riemann tensor. For a large but finite size black hole we expect
the effect
of these higher derivative terms at the horizon to be small
but non-zero, giving rise to 
small modifications of the horizon geometry
and consequently the black hole entropy.   
On the other hand for finite but large charges the statistical
entropy computed from the Cardy formula 
will also receive corrections
which are suppressed by
inverse powers of charges. Thus it would be natural to ask:
Does the agreement between $S_{BH}(Q)$ and
$S_{stat}(Q)$ continue to hold even after taking
into account the effects of higher derivative corrections on the black
hole side, and deviation from the Cardy formula on the
statistical side?

Due to
an inherent ambiguity
in defining the black hole entropy and the statistical
entropy beyond the large charge limit,
the issue involved is more complex than it sounds. The original 
Bekenstein-Hawking entropy was computed in classical general
theory of relativity.
However once we begin including higher
derivative corrections, various string dualities
can map the classical contribution to the effective action 
to the quantum contribution and vice versa.
Thus it no longer makes sense to restrict our
analysis to the classical theory.
A natural choice will be to use 
the one particle
irreducible (1PI) 
effective action of the theory since it respects all
the duality symmetries. However since string
theory has massless particles, the 1PI effective action typically gets
non-local contributions if we go to sufficiently high order in the
derivatives. 
As we shall describe in this review, while for any higher derivative
theory of gravity with a local Lagrangian density there is a well defined
algorithm for computing the entropy of a black hole, 
at present no technique is available for treating theories with
non-local action. This causes a potential problem in defining the
entropy of a black hole in string theory beyond the leading order.
We could circumvent this problem using the
Wilsonian effective action which is
manifestly local, but this does not respect all the duality symmetries
of the theory. 
There is a similar ambiguity 
on the statistical side as well. Since the corrections to the entropy
are suppressed
by inverse powers of various charges, they can be regarded as
finite size corrections. However such corrections are known to
depend crucially on the ensemble we choose to define the
entropy. For example we could use duality invariant microcanonical
or grand canonical ensembles, or use 
duality non-invariant mixed
ensembles where we treat a subset of the charges as we would do
in a microcanonical ensemble and the rest of the charges as we
would do in a grand canonical ensemble\cite{0405146}.

We  hope that by studying explicit examples where
one could compute the corrections to both the statistical entropy
and the black hole entropy, we may be able to resolve the
above mentioned ambiguities and make a more precise
formulation of the relationship between the two entropies.
In order to proceed along these lines we need to open two fronts.
First of all we need to learn how to take into account the effect of
the higher derivative terms on the computation of 
black hole entropy.
But we also need to know how to calculate the statistical
entropy to greater accuracy.
This involves precise computation of the degeneracy of states with
a given set of charges.
In this review we shall address both these issues. The first part of the
review, dealing with the computation of black hole entropy
in the presence of higher derivative terms, will be based on the
entropy function formalism of \cite{0506177,0508042}, -- 
this in turn is an adaptation of a more 
general formalism for computing
black hole entropy in the presence of higher derivative 
terms\cite{9307038}
to the
special case of extremal black holes.
The second part of the review, dealing with precision computation
of statistical entropy for  a class of four dimensional black holes
in $\NN=4$ supersymmetric string theories, will follow the analysis
of
\cite{0605210,0607155,0609109}.
The original formula for the statistical entropy was first proposed
in \cite{9607026}
for the special case of heterotic string theory compactified on
$T^6$, and later extended to more general models in
\cite{0510147,0602254,0603066}.
Various 
alternative
approaches to proving these formul\ae\ have been explored 
in \cite{0505094,0506249,0508174,0506151,0612011}.
We shall not review these different approaches, except
parts of \cite{0505094,0508174} that will be directly relevant
for our approach to the counting problem, and also
part of \cite{0412287} that will be useful in extracting
the asymptotic behaviour of the statistical entropy for large
charges.

In our
analysis we shall try to maintain manifest duality invariance
by using the 1PI effective action on the black hole side and
microcanonical ensemble on the statistical side. 
As we shall see,
to the extent we can compute both sides, the black hole and the
statistical entropy agree even after taking into account higher
derivative corrections.  
The analysis on the statistical side is quite clean, and we find
an explicit algorithm to generate an expansion of the
statistical entropy in inverse powers of charges. The analysis on
the black hole side suffers from the inherent
problem of having to deal with non-local terms, but we carry out
our analysis by including the effect of a duality invariant set of
local four derivative terms in the action, -- the Gauss-Bonnet term.
Although we cannot prove that the other four derivative terms
(including non-local terms) do not contribute to this order, we
find that in various limits where this additional contribution 
can be computed, the result for the black hole entropy agrees with
the one computed using the Gauss-Bonnet term. We believe that
the limitation due to the non-local nature of the 1PI effective
action can be overcome in the future, and 
we shall find a systematic procedure for calculating the black hole
entropy using the 1PI effective action.
On the other hand it is also possible in principle
that we can give up manifest duality invariance and work with
Wilsonian effective action on the black hole side and a mixed
ensemble on the statistical side. This approach has been
advocated in \cite{0405146}.

The rest of the review is organised as follows. In \S\ref{s1} we
develop the entropy function formalism for computing the entropy
of extremal black holes. 
We include in this discussion spherically symmetric black holes
in various dimensions, rotating black holes, theories with Chern-Simons
terms etc., and also describe how the entropy function formalism
provides a simple proof of the attractor phenomenon.
\S\ref{s2} contains application of this
formalism to the computation of entropy of a wide class of extremal
black holes in a wide class of theories. These include in particular
a class of quarter BPS black holes in the $\NN=4$ supersymmetric
four dimensional string theories; these are the black holes for which
we later carry out a detailed comparison between the
black hole entropy and the statistical entropy. In \S\ref{sads3}
we compare our approach, which holds for a general extremal black
hole, with that of \cite{0506176,0508218,0607138,0609074}
where for a special class of extremal 
black holes  -- with an $AdS_3$ factor
in the near horizon geometry --  a more powerful technique for
computing the entropy was developed.   In \S\ref{s3}
we describe the computation of statistical entropy of  a special
class of black holes in $\NN=4$ supersymmetric string theories in
four dimensions, and compare the results with the black hole entropy
computed in \S\ref{s2}. We conclude in \S\ref{sopen} with a list
of open questions. We also speculate on how our degeneracy
formula might be extendable to the $\NN=2$ supersymmetric
string theories. The appendices provide us with some technical 
results which are used mainly in the computation of the statistical
entropy.

We must caution the reader that this is not a complete review of
attractor mechanism or computation of black hole and statistical
entropy. Instead it deals with only a very specific 
approach to computing entropy of extremal black holes
using the entropy function formalism, and computation of the
statistical entropy of a special class of dyons in a special class
of theories. It 
does not even contain all aspects of the entropy function
formalism or the computation of the statistical entropy in $\NN=4$
supersymmetric string theories. However
we have tried to make the review self-contained in the sense that
for the material that does get covered in the review, one does not always
have to go back and
consult the original literature in order to follow the material. 
As a result the
review has become somewhat long; however we hope that this has not
significantly increased the time needed to go through the rest of the
review.

\renewcommand{\theequation}{\thesubsection.\arabic{equation}}

\sectiono{Black Hole Entropy Function and the
Attractor Mechanism} \label{s1}

In this section we shall develop a general method for computing
the entropy of extremal black holes in a theory of gravity with higher
derivative corrections. The first question we need to address is:
How do we define
extremal black holes in a general higher derivative theory of gravity?
For this we shall
take the clue from usual  theories of gravity with two derivative
actions, -- namely study the
properties of extremal black holes in these theories and then identify
certain universal features which can be adopted as the definition of
extremal black holes in a more general class of theories with higher
derivative terms.

\subsectiono{Definition of extremal black holes}
\label{sreiss}

We begin our analysis with the Reissner-Nordstrom solution
describing a spherically symmetric charged black hole
in the usual Einstein-Maxwell theory in four dimensions. This theory
is described by the action
\be\label{e1.1}
\SSS = \int d^4 x\, \sqrt{-\det g}\, \LL, \qquad
\LL = {1\over 16\pi G_N} R -
{1\over 4} F_{\mu\nu} F^{\mu\nu} \, .
\ee
We shall be using the following notations for the Christoffel symbol
and Riemann tensors:
\ben \label{e7btz}
&& \Gamma^\mu_{\nu\rho} = {1\over 2} g^{\mu\sigma}
\left( \p_\nu g_{\sigma\rho} + \p_\rho g_{\sigma\nu}
- \p_\sigma g_{\nu\rho} \right) \nonumber \\
&& R^\mu_{~\nu\rho\sigma} = \p_\rho
\Gamma^\mu_{\nu\sigma} - \p_\sigma
\Gamma^\mu_{\nu\rho} + \Gamma^\mu_{\tau\rho}
\Gamma^\tau_{\nu\sigma} - \Gamma^\mu_{\tau\sigma}
\Gamma^\tau_{\nu\rho} \nonumber \\
&& R_{\nu\sigma} = R^\mu_{~\nu\mu\sigma}, \qquad
R = g^{\nu\sigma}  R_{\nu\sigma}\, .
\een
The 
Reissner-Nordstrom solution in this theory is given by
\ben \label{e1.2}
 ds^2  &=& - (1 - a/\rho) (1 - b/\rho) d\tau^2   + {d\rho^2\over
 (1 -a/\rho) (1 - b/\rho)}   
 + \rho^2 (d\theta^2 + \sin^2\theta d\phi^2) \, , \nonumber \\
 F_{\rho\tau} &=& {q\over 4\pi\rho^2}
 , \qquad F_{\theta\phi} = {p\over 4\pi} \, \sin\theta\, ,
 \een
  where $\rho$, $\theta$, $\phi$ and $\tau$ are the space-time
 coordinates,  $a$ and $b$ are two constants determined from the relation
 \be \label{e1.2.1}
 a+b = 2G_NM, \qquad ab = {G_N\over 4\pi} (q^2 + p^2)\, ,
 \ee
 and $q$, $p$ and $M$ 
 denote the electric and magnetic charges and the mass
 of the black hole respectively.
 If we take $a>b$ then the inner and the outer
 horizon of the black hole are at $r=b$ and at $r=a$ respectively.
 The extremal limit corresponds to choosing 
 \be \label{e1.2.2}
 M^2 = {1\over 4\pi G_N} (q^2 + p^2)\, ,
 \ee
 so that we have
 \be \label{e1.2.3}
 a=b = \sqrt{{G_N\over 4\pi} (q^2 +p^2)}\, .
 \ee
 We now define
\be \label{e1.3}
t = \lambda\, \tau /  a^2, \qquad r=\lambda^{-1} (\rho-a), 
\ee
 where $\lambda$ is an arbitrary constant, and rewrite the
 extremal solution in this new coordinate system. This gives
\ben \label{e1.4}
 ds^2 &=& - {r^2 a^4 \over (a+\lambda\, r)^2} d t^2 + 
 {(a+\lambda\, r)^2\over r^2} dr^2  
 + (a+\lambda\, r)^2 (d\theta^2 +
 \sin^2\theta d\phi^2) \, , \nonumber \\
 F_{rt} &=& {q a^2 \over 4\pi (a + \lambda r)^2}, 
 \qquad F_{\theta\phi} = {p\over 4\pi}\sin\theta\, .
 \een
Finally we take the `near horizon' 
limit $\lambda\to 0$. In this limit the solution takes
the form:
 \ben\label{e1.5}  
 ds^2 &=&  a^2 \left(-r^2 dt^2 + {dr^2\over r^2} \right)+
 a^2 (d\theta^2 + \sin^2\theta d\phi^2) \, , \nonumber \\
 F_{rt} &=& {q\over 4\pi}, \qquad F_{\theta\phi} = {p\over 4\pi}
 \sin\theta\, .
 \een
 The entropy of the black hole, obtained by dividing the area of the
 horizon by $4 G_N$, is
 \be \label{entropyr}
 S_{BH} = {1\over 4}(q^2 + p^2)\, .
 \ee
 
The field configuration 
given in \refb{e1.5} has the following features:
\begin{enumerate}
\item In the limit $\lambda\to 0$ keeping $r$ fixed, the original
coordinate $\rho$ approaches $a$. Thus \refb{e1.5} describes
the field configuration of the black hole near the horizon.

\item 
 Since for any $\lambda$ \refb{e1.4} describes 
 an exact classical solution, in
 the $\lambda\to 0$ limit also we have an exact classical solution
 for all finite $r$, not just for small $r$. 
Indeed, we could have obtained
 \refb{e1.5} by directly solving the equations of motion of the
 Einstein-Maxwell theory without any reference to black holes.
 
 \item The space-time described by \refb{e1.5} splits into a product
 of two spaces. One of them, labelled by $(\theta,\phi)$ describes an
 ordinary two dimensional sphere $S^2$. The other, labelled by $(r,t)$,
 describes a two dimensional 
 space-time known as $AdS_2$. This is a solution
 of two dimensional Einstein gravity with negative cosmological
 constant.
 
\item The background described in \refb{e1.5} has an
$SO(3)$ isometry acting on the sphere $S^2$. This reflects the
spherical symmetry of the original black hole and is present
even in the full black hole solution. The background also has
an SO(2,1) isometry acting on the $AdS_2$ space that was 
not present in the full black hole solution.
This isometry is
 generated by
\be \label{e1.7}
 L_1=\p_t, \quad L_0=t\p_t - r\p_r, \quad
 L_{-1} = {1\over 2} \left( {1\over r^2}+t^2\right) \p_t 
 -t\, r \, \p_r\, .
 \ee
 Not only the metric, but also the gauge field strengths given in
 \refb{e1.5} can be shown to be invariant under the 
 $SO(2,1)\times SO(3)$ transformation.
 \end{enumerate}  
 It turns out that
all known extremal spherically symmetric 
 black holes
 in four dimensions
 with non-singular
horizon  have near horizon geometry
 $AdS_2\times S^2$ and an associated isometry 
 $SO(2,1)\times SO(3)$.
 
 Consider now the effect of adding higher derivative terms in the
 action. In general it is quite difficult to find the full black hole
 solution after taking into account these higher derivative terms.
 However it is natural to postulate that the symmetries of the near
 horizon geometry will not be destroyed by these higher derivative
 terms. This suggests the following postulate:
 
 \noindent{\it In any generally covariant theory of gravity coupled
 to matter fields, the near horizon geometry of a 
 spherically symmetric  extremal black hole in four
 dimensions has $SO(2,1)\times SO(3)$ isometry.}
 
 \noindent We shall take this as the definition of spherically
 symmetric extremal black holes in four dimensions. Although
 we arrived at this definition by analyzing extremal black holes
 in theories with only two derivative terms in the action,
 and possible small modification of the solution due to
 higher derivative terms, we shall extend the definition to
 include even black holes with large curvature at the
 horizon so that the higher derivative terms are as important
 as the two derivative terms.\footnote{Such black holes are
 called small black holes and will be the subject of discussion
 in \S\ref{s2.3} and \S\ref{smallre}.}
 
 This analysis can be generalized to study spherically symmetric
 extremal black holes in other dimensions,
 as well as rotating extremal black holes in four and other
 dimensions. In every known example one finds that the
 near horizon geometry of extremal black holes has an enhanced
 $SO(2,1)$ isometry that is not present in the full black holes
 solution. The full isometry of the near horizon geometry is then
 the product of the $SO(2,1)$ isometry and rotational isometry
 of the full black hole solution. For example
 \begin{enumerate}
 \item The near horizon geometry of
 an extremal spherically symmetric black hole in $D$
 dimensions has $SO(2,1)\times SO(D-1)$ isometry.
 \item The near horizon geometry of
 an extremal rotating black hole in four dimensions has
 $SO(2,1)\times U(1)$ isometry.
 \end{enumerate}
 We shall take these as definitions of the corresponding
 extremal black holes even after inclusion of higher derivative 
 terms.\footnote{In four and five dimensions these postulates 
 have recently been
 proved in \cite{0705.4214}.}
 
 Based on these postulates we shall now develop a general procedure
 for computing the entropy of extremal black holes.
 Our discussion will follow \cite{0506176,0508042,0606244}.

\subsectiono{Spherically symmetric black holes in $D=4$}
\label{s1.1}

Let us consider
a four dimensional theory of gravity coupled
to a set of  abelian gauge fields $A_\mu^{(i)}$
and 
neutral scalar fields $\{\phi_s\}$.  Let $\sqrt{-\det g}\,
\LL$ be the lagrangian density, 
expressed as a function of the metric $g_{\mu\nu}$, the scalar fields 
$\{\phi_s\}$,
the gauge field strengths $F^{(i)}_{\mu\nu}$, 
and covariant derivatives of these fields. We have
not included any antisymmetric rank two tensor field in our list
of fields since such fields can always be dualized to a scalar
field. 
When written in terms of the anti-symmetric tensor field, 
the definition
of the field strength often contains gauge and Lorentz Chern-Simons
terms. However
when written in terms of the dual scalar fields there are no 
Chern-Simons type term in the action. Hence reparametrization
and gauge invariance 
of the action implies that $\LL$ is
manifestly reparametrization  invariant and gauge invariant under the
usual transformation laws of various fields.  Thus $\LL$ 
must be constructed from the scalar fields $\phi_s$, gauge field strengths
$F^{(i)}_{\mu\nu}\equiv \p_\mu A^{(i)}_\nu - \p_\nu A^{(i)}_\mu$,
the inverse metric $g^{\mu\nu}$, the Riemann tensor 
$R_{\mu\nu\rho\sigma}$ and covariant derivatives of these fields.

We consider a spherically symmetric extremal black hole solution
with $SO(2,1)\times SO(3)$ invariant
near horizon geometry.  
The most general field configuration consistent
with this isometry
is of the form:
\ben \label{e1}
&& ds^2\equiv g_{\mu\nu}dx^\mu dx^\nu = v_1\left(-r^2 dt^2+{dr^2\over 
r^2}\right)  +
v_2 \left(d\theta^2+\sin^2\theta d\phi^2\right) \nonumber \\
&& \phi_s =u_s \nonumber \\
&& F^{(i)}_{rt} = e_i, \qquad  F^{(i)}_{\theta\phi} = {p_i \over 4\pi} \,  
\sin\theta\, , 
\een
where $v_1$, $v_2$, $\{u_s\}$, $\{e_i\}$ and $\{p_i\}$ are constants.
For this background the nonvanishing components of the Riemann tensor 
are:
\ben \label{e1a}
R_{\alpha\beta\gamma\delta} &= &-v_1^{-1} (g_{\alpha\gamma} g_{\beta\delta} 
- g_{\alpha\delta} g_{\beta\gamma})\, , \qquad \alpha, \beta, \gamma, 
\delta =r, t\, ,
\nonumber \\
R_{mnpq} &=& v_2^{-1}\,  (g_{mp} g_{nq} - g_{mq} g_{np} )\, , \qquad m,n,p,q
= \theta, \phi \, .
\een
It follows from the general form of the background that the covariant 
derivatives of 
the scalar fields $\phi_s$, the gauge field strengths 
$F^{(i)}_{\mu\nu}$ and the Riemann tensor
$R_{\mu\nu\rho\sigma}$ all vanish for the near horizon geometry. By the general
symmetry consideration 
it follows that the contribution to the equation of motion from
any term in $\LL$ that involves covariant derivatives of the gauge field
strengths, scalars
or the Riemann tensor vanish identically 
for this background and we can restrict our
attention to only those terms which do not involve 
covariant derivatives of these fields.

Let us
denote by $f(\vec u, \vec v, \vec e, \vec p)$ the Lagrangian density 
$\sqrt{-\det g}\,
\LL$ evaluated for
the near horizon geometry \refb{e1} and integrated over the angular 
coordinates:
\be \label{e2}
f(\vec u, \vec v, \vec e, \vec p) = \int d\theta\, d\phi\, \sqrt{-\det 
g}\, \LL\, .
\ee
The scalar and the metric field equations in the near horizon geometry 
correspond to
extremizing $f$ with respect to the variables $\vec u$ and $\vec v$:
\be \label{e3}
{\p f\over \p u_s} =0, \qquad {\p f\over \p v_i} = 0\, .
\ee
Furthermore since $u_s$, $v_1$ and $v_2$ describe the most
general $SO(2,1)\times SO(3)$ invariant
scalar and metric deformations, these are the only independent 
components of the equations
of motion of scalar fields and the metric.

On the other hand the  non-trivial 
components of the gauge field equations
and the Bianchi
identities for the full black hole solution takes the form:
\be \label{e4}
\p_r \left(
{\delta \, \SSS\over \delta F^{(i)}_{rt}}\right) = 0, 
\qquad 
\p_r F^{(i)}_{\theta\phi} = 0\ ,
\ee
where $\SSS=\int d^4 x\sqrt{-\det g}\LL$ is the action.
These equations are of course
automatically satisfied by the near horizon
background \refb{e1}, but we can extract more information from them.
{}From \refb{e4} it follows that
\be \label{e4a}
\int d\theta d\phi \, {\delta
\SSS\over \delta F^{(i)}_{rt}} 
= a_i, \quad \int d\theta d\phi \, F^{(i)}_{\theta\phi}
=b_i\, ,
\ee
where $a_i$ and $b_i$ are $r$ independent constants. Evaluating
these integrals on the near horizon geometry \refb{e1} gives
\be \label{e4b}
a_i = {\p f\over \p e_i}, \qquad b_i = p_i\, .
\ee
On the other hand if we evaluate the integrals in \refb{e4a}
at asymptotic infinity, then $a_i$ and $b_i$ are just the integrals
of electric and magnetic flux at infinity, and hence can be identified
with the electric  and magnetic charges respectively.
{}From this it follows 
that the constants $p_i$ appearing
in \refb{e1} correspond to magnetic charges of the black hole, and
\be \label{e5}
{\p f\over \p e_i} = q_i\, 
\ee
where  $q_i$ denote the electric
charges carried by the black hole.

For fixed $\vec p$ and $\vec q$,
 \refb{e3} and \refb{e5} give a set of equations which are equal in number
to the number of 
unknowns $\vec u$, $\vec v$ and $\vec e$. In a generic case we may
be able to solve these equations completely to determine the background in 
terms of
only the electric and the magnetic charges $\vec q$ and 
$\vec p$.~\footnote{The situation in string theory is not completely 
generic. For example
in $\NN=2$ supersymmetric string theories there is no coupling of the 
hypermultiplet
scalars to the vector multiplet fields or the curvature tensor to lowest 
order in $\alpha'$,
and hence in this 
approximation the function $f$ does not depend on the hypermultiplet
scalars. Thus the equations \refb{e3}, \refb{e5}
 do not fix the values of the hypermultiplet scalars in this approximation.}
This is consistent
with the attractor mechanism for supersymmetric 
background which says that the
near horizon configuration of 
a black hole depends only on the electric and magnetic
charges carried by the black hole and not on the asymptotic values of 
these scalar
fields.  We shall elaborate on this in \S\ref{s2.1}.

Let us define
\be \label{e2a}
 {\EE(\vec u, \vec v, \vec e, \vec q, \vec p) \equiv 2\, 
\pi ( e_i \, q_i - f(\vec u, \vec v, \vec e, \vec p)) }
\ee
The equations \refb{e3},
\refb{e5}
determining $\vec u$, $\vec v$ and $\vec e$ are then given 
by:
\be \label{e3a}
{ {\p \EE \over \p u_s}=0, \quad {\p
\EE \over \p v_1}=0\,, \quad  {\p
\EE \over \p v_2}=0}\, , \quad {\p \EE\over \p e_i}=0\, .
\ee
Thus all the near horizon parameters may be determined
by extremizing a
single function $\EE$.

We shall now turn to the analysis of the 
entropy associated with this black hole\cite{0506177}. A general
formula for the entropy in the presence of higher derivative terms has 
been given in
\cite{9307038,9312023,9403028,9502009}.\footnote{This 
formula for the entropy has been derived for
regular black holes with bifurcate event horizon and not for extremal
black holes. We are defining the entropy of extremal black holes
as the entropy of a non-extremal black hole in the extremal
limit. This allows us to use Wald's formula.}
For a spherically symmetric black hole this formula takes the form
\be \label{eio1}
S_{BH} = -8\pi\, \int_H\,  d\theta \, d\phi\,
{\delta \SSS \over \delta R_{rtrt}}  \, \sqrt{-g_{rr} \, g_{tt}} \, ,
\ee
where $H$ denotes the horizon of the black hole.
In computing $\delta \SSS / \delta R_{\mu\nu\rho\sigma}$ in 
\refb{eio1} we need to
\begin{enumerate}
\item express the action $\SSS$ in terms of
symmetrized covariant derivatives of fields by replacing
anti-symmetric combinations of covariant derivatives in terms
of the Riemann tensor, and then
\item  
treat $R_{\mu\nu\rho\sigma}$ as independent variables.
\end{enumerate}
This formula simplifies enormously 
here since
the covariant derivatives of all the tensors vanish, and as a result
\be \label{eio2}
{\delta \SSS \over \delta R_{rtrt}} = \sqrt{-\det g}
{\p \LL\over \p R_{rtrt}}\, ,
\ee
where in the expression for $\LL$ we need to keep only those terms
which do not involve explicit covariant derivatives, and
${\p\LL/\p R_{\mu\nu\rho\sigma}}$
is defined through the equation
\be \label{ex1}
\delta\, \LL = {\p\LL\over \p R_{\mu\nu\rho\sigma}} \, \delta\, 
R_{\mu\nu\rho\sigma}\, .
\ee
In computing $\delta\LL$
we need to treat the components of the Riemann tensor as independent
variables not related to the metric. Substituting \refb{eio2} into
\refb{eio1} we get simple 
formula for the entropy
\be \label{e6}
S_{BH} = 8\pi\, {\p \LL\over \p R_{rtrt}} \, g_{rr} \, g_{tt} \, A_H
=-8\pi\, v_1^2\, {\p\LL\over \p R_{rtrt}} \, A_H\, ,
\ee
where $A_H$ is the area of the event horizon.

In order to express this in terms of the function $f$ defined
in \refb{e2},  let us denote by $f_\lambda(\vec u,\vec v,\vec e, \vec p)$ 
an expression similar to the right hand side of \refb{e2} except 
that each factor of
$R_{rtrt}$ in the expression of $\LL$ is multiplied by a factor of 
$\lambda$. Then we
have the relation:
\be \label{e7}
\left . {\p f_\lambda
(\vec u,\vec v,\vec e, \vec p) \over \p\lambda} \right|_{\lambda=1}
=  \int d\theta \, d\phi\, \sqrt{-\det g} 
\, R_{\alpha\beta\gamma\delta} \, {\p \LL\over
\p R_{\alpha\beta\gamma\delta} }\, ,
\ee
where the repeated indices $\alpha,\beta,\gamma,\delta$ are summed over 
the coordinates $r$ and $t$.  Now since by symmetry consideration
$(\p \LL / \p R_{\alpha\beta\gamma\delta})$
is proportional to $(g^{\alpha\gamma} g^{\beta\delta} - g^{\alpha\delta} 
g^{\beta\gamma})$, we have
\be \label{e8}
{\p \LL \over \p R_{\alpha\beta\gamma\delta} }= - v_1^2 \, 
(g^{\alpha\gamma} g^{\beta\delta} - g^{\alpha\delta} 
g^{\beta\gamma}) \, {\p \LL \over \p R_{rtrt}}\, .
\ee 
The constant of proportionality has been fixed by 
taking $(\alpha\beta\gamma\delta)=(rtrt)$.
Using \refb{e1a}  and \refb{e8}, and that for a spherically symmetric
background $\p\LL/\p R_{rtrt}$ is independent of the $(\theta,\phi)$
coordinates, we can  rewrite \refb{e7} as
\be \label{e9}
\left . {\p 
f_\lambda(\vec u,\vec v,\vec e, \vec p) \over 
\p\lambda}\right|_{\lambda=1}
= 4\, v_1^2 \,
{\p \LL \over \p R_{rtrt}}\, A_{H}   
\, .
\ee
Substituting this into \refb{e6} gives
\be \label{e10}
S_{BH} = -2\pi\, \left . {\p 
f_\lambda(\vec u,\vec v,\vec e, \vec p) \over \p\lambda} 
\right|_{\lambda=1}\, .
\ee

We shall now try to express
the right hand side of \refb{e10} in terms of derivatives of
$f$ with respect to the
variables $\vec u$, $\vec v$, $\vec e$ and $\vec p$. 
For this let us focus on the $v_1$ dependence of $f_\lambda$.
Since the
expression for $\LL$ is invariant under reparametrization
of the $r,t$ coordinates, every factor of
$R_{rtrt}$ in the expression for $f_\lambda$ must appear in
the combination  $\lambda\, g^{rr} g^{tt} R_{rtrt} = 
-\lambda v_1^{-1}$, every factor of $F^{(i)}_{rt}$ must appear
in the combination $\sqrt{-
g^{rr} g^{tt}}F^{(i)}_{rt}=e_i v_1^{-1}$, and every factor of
$F^{(i)}_{\theta\phi}=p_i/4\pi$ and $\phi_s=u_s$ must appear without
any accompanying power of $v_1$. 
The contribution from all terms which
involve covariant derivatives of $F^{(i)}_{\mu\nu}$,
$R_{\mu\nu\rho\sigma}$ or $\phi_s$ vanish; hence there is no further factor of
$v_1$ coming from contraction of the metric with these derivative operators.
The only other $v_1$ dependence of $f_\lambda(\vec u,\vec v,\vec e, \vec p)$ is
through the overall multiplicative factor of $\sqrt{-\det g}\propto v_1$.  Thus
$f_\lambda(\vec u,\vec v,\vec e, \vec p)$ must be of the form 
\be\label{eio3}
f_\lambda(\vec u,\vec v,\vec e, \vec p)=v_1 g(\vec u,
v_2,\vec p, \lambda v_1^{-1}, \vec e v_1^{-1})\, ,
\ee
for some function $g$. This gives
\be \label{e11}
\lambda {\p f_\lambda(\vec u,\vec v,\vec e, \vec p) \over \p\lambda}  
+ v_1  {\p f_\lambda(\vec u,\vec v,\vec e, \vec p)  \over \p v_1}
+ e_i  {\p f_\lambda(\vec u,\vec v,\vec e, \vec p)   \over \p e_i}
- f_\lambda(\vec u,\vec v,\vec e, \vec p) = 0\, .
\ee
Setting $\lambda=1$ in \refb{e11}, using the equation of motion of $v_1$ 
as given in
\refb{e3}, and substituting the result into eq.\refb{e10} we get
\be \label{e12}
S_{BH} = 2\pi\, \left( e_i \, {\p f\over \p e_i} - f\right) \, .
\ee
This together with \refb{e5} shows
that $S_{BH}(\vec q, \vec p)/2\pi$ may be regarded as the 
Legendre transform of the function $
f(\vec u,\vec v, \vec e, \vec p)$ with respect to the
variables 
$e_i$ after eliminating $\vec u$ and 
$\vec v$ through their equations of motion \refb{e3}. Using
\refb{e2a} we can also express 
\refb{e12} as
\be \label{e5a}
S_{BH} = \EE(\vec u, \vec v, \vec e, \vec q, \vec p) \, ,
\ee
at the extremum \refb{e3a}. This suggests that we call the
function $\EE$ the entropy function\cite{0506177}.

We can take a slightly different viewpoint in which we define
the entropy function $\EE$ as a function of $\vec u$, $\vec v$,
$\vec q$ and $\vec p$
after eliminating the electric field variables $\vec e$ by the
$\p f /\p e_i=q_i$ condition. In this form the entropy function
given in \refb{e2a} will just be $2\pi$ times the Legendre
transform of the function $f$ with respect to the variables 
$\{e_i\}$. We shall continue to use the same symbol $\EE$
for both entropy functions since the second
definition  is obtained from the first simply 
by extremizing the latter with respect to $\{e_i\}$.

Given an action, the entropy function 
formalism reduces the problem of
computing the entropy of an extremal black hole into the problem
of solving a set of algebraic equations. We shall illustrate this by
applying this formalism to extremal Reissner-Nordstrom black hole
in the Maxwell-Einstein theory described by the action
\refb{e1.1}. The most general $SO(2,1)\times SO(3)$ invariant
background in this theory is given by
\ben \label{eio7}
ds^2 &=& v_1\left(-r^2 dt^2+{dr^2\over
r^2}\right)   +
v_2 \left(d\theta^2+\sin^2\theta d\phi^2\right) \nonumber \\
F_{rt}&=& e, \qquad F_{\theta\phi}=p\sin\theta/4\pi \, .
\een
Using \refb{e1.1},  \refb{e1a} we get 
\ben \label{eio8}
f(v_1,v_2,e,p) &\equiv& \int d\theta d\phi \, \sqrt{-\det g}
\, \LL \nonumber \\
&=& 4\pi\, v_1 v_2 \, \left[
{1\over 16\pi G_N} \, \left( -{2\over v_1} + {2\over v_2} \right)
 + {1\over 2} \, v_1^{-2} e^2 - {1\over 2} v_2^{-2}
\left({p\over 4\pi}\right)^2\right]
\, .  
\een
This in turn gives
\ben \label{eio9}
\EE (v_1,v_2, e, q, p) &\equiv& 2\pi (q\, e - f)  \nonumber \\
&=& 2\pi \left[ q \, e 
-{1\over 4 G_N} (2v_1 - 2 v_2)  
- 2\pi\, \, v_2\, v_1^{-1} \, e^2 + 2\pi\, v_1\, v_2^{-1}
\left({p\over 4\pi}\right)^2\right]
\, . 
\een
It is easy to verify that
$\EE$ has an extremum at
\be \label{eio9a}
v_1=v_2 = G_N
\, {q^2+p^2\over 4\pi}\, , \qquad
e={q\over 4\pi}\, .
\ee
Substituting this into the expression for $\EE$ we
get
\be \label{eio10}
S_{BH}\equiv \EE  = {1\over 4} (q^2+p^2) \, .
\ee
Eqs.\refb{eio9a} reproduces \refb{e1.2.3},
\refb{e1.5} and \refb{eio10} reproduces \refb{entropyr}.

Finally we note that although the entropy function formalism
developed in this section gives a simple method for computing
the entropy of an extremal black hole if such a solution exists,
our analysis does not tell us if the full black hole solution,
interpolating between $AdS_2\times S^2$ near horizon
geometry and the asymptotically flat Minkowski space, really exists.
For a general two derivative theory this issue has been addressed
in \cite{0507096} where it was shown that such a solution exists
provided the matrix of second derivatives of the entropy function
with respect to the scalar field values at the horizon is positive
definite at the extremum of the entropy function. Whether there
is a generalization of this result in higher derivative theories is
still an open question.

\subsectiono{Attractor, field redefinition and 
duality transformation} \label{s2.1}

In this section we shall discuss some important consequences of the
results derived in \S\ref{s1.1}.

\begin{enumerate}

\item Since the construction of the function $\EE$ 
involves knowledge of 
only the 
Lagrangian density,  the functional form of
$\EE$ is independent of asymptotic values of 
the moduli scalar 
fields, -- scalar fields which have no potential in flat space-time
and hence can take arbitrary constant values asymptotically.
Thus if the extremization equations \refb{e3a} 
determine all the parameters $\vec u$, $\vec v$, $\vec e$ uniquely
then the value of $\EE$ at 
the extremum and hence the entropy $S_{BH}$ 
is completely independent of 
the asymptotic values of the moduli fields. If on the other hand the 
function $\EE$ has flat directions then only some combinations of
the parameters $\vec u$, $\vec v$, $\vec e$ are
determined by extremizing $\EE$, and the rest
may depend on the asymptotic values of 
the moduli fields. However since $\EE$ is independent of the flat 
directions, it depends only on the combination of parameters which are 
fixed by the extremization equations. As a result the value of $\EE$ at the 
extremum is still independent 
of the asymptotic moduli. This shows that 
the entropy of the black hole is 
independent of the asymptotic values of the moduli fields irrespective of 
whether or not $\EE$ has flat directions. This is a generalization of the usual
attractor mechanism for  black holes in supergravity 
theories\cite{9508072,9602111,9602136}.

This result in particular implies that 
the entropy of an extremal black hole
does not change as we change the asymptotic value of the string
coupling constant from a sufficiently large value where the
black hole description is good to a sufficiently small value where the
microscopic description is expected to be valid. 
This fact has been used to argue that under certain conditions
the statistical entropy of the system, computed at weak string coupling,
should match the black hole entropy even for
non-supersymmetric extremal black holes\cite{0611143}.

\item An arbitrary field redefinition of the metric and the scalar 
fields will induce a redefinition of the parameters $\vec u$, $\vec v$, 
and hence the functional form of $\EE$ will 
change.\footnote{A redefinition of gauge fields preserving the gauge
transformation laws requires adding to the gauge field a gauge
invariant vector field constructed out of other fields and their
covariant derivatives. Since in the $AdS_2\times S^2$ background
all such vector fields vanish, the parameters labeling the gauge field
strengths are not redefined.}
However, since the value 
of $\EE$ at the extremum 
is invariant under non-singular field redefinition,
the entropy is unchanged under a redefinition of the metric and 
other scalar fields. To  see this more explicitly, let us consider a 
reparametrization of $\vec u$ and $\vec v$ of the form:
\be \label{erepar}
\wh u_s = g_s(\vec u, \vec v, \vec e, \vec p), \quad 
\wh v_i = h_i(\vec u, \vec v, 
\vec e, \vec p)\, ,
\ee
for some functions $\{g_s\}, \{h_i\}$.
Then it follows from eqs.\refb{e2}, \refb{e2a} that the new entropy 
function $\wh \EE(\vec{\wh u}, \vec{\wh v}, \vec e, 
\vec q, \vec p)$ is given by:
\be \label{erepar1}
\wh \EE(\vec{\wh u}, \vec{\wh v}, \vec e, \vec q, \vec p)
= \EE(\vec u, \vec v, \vec e, \vec q, \vec p)\, .
\ee
It is now easy to see that eqs.\refb{e3a} are equivalent to:
\be \label{e3bb}
{ {\p \wh \EE \over \p \wh u_s}=0, \quad {\p
\wh \EE \over \p \wh v_1}=0\,, \quad  {\p
\wh \EE \over \p \wh v_2}=0}\, , \quad {\p \wh \EE
\over \p e_i}=0\, .
\ee
Thus the value of $\wh \EE$ evaluated at this 
extremum is equal to the value 
of $\EE$ evaluated at the extremum \refb{e3a}, showing that the
entropy of an extremal black hole remains unchanged under field
redefinition.
This result of course is a consequence of the field 
redefinition invariance of Wald's entropy formula as discussed in 
\cite{9312023}.

\item As is well known, Lagrangian density is not invariant under an 
electric-magnetic duality transformation. However, the function 
$\EE$, being 
Legendre transformation of the Lagrangian density with respect to the 
electric field variables, is invariant under an electric-magnetic duality 
transformation. In other words, if instead of the original Lagrangian 
density $\LL$, we use an equivalent dual Lagrangian density
$\wt \LL$ where some 
of the gauge fields have been dualized, and construct a new entropy
function $\wt \EE(\vec u, \vec v, \vec q, \vec p)$ from this 
new Lagrangian 
density, then $\EE$ and $\wt \EE$ will be related to 
each other by exchange of the
appropriate $q_i$'s and $p_i$'s.

\end{enumerate}

\subsectiono{Spherically symmetric black holes for arbitrary $D$}
\label{s1.2}

The analysis can be generalized to higher dimensional theories as 
follows. 
We begin with a $D$-dimensional field theory of metric, various
$p$-form gauge fields and neutral scalars with lagrangian
density $\LL$. In this section we 
shall assume that the neither the definition of
the field strengths associated with the $p$-form gauge fields, nor
the Lagrangian density has any Chern-Simons terms. Thus the
Lagrangian density will be manifestly invariant under general
coordinate transformation and gauge transformation of the $p$-form
gauge field $B_{\mu_1\cdots \mu_p}$ of the form
\be\label{epf1}
\delta B_{\mu_1\cdots \mu_p} = \p_{[\mu_1}
\Lambda_{\mu_2\cdots \mu_p]}\, .
\ee
The cases where either the Lagrangian density or the definition of a
field strength has a Chern-Simons term will be dealt with
separately in \S\ref{s1.4}.
 
 In
 $D$ space-time dimensions the near horizon geometry
 of a spherically symmetric extremal black hole solution has 
 $SO(2,1)\times SO(D-1)$ isometry. This forces the metric to have
 the form
 $AdS_2\times S^{D-2}$. The relevant fields which can 
take non-trivial
 expectation values near the horizon are scalars $\{\phi_s\}$, metric 
$g_{\mu\nu}$, 
 gauge fields $A^{(i)}_\mu$,  $(D-3)$-form gauge fields 
 $B^{(a)}_{\mu_1\ldots \mu_{D-3}}$, 2-form gauge fields
 $C^{(m)}_{\mu\nu}$ and $(D-2)$-form gauge fields
 $D^{(I)}_{\mu_1\cdots \mu_{D-2}}$.
 If 
 \be\label{epf2}
 H^{(a)}_{\mu_1
\ldots \mu_{D-2}}=\p_{[\mu_1} B^{(a)}_{\mu_2\cdots
\mu_{D-2}]}\, ,
\ee
denotes the field strength 
associated with the  field $B^{(a)}$,
 then the general background consistent 
 with the $SO(2,1)\times SO(D-1)$ 
symmetry of
 the background geometry is of the form:
 \ben \label{e21}
&& ds^2\equiv 
g_{\mu\nu}dx^\mu dx^\nu = v_1\left(-r^2 dt^2+{dr^2\over r^2}\right)  +
v_2 \, d\Omega_{D-2}^2 \nonumber \\
&& \phi_s =u_s, \quad  C^{(m)}_{rt}
= w_m, \quad D^{(I)}_{l_1\cdots l_{D-2}} =
z_I \, \epsilon_{l_1\cdots l_{D-2}}\, \sqrt{\det h^{(D-2)}}\,
/\Omega_{D-2}
\nonumber \\
&& F^{(i)}_{rt} = e_i, \qquad  H^{(a)}_{l_1\cdots l_{D-2}} =
p_a  \, \epsilon_{l_1\cdots l_{D-2}}\, \sqrt{\det h^{(D-2)}}\,
/\Omega_{D-2}\,  ,
\een
 where $v_1$, $v_2$, $\{u_s\}$, $\{w_m\}$, $\{z_I\}$, $\{e_i\}$
 and $\{p_a\}$ are constants parametrizing the background,
 $d\Omega_{D-2}^2\equiv h^{(D-2)}_{ll'} dx^l dx^{l'}$ 
 denotes the line element on the unit $(D-2)$-sphere,
  $\Omega_{D-2}$ denotes the volume of the unit $(D-2)$-sphere,  
 $x^{l_i}$  with $2\le l_i\le (D-1)$
 are coordinates along this sphere and $\epsilon$ denotes
 the totally anti-symmetric symbol with $\eps_{2\ldots(D-1)}=1$. 
 Any other $k$-form field
for $k$ different from
1, 2, $D-3$ or $D-2$ 
  will vanish in this background since
 the only $SO(2,1)\times SO(D-2)$ invariant forms on $AdS_2\times
 S^{D-2}$ are the 2-form corresponding to the
 volume form on $AdS_2$ and the  $(D-2)$-form corresponding to the
volume form on $S^{D-2}$. Even among the ones given above,
the constant $C^{(m)}_{rt}$ background proportional to $w_m$
can be removed by gauge transformation.
 
 We now define
 \be \label{e22}
f(\vec u, \vec v, \vec e, \vec p) = \int dx^2\cdots dx^{D-1}\, \sqrt{-\det 
g}\, \LL\, ,
\ee
as in \refb{e2}. Note that since the lagrangian density depends
on the various $k$-form fields only through their field
strength, $f$ is independent of $\vec w$ and $\vec z$.\footnote{As
we shall see in \S\ref{s1.4}, this
situation will change once we allow Chern-Simons terms.}
Analysis identical to that for $D=4$ now tells us that the constants $p_a$
represent magnetic type charges carried by the black hole, and the 
equations which
determine the values of $\vec u$, $\vec v$ and $\vec e$ are
\be \label{e23}
{\p f\over \p u_s} =0, 
\qquad {\p f\over \p v_i} = 0\, , \qquad {\p f\over \p e_i}=q_i\, ,
\ee
where $q_i$ denote the electric charges carried by the black hole. 
$\vec w$ and $\vec z$ remain undermined.
Also using \refb{e6} which is valid for spherically symmetric
black holes in any dimension, we can show that
the entropy of the black 
hole is given by $2\pi$ times the Legendre transform of $f$:
\be \label{e24}
S_{BH} = 2\pi\, \left( e_i \, {\p f\over \p e_i} - f\right) \, .
\ee
as in \refb{e12}.

As in the case of four dimensional black holes, we can define
\be \label{e2ah}
 {\EE(\vec u, \vec v, \vec e, \vec q, \vec p) \equiv 2\, 
\pi ( e_i \, q_i - f(\vec u, \vec v, \vec e, \vec p)) }
\ee
The equations \refb{e23}
determining $\vec u$, $\vec v$ and $\vec e$ are then given 
by:
\be \label{e3ah}
{ {\p \EE \over \p u_s}=0, \quad {\p
\EE \over \p v_1}=0\,, \quad  {\p
\EE \over \p v_2}=0}\, , \quad {\p \EE\over \p e_i}=0\, .
\ee
Furthermore \refb{e24} shows that
the entropy associated with the black hole is given by:
\be \label{e5ah}
S_{BH} = \EE(\vec u, \vec v, \vec e, \vec q, \vec p) \, ,
\ee
at the extremum \refb{e3ah}.

A useful viewpoint that treats extremal black holes in all dimensions
in one go is to regard the $S^{D-2}$ part of the near horizon geometry
as a compact space and treat the effective field theory governing the
dynamics of the near horizon geometry as two dimensional. 
The effective Lagrangian density of this two dimensional theory 
spanned by $r$ and $t$ is
given by
\be\label{e2e.1}
\sqrt{-\det h}\, \LL^{(2)}=\int_{S^{D-2}} \sqrt{-\det g}\, \LL
\, ,
\ee
where $g_{\mu\nu}$ and $\LL$ denote the original $D$-dimensional
metric and Lagrangian density, and $h_{\alpha\beta}$ and
$\LL^{(2)}$ denote the two dimensional metric and two dimensional
Lagrangian density obtained via dimensional reduction. 
Only non-trivial degrees of freedom in this two dimensional theory
are the metric, gauge fields and scalars coming from the dimensional
reduction of various $D$-dimensional fields on $S^{D-2}$. The
magnetic charges $p_a$ labeling the flux of the $(D-2)$-form
field strengths through $S^{D-2}$ appear as parameters in this
two dimensional theory.
The most general near horizon configuration consistent with
$SO(2,1)$ isometry of $AdS_2$ will have an $AdS_2$ metric,
constant two dimensional gauge field strengths and constant scalars.
By regarding
this as the near horizon geometry
of a two dimensional extremal black hole, we can write
the Wald's formula for the entropy as
\be \label{eio1a}
S_{BH} = -8\pi\, \left.
{\delta \SSS \over \delta R^{(2)}_{rtrt}}  \, \sqrt{-h_{rr} \, h_{tt}}
\right|_{\hbox{Horizon}} \, ,
\ee
where $\SSS\equiv \int dr dt \sqrt{-\det h} \, \LL^{(2)}$ now
is to be regarded as a two dimensional action.
For extremal black holes all covariant derivatives of scalar fields
and field strengths vanish, and we have the analog of \refb{e6}
\be \label{e6a}
S_{BH} = 8\pi\, \left.
{\p \LL^{(2)}
\over \p R^{(2)}_{rtrt}} \, h_{rr} \, h_{tt} \right|_{\hbox{Horizon}}
\, .
\ee
If we now define 
\be \label{e6b}
f = \left. \sqrt{-\det h}\, \LL^{(2)} \right|_{\hbox{Horizon}}
\ee
and 
\be\label{e6c}
\EE = 2\pi (q_i e_i - f)\, ,
\ee
then one can show, following the same procedure as in \S\ref{s1.1},
that the parameters labeling the near horizon background are obtained
by extremizing $\EE$ with respect to these various parameters, and
furthermore that the entropy itself is given by the function $\EE$
evaluated at its extremum.

The arguments of \S\ref{s2.1} can now be used to prove attractor
behaviour of these black holes. We shall see in \S\ref{s1.3} that
the two dimensional viewpoint provides a useful tool for
proving the attractor behaviour of extremal rotating black holes
as well.

\subsectiono{Rotating black holes in $D=4$} \label{s1.3}

In this section we shall describe the construction of the entropy function
for rotating black holes in four dimensions following the
analysis performed in \cite{0606244}.
The results can be
easily generalized to higher dimensions. Early work on attractor
mechanism for rotating black holes has been carried out in
\cite{9602065,9611094,0605139}.

As in \S\ref{s1.1} we begin by considering a general
four dimensional theory of gravity coupled
to a set of  abelian gauge fields $A_\mu^{(i)}$
and
neutral scalar fields $\{\phi_s\}$ with action
\be\label{eact}
\SSS = \int d^4 x\,  \sqrt{-\det g}\,
\LL\, ,
\ee
where   $\sqrt{-\det g}\,
\LL$ is the lagrangian density,
expressed as a function of the metric $g_{\mu\nu}$, the Riemann
tensor $R_{\mu\nu\rho\sigma}$,
the scalar fields
$\{\phi_s\}$,
the gauge field strengths $F^{(i)}_{\mu\nu}=\p_\mu
A_\nu^{(i)} - \p_\nu
A_\mu^{(i)}$,
and covariant derivatives of these fields. In general $\LL$
will contain terms with more than two derivatives.
We now define a rotating extremal black hole solution to be
one whose
near horizon geometry has the symmetries of
$AdS_2\times S^1$, -- this holds 
for known extremal rotating black hole
solutions\cite{9905099,0606244} and has recently been proven
in \cite{0705.4214}.
The most general field configuration consistent
with the $SO(2,1)\times U(1)$ symmetry of $AdS_2\times S^1$
is of the form:\footnote{Our convention for $\alpha$ differs from
the one used in \cite{0606244} by a minus sign. With this
new convention the variable $J$ conjugate to $\alpha$ will represent
angular momentum in the standard convention.}
\bea{eg1rep}
&& ds^2\equiv g_{\mu\nu}dx^\mu dx^\nu = v_1(\theta)
\left(-r^2 dt^2+{dr^2\over
r^2}\right)  + \beta^2 \, d\theta^2+
\beta^2\, v_2(\theta) (d\phi + \alpha r dt)^2 \nonumber \\
&& \phi_s =u_s(\theta) \nonumber \\
&& {1\over 2}\, F^{(i)}_{\mu\nu}dx^\mu\wedge dx^\nu
= (e_i+\alpha b_i(\theta)) dr \wedge dt
+ \p_\theta b_i(\theta) d\theta\wedge (d\phi + \alpha r dt)\, ,
\een
where $\alpha$, $\beta$ and $e_i$ are constants, and $v_1$, $v_2$,
$u_s$ and $b_i$  are functions of $\theta$.
Here $\phi$ is a periodic coordinate with period $2\pi$ and
$\theta$ takes value in the range $0\le\theta\le \pi$.
The SO(2,1) isometry of $AdS_2$ is generated by the Killing
vectors\cite{9905099}:
\be\label{ekill}
L_1=\partial_t, \qquad
L_0=t\partial_t-r\partial_r, \qquad
L_{-1}=(1/2)(1/r^2+t^2)\partial_t - (tr)\partial_r - (\alpha /r)
\partial_\phi\, .
\ee

A simple way to see
the $SO(2,1)\times U(1)$ symmetry of the configuration
\refb{eg1rep} is
as follows.
The $U(1)$ transformation acts as a translation of $\phi$ and is
clearly a symmetry of this configuration.
In order to see the SO(2,1) symmetry of this background
we regard $\phi$ as a compact
direction and interprete this as a theory in three dimensions
labelled by coordinates $\{x^m\}\equiv(r,\theta,t)$ with
metric $\hat
g_{mn}$, gauge fields $a_m^{(i)}$ and $a_m$ and  scalar fields
$\psi$ and $\chi_i$ defined through the relations
\ben\label{e3ddimred}
ds^2 &=& \hat g_{mn} \, dx^m dx^n + \psi (d\phi + a_m dx^m)^2\cr
A^{(i)}_\mu dx^\mu &=& a^{(i)}_m dx^m 
+ \chi_i (d\phi + a_m dx^m)\, .
\een
Besides these we also have scalar fields $\phi_s$ descending down
from four dimensions.
If we denote by $f^{(i)}_{mn}$
and $f_{mn}$ the field strengths associated with the three
dimensional gauge fields $a_m^{(i)}$ and $a_m$ respectively,
then the background \refb{eg1rep} can be interpreted as the following
three dimensional background:
\bea{eg2}
&& \wh{ds}^2\equiv \hat g_{mn}dx^m dx^n = v_1(\theta)
\left(-r^2 dt^2+{dr^2\over
r^2}\right)  + \beta^2 \, d\theta^2 \nonumber \\
&& \phi_s =u_s(\theta), \qquad \psi= \beta^2\, v_2(\theta), \quad
\chi_i =   b_i(\theta)\, , \nonumber \\
&& {1\over 2}\, f^{(i)}_{mn}dx^m\wedge dx^n
= e_i\,  dr \wedge dt ,
\qquad {1\over2}\,
f_{mn}dx^m\wedge dx^n = \alpha \, dr\wedge dt\, .
\een
The $(r,t)$ coordinates now describe an AdS$_2$ space and
this background is manifestly $SO(2,1)$ invariant.    In this
description the Killing vectors take the standard form
\be\label{ekill1}
L_1=\partial_t, \qquad
L_0=t\partial_t-r\partial_r, \qquad
L_{-1}=(1/2)(1/r^2+t^2)\partial_t - (tr)\partial_r  \, .
\ee

Let us now return to the four dimensional viewpoint.
For the configuration given in
\refb{eg1rep} the magnetic charge associated with the $i$th gauge
field is given by
\be\label{eg1.5}
p_i = \int d\theta d\phi F^{(i)}_{\theta\phi} = 2\pi (b_i(\pi)
- b_i(0))\, .
\ee
Since an additive constant in $b_i$ can be absorbed into the
parameters $e_i$, we can set $b_i(0)=-p_i/4\pi$. This, together
with \refb{eg1.5}, now gives
\be\label{eg1.6}
b_i(0)=-{p_i\over 4\pi}, \qquad b_i(\pi) = {p_i\over 4\pi}\, .
\ee
We shall assume that the deformed horizon, labelled by the
coordinates $\theta$ and $\phi$, is a smooth deformation
of the sphere. In particular there should be no conical defects
near $\theta=0,\pi$. We shall further assume that the 
gauge field strengths and scalar fields are also smooth at $\theta=0$,
$\pi$.
This requires
\bea{eg1.4}
v_2(\theta) &=& \theta^2 +\OO(\theta^4) \quad
\hbox{for $\theta\simeq 0$} \nonumber \\
&=& (\pi-\theta )^2 +\OO((\pi-\theta)^4) \quad
\hbox{for $\theta\simeq \pi$}\, ,
\een
\bea{ebcca}
b_i(\theta)&=&-{p_i\over 4\pi} +\OO(\theta^2) \quad
\hbox{for $\theta\simeq 0$} \nonumber \\
&=& {p_i\over 4\pi}
 +\OO((\pi-\theta)^2) \quad
\hbox{for $\theta\simeq \pi$}\, ,
\een
\bea{ebscalar}
u_s(\theta)&=&u_s(0) +\OO(\theta^2) \quad
\hbox{for $\theta\simeq 0$} \nonumber \\
&=& u_s(\pi)
 +\OO((\pi-\theta)^2) \quad
\hbox{for $\theta\simeq \pi$}\, .
\een
Note that the smoothness of the background requires the Taylor series
expansion around $\theta=0,\pi$ to contain only even powers of $\theta$
and $(\pi-\theta)$ respectively. This can be seen by expressing the
solutions near $\theta=0$ or $\theta=\pi$ using the local Cartesian
coordinates $(x,y)=(\sin\theta\cos\phi, \sin\theta\sin\phi)$
and requiring the solution to be non-singular and invariant under
$\phi$ translation in this coordinate system.

Eq.\refb{eg2} and hence
\refb{eg1rep} describes the most general field
configuration
consistent with the $SO(2,1)\times U(1)$
symmetry.
Thus in order to
derive the equations of motion we can evaluate the action on this
background and then extremize the resulting expression with respect
to the parameters labeling the background \refb{eg1rep}. The only
exception to this are the parameters $e_i$ and $\alpha$ labeling
the field strengths. 
{}From the three dimensional viewpoint we see that the background
\refb{eg2} automatically satisfies the equations of motion of the gauge
fields $a^{(i)}_m$ and $a_m$. Thus 
the variation of the action with respect to the
parameters $e_i$ and $\alpha$ need not
vanish, -- instead they give the corresponding conserved
electric charges $q_i$ and the angular momentum $J$ (which can be
regarded as the electric charge associated with the three dimensional
gauge field $a_m$.)

To implement this procedure we define:
\be\label{eg3}
f[\alpha,\beta,\vec e, v_1(\theta), v_2(\theta), \vec u(\theta), \vec
b(\theta)]
= \int d\theta d\phi \sqrt{-\det g}\,  \LL\, .
\ee
Note that $f$ is a function of $\alpha$, $\beta$, $e_i$
and a functional
of $v_1(\theta)$, $v_2(\theta)$,
$u_s(\theta)$ and $b_i(\theta)$. The equations
of motion now correspond
to
\be\label{eg4}
{\p f\over \p\alpha} = J, \quad {\p f\over \p\beta}=0,
\quad {\p f\over \p
e_i} = q_i\, , \quad {\delta f\over \delta v_1(\theta)}=0\, ,
\quad {\delta f\over \delta v_2(\theta)}=0,
\quad {\delta f\over \delta u_s(\theta)}=0,
\quad {\delta f\over \delta b_i(\theta)}=0\, .
\ee
Equivalently, if we define:
\be\label{eg5}
\EE[J,\vec q,\alpha,\beta,\vec e, v_1(\theta),
v_2(\theta), \vec u(\theta), \vec
b(\theta) ]
= 2\pi \left( J\alpha + \vec q\cdot \vec e - f[\alpha,\beta,
\vec e, v_1(\theta), v_2(\theta), \vec u(\theta), \vec
b(\theta)]
\right)\, ,
\ee
then the equations of motion take the form:
\be\label{eg6}
{\p \EE\over \p\alpha} = 0, \quad
{\p \EE\over \p\beta}=0, \quad {\p \EE\over \p
e_i} = 0\, , \quad {\delta \EE\over \delta v_1(\theta)}=0\, ,
\quad {\delta \EE\over \delta v_2(\theta)}=0,
\quad {\delta \EE\over \delta u_s(\theta)}=0,
\quad {\delta \EE\over \delta b_i(\theta)}=0\, .
\ee

These equations are subject to the boundary conditions \refb{eg1.4},
\refb{ebcca}, \refb{ebscalar}. For formal arguments it will be useful to
express the various
functions of $\theta$
appearing here by expanding them as a linear combination
of appropriate basis states which make the constraints \refb{eg1.4},
\refb{ebcca} manifest, and then varying $\EE$ with respect to the
coefficients appearing in this expansion. The natural functions
in terms of which we can expand an arbitrary $\phi$-independent
function on a sphere are the
Legendre polynomials $P_l(\cos\theta)$. We take
\bea{eg7}
&& v_1(\theta) = \sum_{l=0}^\infty \wt v_1(l) \, P_l(\cos\theta)\, ,
\quad v_2(\theta) = \sin^2\theta + \sin^4\theta
\sum_{l=0}^\infty \wt v_2(l) \, P_l(\cos\theta)\, , \nonumber \\
&&
u_s(\theta) = \sum_{l=0}^\infty \wt u_s(l) \, P_l(\cos\theta)\, ,
\quad b_i(\theta) = -{p_i\over 4\pi} \cos\theta
+ \sin^2\theta \sum_{l=0}^\infty \wt b_i(l) \, P_l(\cos\theta)\, .
\nonumber \\
\eea
This expansion explicitly implements the constraints \refb{eg1.4},
\refb{ebcca} and \refb{ebscalar}. Substituting this into \refb{eg5} gives
$\EE$
as a function of $J$, $q_i$, $\alpha$, $\beta$, $e_i$,
$\wt v_1(l)$, $\wt v_2(l)$, $\wt u_s(l)$ and
$\wt b_i(l)$. Thus the equations
\refb{eg6} may now be reexpressed as
 \be\label{eg8}
{\p \EE\over \p\alpha} = 0, \quad
{\p \EE\over \p\beta}=0, \quad {\p \EE\over \p
e_i} = 0\, , \quad {\p \EE\over \p \wt v_1(l)}=0\, ,
\quad {\p \EE\over \p \wt v_2(l)}=0,
\quad {\p \EE\over \p \wt u_s(l)}=0,
\quad {\p \EE\over \p \wt b_i(l)}=0\, .
\ee

Let us now turn to the analysis of the
entropy associated with this black hole.
For this it will be most convenient to regard this configuration
as a two dimensional extremal black hole by regarding the
$\theta$ and $\phi$ directions as compact.
In this interpretation the zero mode of the metric
$\wh g_{\alpha\beta}$ given in \refb{eg2},
with $\alpha,\beta=r,t$, is interpreted
as the two dimensional metric $h_{\alpha\beta}$:
\be\label{enx1}
h_{\alpha\beta} = {1\over 2} \int_0^\pi \, d\theta\, \sin\theta\,
\wh g_{\alpha\beta}\, ,
\ee
whereas all the non-zero modes of $\wh g_{\alpha\beta}$ are
interpreted as massive symmetric rank two tensor fields. This gives
\be\label{erel0}
h_{\alpha\beta}dx^\alpha dx^\beta =
v_1(-r^2 dt^2 + dr^2/r^2)\, , \qquad v_1 \equiv \wt v_1(0)\, .
\ee
Thus
the near horizon configuration, regarded from two dimensions,
involves $AdS_2$ metric, accompanied by
background electric fields $f^{(i)}_{\alpha\beta}$ and
$f_{\alpha\beta}$, a set of massless and massive scalar fields
with vacuum expectation values $\wt u_s(l)$, $\wt v_2(l)$,
$\wt b_i(l)$, and a set of massive
symmetric rank two tensor fields with vacuum expectations values
${\wt v_1(l)} h_{\alpha\beta} /  \wt v_1(0)$.
According to the Wald formula
\cite{9307038,9312023,9403028,9502009}, the entropy of this
black hole is given by:
\be\label{ex-1}
S_{BH} = -8\pi\,   \, {\delta \SSS^{(2)}
\over \delta R^{(2)}_{rtrt}}
\, \sqrt{-h_{rr} \, h_{tt}} \,  ,
\ee
where $R^{(2)}_{\alpha\beta\gamma\delta}$ is the
two dimensional Riemann tensor associated with the metric
$h_{\alpha\beta}$, and $\SSS^{(2)}$ is the general coordinate
invariant action of this two dimensional field theory.
We now note that for this two dimensional configuration
that we have,
the electric
field strengths $f^{(i)}_{\alpha\beta}$ and $f_{\alpha\beta}$ are
proportional to the volume form on $AdS_2$, the scalar fields are
constants and the tensor fields are proportional to the $AdS_2$ metric.
Thus the covariant derivatives of all
gauge and generally covariant tensors which one can
construct out of these two dimensional fields vanish.
{}In this case \refb{ex-1} simplifies to:
\be\label{ex0-}
S_{BH} = -8\pi\,  \sqrt{-\det h}\, {\p \LL^{(2)}
\over \p R^{(2)}_{rtrt}}
\, \sqrt{-h_{rr} \, h_{tt}} \,
\ee
where $\sqrt{-\det h}\, \LL^{(2)}$
is the two dimensional Lagrangian density, related to the four
dimensional Lagrangian density via the formula:
\be\label{enx2}
\sqrt{-\det h}\, \LL^{(2)} = \int d\theta d\phi \sqrt{-\det g}\,
\LL\, .
\ee
Also while computing \refb{ex0-} we set to zero
all terms in $\LL^{(2)}$ which involve covariant derivatives of the
Riemann tensor, gauge field strengths, scalars and the massive tensor
fields.

We can now proceed in a manner
identical to that in \S\ref{s1.1}, \S\ref{s1.2}
to show that the right hand side of \refb{ex0-} is the
entropy function at its extremum. First of all
from  \refb{erel0} it follows that
\be\label{erel1}
R^{(2)}_{rtrt} =  v_1  = \sqrt{-h_{rr} h_{tt}}\,.
\ee
Using this we can express \refb{ex0-} as
\be\label{erel2}
S_{BH} = -8\pi \, \sqrt{-\det h}{\p \LL^{(2)}
\over \p R^{(2)}_{rtrt}} R^{(2)}_{rtrt}\, .
\ee
Let us denote by $\LL^{(2)}_\lambda$ a deformation of
$\LL^{(2)}$ in which we replace all factors of
$R^{(2)}_{\alpha\beta\gamma\delta}$  for
$\alpha,\beta,\gamma,\delta=r, t$ by
$\lambda R^{(2)}_{\alpha\beta\gamma\delta}$, and define
\be\label{erel3}
f^{(2)}_\lambda \equiv \sqrt{-\det h} \, \LL^{(2)}_\lambda\, ,
\ee
evaluated on the near horizon geometry. Then
\be\label{erel4}
\lambda {\p f^{(2)}_\lambda\over \p\lambda} \,
= \sqrt{-\det h} \, R^{(2)}_{\alpha\beta\gamma\delta}
{\p \LL^{(2)}
\over \delta R^{(2)}_{\alpha\beta\gamma\delta}}
= 4\, \sqrt{-\det h} \, R^{(2)}_{rtrt} {\p \LL^{(2)}
\over \p R^{(2)}_{rtrt}}\, .
\ee
Using this \refb{erel2} may be rewritten as
\be\label{erel5}
S_{BH} = -2\pi \lambda {\p f^{(2)}_\lambda\over \p\lambda}\bigg|_{
\lambda=1}\, .
\ee
Next we consider the effect of the scaling
\be\label{erel6}
\lambda \to s\lambda, \quad e_i \to s e_i, \quad \alpha\to s\alpha,
\quad \wt v_1(l)\to s \wt v_1(l)\quad \hbox{for \, \,
$0\le l<\infty$}\, ,
\ee
under which $\lambda\, R^{(2)}_{\alpha\beta\gamma\delta}\to
s^2 \, \lambda\, R^{(2)}_{\alpha\beta\gamma\delta}$.
Since $\LL^{(2)}$ does not involve any
explicit covariant derivatives,   all indices
of $h^{\alpha\beta}$ must contract with the indices in
$f^{(i)}_{\alpha\beta}$, $f_{\alpha\beta}$, $R^{(2)}_{\alpha\beta
\gamma\delta}$ or the indices of the massive
rank two symmetric tensor
fields whose near horizon values are proportional to the parameters
$\wt v_1(l)$. {}From this and
the definition of the parameters $e_i$, $\wt v_1(l)$, and
$\alpha$  it follows that $\LL^{(2)}_\lambda$ remains
invariant under
this scaling, and hence $f^{(2)}_\lambda$ transforms to
$s f^{(2)}_\lambda$, with the overall factor of $s$
coming from the $\sqrt{-\det h}$ factor in the definition of
$f^{(2)}_\lambda$. Thus we have:
\be\label{erel7}
\lambda {\p f^{(2)}_\lambda\over \p\lambda} + e_i
{\p f^{(2)}_\lambda\over \p e_i} + \alpha {\p f^{(2)}_\lambda\over \p
\alpha} + \sum_{l=0}^\infty \wt v_1(l)
{\p f^{(2)}_\lambda\over \p \wt v_1(l)} = f^{(2)}_\lambda\, .
\ee
Now it follows from \refb{eg3}, \refb{enx2} and \refb{erel3}
that
\be\label{erel8}
f[\alpha,\beta,\vec e, v_1(\theta), v_2(\theta), \vec u(\theta), \vec
b(\theta)] = f^{(2)}_{\lambda=1}\, .
\ee
Thus the extremization equations \refb{eg4} implies that
\be\label{erel9}
{\p f^{(2)}_\lambda\over \p e_i} = q_i, \quad
{\p f^{(2)}_\lambda\over \p
\alpha} = J, \quad {\p f^{(2)}_\lambda\over \p \wt v_1(l)}=0\, ,
\qquad \hbox{at $\lambda=1$}\, .
\ee
Hence setting $\lambda=1$ in \refb{erel7} we get
\be\label{erel8.5}
\lambda {\p f^{(2)}_\lambda\over \p\lambda}\bigg|_{\lambda=1}
 =
- e_i q_i - J \alpha + f^{(2)}_{\lambda=1} =
- e_i q_i - J \alpha + f[\alpha,\beta,\vec e, v_1(\theta),
v_2(\theta), \vec u(\theta), \vec
b(\theta)]\, .
\ee
Eqs.\refb{erel5} and the definition \refb{eg5} of the entropy
function  now gives
\be\label{ex2}
S_{BH} = \EE
\ee
at its extremum.

The arguments of \S\ref{s2.1} can now be used to prove attractor
behaviour of these black holes, \i.e.\ the black hole entropy
depends only on the charges $\{q_i, p_i\}$ and the angular
momentum $J$ but not on any other asymptotic data.

For practical computations it is often useful to work with the functions
$v_i(\theta)$, $u_s(\theta)$ and $b_i(\theta)$ instead of their
mode decompositions given in \refb{eg7}. In this case the extremization
of the entropy functional $\EE$ with respect to these functions,
as described in eqs.\refb{eg6}, would
give rise to a set of ordinary differential equations in $\theta$ for
these functions, and the entropy is obtained by evaluating the
entropy function at a solution of these equations. In order to carry
out this procedure we need to carefully keep track of all the boundary
terms that arise in the expression for the entropy function.
This has been discussed in detail in \cite{0606244} where we have also
illustrated the general method by applying it to a specific class
of rotating black holes in string theory studied in
\cite{9505038,9603147,9610013,9909102}.

Finally, one interesting question that arises for rotating black
holes is whether the horizon can have non-spherical topology,
{\it e.g.} the topology of a torus.
Although in two derivative theories the
horizon of a four dimensional black hole is known to have
spherical topology,  once
higher derivative terms are added to the action there may be
other possibilities. Our analysis can be
easily generalized to the case where  the
horizon has the topology of
a torus rather than a sphere.
All we need to do is to take the $\theta$
coordinate to be a periodic variable with period $2\pi$ and expand the
various functions in the basis of periodic functions of $\theta$.
However
if the near horizon geometry is invariant under both $\phi$ and $\theta$
translations, then in the expression for $L_{-1}$ given in \refb{ekill}
we could add a term of the form $-(\gamma/ r)
\p_\theta$, and
the entropy could have an additional
dependence on the charge conjugate to the variable $\gamma$.
This represents the Noether charge
associated with $\theta$ translation, but
does not correspond to a physical charge
from the point of view of the asymptotic observer since the full solution
is not invariant under $\theta$ translation. These are commonly
known as dipole charges. The entropy function method
can also be used to compute entropies of higher dimensional
black holes with
non-spherical horizons where such solutions are known to exist
even for two derivative action\cite{0110260,0408120}.

\subsectiono{Dealing with Chern-Simons terms} \label{s1.4}

The analysis in the previous sections relies on several 
important assumptions about the structure of the Lagrangian density.
In particular we have assumed that

\begin{enumerate}

\item The lagrangian density depends on the metric in a manifestly
covariant manner, namely the only dependence on the metric comes
via the metric, Riemann tensor and covariant derivatives of various
tensor fields, but does not 
have any explicit dependence on
spin connections and Christoffel symbols. 

\item For any $p$-form gauge field $B$ present in the theory, the
covariant field strength has the form $H=dB$ so that $H$ satisfies
the Bianchi identity $dH=0$. If this is not the case, then  a
field configuration of the form given in \refb{e21} will not
automatically satisfy the Bianchi identity, and we shall get additional
constraints on the parameters labeling the near horizon background
by requiring that $H$ satisfies the Bianchi identity.

\item The dependence on the gauge fields $A_\mu^{(i)}$ and more
generally on the $(D-3)$-form gauge fields $B^{(a)}$
appears through
their field strengths. 
Otherwise we shall encounter the following problems:
\begin{enumerate}
\item
If the lagrangian density had any explicit dependence
on the gauge fields then the gauge field equations of motion would not
take the form given in \refb{e4}. \label{aitem}

\item While making the ansatz \refb{e21} we have taken the gauge field
strengths $F^{(i)}_{rt}$ and $H^{(a)}_{l_1\cdots l_{D-2}}$ to be 
invariant under the symmetries of $AdS_2\times S^{D-2}$, but the
gauge fields themselves do not carry the symmetry. As a result
the Lagrangian density evaluated in the background will be
invariant under the symmetries of $AdS_2\times S^{D-2}$ only
if it does not involve explicitly the gauge fields and is a function
only of the field strengths. 
Otherwise the extremization of the
entropy function may not give 
the complete set of independent equations of
motion.
\end{enumerate}

\end{enumerate}

The type of Lagrangian densities which appear in low energy effective
action of string theory often violates one or
more of these conditions. In particular the
Lagrangian density often involves Chern-Simons forms, which are
totally antisymmetric  tensors 
$\Omega_{\mu_1\ldots \mu_n}$ which depend on one or
more lower rank gauge fields  or spin connetions / Christoffel symbols
rather than just on the 
field strengths. As a 
result $\Omega$ is not invariant under the gauge transformation
associated with these lower rank gauge fields. Nevertheless
$\Omega$ has  the special property that
the variation of $\Omega$ under various gauge transformations
are exact forms:
\be\label{edeltaomega}
\delta \Omega_{\mu_1\ldots \mu_n} = \p_{[\mu_1} \chi_{\mu_2\ldots
\mu_n]}\, ,
\ee
for some quantity $\chi$.
As a result $\p_{[\mu_1} \Omega_{\mu_2\ldots \mu_{n+1}]}$ is a
covariant tensor. 

A simple example of such a Chern-Simons term is as follows.
Suppose the theory contains a $p$-form gauge field 
$B^{(1)}_{\mu_1\ldots \mu_p}$ and a $q$-form gauge field
$B^{(2)}_{\mu_1\ldots \mu_q}$, with associated gauge
transformations of the form
\be \label{ecsexample1}
\delta B^{(1)}_{\mu_1\ldots \mu_p} = \p_{[\mu_1}
\Lambda^{(1)}_{\mu_2 \ldots \mu_p]} \qquad 
\delta B^{(2)}_{\mu_1\ldots \mu_q} = \p_{[\mu_1}
\Lambda^{(2)}_{\mu_2 \ldots \mu_q]}\, .
\ee
Then the $(p+q+1)$-form
\be \label{edefcsform}
\Omega_{\mu_1\ldots \mu_{p+q+1}}
= B^{(1)}_{[\mu_1\ldots \mu_p} \, \p_{\mu_{p+1}}
B^{(2)}_{\mu_{p+2}\ldots \mu_{p+q+1}]}
\ee
transforms by a total derivative of the form \refb{edeltaomega}
under the gauge transformation induced by $\Lambda^{(1)}$.
Thus $\Omega_{\mu_1\ldots \mu_{p+q+1}}$ defined in 
\refb{edefcsform} is a Chern-Simons $(p+q+1)$-form. 

The Chern-Simons terms could appear in the expression
for the lagrangian density in two different ways:
\begin{enumerate}
\item The action itself may contain a Chern-Simons
term of the form
\be \label{eactionchern}
\int d^D x \, \epsilon^{\mu_1\ldots \mu_D} \, 
\Omega_{\mu_1\ldots \mu_D} \, .
\ee
Since $\delta\Omega$ is a total derivative, an action of this form is
gauge invariant up to boundary terms. In the presence of
such a term the Lagrangian density
may fail to satisfy our assumptions on three counts. First of
all since the Lagrangian density is not gauge invariant, it may not
be invariant under the symmetries of $AdS_2\times S^{D-2}$ when
evaluated in the near horizon background \refb{e21}. Second, since
the lagrangian density may now depend 
explicitly on the gauge fields and
not just their field strengths, the equations of motion of the gauge
fields may no longer be of the form \refb{e4}. Finally, if the 
Chern-Simons
form explicitly involves Christoffel symbol or spin connection, then even
Wald's formula for the entropy is not directly applicable.\footnote{For
some recent work on application of Wald's formula in the presence of 
Chern-Simons term see \cite{0611141}.}

\item In some theories 
the gauge invariant field strength associated with an
antisymmetric tensor field $B_{\mu_1\ldots \mu_{n-1}}$ 
is given by
\be\label{edeffield}
H_{\mu_1\ldots \mu_n} = \p_{[\mu_1} B_{\mu_2\ldots
\mu_n]} +  \Omega_{\mu_1\ldots \mu_n}
\ee
for some Chern-Simons $n$-form $\Omega$ 
constructed out of lower
dimensional gauge fields and spin connection. 
Under the gauge transformation
\refb{edeltaomega}, $B_{\mu_1\ldots \mu_{n-1}}$ is assigned the
transformation
\be \label{edeltab}
\delta B_{\mu_1\ldots \mu_{n-1}} = - \chi_{\mu_1\ldots
\mu_{n-1}}\, ,
\ee
so that 
\be \label{edeltah}
\delta H_{\mu_1\ldots \mu_n} = 0\, .
\ee
A typical example of such a term is the 3-form field strength associated
with the NS sector 2-form gauge field of heterotic string theory. The
definition of the three form field strength involves both gauge and
Lorentz Chern-Simons 3-forms.
In such cases the Lagrangian density, being a function of 
$H_{\mu_1\ldots \mu_n}$, is invariant under the gauge transformation
\refb{edeltaomega}, \refb{edeltab}. Nevertheless, since the definition
of $H_{\mu_1\ldots \mu_n}$ involves various lower rank gauge
fields and not just their field strengths, the presence of such terms in
the Lagrangian density violates our assumptions on three counts. First
the Bianchi identity of $H_{\mu_1\ldots \mu_n}$, being of the
form
\be \label{ebianchih}
\p_{[\mu_1} H_{\mu_2\ldots \mu_{n+1}]} =
\p_{[\mu_1} \Omega_{\mu_2\ldots \mu_{n+1}]}\, ,
\ee
now could give additional constraints on the near horizon parameters
besides the ones obtained by entropy function extremization condition.
Second since the definition of $H_{\mu_1\ldots \mu_n}$ involves
explicitly lower rank gauge fields, the equations of motion of the gauge
fields may no longer be of the form \refb{e4}.
Finally, if the 
Chern-Simons
form explicitly involves Christoffel symbol or spin connection, then
Wald's formula for the entropy is not directly applicable.
\end{enumerate}
In order to deal with these Chern-Simons terms we shall proceed in
two steps. First we shall show that the second type of Chern-Simons
terms described above, where it appears in the definition
of a field strength, can be transformed to the first type. 
This will involve generalizing the analysis in \cite{0608182}.
We shall then
describe a general procedure for dealing with the first type of 
Chern-Simons term\cite{0601228}. 
In carrying out these manipulations we shall need
to add total derivative terms to the Lagrangian density. Since addition
of such term do not affect the equations of motion we expect that the
entropy computed from the new Lagrangian density will continue
to describe entropy of extremal black holes in the original theory.

The first step is carried out as follows. Suppose in a theory in $D$
dimensions we have an $(n-1)$-form field $B$ whose field strength
$H$ contains a Chern-Simons term as in \refb{edeffield}, but $\LL$
depends on $B$ only through its field strength $H$.
We now introduce a new $(D-n-1)$-form field $\BB_{\mu_1\ldots
\mu_{D-n-1}}$ and define its field strength to be
\be \label{edefhh}
\HH_{\mu_1\ldots
\mu_{D-n}} = \p_{[\mu_1} \BB_{\mu_2\ldots \mu_{D-n}]}\, .
\ee
$\HH_{\mu_1\ldots
\mu_{D-n}}$ is invariant under a gauge transformation
\be \label{edeltabb}
\delta \BB_{\mu_1\ldots
\mu_{D-n-1}} = \p_{[\mu_1} \Lambda_{\mu_2\ldots \mu_{D-n-1}]}\, .
\ee
We now consider a new Lagrangian density
\be \label{enewlagr}
\sqrt{-\det g}\, \wt\LL =  \sqrt{-\det g}\, \LL + 
\epsilon^{\mu_1\ldots \mu_D} \,
(H_{\mu_1\ldots \mu_{n}} 
- \Omega_{\mu_1\ldots \mu_{n}}) \, 
\HH_{\mu_{n+1}\mu_{n+2}\ldots \mu_D}\, ,
\ee
and treat
$H_{\mu_1\ldots \mu_{n}}$ 
and $\BB_{\mu_1\ldots\mu_{D-n-1}}$ as independent fields. 
In this case
the equation of motion of the $\BB_{\mu_1\ldots \mu_{D-n-1}}$
field gives
\be \label{ebbeqr}
\epsilon^{\mu_1\ldots \mu_D} \p_{\mu_1}\,
(H_{\mu_2\ldots \mu_{n+1}} 
- \Omega_{\mu_2\ldots \mu_{n+1}}) =0\, ,
\ee
whose general solution is of the form \refb{edeffield}. On the
other hand the equation of motion of $H_{\mu_1\ldots \mu_n}$
associated with the new action
has the form
\be \label{eheqr}
{\delta \SSS \over \delta H_{\mu_1\ldots \mu_n}}
+ 
\epsilon^{\mu_1\ldots \mu_D}
\, \p_{\mu_{n+1}} \, \BB_{\mu_{n+2}\ldots \mu_D} = 0\, .
\ee
This gives
\be \label{ebbbiancr}
\p_{\mu_1} {\delta \SSS \over \delta H_{\mu_1\ldots \mu_n}} = 0\, ,
\ee
which is the equation of motion of the field 
$B_{\mu_1\ldots \mu_{n-1}}$ in the original theory. 
Furthermore, the equation of motion of
any other field $\psi(x)$ computed from the new action 
$\int \sqrt{-\det g} \wt \LL$ is the same as the one derived
from the original action $\SSS=\int \sqrt{-\det g}\LL$. To see
this note that \refb{enewlagr} gives an equation of motion of $\psi$
of the form:
\be \label{efff}
\left. {\delta \SSS\over \delta\psi(x)}\right|_H -
\int d^D y \eps^{\mu_1\cdots \mu_D} 
{\delta\Omega_{\mu_1\cdots \mu_n}(y)\over \delta
\psi(x)} \HH_{\mu_{n+1}\cdots \mu_D}(y) = 0\, ,
\ee
where the subscript $_H$ denotes that we need to carry out the
functional derivative treating $H_{\mu_1\cdots \mu_n}$ as an
independent field. On the other hand the original equations of
motion derived from the action $\SSS$, where we treat
$B_{\mu_1\ldots \mu_{n-1}}$ as independent field, may be expressed as
\be \label{esss}
\left. {\delta \SSS\over \delta\psi(x)}\right|_H 
+ \int d^D y {\delta \SSS\over \delta H_{\mu_1\cdots \mu_n}}\, 
{\delta\Omega_{\mu_1\cdots \mu_n}(y)\over \delta
\psi(x)}  = 0\, ,
\ee 
where we have taken into account the fact that there may be a hidden
dependence of $\SSS$ on $\psi$ through the Chern-Simons form
$\Omega$ in the definition of $H$.
Using \refb{eheqr} one can verify that 
\refb{efff} and \refb{esss} are
identical.
Thus
\refb{enewlagr} is classically equivalent to the original
Lagrangian density and we can use this new Lagrangian density
to carry out the computation of black hole entropy in this theory.

Since $H_{\mu_1\ldots \mu_n}$ is now an independent field, and
since the field strength $\HH$ is defined as in \refb{edefhh}, we
see that in the new theory the definition of field strengths do not
contain any Chern-Simons term. However the last term in the
Lagrangian density, 
\be\label{elastterm}
- \epsilon^{\mu_1\ldots \mu_D} \,
\Omega_{\mu_1\ldots \mu_{n}} \, 
\HH_{\mu_{n+1}\mu_{n+2}\ldots \mu_D}\, ,
\ee
is a Chern-Simons term. Thus the new Lagrangian density is of
type 1 where the Lagrangian density has an explicit Chern-Simons
term.

Let us now proceed to analyze Lagrangian densities of type 1. We shall
find it useful to use the notation of differential forms rather than
tensors. Chern-Simons terms which appear in string theory Lagrangian
density have one of two forms
\be \label{efirstcs}
B^{(1)}\wedge d B^{(2)}\wedge \ldots d B^{(s)}\, 
\ee
or
\be \label{esecondcs}
dB^{(1)}\wedge d B^{(2)}\wedge \ldots d B^{(s)} \wedge 
\Omega_{3L}\, 
\ee
where $B^{(i)}$ are $r_i$-form fields with associated $(r_i-1)$-form
gauge transformations as in \refb{ecsexample1}, 
and $\Omega_{3L}$ is the
Lorentz Chern-Simons $3$-form.\footnote{We are assuming that the
relevant rank 1 gauge fields are abelian so that their Chern-Simons
terms are of the form \refb{efirstcs}.}
The first 
term is manifestly invariant under
the gauge transformations associated with the $B^{(2)}$, $B^{(3)}$,
$\ldots$ $B^{(s)}$ fields, but fails to be invariant under the gauge
transformation associated with the $B^{(1)}$ field. However by
adding a total derivative term to the Lagrangian density
we can transfer the exterior
derivative from any of the other $B^{(i)}$ fields to $B^{(1)}$.
In this case the term will be manifestly invariant under the gauge
transformation associated with $B^{(1)}$, but will fail to be
invariant under one of the other gauge transformations. The
second term \refb{esecondcs} is manifestly invariant under all the
$B^{(i)}$ gauge transformations, but fails to be invariant under
the local Lorentz transformation. Again by adding a 
total derivative term
we can transfer the exterior
deivative from one of the $B^{(i)}$'s to the
Lorentz Chern-Simons term. The resulting Lagrangian density
will be manifestly
general coordinate and local Lorentz invariant since $d\Omega_{3L}$
transforms covariantly under these transformations, 
but will fail to be invariant
under one of the $B^{(i)}$ gauge transformations.

Our proposal for dealing with the terms given in \refb{efirstcs}
and \refb{esecondcs} is to dimensionally reduce the theory to
two dimensions by regarding the sphere $S^{D-2}$ as a compact
direction, express the resulting action as the integral of a covariant
Lagrangian density in two dimensions spanned by the $r$ and
$t$ coordinates and then calculate its
contribution to the entropy function in the usual manner. 
After dimensional reduction the magnetic flux $p_a$ through
$S^{D-2}$ will appear as parameters labeling the two dimensional
theory.
Since we are interested in only $SO(D-1)$ invariant
field configuration, the dimensional reduction is a straightforward
process except in cases where the Lagrangian density, evaluated in
the $SO(D-1)$ invariant background, has a
term that is not manifestly $SO(D-1)$ invariant. This would happen
if the Lagrangian density either
contains a Lorentz Chern-Simons term which, evaluated
for the sphere metric, is not manifestly $SO(D-1)$ invariant, 
or depends explicitly on a
$B^{(i)}$ whose field strength $dB^{(i)}$ has a non-zero flux
through $S^{D-2}$ since in this case $B^{(i)}$ itself does
not remain invariant under an $SO(D-1)$ rotation. 
Our strategy will be to avoid these terms to
whatever extent possible by adding total derivative to the Lagrangian
density {\it before dimensional reduction} to transfer the
derivatives in appropriate places. Thus if in \refb{efirstcs} and/or
\refb{esecondcs} there is even a single $B^{(i)}$ which does not
have an associated flux through $S^{D-2}$, we can take the
Lagrangian density in a form where that particular $B^{(i)}$
appears without a derivative, and every other factor has a
manifestly covariant form. 
In this case the Lagrangian density evaluated for a generic $SO(D-1)$
invariant background will have a manifestly $SO(D-1)$ invariant
form. 
Thus the only cases where the dimensional
reduction is complicated is the one where all the $B^{(i)}$'s have
flux through $S^{D-2}$. This requires all the $B^{(i)}$'s appearing
in \refb{efirstcs} and/or \refb{esecondcs} to be $(D-3)$-form so that
their field strengths are $(D-2)$-forms. Since the Lagrangian density
must be a $D$-form, for \refb{efirstcs} this gives
\be \label{efcond}
s(D-2) -1 = D \qquad \hbox{\i.e.} \qquad s = {D+1\over D-2}\, .
\ee
On the other hand for \refb{esecondcs} this gives
\be \label{escond}
s(D-2) + 3 = D \qquad \hbox{\i.e.} \qquad s = {D-3\over D-2}\, .
\ee

First let us deal with the case \refb{efcond}. Since $s$ must be an integer,
the only possible cases are $D=3$, $s=4$ and $D=5$, $s=2$. 
For simplicity we shall explain how to deal with the second case;
the analysis of the first case will proceed in an identical manner.
The relevant term in the Lagrangian density is proportional to
\be\label{ecs10}
B^{(1)}\wedge dB^{(2)}\, ,
\ee
where $B^{(1)}$ and $B^{(2)}$ are 2-form fields.
Note that $B^{(1)}$ and $B^{(2)}$ must be different fields since
$B\wedge dB$ is a total derivative for any even form field $B$.
Suppose $dB^{(1)}$ and $dB^{(2)}$ have flux $p_1$ and $p_2$
through $S^3$. We now define new 2-form fields
\be \label{ecs11}
C^{(1)} = {p_2 B^{(1)} - p_1 B^{(2)}\over \sqrt{p_1^2+p_2^2}}, 
\qquad
C^{(2)} = {p_1 B^{(1)} + p_2 B^{(2)}\over \sqrt{p_1^2+p_2^2}}\, .
\ee
In terms of these fields \refb{ecs10} can be wrtten
as
\be \label{ecs12}
C^{(1)}\wedge dC^{(2)}
\ee
plus total derivative terms. Furthermore the field $C^{(1)}$
has no flux through $S^3$ and $C^{(2)}$ has a flux proportional
to $\sqrt{p_1^2 + p_2^2}$. Since the Lagrangian density does not
involve $C^{(2)}$ explicitly,
we can carry out the dimensional
reduction of this term in a straightforward fashion, and get a
term proportional to
\be \label{ecs13}
\sqrt{p_1^2 + p_2^2} \, \, C^{(1)}\, ,
\ee
in the two dimensional theory. Since the 2-form
field $C^{(1)}$ can be regarded
as a scalar field density
in the two dimensional theory, \refb{ecs13} has
a manifestly covariant form in two dimensions and we can use
entropy function formalism to analyze extremal black hole solutions
in this theory. 

We note in passing that \refb{ecs13}
is the only term in the two dimensional Lagrangian
density which depends explicitly on $C^{(1)}$; the rest of the 
Lagrangian density depends on $C^{(1)}$ through $dC^{(1)}$
and hence vanishes in two dimensions.
Requiring the action to be stationary
with respect to $C^{(1)}$ then gives $p_1=p_2=0$. This shows
that in the presence of a term of the form \refb{ecs10}
we cannot have an extremal black hole solution with magnetic
charges associated with the $B^{(1)}$ and $B^{(2)}$ fields.

We now turn to a discussion of terms of the form \refb{esecondcs}
for which the only problematic case is \refb{escond}. Requiring $s$
to be integer gives $s=0$, $D=3$ as the only case. 
This corresponds to
the presence of a gravitational 
Chern-Simons term in the three dimensional
theory\cite{djw1,djw2}:
\be \label{elcs}
\sqrt{-\det g} \LL^{(3)}_{CS} = K\, 
\epsilon^{\mu\nu\tau} \left[{1\over 2}
\wh\Gamma^\rho_{\mu\sigma} 
\p_\nu \wh\Gamma^\sigma_{\tau \rho} + {1\over 3}
\wh\Gamma^\rho_{\mu\sigma} \wh\Gamma^\sigma_{\nu\eta} 
\wh\Gamma^\eta_{\tau\rho}\right]\, ,
\ee
where $\wh\Gamma^\mu_{\nu\rho}$ denotes the Christoffel
symbol and $K$ is a constant. 
To deal with this term, we regard the horizon $S^1$ as a compact
direction and carry out the dimensional reduction of this term
by taking the ansatz:
\be \label{elcs1}
g_{\mu\nu} dx^\mu dx^\nu = \phi \left[ g^{(2)}_{\alpha\beta} 
dx^\alpha dx^\beta + (dy + a_\alpha 
dx^\alpha)^2\right]\, .
\ee
Here $g^{(2)}_{\alpha\beta}$ ($0\le \alpha,\beta\le 1$) denotes a 
two dimensional metric, $a_\alpha$ denotes a two dimensional 
gauge  field and $\phi$ denotes a 
two dimensional scalar field. The $y$ coordinate labeling the
horizon $S^1$ is taken to have
period $2\pi$. In terms of these 
two dimensional fields the 
lagrangian density \refb{elcs}, after dimensional reduction to
two dimensions by integration over the $y$ coordinate, 
takes the form\cite{0305117,0601228}:
\be \label{elcs2}
K \, \pi \, \left[
 {1\over 2} R^{(2)} \varepsilon^{\alpha\beta} f_{\alpha\beta}
+{1\over 2} \varepsilon^{\alpha\beta} f_{\alpha\gamma} 
f^{\gamma\delta} 
f_{\delta\beta} \right]\, ,
\ee
plus total derivative terms. Here
$f_{\alpha\beta}=\p_\alpha a_\beta - \p_\beta a_\alpha$ is
the field strength associated with the two dimensional gauge field
$a_\alpha$,
$R^{(2)}$ denotes the Ricci scalar constructed out of the
two dimensional metric $g^{(2)}_{\alpha\beta}$ and
$\varepsilon^{\alpha\beta}$ is the totally anti-symmetric symbol
with $\epsilon^{01}=1$.
Since the Lagrangian density \refb{elcs2} has a manifestly covariant
form in two dimensions, we can apply the entropy function
formalism on this lagrangian density. This will be illustrated in
more detail in the
context of BTZ black holes in \S \ref{s2.4}.

This concludes our discussion on Chern-Simons terms in the context
of spherically symmetric black holes in arbitrary dimensions. A
similar analysis may be carried out for rotating black holes, but
we shall not discuss this case here.

\sectiono{Explicit Computation of Black Hole Entropy} \label{s2}

In this section we shall illustrate the entropy function formalism of
\S\ref{s1} by using it
to calculate the entropy of extremal black holes in a
variety of theories. Many of these results can also be derived from
other methods; in each of these cases the result obtained using
entropy function method (naturally) agrees with the ones obtained
by other methods. The analysis of this section will serve the
twin purpose of illustrating the entropy function formalism and
deriving specific results for black hole entropy  which will later be
compared with the statistical entropy in \S\ref{s3}.

\subsectiono{Black holes in $\NN=4$ supersymmetric theories in 
$D=4$} \label{s2.2}

There are various string compactifications which lead to $\NN=4$
supersymmetric theories in four dimensions. These theories have 
many scalar fields, known as moduli fields,
whose potential vanishes identically by the requirement
of supersymmetry. Thus they can take arbitrary vacuum expectation
values (vev), and the space of vev of these scalar fields
describe the moduli space. At a generic point in the moduli space
the requirement of $\NN=4$
supersymmetry completely determines 
the massless field content of the theory in terms of a single integer
$r\ge 6$ which labels the number
of $U(1)$ gauge fields in the theory. In particular the massless bosonic
fields are the string metric 
$G_{\mu\nu}$, $r$ abelian gauge fields
$A_\mu^{(i)}$ ($i=1,\ldots r$), a complex scalar field $a+iS$ taking
value in the upper half plane, and a 
set of $r\times r$ matrix valued scalar fields $M$ subject to the 
constraint:
\be \label{eag1}
MLM^T=L, \quad M^T=M \, ,
\ee
where $L$ is a matrix with 6 eigenvalues $+1$
and $(r-6)$  eigenvalues $-1$.
For $r\ge 12$, a convenient choice of $L$ is
\be \label{echoicel}
L = \pmatrix{0_6 & I_6 & \cr I_6 & 0_6 & 
\cr   &  & -I_{r-12}}\, ,
\ee
where $I_k$ denotes a $k\times k$ identity matrix and $0_k$
denote $k\times k$ zero matrix. The canonical metric $g_{\mu\nu}$
is related to the string metric $G_{\mu\nu}$ via the relation
$G_{\mu\nu}=Sg_{\mu\nu}$.

Our analysis in this section will mostly follow 
\cite{0508042,0609109}.

\subsubsection{Supergravity approximation} \label{sugra}

Requirement of $\NN=4$ supersymmetry also fixes all the terms
in the action containing at most two derivatives.
The part of the action containing the massless bosonic fields is
given by
\ben \label{eag3}
\SSS &=& {1\over 2\pi\alpha'}\, \int d^4 x \, 
\sqrt{-\det G} \,  S\, 
\left[ R_G + {1\over S^2}\,
G^{\mu\nu}  (\p_\mu S \p_\nu S -{1\over 2} 
\p_\mu a \p_\nu a) +{1\over 8}
G^{\mu\nu} Tr(\p_\mu M L \p_\nu M L) \right. \nonumber \\
&& \left. -  G^{\mu\mu'} G^{\nu\nu'}\,
F^{(i)}_{\mu\nu} (LML)_{ij} F^{(j)}_{\mu'\nu'}  - {a\over S}
G^{\mu\mu'} G^{\nu\nu'}\,
F^{(i)}_{\mu\nu} L_{ij} \wt F^{(j)}_{\mu'\nu'}  \right]\, .
\een
Note that this action has an
$SO(6,r-6)$ symmetry acting
on $M$ and $F^{(i)}_{\mu\nu}$:
\be \label{eag8c}
M\to \Omega M \Omega^T, \quad F^{(i)}_{\mu\nu}\to \Omega_{ij} 
F^{(j)}_{\mu\nu}\, ,
\ee   
where $\Omega$ is an $r\times r$ matrix satisfying
\be \label{eag9}
\Omega^T L \Omega=L\, .
\ee
This corresponds to the continuous T-duality 
symmetry of the supergravity
action. As will be discussed in \S\ref{sdual}, an appropriate
discrete subgroup of this is an exact symmetry of the full string
theory.

In this theory we look for a spherically symmetric extremal
black hole solution carrying arbitrary electric charges $q_i$
and magnetic charges $p_i$ for $i=1,\cdots r$.
{}Following the analysis of   \S\ref{s1.1} we look for a
near horizon field configuration of the form:
\ben \label{ehor}
 {ds^2  = {\alpha'\over 16} v_1\left(-r^2 dt^2+{dr^2\over 
r^2}\right)   
+ {\alpha'\over 16}
v_2 (d\theta^2+\sin^2\theta\, d\phi^2) \, , }\nonumber \\ 
 S =u_S, \qquad a=u_a , \qquad  {M_{ij} = u_{Mij} }\nonumber \\
  {F^{(i)}_{rt} = {\sqrt{\alpha'}\over 4}\, e_i,} \quad  
 {F^{(i)}_{\theta\phi}={p_i\sqrt{\alpha'}\over 16\pi}} \, \sin\theta\, ,
 \een
 where, for later convenience, we have included additional factors
 of $\alpha'/16$ multiplying $v_1$ and $v_2$ and additional
 factors of $\sqrt{\alpha'}/4$ multiplying $e_i$ and 
 $p_i$.\footnote{We shall use the convention that the coordinates
 $r,t,\theta,\phi$ and all the scalar fields are dimensionless, the
 gauge fields have dimension of length and the metric has dimension
 of length$^2$. With this convention the near horizon parameters
 $v_1$, $v_2$, $u_a$, $u_S$,
 $u_{Mij}$, $e_i$ and $p_j$ are dimensionless.}
 The effect
 of this will be to change the definition of the electric 
 and magnetic charges. Eq.\refb{ehor} agrees with the corresponding
 equations in \cite{0508042} for $\alpha'=16$.
Substituting \refb{ehor} into \refb{eag3} and using \refb{e2} we get
\ben \label{eag5}
&& f(u_S, u_a, u_M, \vec v, \vec e, \vec p)
\equiv \int d\theta d\phi \, \sqrt{-\det G} \, \LL
\nonumber \\ 
&=& 
{1\over 8} \, v_1 \, v_2
\, u_S  \left[ -{2\over v_1} +{2\over v_2} + 
 {2\over v_1^2} e_i (Lu_M L)_{ij} e_j - {1\over 8\pi^2 v_2^2}  p_i 
 (Lu_ML)_{ij} p_j + { u_a\over \pi u_S v_1 v_2}  e_i L_{ij} p_j
 \right] \, . \nonumber \\
\een
Eq.\refb{e2a} now gives
\ben \label{eag7pre}
\EE(u_S, u_a, u_M, \vec v, \vec e,
\vec q, \vec p) &\equiv& 2\pi \left( e_i q_i 
- f(u_S, u_a, u_M, \vec v, \vec e, \vec p) \right) \nonumber \\
&=& 2\pi \Bigg[ e_i q_i - {1\over 8} \, v_1 \, v_2
\, u_S  \bigg\{ -{2\over v_1} +{2\over v_2} + 
 {2\over v_1^2} e_i (Lu_M L)_{ij} e_j \nonumber \\
 && \qquad - {1\over 8\pi^2 v_2^2}  p_i 
 (Lu_ML)_{ij} p_j + { u_a\over \pi u_S v_1 v_2}  e_i L_{ij} p_j
 \bigg\} \Bigg]\, .
 \een
Eliminating $e_i$ from 
\refb{eag7pre} using the equation $\p\EE/\p e_i=0$ we get:
\ben \label{eag7}
\EE(u_S, u_a, u_M, \vec v, \vec q, \vec p) 
&=& 2\pi \bigg[ {u_S\over 4} (v_2 - v_1)
+{v_1\over v_2 u_S} \, q^T u_M q 
+{v_1\over 64\pi^2 v_2 u_S} (u_S^2 + u_a^2)
p^T L u_M L p \nonumber \\
&& -{v_1\over 4 \pi v_2 u_S}  \, u_a\, q^T u_M L p\bigg]\,.
\een
We can simplify the formul\ae\ 
by defining new charge vectors:
\be \label{eag8a}
Q_i = 2 q_i, \qquad P_i = {1\over 4\pi}\, L_{ij} p_j\, .
\ee
In terms of $\vec Q$ and $\vec P$
the entropy function $\EE$ is given by:
\be \label{eag8b}
\EE = {\pi\over 2} \bigg[ u_S(v_2 - v_1) +{v_1\over v_2 u_S} 
\left( Q^T u_M 
Q +
 (u_S^2 + u_a^2) \, P^T u_M P - {2 }
\, u_a \, Q^T u_M P \right)\bigg]\, .
\ee

We now need to find the extremum of $\EE$ with respect to $u_S$, $u_a$,
$u_{Mij}$, $v_1$ and $v_2$. In general this leads to a complicated set
of equations. However we can simplify the analysis by noting that 
\refb{eag8c} induces the following transformation 
on the various parameters:
\ben \label{eag9a}
e_i\to \Omega_{ij} e_j, \qquad  p_i\to \Omega_{ij} p_j, \qquad u_M\to
\Omega u_M \Omega^T\, , \nonumber \\
q_i\to (\Omega^T)^{-1}_{ij} q_j \, , \qquad
Q_i\to (\Omega^T)^{-1}_{ij} Q_j, \qquad P_i \to 
(\Omega^T)^{-1}_{ij} P_j\, .
\een
The entropy function \refb{eag8b} is invariant under these 
transformations.
Since at its extremum with respect to $u_{Mij}$
the entropy function depends only on $\vec P$, $\vec Q$, $v_1$, $v_2$, 
$u_S$ and $u_a$ it must be a function
of  the $SO(6,r-6)$ invariant combinations:
\be \label{eag10}
Q^2 = Q_i L_{ij} Q_j, \quad P^2 = P_i L_{ij} P_j, \quad 
Q\cdot P = Q_i L_{ij} P_j\, ,
\ee
besides $v_1$, $v_2$,
$u_S$ and $u_a$.
Let us for definiteness take $Q^2>0$, $P^2>0$, 
and $(Q\cdot P)^2<Q^2 P^2$. In that case with the help
of an $SO(6,r-6)$ transformation we can make 
\be \label{eag11}
(I_r-L)_{ij}Q_j=0, \quad (I_r-L)_{ij}P_j=0\, ,
\ee
where $I_r$ denotes the $r\times r$ identity matrix.
This
is most easily seen by diagonalizing 
$L$ to the form $\pmatrix{I_6& \cr & -I_{r-6}}$.
In this case $\vec Q$ and $\vec P$ satisfying \refb{eag11}
will have 
\be\label{eqipi}
Q_i=0, \quad P_i = 0, \quad \hbox{for $7\le i\le r$}\, .
\ee
We shall now show
that for $\vec P$ and 
$\vec Q$ satisfying this condition, every
term in \refb{eag8b} is extremized 
with respect to $u_{M}$ for
\be \label{eag12}
u_M  = I_r\, .
\ee
Clearly a variation $\delta u_{Mij}$
with either $i$ or $j$ in the range $[7,r]$ 
will give vanishing contribution to each term in
$\delta \EE$ computed from \refb{eag8b}. 
On the other hand due to the constraint \refb{eag1} on $M$, any 
variation $\delta M_{ij}$ (and hence 
$\delta u_{Mij}$) with $1\le i,j\le 6$ must vanish, since in this
subspace \refb{eag1} requires $M$ to be both symmetric and 
orthogonal.
Thus each term in $\delta \EE$ vanishes under all the 
allowed variations of $u_M$.

\refb{eag12} is not the only possible value of $u_M$ that
extremizes $\EE$. 
Any $u_M$ related to \refb{eag12} by an $SO(6,r-6)$ transformation
that preserves the vectors $\vec Q$ and $\vec P$ will extremize
$\EE$.
Thus there is  a family of  extrema
representing flat directions of $\EE$. However
as we have argued in \S\ref{s2.1}, the value of the entropy is
independent of the choice of $u_M$.

We  note in passing that the entropy function 
\refb{eag8b} is also invariant under continuous
S-duality transformation
\be\label{esdtrs}
\pmatrix{Q'\cr P'}=\pmatrix{ m & 
n\cr r & s} \pmatrix{Q\cr P}, \quad
u_a'+ i u_S' = {m (u_a + i u_S) + n\over 
r (u_a + i u_S) + s},  \quad v_i'=  {u_S \over u_S'}  v_i \, ,
\ee
where $m$, $n$, $r$, $s$ are real numbers satisfying
$ms - nr=1$.
This is a reflection of the continuous
S-duality covariance of the equations
of motion derived from the action \refb{eag3}. We shall not make
use of this symmetry in our analysis. However a discrete subgroup
of this, to be introduced in \S\ref{sdual}, is an exact symmetry of
string theory and will play an important role in our analysis.

Substituting \refb{eag12} into \refb{eag8b} and using \refb{eag10},
\refb{eag11}, we get:
\be \label{eag13}
\EE= {\pi\over 2} \left[ u_S(v_2-v_1) 
+{v_1\over v_2} \left\{{Q^2\over u_S} + 
{P^2\over u_S} (u_S^2 + u_a^2)
- 2\, {u_a\over u_S}\, Q\cdot P \right\} \right]\, .
\ee
Note that we have expressed the right hand side of this 
equation in an $SO(2,2)$ invariant form.
Written in this manner, eq.\refb{eag13} is valid
for general $\vec P$, $\vec Q$ satisfying 
\be \label{erg1}
P^2>0, \quad Q^2>0, \quad (Q\cdot P)^2<Q^2P^2\, .
\ee
 
It remains to extremize $\EE$ 
with respect to $v_1$, $v_2$, $u_S$ and $u_a$.
Extremization with respect to $v_1$ and $v_2$ give:
\be \label{es10}
v_1=v_2 = u_S^{-2} \, \left({Q^2} +
{P^2} (u_S^2 + u_a^2) - 2 u_a\, Q\cdot P\right)\, .
\ee
Substituting this into \refb{eag13} gives:
\be \label{es11copy}
\EE= {\pi\over 2}  
{1\over u_S}\, \left\{{Q^2} - 2\, {u_a}\, Q\cdot P
+
{P^2} (u_S^2 + u_a^2)\right\} 
\, .
\ee
Finally extremizing this with respect to $u_a$, $u_S$ we get
\be \label{eag14}
u_S= {\sqrt{Q^2 P^2 - (Q\cdot P)^2}\over P^2} \, ,
\qquad u_a = 
{Q\cdot P\over P^2}, \qquad
v_1=v_2= 2 P^2  \, .
\ee
The black hole entropy, given by the value of $\EE$ for this configuration,
is 
\be \label{eag15}
S_{BH} = \pi \, \sqrt{Q^2 P^2 - (Q\cdot P)^2}\, .
\ee

Although this formula has been derived under the condition
$P^2>0$, $Q^2>0$, $P^2 Q^2 > (Q\cdot P)^2$, the final result for
the entropy also holds for arbitrary $P^2$, $Q^2$ as long as
$P^2 Q^2 > (Q\cdot P)^2$. 
Extremal black holes with $P^2 Q^2 > (Q\cdot P)^2$ are known to be
supersymmetric although we cannot see this in our analysis. The
entropy function formalism also allows us to calculate the
entropy of extremal black holes with
$P^2 Q^2 < (Q\cdot P)^2$. We shall not go through the analysis here,
but just quote the final result:
\be \label{eag15ns}
S_{BH} = \pi \, \sqrt{(Q\cdot P)^2-Q^2 P^2}\, .
\ee

\subsubsection{A special class of $\NN=4$ supersymmetric string
theories} \label{sphys}

Our goal is to study the effect of higher derivative terms
in the action on the entropy function. However, 
unlike in the case of two
derivative terms, the higher derivative corrections do depend on
the details of the 
string compactification that gives rise to the theory. Thus
in order to study the effect of higher derivative corrections we need
to consider specific theories. We shall restrict our analysis to
a  class of string theories where we begin with
type IIB string theory on $\MM\times \wt 
S^1\times S^1$ where
$\MM$ is either K3 or $T^4$ and $\wt S^1$ and $S^1$ are two
circles, and 
take an orbifold of this theory by a $\ZZZ_N$ symmetry
group. The generator $g$ of the $\ZZZ_N$ group involves
$1/N$ unit of shift along the circle $S^1$ 
together with an order
$N$ transformation
$\wt g$ in $\MM$. $\wt g$ is chosen so that it commutes with
the $\NN=4$ supersymmetry generators of the parent theory.
Thus the final
theory has $\NN=4$ supersymmetry. 
Various properties of $\wt g$ coming from this requirement have
been discussed in appendix \ref{s0}. In particular
this requires $\MM/\wt g$
to be an orbifold of $SU(2)$ holonomy.

The description of the theory given above will be referred to as
the first
description of the theory. Another useful description is
obtained by
making an S-duality transformation in the type IIB theory on
$\MM\times \wt S^1\times S^1$ that
exchanges the NS 5-branes with D5-branes and fundamental
strings with D-strings, followed
by an $R\to 1/R$ duality on the circle $\wt S^1$ that takes type IIB
theory on $\wt S^1$ to type IIA theory on the dual circle
$\wh S^1$, and then using six dimensional string-string
duality to relate this to a heterotic string theory on $T^4\times \wh
S^1
\times  S^1$ for 
$\MM=K3$\cite{9410167,9501030,9503124,9504027,9504047} 
and type IIA string theory on
$T^4\times \wh S^1
\times  S^1$ for $\MM=T^4$\cite{9508064}. 
Under this duality the transformation
$\wt g$ gets mapped to a transformation $\wh g$ that acts only as
a shift on the right-moving degrees of freedom on the world-sheet
and as a shift plus rotation on the left-moving degrees of freedom.
In the final theory, obtained by taking the orbifold of heterotic or
type IIA
string theory on $T^4\times \wh
S^1\times  S^1$ by a $1/N$ unit
of
shift along $S^1$ together with the transformation $\wh g$, all
the space-time supersymmetries come from the right-moving sector
of the world-sheet. We shall call this the second description of the
theory. 

The heterotic models were first discovered in the analysis of
\cite{9505054,9506048,9507050,9508144,9508154}. The type II
versions were analyzed in \cite{9508064}.

At the level of two derivative terms, the effective action of
each of these theories is given by \refb{eag3} for appropriate choice
of $r$ determined from the details of compactification.
The precise expression for $r$ is given by
\be \label{edefr}
r = 2 \, k + 8\, ,
\ee
where
$k$ has been computed in appendix \ref{ssiegel}
(eq.\refb{ekvalue})  and is equal to half the number of $\wt g$
invariant harmonic $(1,1)$ forms on $\MM$.
Explicit form of $k$ for special cases can be found in \refb{ekssh},
\refb{ekssii}.

In order to facilitate later analysis where we compare the black
hole entropy with the statistical entropy, it will be useful
to know the correspondence between the various fields and charges
appearing in \refb{eag3} with the physical fields and charges
in string theory.
First of all the field $\tau=a+iS$ corresponds to the
complex structure modulus of the torus $\wt S^1\times S^1$ in the
first description. By following the duality chain carefully one can see
that it represents the usual axion-dilaton field in the second
description, -- $a$ being the scalar field obtained by dualizing
the anti-symmetric tensor field in the NSNS sector, and $S$ being
$e^{-2\Phi}$ where $\Phi$ denotes the dilaton field. 
The matrix valued scalar field $M$ encodes information about
the shape and size of the compact space $T^4\times \wh
S^1\times 
S^1$ and the components of the NSNS sector 2-form field
along $T^4\times \wh S^1\times S^1$ in the second description.
Finally 
in the second description the gauge fields appearing in the action
\refb{eag3} can be related directly to the ones coming from
the dimensional reduction of
the  ten dimensional metric, NSNS sector
anti-symmetric tensor field and gauge fields, without any further
electric-magnetic duality transformation. As a result the elementary
string states in this description carry electric charge vector $\vec Q$ and
the various solitons carry magnetic charge vector $\vec P$. 
We shall carry on the rest of the discussion in this section in the
second description, but following the chain of dualities relating
the two descriptions one can easily work out the interpretation of
various quantities in the first description.

Let $x^4$ and $x^5$ denote the coordinates along $\wh S^1$
and $S^1$ respectively,
both normalized to have period
$2\pi\sqrt{\alpha'}$ after the $\ZZZ_N$ orbifolding, 
and let $x^\mu$ ($0\le
\mu\le 3$) denote the non-compact coordinates.
For most of our analysis it will be 
useful to study in detail a subsector
of the theory in which we include only those gauge fields 
which are associated with the $4\mu$ and $5\mu$
components of the metric and the
anti-symmetric tensor field, 
only those components of $M$ which encode information
about the
components of the metric and the
anti-symmetric tensor field
along $\wh S^1\times S^1$, the axion-dilaton field,
and the four dimensional metric. In this subsector there are
four gauge fields $A^{(i)}_\mu$ ($1\le i\le 4$) and 
a $4\times 4$ matrix valued field $M$ satisfying
\be\label{enewm}
M^T = M, \quad M L M^T = L, \quad L\equiv\pmatrix{0 & I_2\cr
I_2 & 0}\, .
\ee
The fields $A^{(i)}_\mu$ and $M$
are related to the ten dimensional string metric $G_{MN}$ and
2-form field $B_{MN}$ via the relations\cite{9207016,9402002}:
\ben \label{etenfour}
&& \wh G_{mn} \equiv G^{(10)}_{mn}, \quad 
\wh B_{mn} \equiv B^{(10)}_{mn}\, ,  
\qquad 
m,n=4,5\, , 
\nonumber  \\ 
&& M =  
\pmatrix{ \wh G^{-1} & \wh G^{-1} B \cr -\wh B \wh G^{-1} & \wh 
G - \wh B \wh G^{-1} \wh B} \nonumber \\
&& A^{(m-3)}_\mu = {1\over 2} (\wh G^{-1})^{mn} 
G^{(10)}_{m\mu} , \quad
A^{(m-1)}_\mu = {1\over 2} B^{(10)}_{m\mu} - 
\wh B_{mn} A^{(m-3)}_\mu, \nonumber \\
&& \qquad  4\le m,n\le 5, \quad 
0\le \mu, \nu \le 3 \, . 
\een
The Lagrangian density involving the axion-dilaton field, the four
gauge fields and the $4\times 4$ matrix valued scalar field $M$
has a form identical to the one given in \refb{eag3} with $L$ given as
in \refb{enewm}.
In fact this is a consistent truncation of the full $\NN=4$ 
supergravity theory.

With the normalization convention we have used for the charges
$\vec P$ and $\vec Q$, and the sign conventions described in appendix 
\ref{s3.1}, a state with $\wh n$ unit of momentum and
$-\wh w$ unit of winding along $\wh S^1$, $n'$ unit of
momentum and
$-w'$ unit of winding along $S^1$,
$\wh N$ unit of 
Kaluza-Klein monopole charge\cite{grossperry,sorkin} 
associated with 
$\wh S^1$, 
$-\wh W$ unit of NS 5-brane wrapped along 
$T^4\times S^1$,
$N'$ unit of 
Kaluza-Klein monopole charge associated with 
$S^1$ and
$W'$ unit of NS 5-brane wrapped along $T^4\times \wh S^1$
describes a four dimensional charge vector of
the form\footnote{A Kaluza-Klein monopole associated with
a circle denotes a Taub-NUT space whose asymptotic geometry
is the product of this circle 
and the three non-compact spatial directions. Also an NS 5-brane wrapped
along $T^4\times S^1$ will be said to carry one unit of H-monopole
charge associated with $\wh S^1$ and an NS 5-brane wrapped along
$T^4\times \wh S^1$ will be said to carry $-1$ unit of H-monopole
charge associated with $S^1$\cite{9211056}.}
\be\label{e2dcharge}
Q=\pmatrix{\wh n\cr n'\cr \wh w\cr w'}, \qquad 
P = \pmatrix{\wh W\cr W'\cr
\wh N\cr N'}\, .
\ee
Thus we have
\be\label{echsq}
Q^2 = 2(\wh n \wh w + n'w'), \qquad P^2 = 2(\wh N \wh W
+N'W'), \qquad
P\cdot Q = \wh N \wh n + \wh W \wh w+ N' n' + W' w'\, .
\ee
We shall denote by $\VV$ the subspace spanned by charge vectors of the 
form \refb{e2dcharge}.

The sign conventions for various charges have been described in detail
in appendix \ref{s3.1}. Here 
we shall say a few words about the normalization
of the various charges appearing in \refb{e2dcharge}. First of all
units of momentum along $S^1$ and $\wh S^1$ will be taken to be
$1/\sqrt{\alpha'}$. We take $\wh S^1$ to have coordinate radius
$2\pi\sqrt{\alpha'}$ and $S^1$ to have coordinate radius 
$2\pi\sqrt{\alpha'}N$ before orbifolding. Thus after orbifolding
$S^1/\ZZZ_N$ has coordinate radius $2\pi\sqrt{\alpha'}$, and
various fields satisfy $\wh g$ twisted boundary condition under a
translation by $2\pi\sqrt{\alpha'}$ along $S^1$. As a result the
momentum along $S^1$ is quantized in units of 
$1/(N\sqrt{\alpha'})$. 
Similar conventions are followed
for all other quantum numbers. One unit of winding
along $S^1$ will refer to a  state such that as we go once
around the string, its coordinate along $S^1$ shifts by $2\pi 
\sqrt{\alpha'}$. This represents a twisted sector state.
An untwisted sector state whose coordinate along $S^1$
changes by multiples of $2\pi \sqrt{\alpha'}N$ 
will carry winding charge 
$w'$ in
multiples of $N$.
A single H-monopole associated with $S^1$, with $W'=1$,
will correspond to an
array of NS 5-branes wrapped on $\wh S^1\times T^4$
and placed at
intervals of $2\pi  \sqrt{\alpha'}$ along $S^1$. Finally the original 
Kaluza-Klein monopole represented by a
Taub-NUT
space with an asymptotic circle of radius $N\sqrt{\alpha'}$, 
after the orbifolding,
will develop a $\ZZZ_N$ singularity at its centre and has to be 
regarded as carrying $N$ units of  Kaluza-Klein
monopole charge associated with $S^1$. 
Thus the Kaluza-Klein monopole charge $N'$ will be quantized
in units of $N$. The charges $\wh n$, $\wh w$, $\wh N$ and $\wh W$
are all quantized in integer units since $\wh S^1$ has period
$2\pi\sqrt{\alpha'}$ and the orbifold group does not act on $\wh S^1$.
Similar convention must
also be followed in the definition of various fields. For example 
in defining the matrix valued field $M$ and the gauge fields
$A_\mu^{(i)}$ via eq.\refb{etenfour},
the coordinates $x^4$ and $x^5$ must be chosen so that 
$x^4$ has period $2\pi\sqrt{\alpha'}$ and $x^5$ has period
$2\pi\sqrt{\alpha'}N$ 
before orbifolding.

We must also follow the same 
convention in identifying
fields in the first description. For example if the physical radii of
$S^1$ and $\wt S^1$ are $R_0$ and $\wt R$ before orbifolding, then
the
field $\tau=a+iS$ has to be regarded as 
the complex structure modulus of the  torus 
$\wt S^1\times  S^1$ with $S^1$ direction regarded as having
period $2\pi R_0/N$. Thus
we shall have
 $\sqrt{a^2+S^2} = R_0/(N\wt R)$, and
$\tan^{-1}(a/S)$ will be
given by the angle between the two circles.

\subsubsection{Duality symmetries} \label{sdual}

The $\NN=4$ supersymmetric string 
theories discussed here have T- and
S-duality symmetries induced from the duality symmetries of the
parent theory before orbifolding. Since classification
of duality symmetries into T- and S-dualities depends on the
description of the system we are using, we shall follow the
convention that unless
  mentioned otherwise, whenever we refer to T- or S-duality
  symmetry of the theory we shall imply 
  T- or S-duality symmetry in the
  second description. Similarly whenever we mention 
  electric or magnetic
  charges we shall imply electric or magnetic
  charges in the
  second description. 
With this convention a general T-duality transformation
acts non-trivially 
on the charges and the matrix valued scalar field $M$ as:
\be \label{tdual1}
M\to \Omega M \Omega^T, \quad Q\to (\Omega^T)^{-1} 
Q, \quad
P\to (\Omega^T)^{-1} P\, ,
\ee
where $\Omega$ is an $r\times r$ matrix that preserves the charge
lattice and satisfies
\be \label{tdual2}
\Omega L \Omega^T = L\, ,
\ee
$L$ being a matrix with 6 eigenvalues $+1$ and $(r-6)$ eigenvalues
$-1$. Since $L^2=1$ it follows from \refb{tdual2} that
$\Omega^T L\Omega=L$.

Since much of our analysis will involve states with
electric and magnetic charges of the form given in 
\refb{e2dcharge}, we shall explicitly determine the part
of the T-duality group that acts on this subspace. This is
the T-duality group associated with the torus 
$\wh S^1\times S^1$ in the second description.
Before taking the $\ZZZ_N$
orbifold this T-duality group was 
$SL(2,\ZZZ)\times SL(2,\ZZZ)$. Taking the orbifold preserves
a subgroup of this group which commutes with the
orbifold action, \i.e.\ commutes with translation by
$2\pi\sqrt{\alpha'}$
along $S^1$ up to translations by $2\pi\sqrt{\alpha'}N$ and
$2\pi\sqrt{\alpha'}$ along $S^1$ and $\wh S^1$ respectively. 
This turns out to be isomorphic to the group $\Gamma_1(N)
\times \Gamma_1(N)$ where $\Gamma_1(N)$ consists
of $2\times 2$ matrices of the form $\pmatrix{a & b\cr c & d}$
satisfying
\be \label{edefgamma1}
ad-bc
=1, \quad a,d \in N\ZZZ+1, \quad b\in \ZZZ, \quad 
c \in N\ZZZ\, .
\ee
The matrix $\Omega$ is expressed in terms of the pair of
$\Gamma_1(N)$ matrices $\pmatrix{a & b\cr c & d}$ and
$\pmatrix{p & q\cr r & s}$ as
\be \label{efulltduality}
(\Omega^T)^{-1}= 
\pmatrix{d & -c & 0 & 0\cr -b & a & 0 & 0\cr 
0 & 0 & a & b\cr 0 & 0 & c & d}
\pmatrix{p & 0 & 0 & -q\cr 0 & p & q & 0\cr
0 & r & s & 0\cr -r & 0 & 0 & s}\, .
\ee
It is straightforward to verify that this matrix
satisfies \refb{tdual2} and preserves the charge
quantization laws described below 
\refb{echsq}.

If in the second description the theory is an
asymmetric orbifold of heterotic string theory then there is an
additional $\ZZZ_2$ duality symmetry represented by the
matrix
\be \label{esemidirect}
\pmatrix{0 & 0 & 1 & 0\cr 0 & 1 & 0 & 0\cr
1 & 0 & 0 & 0\cr 0 & 0 & 0 & 1}\, .
\ee
Physically this represents the
effect of $R\to 1/R$ duality on the circle $\wh S^1$.

The S-duality symmetry of the theory is best described in the
first description where it corresponds to the global diffeomorphism
symmetry associated with the torus $\wt S^1\times S^1$. Before
orbifolding this symmetry is $SL(2,\ZZZ)$. However 
only those elements of 
$SL(2,\ZZZ)$ which commute with the orbifold group
generator $g$ are true
symmetries of the theory. This again gives rise to the group 
$\Gamma_1(N)$ consisting of $2\times 2$ matrices 
$\pmatrix{\alpha & \beta\cr \gamma & \delta}$ satisfying the constraints:
\be \label{tdual3}
\alpha\delta-\beta\gamma
=1, \quad \alpha,\delta \in N\ZZZ+1, \quad \beta\in \ZZZ, \quad 
\gamma \in N\ZZZ\, .
\ee
Its action on the axion-dilaton modulus $\tau = a + i S$ and the
charges is given by
\be \label{tdual4}
\tau \to {\alpha\tau + \beta\over 
\gamma\tau +\delta}, \qquad \pmatrix{Q\cr P}
\to \pmatrix{\alpha & \beta\cr \gamma & \delta} \pmatrix{Q\cr P}\, .
\ee
The entropy function \refb{eag13} obtained in the leading 
supergravity approximation is invariant under both these symmetries.

\subsubsection{Corrections due to Gauss-Bonnet terms} 
\label{sgauss}

In the first description of the theory the axion-dilaton field $\tau
=a+iS$ has the interpretation of the complex structure modulus
of the torus $\wt S^1\times  S^1$. As discussed in appendix
\ref{ssgauss}, one loop effective action in this
theory contains a term of the form\cite{rzwiebach,9610237}:
\be\label{ecorgaus}
\Delta \SSS = \int d^4 x\, \sqrt{-\det g} \, \Delta \LL\, ,
\ee
where
\be \label{ecorg1}
\Delta\LL = \phi(a,S)\, 
\left\{ R_{\mu\nu\rho\sigma} R^{\mu\nu\rho\sigma}
- 4 R_{\mu\nu} R^{\mu\nu}
+ R^2
\right\} \, .
\ee
Here $R_{\mu\nu\rho\sigma}$ is the Riemann tensor constructed
from the canonical metric $g_{\mu\nu}$:
\be \label{edefgmunu}
g_{\mu\nu} = S \, G_{\mu\nu}\, .
\ee
The function $\phi(a,S)$ appearing in \refb{ecorg1} 
was originally
computed in \cite{0609109} using the formalism developed in
\cite{9708062}. This calculation has been reproduced in
appendix \ref{ssgauss} and the result is
\be\label{eh10bint}
\phi(a,S) = - {1\over 64\pi^2} \, \left( (k+2) \ln S 
+ \ln g(a+iS) + \ln g(-a+iS)\right)
+\hbox{constant}\, ,
\ee
where, as mentioned below  \refb{edefr},  $k$ is equal to half
the number of $\wt g$ invariant harmonic (1,1) forms on $\MM$,
and $g(\tau)$, computed in \refb{enn13}, is given by:
\be\label{enn13pre}
g(\tau) = e^{2\pi i \wh\alpha\tau}\,
\prod_{n=1}^\infty \prod_{r=0}^{N-1}
 \left( 1 - e^{2\pi i r/N}
e^{2\pi i n\tau}\right)^{s_{r}}\, .
\ee
Here $s_r$ counts the number of harmonic $p$-forms of $\MM$
with $\wt g$ eigenvalue $e^{2\pi i r/N}$ weighted by $(-1)^p$ and
$\wh\alpha$, given in \refb{enn9d}, 
is the Euler character of $\MM$ divided
by 24.
Thus we have
\be \label{ealphavalue}
\wh\alpha = \cases{1 \quad \hbox{for} \quad \MM=K3\cr
0 \quad \hbox{for} \quad \MM=T^4}\, .
\ee
Since in the second description of the theory $\MM=K3$ corresponds
to an orbifold of heterotic string theory on 
$T^4\times \wh S^1\times  S^1$ and
$\MM=T^4$ corresponds
to an orbifold of type II string theory on 
$T^4\times \wh S^1\times  S^1$, we see that in this description
\be \label{ealphafin}
\wh\alpha = \cases{1 \quad \hbox{for heterotic}\cr
0 \quad \hbox{for type II}}\, .
\ee
For special cases explicit expressions for $g(\tau)$
in terms of Dedekind eta function can be found in
in \refb{egtauh}, \refb{egtauii}. 

As shown in \refb{egrhotrs},
under a duality transformation
$g(\tau)$ transforms as 
\be \label{egrhotrspre}
g((a\tau+b)(c\tau +d)^{-1}) = (c\tau + d)^{k+2} g(\tau)\, .
\ee
Using this one can
show that $\phi(a,S)$ is
manifestly invariant under the S-duality 
transformation \refb{tdual4}. Since $a$ and $S$ do not
transform under a T-duality transformation of the form given
in \refb{eag8c}, this shows that \refb{ecorg1} is
invariant under both S- and T-duality transformations. Note that
without the $\ln S$ term in its definition, $\phi(a,S)$ would not
have been 
S-duality invariant. 
This
is the only term in $\phi(a,S)$ that cannot be written as a sum of
a holomorphic and an anti-holomorphic term, and has been called
the holomorphic anomaly\cite{9302103,9307158,9309140}.

The effect of this additional term \refb{ecorg1} in the Lagrangian
density
gives correction to the entropy of the black hole. This correction was
first studied in \cite{9711053} using an Euclidean action
formalism; here we describe a systematic method for calculating
this correction using the entropy function formalism.
Using the definition of the entropy function it is easy to calculate
the correction to the entropy function due to this additional term.
It is given by
\be \label{es7}
 \Delta \EE = -2\pi\, \int d\theta d\phi \, \sqrt{-\det g} \,
 \Delta \LL  = 64\pi^2 
\phi(u_a, u_S)\, .
\ee
Together with \refb{eag8b} this gives
\ben \label{es8}
\EE &=& {\pi\over 2} \bigg[ u_S(v_2 - v_1) 
+{v_1\over v_2 u_S} \left\{ Q^T 
u_M Q +
(u_S^2 + u_a^2) \, P^T u_M P \right. \nonumber \\
&& \left. - {2 } 
\, u_a \, Q^T u_M P \right\} + 128\, \pi \, \phi(u_a, u_S)\bigg]\, .
\een
Since the extra term is independent of $u_M$, $v_1$ and $v_2$,
the extremization of $\EE$ with respect to these variables can be
carried out as before without any change. This gives, for $P^2>0$,
$Q^2>0$, $P^2 Q^2 > (Q\cdot P)^2$,
\be \label{es11}
\EE= {\pi\over 2} \bigg[ {1\over u_S}\, \left\{
{Q^2} - 2\, {u_a}\, Q\cdot P
+
{P^2} (u_S^2 + u_a^2)\right\}
+ 128 \pi \phi(u_a, u_S)\bigg]\, .
\ee
The values of $u_a$ and $u_S$ at the horizon are 
determined by extremizing 
$\EE$ with respect to $u_a$ and $u_S$. This gives:
\ben \label{es12}
P^2 u_a - Q\cdot P + 64 \, \pi \, u_S \, {\p\phi\over \p u_a} = 0 \, ,
\nonumber \\
-{1\over u_S^2}  \left({Q^2} - 2\, {u_a}\, Q\cdot P
+
{P^2} u_a^2\right) + P^2  + 128 \, \pi \, {\p\phi\over \p u_S} = 0\, .
\een
Finally the value of $\EE$ evaluated at the solution to eqs.\refb{es12}
gives the entropy of the black hole. As mentioned earlier, these
black holes are expected to be supersymmetric. Eqs.\refb{es11},
\refb{es12} first appeared in \cite{9906094} in the context
of $\NN=2$ supergravity theories.

Although it is difficult to solve the extremization equations \refb{es12}
analytically, we can solve it iteratively. In particular, at the level of
four derivative terms in the action, we are interested in corrections to
the entropy which are suppressed compared to the leading contribution
\refb{eag15} by two powers of various charges, 
\i.e.\ terms which remain invariant under a
simultaneous rescaling of all the charges.
For this we can simply evaluate the modified entropy function at 
the leading order solution \refb{eag14} 
for $u_S$ and $u_a$.  This gives the following expression for the
black hole entropy:
\be \label{ebhexplicit}
S_{BH} = \pi \, \sqrt{Q^2 P^2 - (Q\cdot P)^2}
+64\, \pi^2\, \phi\left({Q\cdot P\over P^2},
{\sqrt{Q^2 P^2 - (Q\cdot P)^2}\over P^2}  \right) +\cdots\, 
\ee
where $\cdots$ denote correction terms which are suppressed by inverse
powers of charges.

Although \refb{ebhexplicit} give the correction to the black hole
entropy due to the Gauss-Bonnet term, we should note that the effective
action of string theory contains other four derivative terms besides the
Gauss-Bonnet term. In principle their contribution to the entropy will
be of the same order as that of the Gauss-Bonnet term. Thus one could
question the significance of the result given in \refb{ebhexplicit}.
At present there is no completely satisfactory 
answer to this question, but we shall 
try to summarize our current understanding of the situation.
In order to set up the
background for this analysis
we shall first study \refb{es11}, \refb{es12}
in a particular limit. 
Let us consider a range of charges
where the electric
charge $\vec Q$ is much larger than the magnetic charge 
$\vec P$.
In this case the leading order result
\refb{eag14} shows that $u_S$ is large at the horizon
and hence the string loop corrections in the second
description, involving inverse powers
of $u_S$, should be small. Thus we can expect that in 
the second description
we only need
to include corrections to the effective action at string tree level.
In the particular context of the Gauss-Bonnet term, this corresponds
to replacing $\phi(u_a,u_S)$
in \refb{es11}, \refb{es12} by its expression for large $u_S$.
Using \refb{eh10bint},
\refb{enn13pre} we see that for large $u_S$
\be\label{elarge}
\phi(u_a,u_S) \simeq {1\over 16\pi}\, \wh\alpha \, u_S\, .
\ee
Substituting this in \refb{es11} we get
\be \label{es11spec}
\EE= {\pi\over 2} \bigg[ {1\over u_S}\, \left({Q^2} - 2\, {u_a}\, Q\cdot P
+
{P^2} (u_S^2 + u_a^2)\right) 
+  8\, \wh\alpha \, u_S \bigg]\, .
\ee
Extremization of this function with respect to $u_S$ and $u_a$ can now
be carried out analytically and, using \refb{es10},
yields the answer
\ben \label{eag16}
&& u_S= \sqrt{Q^2 P^2 - (Q\cdot P)^2\over P^2 (P^2+8\wh\alpha)} \, ,
\qquad u_a = 
{Q\cdot P\over P^2}\, ,
\nonumber \\
&& v_1=v_2 =  2 P^2+8\wh\alpha\, , 
\een
and
\be \label{eag17}
S_{BH}=\EE = \pi \sqrt{ (P^2+8\wh\alpha) (Q^2 P^2 - (Q\cdot P)^2)\over
P^2}\, .
\ee

For later use in \S\ref{stu} and \S\ref{sads.2}
we shall write down the solution \refb{eag16} for
a special class of black holes for which   
\be \label{echargepremore}
Q = \pmatrix{\wh n\cr 0\cr  \wh w\cr 0}, \qquad
P = \pmatrix{ 0 \cr W'\cr 0 \cr N'}\, .
\ee
For definiteness we shall take
\be \label{erg2}
N',W'>0, \quad \wh n, \wh w < 0\, ,
\ee
so that \refb{erg1} is satisfied.
Let us further assume that $\wh n\wh w>> N'W'$ so that $u_S$
at the horizon is large and hence $\phi(u_a,u_S)$ can be
approximated as in \refb{elarge}.
In this case \refb{eag16}, \refb{eag17} take the form:
\ben \label{eag16ag}
&& u_S= \sqrt{\wh n \wh w\over N'W'+4\wh\alpha} \, ,
\qquad u_a = 0\, ,
\nonumber \\
&& v_1=v_2 =  4(N'W'+2\wh\alpha) \, .
\een
The solution for $M$ is also easy to calculate by extremizing
\refb{eag8b} and we get
\be \label{emvaluer}
M = \pmatrix{\wh w/\wh n &&& \cr & N'/W' && \cr
&& \wh n/\wh w & \cr &&& W'/N'}\, .
\ee
Comparing with eq.\refb{etenfour} we see that if $\wh R$ and
$R$ denote the radii of $\wh S^1$ and $S^1$ measured in the
string metric after $\ZZZ_N$ orbifolding then
\be \label{erv}
\wh R = \sqrt{\wh n/\wh w}, \qquad R=\sqrt{W'/N'}\, .
\ee
Finally \refb{eag17} gives
\be \label{eag17spec}
S_{BH}= 2\pi \sqrt{ \wh n\wh w(N'W'+4\wh\alpha)}
\, .
\ee
We shall see in \S\ref{sads3} (see eq.\refb{efe} and the 
discussion below) that 
\refb{eag17spec} is the exact answer for the entropy in the
$\wh n\wh w>> N'W'$ limit.  More generally one can show that
\refb{eag17} is exact in the limit $\sqrt{Q^2P^2-(Q\cdot P)^2}
>> P^2$.

Let us now turn to the question of validity of
\refb{ebhexplicit}.  The
question is: how does the formula get corrected by other four
derivative terms? To this end we make the 
following observations:
\begin{itemize}
\item A simple scaling argument shows that the 
contribution to the entropy from
the four derivative terms must remain invariant when
all charges are scaled by a common parameter. This is manifestly
true for the contribution from the Gauss-Bonnet term.  The
contribution from the other four derivative terms must also satisfy
this constraint. 
\item Since the answer for the
entropy after inclusion of Gauss-Bonnet term is duality invariant, 
the additional contribution due to the other four derivative terms
must be duality invariant by itself. 
\item For $\sqrt{Q^2P^2-(Q\cdot P)^2}
>> P^2$ the additional
contribution must vanish since
we know that  
in this limit the correction to the entropy
due to the Gauss-Bonnet term captures the complete
answer. 
\end{itemize}
This gives strong constraints on the contribution from the additional
four derivative
terms. Having this contribution vanish is consistent with
all these constraints.
We would like to believe that these, together with some
other constraints, can be used to argue that the additional term
actually vanishes, but we do not have such a proof as of now.
However we shall see in \S\ref{s3.3} that 
\refb{ebhexplicit} agrees perfectly
with the first non-leading correction to the statistical entropy.

\subsubsection{Non-supersymmetric extremal black holes}
\label{snsn=4}

In the supergravity approximation 
non-supersymmetric black holes
arise for $P^2 Q^2< (Q\cdot P)^2$, and the entropy of the
corresponding black hole has been given in \refb{eag15ns}.
Let us consider a special class of such black holes with charge
vector given in \refb{echargepremore}, for the range
\be \label{erg3}
N', W',\wh n > 0, \qquad \wh w<0\, .
\ee
This differs from \refb{erg1} by a flip $\wh n\to -\wh n$.
Extremization of $\EE$ can be done simply by flipping,
in the solution corresponding to \refb{echargepremore}, 
\refb{erg2}, the
sign of the electric field conjugate to $\wh n$ keeping every
other near horizon parameter unchanged.
Thus the solution and the black hole entropy for $\wh n>0$ are
given by eqs.\refb{eag16ag}-\refb{eag17spec} 
with $\wh n$ replaced by $-\wh n$.
In particular the black hole entropy is given by
\be \label{ensns1}
S_{BH}^{ns} = 2\pi \sqrt{|\wh n\wh w| (N'W'+4\wh\alpha)}\, .
\ee

\subsectiono{Black holes in $\NN=2$  supersymmetric theories in 
$D=4$} \label{sn=2}

In this section we shall apply the entropy function formalism to calculate
the entropy of extremal black holes
in a more general class of theories in four dimensions,
namely $\NN=2$ supergravity theories. Our analysis will follow
\cite{0603149}.

\subsubsection{General considerations} \label{sgen}

The off-shell formulation of $\NN=2$ supergravity action in four
dimensions was developed in 
\cite{r1,r2,r3,r4,r5,r6,r7,9602060,9603191}. Here
we shall review this formulation following 
the notation of \cite{0007195}.
The basic bosonic fields in the theory are a set of $(N+1)$ 
complex scalar fields
$X^I$ with $0\le I\le N$
(of which one can be gauged away using a scaling 
symmetry), $(N+1)$ gauge fields $A_\mu^I$ and
the metric $g_{\mu\nu}$. Besides this the theory contains
several non-dynamical fields. 
For the black hole solution we shall study, many of these fields
can be set to zero using certain symmetries of the action and
constraints. The relevant field which takes non-zero value near
the horizon  is a complex 
anti-self-dual antisymmetric
tensor field $T^-_{\mu\nu}$.
The lagrangian density $\LL$ of the theory involving these fields is
determined completely in terms of a single holomorphic function
$F(\vec X, \wh A)$  of the scalars $X^I$ and an auxiliary variable
$\wh A$, satisfying
\be \label{eprepo}
F(\lambda \vec X, \lambda^2 \wh A) = \lambda^2 \, 
F(\vec X, \wh A)\, .
\ee
The expression for $\LL$ in terms of this function $F$
has been reviewed in \cite{0007195}; 
for brevity we shall not
reproduce it here. 

In order to facilitate comparison with the results obtained by other
approaches, {\it e.g.} in \cite{0007195}, we shall use a different
normalization convention for the charges than the one used so far.
We introduce charges $\wt q_I$, $\wt p^I$ related to the charges
$q_I$, $p^I$ of the earlier convention via the relations:
\be \label{econv1}
\wt q_I = - 2 \, q_I, \qquad \wt p^I = {p^I\over 4\pi}\, .
\ee
With this normalization convention
a general extremal black hole in this theory has a near horizon geometry
of the form:
\ben \label{e1ns}
&& ds^2 = v_1 ( - r^2 dt^2 + dr^2 / r^2) + v_2 (d\theta^2
+ \sin^2\theta d\phi^2) \nonumber \\
&& F^I_{rt} = e_I, \quad F^I_{\theta\phi}=  \wt p^I
\, \sin\theta, \quad 
X^I = x^I, \quad
T^-_{rt}= v_1\, w \, .
\een
Using the known Lagrangian density $\LL$, we can calculate 
the entropy
function associated with this black hole:
\be \label{e8ns}
\EE(v_1, v_2, w, \vec x,
\vec e, \vec {\wt q}, \vec {\wt p}) = 2\pi
\left( -{1\over 2}\,
\wt q_I e^I - \int d\theta d\phi \, 
\sqrt{-\det g} \, \LL\right)\, .
\ee
The result is\cite{0603149}:
\ben \label{e9ns}
\EE &=& -\pi \wt q_I e^I - \pi \, g(v_1, v_2, w, \vec x, 
  \vec e, \vec {\wt p}) \nonumber \\
g(v_1, v_2, w, \vec x, 
  \vec e, \vec {\wt p}) &=&  
  v_1 v_2 \Bigg[ i (v_1^{-1} - v_2^{-1})
(x^I \bF_I - \bx^I F_I) \nonumber \\
&& -\Big\{
{i\over 4} v_1^{-2} F_{IJ}
(e^I - i v_1 v_2^{-1} \wt p^I - {1\over 2} \bx^I v_1 w) 
(e^J - i v_1 v_2^{-1} \wt p^J - {1\over 2} \bx^J v_1 w)
 + h.c. \Big\} 
 \nonumber \\
 &&  -\Big\{ {i\over 4} v_1^{-1} w \bF_I (e^I - 
 i v_1 v_2^{-1} \wt p^I - {1\over 2} \bx^I v_1 w) + h.c.\Big\}
 \nonumber \\
 && +\Big\{{i\over 8} \bw^2 F + h.c.\Big\}
 + 8 \, i \,
 \bw w\Big( - v_1^{-1} - v_2^{-1} + {1\over 8} \bw w
 \Big)
 \Big(F_{\wh A} - \bF_{\wh A} \Big)  \nonumber \\
 && + 64\, i \, (v_1^{-1} - v_2^{-1})^2
  \Big(F_{\wh A} - \bF_{\wh A} \Big) \Bigg]_{\wh A=-4w^2}
   \, ,
  \nonumber \\
\een
where
\be \label{e4ns}
F_I = {\p F\over \p x^I}, \quad F_{\wh A} = {\p F\over \p
\wh A}, \quad F_{IJ} ={\p^2 F\over \p x^I \p x^J},
\quad F_{\wh A I} = {\p^2 F\over \p x^I \p \wh A},
\quad F_{\wh A \wh A} =
{\p^2 F\over \p \wh A^2}\, ,
\ee
and bar denotes complex conjugation.
This entropy function has a scale invariance
\be \label{e10ns}
x^I \to \lambda x^I, \quad v_i \to \lambda^{-1}\bar\lambda^{-1}
v_i, \quad e^I \to e^I. \quad w\to \lambda w, \quad
\wt q_I\to \wt q_I, \quad \wt p^I\to \wt p^I\, .
\ee
This descends from the invariance of the lagrangian density
of $\NN=2$ supergravity theories
under local scale transformation,
and is usually eliminated by 
using some gauge fixing
condition. We shall however find it convenient to work 
with the gauge
invariant equations of motion obtained by extremizing \refb{e9ns}
with respect to $v_1$, $v_2$, $w$, $\vec x$ and $\vec e$. 

It can be easily seen that the  entropy function is extremized
if we set 
\be \label{e11bns} 
v_1 = v_2 = {16\over \bw w} \, ,
\ee
\be \label{e11ns}
e^I - i v_1 v_2^{-1} \wt p^I - {1\over 2} \bx^I v_1 w = 0
\ee
\be \label{e11ans}
(\bw^{-1}\bF_I - w^{-1} F_I) = -{i\over 4} \, \wt q_I \, .
\ee
Taking the real and imaginary parts of eq.\refb{e11} gives
\be \label{e11cns}
e^I = 4 (\bw^{-1}\bx^I + w^{-1} x^I) \, ,
\ee
and
\be \label{e11dns}
(\bw^{-1}\bx^I - w^{-1} x^I) = - {1\over 4} \, i\, \wt p^I\, .
\ee 
The black hole entropy computed by evaluating the entropy 
function in this background is given by
\be \label{e12ns}
S_{BH} = 2\pi\left[
-{1\over 2}\, \vec {\wt q} \cdot \vec e -16 \, i \, (w^{-2} F - 
\bw^{-2} \bF) \right]
\, .
\ee
If we choose $w$=constant gauge
(which corresponds to $\wh A$=$-4 w^2$=constant), 
then eqs.\refb{e11bns}-\refb{e11dns}
describe the usual supersymmetric
attractor equations for the near horizon
geometry of extremal black holes, and \refb{e12ns}
gives the expression for the
entropy of these black holes as written down in 
\cite{9801081,9812082,9904005,9906094,
9910179,0007195,0009234,0012232,0405146}. For example
\refb{e12ns} shows that in the gauge  $w=$real constant,
the Legendre transform of the black hole entropy
with respect
to the electric charges $\wt q_I$ is proportional to the
imaginary part of the prepotential $F$. Furthermore
eqs.\refb{e11bns}, \refb{e11ns} shows that the argument
$x^I$ of the prepotential is proportional to $e^I+i\wt
p^I$, \i.e. its
real part is the variable conjugate to the electric charge $\wt q_I$ and
its imaginary part is the magnetic charge $\wt p^I$. These 
observations were originally made in \cite{0405146}.

The attractor equations \refb{e11bns}-\refb{e11dns}
provide sufficient but not
necessary conditions for extremizing the entropy function.
We can have other near horizon configurations
which extremize the entropy function but do not satisfy
eqs.\refb{e11bns}-\refb{e11dns}. These will typically describe
non-supersymmetric extremal black holes.

\subsubsection{S-T-U model} \label{stu}

A special $\NN=2$ supergravity theory which will be of
interest to us is a model with four complex scalars
$X^0,\ldots X^3$, described
by the prepotential 
\be \label{e13ns}
F(X^0, X^1, X^2, X^3, \wh A) = - {X^1  X^2 X^3\over X^0}
- {\wh\alpha\over 64} \wh A \, {X^1\over  X^0}\, ,
\ee
where $\wh\alpha$ is a constant. The scalar fields take value
in the range
\be \label{en=2range}
Im\left({X^a\over X^0}\right) >0 \quad \hbox{for} \quad
a=1,2,3\, .
\ee
For $\wh\alpha=0$ this theory describes a subsector of the $\NN=4$
supergravity theory with two derivative terms as
described in eqs.\refb{enewm}-\refb{echsq}.
In this case the dilaton-axion pair
$(S,a)$ appearing in \refb{eag3} correspond respectively 
to the imaginary and real
parts of $X^1/X^0$:
\be \label{edilrel}
{X^1\over X^0} = a + i S\, .
\ee
Various components of the $4\times 4$ matrix valued field $M$
can be identified with various combinations of the four real fields
associated with $X^2/X^0$ and $X^3/X^0$. In particular
in terms of the components $\wh G_{mn}$ and $\wh B_{mn}$
($4\le m,n\le 5$) appearing in \refb{etenfour} we have
\be\label{exmrel}
{X^2\over X^0} = \wh B_{45} + i\sqrt{\det \wh G}, \qquad
{X^3\over X^0} = (\wh G_{45} + i\wh G_{44}) / \sqrt{\det\wh G}
\, .
\ee
These relations take
simple form if the fields $X^2/X^0$ and $X^3/X^0$ are purely
imaginary. In this case the corresponding matrix $M$ is diagonal.
If we parametrize $X^2/X^0$ and $X^3/X^0$ as
\be \label{expar}
{X^2\over X^0} =  i R \wh R, \qquad {X^3\over X^0}=i {\wh R\over R}
\ee
then
\be \label{emvalue}
M = \pmatrix{\wh R^{-2} & 0 & 0 & 0\cr 0 & R^{-2} & 0 & 0\cr
0 & 0 & \wh R^2 & 0 \cr 0 & 0 & 0 & R^2}\, .
\ee
As can be seen from \refb{exmrel} and \refb{expar},
in this case $\wh R$ and $R$ have 
the interpretation of the radii of
$\wh S^1$ and $S^1/\ZZZ_N$ measured in the string metric.
Finally the four sets of gauge
field strengths $F^{(i)}_{\mu\nu}$ are related to the four sets of
gauge field strengths $\FF^I_{\mu\nu}$ of the $\NN=2$ supergravity
theory under study via a complicated set of duality transformations. 
The relations between the gauge fields
are best summarized by relating the electric and magnetic
charges $\wt q_I$, $\wt p^I$ of the $\NN=2$ supergravity theory with
the charges $Q_i$ and $P_i$ appearing in \S 
\ref{s2.2}:\footnote{This differs from the identification made
in \cite{0603149} by a transformation $(Q_1,Q_2,Q_3,Q_4)
\to (Q_4,Q_3,Q_2,Q_1)$, $(P_1,P_2,P_3,P_4)\to
(P_4,P_3,P_2,P_1)$.}
\ben \label{e15ns}
&& Q_4 =   \wt q_2, \quad Q_3 = - \wt p^1, \quad Q_2 =  \wt q_3, 
\quad Q_1 =  \wt q_0\, , \nonumber \\
&& P_4 = \wt p^3, \quad P_3 = \wt p^0, \quad P_2 = \wt p^2, \quad
P_1 =   \wt q_1\, .
\een
After eliminating the electric field variables $e^I$ from \refb{e9ns}
using the $\p\EE /\p e^I=0$ equations and fixing an appropriate
`gauge' for the symmetry \refb{e10ns}, the entropy function
\refb{e9ns} of the S-T-U model
reduces to the one given in \refb{eag8b}.

We have seen in \S\ref{s2.2} that the 
T-duality invariant combination
of the charges are
\be\label{edefagain}
Q^2 = 2(Q_4 Q_2 + Q_3 Q_1), \quad P^2=2(P_4 P_2 + P_3 P_1),
\quad Q\cdot P = (Q_4 P_2 + Q_3 P_1 + Q_2 P_4 + Q_1 P_3)
\, .
\ee
One can explicitly verify that for 
\be \label{erange}
P^2Q^2>(Q\cdot P)^2, \quad P_2, P_4>0, \quad Q_1,Q_3<0, 
\quad P_3=0
\ee
the supersymmetric
attractor equations \refb{e11bns}-\refb{e11dns} can be solved
by setting\cite{9812082,0603149}\footnote{Here 
$w=1$ is gauge condition.}
\ben \label{ew1nspre} &&
x^0 = -{1\over 8} {Q_3 P^2\over 
\sqrt{
 P^2 Q^2 - (P\cdot Q)^2}} \nonumber \\
&& {x^1\over x^0} = -{P\cdot Q\over P^2} + i {\sqrt{P^2 Q^2 -
(P\cdot Q)^2} \over P^2 } \nonumber \\
&&{x^2\over x^0} = -{1\over 2 Q_3 P_4} (Q_3 P_1 + Q_4 P_2 - P_4
Q_2) - i {P_2\over Q_3}{\sqrt{P^2
Q^2 - (P\cdot Q)^2} \over P^2 } \nonumber \\
&&{x^3\over x^0} = -{1\over 2 Q_3 P_2} (Q_3 P_1 - Q_4 P_2 + P_4
Q_2) - i {P_4\over Q_3}{\sqrt{P^2 Q^2 - (P\cdot Q)^2} \over P^2
} \nonumber \\
&& v_1 = v_2 =16, \qquad e^I = 8 \, Re(x^I) \quad \hbox{for} \quad
0\le I\le 3, \qquad w=1\, ,
 \een
Evaluating
the entropy function at the solution and rewriting the result
in terms of the duality invariant combinations $P^2$,
$Q^2$ and $Q\cdot P$, we get back the result
\be \label{en2res}
\EE = \pi \sqrt{Q^2 P^2 - (Q\cdot P)^2}\, ,
\ee
in agreement with \refb{eag15}.
The parameters $v_1$, $v_2$ given in \refb{ew1nspre}
do not agree with the ones given in \refb{es10} even though
the two theories are supposed to be equivalent.
This can be traced to the fact that in the $w=1$ gauge being used
here, the metric in the two descriptions
are related by a field dependent scale transformation.

Finally note that although the expression \refb{en2res} for the
entropy was derived for a special choice of the charges ($P_3=0$),
T-duality invariance guarantees that \refb{en2res} continues to
hold even when we deform $P_3$ away from 0.

We shall now consider the effect of switching on the 
constant $\wh\alpha$. In the $\NN=2$ supergravity theory this
gives rise to various higher derivative terms in the action. Of
them is a term proportional to $S$ times the square of the Riemann
tensor, and the coefficient is identical to the one appearing
in \refb{ecorg1} for $\phi(a,S)$ given in \refb{elarge}. However
the two actions are not physically equivalent \i.e.\ they cannot be
related by a field redefinition even at the level of
four derivative terms. If we solve the attractor
equations \refb{e11bns}-\refb{e11dns} and calculate the entropy
by evaluating the entropy function at the solution, we 
get
\ben \label{ew1ns} &&
x^0 = -{1\over 8} Q_3 \sqrt{P^2 (P^2+8 \wh\alpha) \over
 P^2 Q^2 - (P\cdot Q)^2} \nonumber \\
&& {x^1\over x^0} = -{P\cdot Q\over P^2} + i \sqrt{P^2 Q^2 -
(P\cdot Q)^2 \over P^2 (P^2+8 \wh\alpha)} \nonumber \\
&&{x^2\over x^0} = -{1\over 2 Q_3 P_4} (Q_3 P_1 + Q_4 P_2 - P_4
Q_2) - i {P_2\over Q_3}\sqrt{P^2
Q^2 - (P\cdot Q)^2 \over P^2 (P^2+8 \wh\alpha)} \nonumber \\
&&{x^3\over x^0} = -{1\over 2 Q_3 P_2} (Q_3 P_1 - Q_4 P_2 + P_4
Q_2) - i {P_4\over Q_3}\sqrt{P^2 Q^2 - (P\cdot Q)^2 \over P^2
(P^2+8 \wh\alpha)} \nonumber \\
&& v_1 = v_2 =16, \qquad e^I = 8 \, Re(x^I) \quad \hbox{for} \quad
0\le I\le 3, \qquad w=1\, ,
 \een
 \be \label{ew2ns}
 S_{BH} = \pi \sqrt{P^2 Q^2 - (P\cdot Q)^2} \sqrt{1 + 
 {8 \wh\alpha\over
 P^2}}\, .
 \ee
 Surprisingly, the result for the entropy agrees with the one given in 
\refb{eag17}. 

For 
a special class of black holes for which   
\be \label{echargepreagain}
Q = \pmatrix{\wh n\cr   0 \cr \wh w\cr 0}, \qquad
P = \pmatrix{0 \cr W'\cr 0 \cr N'}\, ,
\ee
\be \label{enewrange}
N',W'>0, \qquad \wh n, \wh w < 0\, ,
\ee
the solution \refb{ew1ns} gives
\be\label{exvals}
{x^1\over x^0} = i \sqrt{\wh n\wh w\over N'W'+4\wh\alpha}, 
\qquad
{x^2\over x^0} = -i {W'\over \wh w}
\sqrt{\wh n\wh w\over N'W'+4\wh\alpha}, \qquad 
{x^3\over x^0} =
-i {N'\over \wh w}\sqrt{\wh n\wh w\over N'W'+4\wh\alpha}\, .
\ee
Using \refb{edilrel} and \refb{expar}
we get
\be \label{efinx}
\wh R = \sqrt{\wh n\over \wh w}\sqrt{N'W'\over N'W'+4\wh\alpha}, 
\qquad 
R = \sqrt{W'\over N'}\, , \qquad
u_S = \sqrt{\wh n\wh w\over N'W'+4\wh\alpha}\, .
\ee
Finally \refb{ew2ns} gives
\be\label{ebhfin}
S_{BH} = 2\pi\sqrt{\wh n \wh w (N'W'+4\wh\alpha)}\, .
\ee
For nonvanishing $\wh\alpha$ the
solution \refb{efinx} differs from the solution
given in \refb{eag16ag}, \refb{erv} for 
$\NN=4$ supergravity theory with
Gauss-Bonnet term,  but the formul\ae\ 
\refb{eag17spec} and \refb{ebhfin} for the entropy continue
to agree\cite{0506251,0508042}.

We would like to note that neither the Gauss-Bonnet correction
given in \refb{ecorg1}, nor the correction to the prepotential 
proportional to $\wh\alpha$ given in
\refb{e13ns}, describes the complete set of four derivative corrections
in tree level string theory. Nevertheless we shall argue in
\S \ref{sads3} that the result \refb{eag17spec} or
\refb{ebhfin} does not change
after inclusion of additional terms in the effective action at string tree
level, \i.e.\ for large $S$. Translated to a condition on the charges, this
means that the result \refb{eag17spec} or \refb{ebhfin}
become a good approximation in the limit
when the electric charges are much larger than the magnetic charges.

Given this surprising agreement between the $\NN=4$ and $\NN=2$
results one might ask whether it is possible to also reproduce the
more general results \refb{es11}, \refb{es12} from the
$\NN=2$ viewpoint. This cannot be done completely
rigorously since  due to the presence of holomorphic
anomaly term proportional to $\ln S$ in the coefficient 
\refb{eh10bint} of the
curvature squared term there is no known generalization of this
term into an $\NN=2$ supersymmentric lagrangian.
Nevertheless the form of these corrections were guessed in
\cite{9906094} by first examining the contribution from the rest of the
terms in $\phi(a,S)$ and then requiring the result to be duality
invariant.

\subsubsection{Non-supersymmetric extremal black holes}
\label{snsn=2}

We can also construct non-supersymmetric extremal black holes
in this theory by directly solving the equations corresponding to
extremization of $\EE$ rather than solving the attractor equations
\refb{e11bns}-\refb{e11dns}. Since for $\wh \alpha=0$ the theory
is equivalent to the $\NN=4$ supergravity theory described in
\S\ref{sugra}, a convenient starting point for constructing
non-supersymmetric solution is to start with the
non-supersymmetric solution described in \S\ref{snsn=4} for
$\wh\alpha=0$. Thus we consider the charge vectors of the form
given in \refb{echargepreagain} with $\wh n$, $\wh w$,
$N'$, $W'$ in the range
\be \label{erg5}
N',W', \wh n >0, \quad \wh w<0\, .
\ee
The solution and the entropy for $\wh\alpha=0$ are obtained
simply by replacing $\wh n\to -\wh n$ and setting $\wh\alpha=0$
in eqs.\refb{eag16ag}-\refb{eag17spec}. 
Using the relations \refb{edilrel}, \refb{expar} and 
\refb{emvalue} between the scalar
fields in the $\NN=4$ description and the $\NN=2$ description,
we get
\be \label{ennss1}
{x^1\over x^0} = u_a + i u_S =i\sqrt{|\wh n\wh w|\over N'W'},
\quad {x^2\over x^0} = i R\wh R = i\sqrt{\left|{\wh n W'\over
\wh w N'}\right|}, \quad {x^3\over x^0}
= i{\wh R\over R} = i \sqrt{\left| {\wh n N'\over 
\wh w W'}\right|}\, ,
\ee
and
\be \label{ennnss2}
S_{BH} = 2\pi \sqrt{|\wh n\wh w| N'W'}\, .
\ee
Note that we cannot use \refb{eag16ag} to determine the
values of $v_1$ and $v_2$ in the $\NN=2$ description since
they are gauge dependent
in this description.  Instead
 $v_1$, $v_2$ and $w$  are
found by directly
solving the extremization equations of the $\NN=2$ theory
after fixing an appropriate gauge.
Beginning with this leading order solution, we can
solve the extremization equations for
$\EE$ given in \refb{e9ns} in a power series expansion in
$\wh\alpha$.  We shall not describe the details of the calculation
but only give the final result for
the entropy calculated by this method\cite{0603149}:
\be \label{ensns2}
S_{BH}^{ns} = 2\pi \sqrt{|\wh n\wh w|N'W'} 
(1 + 80 u  - 3712 u^2  - 243712 u^3 
- 18325504 u^4  - 9538502656 u^5 +\OO(u^6))\, ,
\ee
where
\be \label{edefu}
u = {\wh\alpha\over 128 N'W'}\, .
\ee

\subsectiono{Small black holes} \label{s2.3}

In this section we shall focus on a special subset of 
the black holes described in \S \ref{s2.2} and 
\S \ref{stu}, -- those
with vanishing magnetic charge $\vec P$. 
These black holes, being purely electrically charged, have the same
quantum numbers as the elementary string excitations in the
second description of these theories.
In particular the extremal supersymmetric black holes should
correspond to BPS states\cite{DH,DGHR} in the spectrum
of elementary string. 
A simple class of such states, corresponding to $\vec P=0$ 
and $\vec Q$ of the form given in \refb{echargepreagain} 
with $\wh n,\wh w<0$,
represent BPS elementary string states carrying $\wh n$ units
of momentum and $-\wh w$ units of winding along $\wh S^1$.
In the light-cone gauge Green-Schwarz formulation of the fundamental
string world-sheet theory, we denote by $L_0'$ and $\bar L_0'$ the
vacuum and oscillator contribution to the zero modes $L_0$
and $\bar L_0$ of the left- and
the right-handed Virasoro generators. On the other hand the contributions 
to $L_0$ and $\bar L_0$ from the momentum and winding 
along $\wh 
S^1$  are given by ${1/4} (\wh n 
\sqrt{\alpha'} / \wh R \mp \wh w\wh R/\sqrt{\alpha'})^2$, where $\wh R$ is 
the radius of $\wh S^1$. Thus the 
contribution to $L_0-\bar L_0$ from these charges is given by $-\wh n\wh w$,
and the left-right level matching condition tells us that these states
must have $L_0'-\bar L_0'=\wh n\wh w$.  
On the other hand since space-time supersymmetry originates
in the right-moving sector of the world-sheet theory,
the BPS condition tells us that $\bar L_0'$
must vanish. Thus we have $L_0'=\wh n\wh w$.
For a more general electric charge vector $\vec Q$,
by matching the quantum numbers of the black
hole with that of the elementary string and imposing the
left-right level matching condition one can easily verify
that a supersymmetric black hole with charge $\vec Q$ would correspond
to elementary string excitations where the right-moving oscillators
are in the  $\bar L_0'=0$ state
and the left-moving oscillators are excited to a level 
$L_0'=Q^2/2$\cite{9504147,9506200,9712150}. The 
degeneracy $d(Q)$
of these states for large $Q^2$
can be computed using the Cardy formula \refb{e0.4}.
Since for the heterotic string theory $c_L=24$ and for the type II
string theory $c_L=12$, we have
\be \label{estatsmall}
S_{stat}(Q)\equiv \ln d(Q) \simeq 4\pi \, \sqrt{Q^2/2}\, 
\ee
for heterotic string theory, and
\be \label{estatsmallii}
S_{stat}(Q)\equiv \ln d(Q) \simeq 2\sqrt 2\, \pi \sqrt{Q^2/2}\, 
\ee
for type II string theory.
We would like to test if the entropy of the corresponding black
hole solution, computed by extremizing the entropy funcion,
agrees with this statistical
entropy. This is the problem we shall now address.
The analysis will have two parts. First we shall use symmetry
principles to argue that the black hole entropy, if non-zero,
will have the same dependence on the charges as in 
\refb{estatsmall}, 
\refb{estatsmallii} \cite{9504147,9506200}. 
Then we shall describe
computation of the overall coefficient\cite{0409148}.

Since the black hole carries zero magnetic charges,
the near horizon background of the black hole is of the form
\ben \label{ehorsmall}
ds^2  = v_1\left(-r^2 dt^2+{dr^2\over 
r^2}\right)   
+
v_2 (d\theta^2+\sin^2\theta\, d\phi^2) \, , \nonumber \\ 
S =u_S, \qquad a=u_a , \qquad  M_{ij} = u_{Mij} \nonumber \\
  F^{(i)}_{rt} = e_i, \quad  
 F^{(i)}_{\theta\phi}=0 \, ,
 \een
and in the leading
supergravity approximation the function
$f$ given in \refb{eag5} becomes:
\ben \label{eag5small}
&& f(u_S, u_a, u_M, \vec v, \vec e)
\equiv \int d\theta d\phi \, \sqrt{-\det G} \, \LL
\nonumber \\ 
&=& 
{1\over 8} \, v_1 \, v_2
\, u_S  \left[ -{2\over v_1} +{2\over v_2} + 
 {2\over v_1^2} e_i (Lu_M L)_{ij} e_j 
 \right] 
 \, . 
\een
It is easy to see that the entropy function computed from this
function $f$ has no non-trivial extremum. Indeed, if we set $\vec P=0$
in \refb{eag14} we get singular solution, and the entropy given in
\refb{eag15} vanishes.

It is in principle possible that once higher derivative and/or string loop
corrections to the supergravity action are included, we might get a
non-singular  solution. However since the leading solution is singular,
these corrections are no longer small, and we may have to include
all possible corrections to the effective action. The goal of this section
will be to analyze the effect of these corrections in detail and 
see if we can extract
some useful information about the entropy of such black holes.

In order to get some insight into the problem we 
note from \refb{eag14} that if we naively take $\vec P\to 0$ in these
equations then $u_S\to\infty$ whereas $v_i\to 0$. Since $u_S$ denotes
the inverse string coupling square this implies that the string coupling
constant becomes small at the horizon. Thus we could expect that the
string loop corrections are not important. On the other hand since the
curvatures of $AdS_2$ and $S^2$ measured in the string metric are
inversely proportional to $v_1$ and $v_2$ respectively, they become
large in this limit and we expect that
higher derivative corrections to the action become important near the
horizon of the small black hole. Thus we should in principle include
all tree level higher derivative corrections. 
For the time being we shall proceed
with this ansatz and study the effect of tree level higher derivative
corrections to the black hole solution. Later we shall verify that the
solution obtained with this assumption is self-consistent, namely that
the effect of string loop corrections on these solutions are indeed small.

The effective action of the tree level string theory, with all the
Ramond-Ramond (RR) fields set to zero,  has two
important properties which will be important for our analysis:
\begin{enumerate}
\item The full tree level effective action of  string theory is
invariant under a continuous $SO(6,r-6)$ T-duality symmetry. As a result
the complete entropy function 
\be \label{entropysmall}
\EE = 2\pi \left[ {1\over 2} e_i Q_i - f(u_S, u_a, u_M, \vec v, \vec e)
\right]\, 
\ee
computed with the tree level effective
action of  string theory
will be invariant under the transformation
\be \label{eag9asmall}
e_i\to \Omega_{ij} e_j, \qquad u_M\to
\Omega u_M \Omega^T, \qquad
Q_i\to (\Omega^T)^{-1}_{ij} Q_j \, .
\ee
\item The tree level effective action picks up a constant multiplicative
factor $\lambda$ under $S\to \lambda S$, $a\to \lambda a$. As
a result
\be \label{efscaling}
f(\lambda u_S, \lambda u_a, u_M, \vec v, \vec e)
= \lambda \, f(u_S, u_a, u_M, \vec v, \vec e)\, ,
\ee
and
\be \label{eescaling}
\EE \to \lambda \EE \quad \hbox{under} \quad 
Q_i\to \lambda Q_i, \quad u_S\to \lambda u_S, \quad u_a\to \lambda u_a
\, .
\ee
\end{enumerate} 
These two properties imply that the black hole entropy 
$S_{BH}(Q)$, obtained by extremizing the entropy function
with respect to the
variables $e_i$, $v_1$, $v_2$, $u_{Mij}$, $u_S$ and $u_a$, 
has the following two properties:
\begin{enumerate}
\item $S_{BH}$ depends on $\vec Q$ only through 
the duality invariant combination
$Q^2\equiv Q_i L_{ij} Q_j$.
\item $S_{BH}$ has the scaling property
\be \label{bhscaling}
S_{BH} \to \lambda \, S_{BH} \quad \hbox{under}
\quad Q_i\to \lambda Q_i\, .
\ee
\end{enumerate}
This gives
\be \label{efinsmall}
S_{BH} = C \, \sqrt{Q^2}\, ,
\ee
for some constant $C$. This agrees with the expression for the statistical
entropy given in \refb{estatsmall}, \refb{estatsmallii}
 up to an overall normalization
constant $C$ which cannot be determined from this simple scaling
argument. 

Eq.\refb{efinsmall} can in fact be derived even
without assuming an $AdS_2\times S^{D-2}$ 
near horizon geometry\cite{9504147,9506200}. The main additional 
complication in this analysis is that in absence of a near horizon $AdS_2$ 
geometry and the associated attractor mechanism we can no longer assume 
that the entropy is independent of the asymptotic moduli; we have to prove 
this by explicitly examining the supergravity solution describing the near 
horizon geometry of small black holes. This was carried out in 
\cite{9504147, 9506200,0505122}.

The same scaling argument also tells us that at the extremum
\be \label{eparscaling}
v_1=c_1, \quad v_2=c_2, \quad e_i = c_3\, L_{ij} Q_j/\sqrt{Q^2}, 
\quad
u_S=c_4 \sqrt{Q^2}\, ,
\ee
where $c_i$'s are numerical constants independent of $\vec Q$. This
shows that $u_S$ is large for large $Q^2$ and hence string loop
corrections are indeed small near the horizon. On the other hand
since $v_1$, $v_2$ and $e_i$
are of order unity, higher derivative corrections
are important, and we must include all tree level higher derivative
corrections to the action to get a reliable estimate of the constant $C$
appearing in \refb{efinsmall}.
Nevertheless it is instructive to see what value of $C$ we get just by
including the Gauss-Bonnet term in the action. For this we
set $Q\cdot P=0$ and then take the $P^2\to 0$ limit of
\refb{eag17} (or equivalently \refb{ew2ns} if we want to do this
computation in $\NN=2$ supersymmetric S-T-U model with the
prepotential given in \refb{e13ns})
which was derived for
the case when electric charges are 
large compared to magnetic charges. 
This gives\cite{0409148,0410076,
0411255,0411272,0501014}
\be \label{eentsmall}
S_{BH} = 4\pi \sqrt{\wh\alpha\, Q^2/2}\, .
\ee

Using \refb{ealphafin}
we see that
for heterotic string compactification, the constant $\wh\alpha=1$. Thus
the 
result for the entropy calculated with the Gauss-Bonnet
term agrees exactly with the statistical entropy given in
\refb{estatsmall}. On the other hand
for type II string theories the constant $\wh\alpha$ vanishes, showing
that the entropy vanishes to this order. This is in disagreement with the
statistical entropy \refb{estatsmallii}.
We should keep in mind however that at this point there is no reason to
expect that including just the Gauss-Bonnet correction will give the
correct value of the constant $C$. Hence at this stage
neither the agreement for
the heterotic string theory, nor the disagreement for the type II string
theory should be taken seriously. We shall return to a more detailed
discussion on this point in \S \ref{smallre}.

The scaling analysis carried out here can 
be easily generalized to higher
dimensional small black holes\cite{9506200,0505122}. 
It can also be generalized to
include elementary string states 
carrying angular momentum\cite{0611166} assuming that the
near horizon geometry of such an object 
has the structure of a black ring
with $AdS_2\times S^1\times S^{d-3}$ near horizon 
geometry\cite{0506215,0511120}.

There is one subtle point about the small black hole that requires
special mention. According to the analysis described in this
section, the near horizon geometry of a small black hole
is given by $AdS_2\times S^2$. As seen from
\refb{ehorsmall}, there is an electric flux through
$AdS_2$, but there is no flux through $S^2$. This creates
a puzzle. If the background given in \refb{ehorsmall} is
what couples to the sigma model describing string
propagation in this background, then the sigma model will
be a direct sum of two conformal field theories, -- one associated
with $AdS_2$ together with the electric flux through it, and the
other associated with $S^2$ without any flux. However it is
well known that the sigma model associated with $S^2$ does
not give rise to a conformal field theory; instead under the 
renormalization group flow the radius of $S^2$ goes to zero
in the infrared. This would seem to contradict the result that
the near horizon geometry of the black hole is given by
\refb{ehorsmall}. 
The only consistent resolution to this puzzle seems to be that the
metric and other field variables in terms of which the solution
takes the form \refb{ehorsmall} are not the ones which couple
directly to the world-sheet sigma model.\footnote{See
refs.\cite{0611062,0707.3818,0707.4303,0708.0016} for further insight into 
this issue.}
Instead the fields which
couple to the sigma model must be related to the ones 
appearing in \refb{ehorsmall} by an appropriate field redefinition
which becomes singular when evaluated on the solution of the
entropy function extremization equations. 
For example the sigma model metric could be
related to the one in \refb{ehorsmall} by multiplication by
a T-duality invariant
function of the gauge field strengths and the matrix $M$, such that
this factor vanishes at the solution. In that case although the
metric appearing in \refb{ehorsmall} is finite at the solution, the
metric that couples to the sigma model would correspond to a sphere
of zero radius in accordance with known results.

An interesting  problem is to construct the conformal field theory
describing the near horizon geometry of a small black hole. As
should be clear from the above discussion, we do not expect
this to be a sum of two decoupled conformal field theories; instead
it must
be described as a single conformal field theory involving the four
non-compact space time coordinates as well as the internal coordinates
responsible for the electric charge of the black hole.
Indeed, in order to argue that the entropy of the black hole has the form
given in \refb{efinsmall}, it is not necessary to assume a 
near horizon geometry of the form \refb{ehorsmall}; it is enough
to assume the existence of a non-singular conformal field theory
associated with the near horizon geometry 
of the black hole\cite{9504147,9506200,9712150}. Furthermore
this argument can be generalized to small black holes in higher
dimensions as well.
Various proposals for this conformal field theory have recently been made
in \cite{0611062,0707.3818,0707.4303,0708.0016}.

Finally, note that the above analysis holds 
only for the Neveu-Schwarz
formulation of the conformal field theory. If instead we use the 
light-cone gauge Green-Schwarz formulation of the theory then
the world-sheet fermions transform in a spinor representation of
the tangent space group of the target manifold, and as a result the
conformal field theories associated with the $S^2$ part does not
decouple from the rest of the conformal field theory. Thus in this
case there is no argument showing that the $\sigma$-model
target space manifold cannot be $AdS_2\times S^2$ (or, as we shall
discuss in \S\ref{smallre},  $AdS_3
\times S^2$).

\subsectiono{Extremal BTZ black holes with gauge and gravitational
Chern-Simons terms} \label{s2.4}

BTZ solution describes a rotating black hole in three 
dimensional theory of gravity with negative cosmological 
constant\cite{9204099} and often
appears as a factor in the near horizon geometry of
higher dimensional black holes in string 
theory\cite{9711138,9712251,9901050}. 
For this
reason
it has provided us with a useful 
tool for relating black hole entropy to the degeneracy of 
microstates of the black hole, both in three dimensional
theories of gravity and also in string theory\cite{9712251,brown}. 
In this section we shall apply the entropy function method to compute
the entropy of a BTZ black hole in a theory of three dimensional
gravity coupled to a set of abelian gauge fields and neutral
scalar fields, with an arbitrary general coordinate
invariant and gauge invariant action.
In particular we shall include 
Lorentz Chern-Simons terms of the form
described in \refb{elcs} and gauge Chern-Simons terms. 
In the spirit of the discussion in
\S \ref{s1.3} this will be done by treating the angular
coordinate along which the brane rotates as a compact direction.
The analysis presented here will be a generalization
of the one given in \cite{0601228} where we considered the case
of a purely gravitational theory, and will borrow insights from
the analysis
of \cite{0608044,0701176}. 
Computation of the entropy of
BTZ black holes in a general higher derivative theory of gravity
without Chern-Simons terms has been carried out previously in
\cite{9909061}. Entropy of BTZ black holes has also been
analyzed using
the Euclidean action formalism in 
\cite{9804085,0506176,0508218,0509148}.

Let us consider a three dimensional theory of gravity with
metric $G_{MN}$ ($0\le M,N\le 2$), $U(1)$ gauge fields
$A_M^{(i)}$ $(1\le i\le n_1)$ and neutral scalar fields
$\{\phi_s\}$ $(1\le s\le n_2)$ and a general 
action of the form:\footnote{We could
add any number of scalar
fields without changing the final result since they must be frozen
to constant values in order to comply with the homogeneity of the
BTZ configuration.}
\be \label{e1btz}
S = \int d^3 x \sqrt{-\det G}\left[ \LL^{(3)}_0
+ \LL^{(3)}_1 + \LL^{(3)}_2\right]\, .
\ee
Here
$\LL^{(3)}_0$ denotes an arbitrary scalar function of the metric, 
scalar fields, gauge field strengths $F^{(i)}_{MN}=\p_M A^{(i)}_N
-\p_N A^{(i)}_M$, 
the 
Riemann tensor and covariant derivatives of these
quantities.  $\sqrt{-\det G}\, \LL^{(3)}_1$ denotes 
the gravitational Chern-Simons term:
\be \label{e1abtz}
\sqrt{-\det G}\, \LL^{(3)}_1 = K \, \Omega_3(\wh\Gamma) \, ,
\ee
where
$K$ is a 
constant, $\wh\Gamma$ is the Christoffel connection constructed out of 
the metric $G_{MN}$ and
\be \label{e2btz}
\Omega_3(\wh\Gamma) = \epsilon^{MNP} \left[{1\over 2}
\wh\Gamma^R_{MS} \p_N \wh\Gamma^S_{PR} + {1\over 3}
\wh\Gamma^R_{MS} \wh\Gamma^S_{NT} 
\wh\Gamma^T_{PR}\right]\, .
\ee
$\epsilon$ is the
totally anti-symmetric symbol with $\epsilon^{012}=1$.
Finally $\sqrt{-\det G} \, \LL_2^{(3)}$ denotes a gauge Chern-Simons
term of the form:
\be \label{egcs1}
\sqrt{-\det G} \, \LL_2^{(3)} = {1\over 2}\,
C_{ij} \, \epsilon^{MNP}\, 
A^{(i)}_M \, F^{(j)}_{NP}\, ,
\ee
for some constants $C_{ij}=C_{ji}$.

We shall consider field configurations where one of the coordinates (say 
$y\equiv x^2$) is compact with 
period $2\pi$ and the metric is independent 
of this compact direction. In this case we can define two 
dimensional fields through the relation:\footnote{The two dimensional
view of the three dimensional black holes was first discussed in
\cite{9304068}.}
\ben \label{e3btz}
G_{MN} dx^M dx^N &=& \phi \left[ g_{\mu\nu} 
dx^\mu dx^\nu + (dy + a_\mu 
dx^\mu)^2\right]\, , \qquad 0\le\mu,\nu \le 1\nonumber \\
A^{(i)}_M dx^M &=& \chi^{(i)} (dy + a_\mu 
dx^\mu) + a^{(i)}_\mu dx^\mu\, .
\een
Here $g_{\mu\nu}$ denotes a 
two dimensional metric, $a_\mu$ and $a^{(i)}_\mu$
denote two dimensional 
gauge  fields and $\phi$ and $\chi^{(i)}$ denote 
two dimensional scalar fields. In terms of these 
two dimensional fields 
\be \label{ebtbt1}
{1\over 2} F^{(i)}_{MN} dx^M \wedge dx^N =
d\chi^{(i)} \wedge (dy + a_\mu 
dx^\mu) +{1\over 2} \, \left( \chi^{(i)} f_{\mu\nu}
+ f^{(i)}_{\mu\nu}\right) dx^\mu\wedge dx^\nu\, ,
\ee
where
\be \label{ebtbt2}
f^{(i)}_{\mu\nu} = \p_\mu a^{(i)}_\nu - \p_\nu a^{(i)}_\mu\, ,
\ee
and
\be \label{e6btz}
f_{\mu\nu} = \p_\mu a_\nu - \p_\nu a_\mu\, .
\ee
The 
action takes the form:
\be \label{e4btz}
S = \int d^2 x \sqrt{-\det g}\left[ \LL^{(2)}_0 + \LL^{(2)}_1 
+ \LL^{(2)}_2\right]
\ee
where
\be \label{e5btz}
\sqrt{-\det g} \, \LL^{(2)}_0 = \int dy \sqrt{-\det G} \,
\LL^{(3)}_0
= 2\pi \sqrt{-\det G} \, \LL^{(3)}_0\, ,
\ee
\be \label{e5abtz}
\sqrt{-\det g} \, \LL^{(2)}_1 = K \, \pi \, \left[
 {1\over 2} R \varepsilon^{\mu\nu} f_{\mu\nu}
+{1\over 2} \varepsilon^{\mu\nu} f_{\mu\tau} f^{\tau\sigma} 
f_{\sigma\nu} \right]\, .
\ee
and
\be \label{egcs}
\sqrt{-\det g} \, \LL^{(2)}_2  = C_{ij}\, \varepsilon^{\mu\nu}\,
\left( \chi^{(i)} f^{(j)}_{\mu\nu} +{1\over 2}
\chi^{(i)} \, \chi^{(j)}\, f_{\mu\nu}\right)\, .
\ee
Here
$R$ is the scalar curvature of the two dimensional metric
$g_{\mu\nu}$ and 
$\varepsilon^{\mu\nu}$ is the totally antisymmetric symbol with
$\varepsilon^{01}=1$.
\refb{e5btz} is a straightforward dimensional reduction of the
$\LL_0^{(3)}$ term.
\refb{e5abtz}
comes from dimensional reduction of the gravitational
Chern-Simons term after
throwing away total derivative terms and was 
worked out in 
\cite{0305117}. 
\refb{egcs} comes from dimensional reduction of \refb{egcs1}
after throwing away total derivative terms.
\refb{e5abtz}, \refb{egcs} show that although the 
Chern-Simons terms cannot be expressed in a 
manifestly covariant form in three dimensions, they do reduce to
manifestly covariant expressions in two dimensions.

We shall define
a general extremal black hole in the two dimensional theory 
to 
be the one whose near horizon geometry is $AdS_2$ and for 
which the scalar 
fields $\phi$, $\{\phi_s\}$ and $\{\chi^{(i)}\}$ and 
the gauge field strengths $f_{\mu\nu}$ and
$f^{(i)}_{\mu\nu}$  are invariant under 
the $SO(2,1)$ isometry of the $AdS_2$ background. 
The most general near 
horizon background consistent with this requirement is
\ben \label{e8btz}
&& g_{\mu\nu} dx^\mu dx^\nu = v\left(-r^2 dt^2 + {dr^2\over r^2}\right), 
\qquad f_{rt} = e, \qquad \phi = u\, , \nonumber \\
&& \phi_s = u_s, \quad f^{(i)}_{rt} = e_i, \quad \chi^{(i)}=w_i\, ,
\een
where $v$, $e$, $u$, $\{u_s\}$, $\{e_i\}$ and $\{w_i\}$ 
are constants. This corresponds to a
three dimensional configuration of the form
\ben \label{especthree}
&& G_{MN} dx^M dx^N = v \, u\, (-r^2 dt^2 + {dr^2\over r^2})
+ u \, (dy + erdt)^2 \nonumber \\
&& A^{(i)}_M dx^M = w_i  \, (dy + erdt) + e_i r dt, \qquad \phi_s=u_s
\nonumber \\
&& {1\over 2} F^{(i)}_{MN} dx^M \wedge dx^N
= (e_i + w_i e) \, dr \wedge dt\, .
\een
Following the procedure described in \S\ref{s1} we define
\be \label{e9btz}
f(u, v, e, \vec u, \vec e, \vec w) 
= \sqrt{-\det g}\,  (\LL^{(2)}_0+\LL^{(2)}_1
+ \LL^{(2)}_2)\, ,
\ee
evaluated in the background \refb{e8btz}, and
\be \label{e10btz}
\EE(u, v, e, \vec u, \vec e, \vec w, q, \vec q) = 
2\pi (eq + e_i q_i - f(u, v, e, \vec u, \vec e, \vec w))\, .
\ee
The near horizon values of $u$, $v$ and $e$, $\{u_s\}$,
$\{e_i\}$ and $\{w_i\}$ for an extremal 
black hole with electric charges $q$, $\{q_i\}$  are 
obtained by extremizing the 
entropy 
function $\EE$ with respect to these variables. 
Furthermore, Wald's entropy 
for this black hole is given by the value of the function $\EE$ at this 
extremum. 

Using eqs.\refb{e5btz}, \refb{e5abtz}, \refb{egcs} 
and \refb{e8btz},
\refb{e9btz} we see that
for the theory considered here,
\be \label{e10abtz}
f(u, v, e, \vec u, \vec e, \vec w) = f_0(u,v,e,\vec u, \vec e +  e\vec w) 
+ \pi\, K\, (2\, e \,
v^{-1} - e^3\, v^{-2}) - 2 \, C_{ij}\, \left(w_i e_j +{1\over 2} e w_i
w_j\right)\, ,  
\ee
where
\be \label{e10bbtz}
f_0(u, v, e,\vec u, \vec e + e\vec w) 
= {2\pi}\, \sqrt{-\det G}\,  \LL^{(3)}_0 \, 
\ee
evaluated in the background \refb{especthree}.
Note that $f_0$ depends on $\vec e$ and $\vec w$ only through
the combination $\vec e + e\vec w$ since this is the combination
that enters the expression for $F^{(i)}_{MN}$ 
given in \refb{especthree}. Substituting \refb{e10abtz} into
\refb{e10btz} we get
\ben \label{ebtbtone}
\EE &=& 2\pi \Bigg[ eq + e_i q_i - 
f_0(u,v,e,\vec u, \vec e +  e\vec w) 
- \pi\, K\, (2\, e \,
v^{-1} - e^3\, v^{-2}) \nonumber \\
&& \qquad \qquad + 2 \, C_{ij}\, \left(w_i e_j +{1\over 2} e w_i
w_j\right) \Bigg]\, .
\een

We shall carry out the extremizaton of $\EE$ in stages. First we
shall eliminate $w_i$ and $e_i$ using their equations of
motion. Extremization of $\EE$ with respect to $w_i$ gives
\be \label{ebtbttwo}
-e {\p f_0\over \p e_i} + 2 \, C_{ij}\, (e_j + e w_j) = 0\, .
\ee
Now the terms in $f_0$ involving $\vec e+ e\, \vec w$ must involve
quadratic and higher powers of $e_i + e w_i$ since these 
come from terms in $\LL^{(3)}_0$ involving the gauge fields
$F^{(i)}_{MN}$. Thus $\p f_0/\p e_i$ vanishes for
$\vec e + e\, \vec w=0$, This shows that \refb{ebtbttwo} can
be solved by choosing
\be \label{ebtbtthree}
e_i + e\, w_i = 0\, .
\ee
Extremization of $\EE$ with respect to $e_i$ now gives
\be \label{ebtbtfour}
w_i = -{1\over 2} C^{-1}_{ij} \, q_j\, ,
\ee
assuming that $C$ is invertible as a matrix. Substituting
\refb{ebtbtthree} and \refb{ebtbtfour} into the expression
\refb{ebtbtone} for $\EE$ we now get
\be \label{ebtbtfive}
\EE = 2\pi \, \left[ e\left( q - {1\over 4}\, C^{-1}_{ij}
\, q_i q_j\right) - f_0 ( u, v, e, \vec u, \vec 0)
- \pi\, K\, (2\, e \,
v^{-1} - e^3\, v^{-2}) \right]\, .
\ee
Eq.\refb{ebtbtthree} together with \refb{especthree} tells us that
the three dimensional gauge field strengths
$F^{(i)}_{MN}$ vanish, although the gauge field components
$A^{(i)}_y$ have
non-zero constant values proportional to $w_i$.

We now turn to the extremization of $\EE$ with respect to $u$,
$v$, $e$ and $\vec u$.
It turns out that if  $v$ and $e$ satisfy the relation
\be \label{e20abtz}
v = e^2\, 
\ee
then the background \refb{especthree} 
actually describes a locally $AdS_3$
space-time. Since a local $AdS_3$ space has a higher degree of isometry
than a local $AdS_2$ space, setting $v=e^2$ is a consistent truncation
of the theory and hence we can extremize the entropy function within
this class of configurations.  This corresponds to a three dimensional
field configuration of the form:
\ben \label{ethrbtz}
&& G_{MN} dx^M dx^N = u \, e^2\, \left(-r^2 dt^2 
+ {dr^2\over r^2}\right) + u \, (dy + e\, r\, dt)^2 \nonumber \\
&& \phi_s = u_s, \qquad {1\over 2} F^{(i)}_{MN} dx^M\wedge
dx^N = 0\, .
\een
By making a coordinate change $y = e \, z$ we can see that the metric
depends only on the combination $u e^2$. Thus all the scalars
constructed out of the metric and the Riemann tensor must be a
function of this combination only. The entropy function will still
depend on $e$ and $ue^2$ since the new coordinate $z$ will have $e$
dependent period.
We shall  proceed by choosing $e$ and 
\be \label{e21btz}
l = 2\sqrt{ue^2}\, ,
\ee 
as independent
variables.  Since $\LL_0^{(3)}$ is a scalar, and hence is a function
of the combination $ue^2$ only, we can define a function $h(l,
\vec u)$
via the relation
\be \label{en2}
h(l,\vec u) = \LL^{(3)}_0
\ee
evaluated in the background \refb{ethrbtz}. 
Eq.\refb{ethrbtz} now gives
\be \label{e22btz}
f_0 \equiv 2\pi \, \sqrt{-\det G} \,
\LL^{(3)}_0  =   {1\over |e|} g(l,\vec u)\, ,
\ee
where 
\be \label{en6btz}
g(l,\vec u) = {\pi \, l^3\, h(l, \vec u)\over 4}\, .
\ee
 Plugging all these results back into eq.\refb{e10btz} we now
 get
\be \label{e23btz}
\EE = 2\pi \left( \wh q\, e - {1\over |e|} g(l,\vec u) - {\pi \, K\over e}
\right)\, ,
\ee
where
\be \label{eddqhat}
\wh q = q - {1\over 4}\, C^{-1}_{ij} q_i q_j\, .
\ee
We need to extremize $\EE$ with respect to $l$, $u_s$ 
and $e$. The
extremization with respect to $l$ and $u_s$ clearly 
requires extremization of
$g(l,\vec u)$ with respect to $l$ and $u_s$. Defining
\be \label{ecdefbtz}
C = -{1\over \pi} g(l, \vec u)
\ee
at the extremum of $g$ we get
\be \label{e24btz}
\EE = 2\pi \left( \wh q\, e + {\pi\, C\over |e|} - {\pi\, K\over e} 
\right) \, .
\ee
We shall assume that $C\ge |K|$. 
Extremizing \refb{e24btz} with respect to $e$ we now get:
\ben \label{eevaluebtz}
e &=& \sqrt{\pi (C-K)\over \wh q} \quad \hbox{for} \, \wh q>0\, ,
\nonumber \\
&=& \sqrt{\pi (C+K)\over |\wh q|} \quad \hbox{for} \, \wh
q<0\, .
\een
Furthermore,  at the extremum,
\ben \label{e25btz}
\EE &&= 2\pi \sqrt{ c_R \, \wh q\over 6} \quad \hbox{for} \, \wh
q>0\, ,
\nonumber \\
&&= 2\pi \sqrt{c_L \, |\wh q|\over 6} \quad \hbox{for} \, \wh q<0\, ,
\een
where we have defined
\be \label{e26btz}
c_L = 24\, \pi\, (C+K)\, , \qquad c_R = 
24\, \pi\, (C-K) \, .
\ee
\refb{e25btz} gives the entropy of extremal BTZ black hole
as a function of the charges $q$ and $\{q_i\}$. Physically $q_i$ are
the charges conjugate to the gauge fields $A_M^{(i)}$ while
$q$ labels the angular momentum of the black hole.

Note that the Chern-Simons term plays no role in the determination
of the parameters $l$, $\vec u$ and
\be \label{e26abtz}
c_L+c_R = 48\, \pi \, C = -48\, g(l,\vec u)\, .
\ee
This is a reflection of the fact that in three dimensions the effect of 
the Chern-Simons term on the equations of motion involves covariant 
derivative of the Ricci tensor\cite{0305117} which vanishes for BTZ 
solution. On the other hand 
\be \label{e27btz}
c_L - c_R = 48\, \pi \, K \, 
\ee
is insensitive to the detailed structure of the higher derivative
terms and is determined completely by the coefficient of the 
Chern-Simons term. This is a consequence of 
the fact that $c_L-c_R$ is
determined by the parity odd part of the action evaluated
on the near horizon geometry of the BTZ black hole, 
and this contribution 
comes solely from the Chern-Simons term.

\sectiono{Black Holes with an $AdS_3$ Factor in the Near Horizon 
Geometry} \label{sads3}

For some extremal black holes in string theory the $ AdS_2$
component of the near horizon geometry, together with an internal
circle, describes a locally $ AdS_3$ space. More accurately the
near horizon geometry of these extremal black holes correspond to
that of extremal BTZ black holes of the type discussed in 
\S\ref{s2.4}
with the momentum along the internal circle representing the angular
momentum of the black hole. In such
situations the enhanced isometry
group of the $ AdS_3$ space allows us to get a more detailed
information about the entropy of the system and prove certain
non-renormalization
theorems\cite{0508218,0506176,0607138,0609074} for the
entropy of supersymmetric as well as non-supersymmetric black holes.
In this section we will outline these arguments and
carry out a comparison between the two approaches
when both methods are available.
Our discussion will follow 
closely the one given in \cite{0611143}.

We shall divide the discussion into three parts. In \S
\ref{sads.1}
we shall describe the origin of the $AdS_3$ factor and the
information it provides for the black hole entropy. In 
\S \ref{sads.2} 
we shall discuss consequences of this result for a specific
class of black holes described in \S \ref{s2.2}-\ref{s2.3}.
In \S \ref{sads.3} 
we shall discuss possible limitations of this
approach.

\subsectiono{Origin and consequences of $AdS_3$ factor}
\label{sads.1}

We begin by reviewing the origin of the $ AdS_3$ geometry. For
this we focus on the $ AdS_2$ part of the near horizon geometry
together with the electric flux through it. By choosing the basis of
gauge fields appropriately we can arrange that only one gauge field
has non-vanishing electric field along the $ AdS_2$; let us
denote this gauge field strength by $F_{\mu\nu}=\p_\mu A_\nu -\p_\nu
A_\mu$. Then the relevant part of the near horizon background takes
the form: 
\be \label{ecomp1} 
ds^2 \equiv g_{\alpha\beta}dx^\alpha
dx^\beta = v (-r^2 dt^2 + r^{-2} dr^2), \quad F_{rt} = e\, . 
\ee
We shall assume that there is an appropriate duality frame in
which we can regard the gauge field component $A_\mu$ as coming from
the component of a three dimensional metric along certain internal
circle labelled by a coordinate $y$. Let us use the convention in
which the two dimensional metric $g_{\alpha\beta}$ is related to the
three dimensional metric $G_{MN}$ in the $r$, $t$, $y$ space via the
dimensional reduction formula given in \refb{e3btz}.
In this case the solution \refb{ecomp1} 
has the same structure as the one
described in \S \ref{s2.4}. In particular if we choose
 \be\label{ecomp4} 
 v = e^2\, , 
 \ee 
then the
three dimensional metric in the $r$-$t$-$y$ plane describes a locally 
$ AdS_3$ space. More precisely, due to the compact nature
of the coordinate $y$ it becomes the quotient of 
the $ AdS_3$ space
by a translation by $2\pi$ along $y$. The effect of taking this
quotient is to break the $SO(2,2)$ isometry group of $ AdS_3$ to
$SO(2,1)\times U(1)$\cite{9803231}, -- 
the symmetries of an $ AdS_2\times S^1$
manifold. Since the physical radius of the $y$ circle is given by 
$\sqrt{G_{yy}}=\sqrt{u}$, we expect that the effect of this symmetry
breaking will be small for large $u$.

Let us for the time being ignore the effect of this symmetry
breaking and suppose that the background has full symmetries of the
$ AdS_3$ space. In this case we expect that the dynamics of the
theory in this background will be governed by an effective three
dimensional action, obtained by treating all the other directions,
including the angular coordinates
labeling the non-compact part of space, as compact. As described
in \refb{e1btz} the
effective Lagrangian density will have a piece $\LL_0^{(3)}$ which
is a scalar function of the metric, Riemann tensor and covariant
derivatives of the Riemann tensor and a gravitational Chern-Simons
term with coefficient $K$.
The resulting entropy will also have the same form
as \refb{e25btz}.
The quantum number $q$ now has the
interpretation of electric charge associated with the gauge field
$A_\mu$ rather than angular momentum. We shall, for
simplicity, ignore the presence of additional gauge fields (with
vanishing electric field but nonvanishing Wilson lines) so that
$\wh q$ appearing in \refb{e25btz} is equal to $q$.  

Although the above analysis gives a general form of the entropy of
a black hole with an $AdS_3$ factor in the near horizon geometry,
this analysis by itself does not determine the
constants $c_L$ and $c_R$ appearing in (\ref{e26btz}). 
While $(c_L-c_R)$ can be determined in terms of the coefficient
of the gravitational Chern-Simons term via eq.\refb{e27btz}, 
the computation of $c_L+c_R$ via eq.\refb{e26abtz} will, 
in general,
require detailed knowledge of higher derivative terms in the action. 
However the situation simplifies if the underlying three dimensional
theory has at least (0,4) 
supersymmetry, -- in this case using AdS/CFT 
correspondence\cite{9711200,9802109,9802150}
one can
also determine
$(c_L+c_R)$ in terms of the
coefficient of a Chern-Simons term in the 
action\cite{0506176,0508218,0509148}.
The argument proceeds
as follows. 
The constants $c_L$ and $c_R$ given in (\ref{e26btz})
can be interpreted
as the left- and right-moving central charges of the two dimensional
CFT living on the boundary of the
$ AdS_3$\cite{9806087,0506176,0508218,0509148}.
If
the boundary theory happens to have $(0,4)$ supersymmetry, then
the central charge $c_R$ is related to the central charge of an
$SU(2)_R$ current algebra which is also a part of the $(0,4)$
supersymmetry algebra. Associated with the $SU(2)_R$ currents there
will be $SU(2)$ gauge fields in the bulk which typically arise
from the dimensional reduction of the full string theory on the
transverse sphere and the central charge of
the $SU(2)_R$ current algebra will be determined in terms of the
coefficient of the gauge Chern-Simons term in the bulk theory. This
determines $c_R$ in terms of the coefficient of the gauge Chern-Simons
term in the bulk theory\cite{0506176,0508218}.
On the other hand we have already seen in \refb{e27btz} that
 $c_L-c_R$ is determined in terms of the
coefficient $K$ of the
gravitational Chern-Simons term.
Since both $c_L$ and $c_R$ are determined in terms of the
coefficients of the Chern-Simons term in the bulk theory, they do
not receive any higher derivative corrections. This completely
determines the entropy from (\ref{e25btz}). 

This result is somewhat surprising from the point of view
of the bulk theory, since for a given three dimensional theory of gravity
the entropy does have non-trivial dependence on all the higher derivative
terms. Thus one could wonder how the dependence of the entropy on these
higher derivative terms disappears by imposing the requirement of (0,4)
supersymmetry.
There is however a simple explanation of this fact even in the bulk theory:
(0,4)
supersymmetry prevents the addition of any higher derivative terms in the
supergravity action (except those which can be removed by field
redefinition\footnote{As has been discussed in 
\cite{0706.3359}, in a
general theory of gravity in three dimensions with negative
cosmological constant we can find explicit field redefinition that
removes all the higher derivative corrections. However this
does not rule out the possibility that the cosmological constant
is renormalized during this field redefinition, -- we need additional
input from supersymmetry to establish this. Furthermore as far
as we can see this argument holds only if the original action was
local, containing powers of Riemann tensor and their covariant 
derivatives. In contrast our argument based of $AdS_3/CFT$
 correspondence applies to the full quantum corrected effective action
 including non-local terms as long as we have a 
 global $AdS_3$ space.}) 
and hence the entropy computed using the three dimensional
supergravity theory with the Chern-Simons terms
is the exact answer.

This non-renormalization theorem may be proved as 
follows\cite{0705.0735}.
In AdS/CFT correspondence the boundary operators dual to the fields
in the supergravity multiplet are just the superconformal currents
associated with the (0,4) supersymmetry algebra. The
correlation functions of these operators in the boundary theory are
determined completely in terms of the central charges $c_L$, $c_R$ of
the left-moving Virasoro algebra and the right-moving super-Virasoro
algebra. Of these $c_R$ is related to the central charge $k_R$ of the
right-moving SU(2) currents which form the R-symmetry currents of the
super-Virasoro algebra and hence to the coefficient of the
Chern-Simons term of the associated SU(2) gauge fields in the bulk theory.
On the other hand $c_L-c_R$ is determined in terms of the coefficient of
the gravitational Chern-Simons term in the bulk theory. 
Thus the  knowledge
of the gauge and gravitational 
Chern-Simons terms in the bulk theory determines
all the correlation functions of (0,4) superconformal currents in the
boundary theory.
Since by AdS/CFT correspondence\cite{9711200}
these correlation functions in the
boundary theory determine completely the boundary S-matrix of the
supergravity fields\cite{9802109,9802150}, we conclude that the
coefficients of the gauge and gravitational Chern-Simons terms in
the bulk theory determine completely the boundary S-matrix elements
in this theory.

Now the boundary S-matrix elements are the
only perturbative observables of the bulk theory. Thus we expect that 
two different theories with the same boundary S-matrix must be related
by a field redefinition. 
Combining this with the observation made in
the last paragraph we see that
two different gravity theories, both with (0,4) supersymmetry
and the same coefficients of the gauge and gravitational Chern-Simons
terms, 
must be related by field redefinition.
Put another way, once we have constructed a classical supergravity
theory with (0,4) supersymmetry and given coefficients of the
Chern-Simons terms, there cannot be any higher derivative corrections
to the action involving fields in the gravity supermultiplet
except for those which can be removed by
field redefinition.
The non-renormalization of the entropy of the BTZ
black hole then follows trivially from this fact.
The complete theory in the bulk
of course will have other matter multiplets whose action will receive
higher derivative corrections. However
since restriction to the
fields in the gravity supermultiplet
provides a consistent truncation of the theory, and since the BTZ
black hole is embedded in this subsector, its entropy will not be affected
by these additional higher derivative terms.

We shall now describe this unique (0,4) supergravity
action and compute
the constants $c_L$ and $c_R$ from this action.
The action was
constructed in \cite{9904010,9904068}
(generalizing earlier work of \cite{ach1,ach2,0610077,wit2}
for supergravity actions based on $Osp(p|2;R)\times Osp(q|2;R)$
group)
by regarding the supergravity
as a gauge theory based on $SU(1,1)\times SU(1,1|2)$ algebra.
If $\Gamma_L$ and $\Gamma_R$ denote the (super-)connections
in the $SU(1,1)$ and $SU(1,1|2)$ algebras respectively, then the
action is taken to be a Chern-Simons action of the form:
\ben \label{esupercs}
\SSS &=& -a_L\, \int d^3 x \left[ Tr( \Gamma_L \wedge 
d\Gamma_L
+ {2\over 3} \Gamma_L\wedge \Gamma_L \wedge \Gamma_L\right]
\nonumber \\
&& + a_R \, 
\int d^3 x \left[ Str( \Gamma_R \wedge 
d\Gamma_R
+ {2\over 3} \Gamma_R\wedge \Gamma_R 
\wedge \Gamma_R\right]\, ,
\een
where $a_L$ and $a_R$ are constants. 
Note that the usual metric degrees of freedom are 
encoded in the connections
$\Gamma_L$ and $\Gamma_R$.
Thus there is no obvious  way to add $SU(1,1)\times SU(1,1|2)$
invariant higher derivative terms in the
action involving the field strengths associated
with the connections $\Gamma_L$ and $\Gamma_R$. 
{}From this viewpoint also
it is natural that the supergravity action
does not receive any higher derivative corrections.

The bosonic fields of this theory include the metric $G_{MN}$
and an
SU(2) gauge field ${\bf A}_M$ ($0\le M\le 2$), represented as a 
linear combination of $2\times 2$ anti-hermitian matrices.
After expressing the action
in the component notation and eliminating auxiliary fields using
their equations of motion as in 
\cite{9904010,9904068,ach1,ach2,0610077}
we arrive at the action 
\be \label{ebosonic}
\SSS= \int d^3 x\left[\sqrt{-\det G} \left[ R
+ 2 m^2 \right] +  K \, \Omega_3(\wh\Gamma) 
- {k_R\over 4\pi}\, 
\epsilon^{MNP} Tr\left({\bf A}_M
\p_N {\bf A}_P + {2\over 3} {\bf A}_M {\bf A}_N {\bf A}_P\right)\right]
\ee
where
\be \label{ecarel}
{1\over m} = {1\over 2} (a_R+a_L), 
\qquad K = {1\over 2} (a_L-a_R)\, ,
\ee
\be \label{edefkr}
k_R = 4\pi a_R = 4\pi \, \left({1\over m} - K\right) \, ,
\ee
and $\Omega_3(\wh\Gamma)$ is the
gravitational Chern-Simons term defined in \refb{e2btz}.

We shall now compute the constants $c_L$ and $c_R$ in this
theory by using the general results of \S\ref{s2.4}.
We have in this theory
\be\label{ehlspec}
h(l) = (- 6 l^{-2} +2 m^2)\, ,
\ee
\be \label{eglspec}
g(l) = {\pi\over 4} l^3 \, (- 6 l^{-2} + 2 m^2)\, ,
\ee
\be \label{el0spec}
l_0 = {1\over m}\, ,
\ee
\be \label{ecspec}
C = {1\over m}
\ee
and 
\be \label{ellcr}
c_L = 24\, \pi \, \left({1\over m} + K\right) = 24\, \pi \,
a_L, \qquad
c_R = 24\, \pi \, \left({1\over m} - K\right) = 24\, \pi \, a_R\, ,
\ee
where in \refb{ellcr} we have used \refb{ecarel}.
Using \refb{edefkr} we get
\be \label{ealar}
c_R = {6\, k_R}, \qquad c_L =  48\, \pi\, K + 6 \, k_R\, .
\ee
This gives the expressions for $c_L$ and $c_R$ in terms of the
coefficients $k_R$ and $K$ of the gauge and gravitational Chern-Simons
terms. By the argument outlined earlier, this result will not be
modified by higher derivative corrections in the theory.

The argument given above can be easily generalized to the cases
where the theory contains additional U(1) gauge fields with
non-degenerate Chern-Simons terms  and the black hole is charged
under these gauge fields. The analysis of \S\ref{s2.4} shows that the
expressions for $c_L$ and $c_R$ are independent of the action
involving these gauge fields and hence will continue to be
given by eqs.\refb{ealar}. The quantity $\wh q$ appearing in
the expression \refb{e25btz}
the black hole entropy will however depend on the additional gauge
charges via eq.\refb{eddqhat}.

The results for black hole entropy computed using the general
arguments outlined above 
have been verified by explicit computation in heterotic string theory
after including all tree level four derivative 
corrections\cite{0607094,0608182} and
also in five dimensional supergravity 
theories\cite{0702072,0703087,0703099} 
with curvature squared
corrections\cite{0611329}.

\subsectiono{Applications to black holes
in string theory} \label{sads.2}

We shall now apply the observations made in
\S \ref{sads.1} to the study
of black holes discussed in \S \ref{sphys}-\S\ref{snsn=4}. 
We shall use the
second description of the theory, -- as a $\ZZZ_N$ orbifold of heterotic
or type II string theory on $T^4\times \wh S^1\times S^1$, -- and 
consider a black hole with $\wh n$ units of momentum and
$-\wh w$ units of fundamental string winding charge along $\wh S^1$
and $N'$ units of Kaluza-Klein monopole charge and $-W'$ units of
H-monopole charge associated with the circle $S^1$. From 
\refb{e2dcharge} we see that such a state has
\be \label{echargepre}
Q = \pmatrix{\wh n\cr   0 \cr \wh w\cr 0}, \qquad
P = \pmatrix{0 \cr W'\cr 0 \cr N'}
\ee
giving
\be \label{echarge}
Q^2 = 2\wh n\wh w, \qquad P^2 = 2 \, W' \, N', \qquad Q\cdot P=0
\, .
\ee

In order to apply the formalism described in \S \ref{sads.1} we
need to ensure that the only electric field carried by the
solution comes from the dimensional reduction of the metric along a
compact circle. In this case however there are two sets of
electric fields corresponding to the dimensional reduction of the
metric along $\wh S^1$ and the dimensionl reduction of the
NS sector 2-form field along $\wh S^1$. To avoid this problem
we take the heterotic or type II string theory on 
$T^4\times S^1/\ZZZ_N$ and dualize the 2-form
field $B_{MN}$ in the five remaining dimensions into a gauge
field $\BB_M$. We then compactify the resulting theory on
$\wh S^1$. In this description 
the fundamental string
winding charge $\wh w$ along $\wh S^1$ 
appears as a magnetic charge of the four dimensional gauge
field $\BB_\mu$. As a result
the quantum numbers
$N'$, $W'$ and $\wh w$ appear as magnetic charges whereas
the quantum number $\wh n$ appears as electric charge.

As in \S\ref{sads.1}
we proceed by making the ansatz
that the $AdS_2$ factor in the near horizon geometry of the
black hole combines with $\wh S^1$ to produce an
$AdS_3$ factor. If this ansatz leads to a finite size $AdS_3$ then our
ansatz is self-consistent. 
In this case we can regard the near horizon geometry of the black hole
as that of an extremal BTZ black hole in a three dimensional theory
of gravity, obtained by compactifying the full string theory on
$(T^4\times S^1)/\ZZZ_N\times S^2$.
The three dimensional theory obtained this way turns out to
have a (0,4) supersymmetry and associated  SU(2) gauge fields which
come from the isometries of $S^2$. Thus due to the arguments outlined
in \S \ref{sads.1}, the central charges $c_L$ and $c_R$ appearing
in the formula for the entropy can be calculated from the knowledge
of the Lorentz and SU(2) Chern-Simons terms. The coefficients of the
Chern-Simons terms in turn can be calculated by carefully keeping track
of various non-covariant terms appearing in the original
Lagrangian density as well those appearing in the process of
dimensional reduction. The final result for $N',W'>0$, $\wh w<0$ 
is\cite{0506176,0508218,0609074}
\ben \label{ecint}
&& c_R = 
- 6 ( N'W'\wh w+ 2\wh \alpha \wh w ), \nonumber \\
&&
c_L = 
- 6 ( N'W'\wh w+ 4\wh \alpha \wh w ) \, ,
\een
where $\wh\alpha=1$ for heterotic string theory and 0 for type II string
theory. On the other hand with the convention we have chosen the
quantity $q$ appearing in \refb{e25btz} is given by
\be \label{eqident}
q = \wh n\, .
\ee
Thus if we take the supersymmetric configuration
\be \label{erangenw}
N', W'>0, \qquad  \wh n, \wh w < 0\, ,
\ee
then from \refb{e25btz} we get
\be \label{efe}
S_{BH} = 2\pi \sqrt{(N'W'+4\wh\alpha) \, \wh w \, \wh n}\, .
\ee
This agrees with \refb{eag17spec} and
\refb{ebhfin}.
By switching on the electric charges associated with various
gauge fields in the way described in \S\ref{s2.4}  
and using T-duality
invariance one can in fact
argue that the more general result \refb{eag17} 
(and the corresponding result \refb{ew2ns} in $\NN=2$
supergravity theory) is exact in the limit 
$\sqrt{Q^2P^2 - (Q\cdot P)^2}>>P^2$. 
This exact
agreement for supersymmetric black holes
is somewhat mysterious since neither the Gauss-Bonnet
term nor the $\NN=2$ supergravity analysis described in \S
\ref{sn=2} captures the complete set of terms even at the four
derivative level.
Neither do they satisfy the condition which led to \refb{efe},
namely neither of these theories come from dimensional
reduction on $\wh S^1$ of a supersymmetric action in one
higher dimension.

For the non-supersymmetric configuration
\be \label{erg7}
N', W', \wh n>0, \qquad \wh w < 0\, ,
\ee
we get
\be \label{ensns3}
S_{BH}^{ns} = 2\pi \sqrt{(N'W'+2\wh\alpha)|\wh n\wh w|}\, .
\ee
This does not agree with either \refb{ensns1} or \refb{ensns2}.

As long as the near horizon geometry of the black hole
is given by a locally $AdS_3$ space, the results \refb{efe},
\refb{ensns3} are exact. Indeed, this has now been verified
by explicit computation of the black hole 
entropy\cite{0608182,0607094}
keeping all the four derivative terms in the effective 
action\cite{tseyt,hull} and
comparing the result for extremal black hole entropy with
the expansion of \refb{efe}, \refb{ensns3} up to first
non-leading order in $1/|N'W'|$.
In \S \ref{sads.3} we shall examine under what condition
this approximation breaks down. 

\subsectiono{Limitations of $AdS_3$ based approach} \label{sads.3}

Clearly the existence of an $ AdS_3$ factor in the near horizon
geometry gives us results which are much stronger than the ones
which can be derived based on the existence of only an $ AdS_2$
factor. 
In this section we shall discuss the approximation under which
\refb{e25btz} holds and possible corrections to this formula.

The main underlying assumption behind \refb{e25btz} is that
the black hole solution is described by an effective three dimensional
theory of gravity with a generally covariant action in three
dimensions of the form \refb{e1btz}. 
In this case we can look for solutions in this three dimensional
theory
with SO(2,2) isometry which corresponds to an $AdS_3$ space.
However  this SO(2,2) isometry of the near horizon background
is only an approximate symmetry since due to the compactness of
the angular coordinate of the BTZ black hole the actual space is
a quotient of the $AdS_3$ space by the group of $2\pi$ translation
along this coordinate. The true symmetry of the background is a
subgroup of SO(2,2) that commutes with the translation and this is
simply SO(2,1)$\times$U(1), -- the product of the isometry group
of $AdS_2$ and the group of translations along the compact direction.  
Let us denote the coordinate along this compact direction
by $y$.
As long as 
the physical radius of the $y$ coordinate is large we expect that the
effect of the breaking of SO(2,2) isometry will be small 
and \refb{e25btz}
will be valid. However
if this radius
is of order unity, then
the SO(2,2) symmetry of $ AdS_3$ to be broken
strongly and it will be more appropriate to regard the background
as a two dimensional background by dimensionally reducing the
theory along $y$. 
The effective two dimensional action governing
the dynamics in $ AdS_2$ space, besides having a `local' piece of the
form (\ref{e1btz}) with three dimensional general
coordinate invariance, contains additional terms which cannot be
written as dimensional reduction of a generally covariant three
dimensional action.\footnote{This can be seen even in ordinary
Kaluza-Klein compactification of flat space-time. If the
space-time contains a compact circle of radius $R$, then the
quantum effective action will typically involve terms with
complicated dependence on $R$. Since
the dimensional reduction of
a higher dimensional generally covariant action on a 
circle of radius $R$ produces a Lagrangian density 
proportional to $R$, not all the terms in the quantum effective action can 
be viewed as coming from the dimensional reduction of
a higher dimensional generally covariant action.}
There are various sources of these additional
terms, {\it e.g.} due to the quantization of the momenta along the
$y$ direction, contribution to the effective action from various
euclidean branes wrapping the $y$ circle, etc. In the presence of
such terms there will be additional contribution to the entropy
which are not of the form (\ref{e25btz}). These additional
corrections can be interpreted as due to the corrections to the full
string theory partition function on thermal
$ AdS_3$\cite{0607138,0609074} or equivalently as
corrections to the Cardy formula in the CFT living on the boundary
of $ AdS_3$, but there is no simple way to calculate these
corrections without knowing the details of this CFT.

We will illustrate this in the context of the black holes
discussed in \S\ref{sads.2}. 
As can be seen from eqs.\refb{eag16ag}, \refb{erv} for 
$\wh\alpha=0$,
in the leading
supergravity approximation the near horizon values of the radii $\wh R$
and $R$ of $\wh S^1$ and $ S^1$ and field $S$
representing square of the inverse string coupling are given by 
\be\label{ecomp11} 
\wh R =
\sqrt{\left|\wh n\over \wh w\right|}, \qquad 
R = \sqrt{\left| W'\over
N'\right|}, \qquad 
u_S = \sqrt{\left| {\wh n\wh w\over  N'  
W'}\right|}\, . \ee 
{}From this we see that if we take  $|\wh 
n|$ large keeping
the other charges fixed, the radius $\wh R$ of the circle $ \wh
S^1$ becomes
large. Thus we expect that in this limit the $SO(2,2)$
isometry of the near horizon geometry will be a good approximation
and the entropy will have the
form given in (\ref{efe}) even after inclusion of higher
derivative corrections. However when all charges are of the same
order then the radius of $\wh S^1$ becomes of order unity
and the higher derivative corrections to the action will
contain terms which cannot be regarded as the dimensional reduction
of a three dimensional general coordinate invariant action of the
form given in (\ref{e1abtz}). Consequently
the higher derivative corrections
to the entropy will cease to be of the form given in (\ref{efe}).

This can be seen explicitly by taking into account the effect of the
four derivative Gauss-Bonnet term in the four dimensional effective
action describing heterotic string compactification on $ T^4\times
\wh S^1\times S^1$. 
An expression for the entropy of a black hole in the presence
of a Gauss-Bonnet term  has been given in 
\refb{ebhexplicit}. For large $|\wh n|$, $u_S$ given in
\refb{ecomp11} is large. In this case we can approximate
$\phi(u_a, u_S)$ by its large $u_S$ limit given in 
\refb{elarge} and
the corresponding entropy \refb{eag17spec} agrees with 
the result \refb{efe}.
However
if all the charges are of the same order,
then $u_S$ given in \refb{ecomp11} 
is of order unity and we
cannot approximate $\phi(u_a,u_S)$ by its large $u_S$ limit. 
Instead we need to use the complete expression given in
\refb{ebhexplicit} with $Q^2$, $P^2$ and $Q\cdot P$ given in
\refb{echarge}:
\be \label{ebbex}
S_{BH} \simeq 2\pi\sqrt{\wh n\wh w N' W'} 
+ 64\pi^2 \phi\left(0, \sqrt{\left| {\wh n\wh w\over  N'  
W'}\right|}\right)\, .
\ee
For the specific case of heterotic string theory
on $T^4\times\wh S^1\times S^1$ 
we get from \refb{eh10bint}, \refb{enn13pre} for the case $\MM=K3$,
$N=1$,
\be \label{ecomp14} 
\phi(a,S) = -{3\over
16\pi^2} \ln \left(2 S |\eta(a+iS)|^4\right)\, , 
\ee 
up to an additive constant.
This gives 
\be\label{ecomp15} 
\Delta S_{BH} \simeq 64 \, \pi^2 \,
\phi(0,u_S)|_{u_S=\sqrt{\left| {\wh n\wh w/  N'  W'}\right|}} = -12 \,
\ln\left[ 2\sqrt{\left| {\wh n\wh w\over  N' W'}\right|} 
\eta\left( i
\sqrt{\left| {\wh n\wh w\over  N' W'}\right|}\right)^4\right]\, . \ee
This clearly has a complicated $\wh n$-dependence and is not
in agreement with the simple form \refb{efe}.
 
It is instructive to study the origin of the terms which break the
SO(2,2) symmetry of $ AdS_3$ in this specific
example. First of all (\ref{ecomp15}) contains a
correction term proportional to $\ln S\sim \ln\left|
{\wh n\wh w\over N' W'}\right|$.
This can be traced to the effect of replacing the continuous
integral over the momentum along $ S^1$ by a discrete sum. There are
also additional corrections involving powers of $e^{-2\pi u_S}$. These
can be traced to the effect of Euclidean 5-branes wrapped on
$ K3\times \wh
S^1\times S^1$\cite{9610237}. Since the 5-brane
has one of its legs along $\wh S^1$, it breaks the SO(3,1) isometry of
Euclidean $ AdS_3$.

The above example also illustrates the basic difference between the
approximation schemes used by the $ AdS_3$ and $ AdS_2$ based
approaches. The $ AdS_3$ based approach is useful when we take
the momentum along the $ AdS_3$ circle $S^1$ to be large keeping
the other charges fixed. In this limit the size of $S^1$ becomes
large (see eq.(\ref{ecomp11})) and hence the $SO(2,2)$ symmetry of
$ AdS_3$ is broken weakly. As a result the entropy has the form
(\ref{e25btz}). In the CFT living on the boundary of $ AdS_3$,
this corresponds to a state with large $L_0$ (or $\bar L_0$)
eigenvalue, keeping the central charge fixed. This is precisely the
limit in which the Cardy formula for the degeneracy of states is
valid. On the other hand the $ AdS_2$ based approach is useful if
all the charges are large since in this limit the $ AdS_2$ has
small curvature, and we can use the derivative expansion of the
effective action to find a systematic expansion of the entropy and
the entropy function in inverse powers of charges.

\subsectiono{Small black hole revisited} \label{smallre}

Given this understanding of the range of validity of the
$AdS_3$ based approach, let us now return to the case of
small black holes. In this case we have $N'=0$, $W'=0$.
Eq.\refb{efe} shows that for $\wh n\wh w>0$, \i.e.\ for
supersymmetric small black holes, the entropy
is given by
\be \label{ere1}
S_{BH} = 4\pi \sqrt{\wh\alpha \wh w\wh n}\, ,
\ee
if we take the large $\wh n$ limit at fixed $\wh w$. In fact
since we have already argued earlier that for large charges the 
result for the entropy depends only on the combination
$Q^2 = 2\wh n\wh w$, \refb{ere1} must be valid for large
$\wh n\wh w$. For heterotic string theory $\wh\alpha=1$ and
the result is in perfect agreement with the statistical entropy
\refb{estatsmall}. However for type II string theory $\wh\alpha=0$
and the result is in disagreement with the statistical entropy
formula \refb{estatsmallii}.

The origin of this discrepancy for type II string theories is not
completely clear at this stage. Since \refb{ere1} gives $S_{BH}=0$,
the most conservative point of view would be that one cannot find a
solution to the equations of motion with the ansatz that there is an
underlying locally $AdS_3$ factor. This still leaves open the
possibility that type II string theory admits a
small black
hole solution 
whose near horizon geometry has an $AdS_2$ factor, and finite
entropy which can be computed using the entropy function method
after taking into account all the higher derivative corrections to the
tree level effective action. A different approach to this problem
has been described in 
\cite{0611062,0707.3818,0707.4303,0708.0016}.

One can also carry out a similar analysis for non-supersymmetric
small black holes. If we set $N'=W'=0$ in eq.\refb{ensns3} we get
\be \label{esmallns}
S_{BH}^{ns} = 2\sqrt 2\pi \sqrt{\wh\alpha|\wh n\wh w|}\, .
\ee
For $\wh\alpha=1$ this agrees with the statistical entropy of these
black holes computed from the spectrum of elementary string
states carrying right-moving excitations with 
$\bar L_0=|\wh n\wh w|$ and no left-moving excitations. However
for $\wh\alpha=0$ the result again fails to agree with the statistical
entropy of small non-supersymmetric
black holes in type II string theory which is given by
$2\sqrt 2\pi\sqrt{|\wh n\wh w|}$.

\sectiono{Precision Counting of Dyon States} \label{s3}

In this section we shall compute the statistical entropy of a
class of dyonic
black holes in the theories
described in \S \ref{sphys}. 
We shall carry out our analysis by first 
counting
states of a 
configuration carrying specific charge vectors 
$(\vec Q, \vec P)$ in a specific corner of the 
moduli space and
then extend the results to more general 
charges. 
In  \S\ref{s3.s} we shall describe
this specific microscopic configuration. Also,
in order to guide the
reader through the rest of the section, we shall summarize in
this section the results of \S\ref{s3.2}-\S\ref{smulti}.
In
\S\ref{s3.2} we carry out the computation of BPS states for
the microscopic configuration described in \S\ref{s3.s}.
 In \S\ref{sadditional} we describe how our results
can be extended to more general charge vectors.
In 
\S\ref{smarginal} we discuss how the spectrum of the theory could
change discontinuously as we move across walls of marginal stability
in the moduli space, and use these results to determine the region
of moduli space in which our formula for the degeneracy of dyons
remains valid.  In \S\ref{sdualtwo} we study T- and 
S-duality transformation
properties of the degeneracy formula, and use the requirement of
duality invariance to determine how the degeneracy changes as we
move across a wall of marginal stability.
In \S\ref{s3.3} we calculate the statistical
entropy of the system, -- given by the logarithm of the degeneracy
of states, -- by expanding the result of \S\ref{s3.2} 
in a series expansion
in inverse powers of charges and compare the results with the results
for black hole entropy given in \S\ref{sgauss}.  Finally in
\S\ref{smulti} we demonstrate how the change in the degeneracy
across walls of marginal stability can be related to (dis)appearance
of 2-centered black holes as we cross the marginal stability 
walls\cite{0005049,0010222,0101135,0206072,0304094,0702146,
0705.2564,0706.3193}.

Throughout this
section we shall set $\alpha'=1$ unless mentioned otherwise.
Our counting of states will follow \cite{0605210,0607155,0609109}.
The original formula for the degeneracy was first proposed
in \cite{9607026}
for the special case of heterotic string theory compactified on
$T^6$, and extended to more general models in
\cite{0510147,0602254,0603066}.
Various 
alternative
approaches to proving these formul\ae\ have been explored 
in \cite{0505094,0506249,0508174,0506151,0612011}.

\subsectiono{Summary of the results} \label{s3.s}

As in \S\ref{sphys} we consider  
type IIB string theory on $\MM\times \wt 
S^1\times  S^1$ where
$\MM$ is either K3 or $T^4$, and 
mod out this theory by a $\ZZZ_N$ symmetry
group generated by a transformation $g$ that involves
$1/N$ unit of shift along the circle $S^1$ 
together with an order
$N$ transformation
$\wt g$ in $\MM$. $\wt g$ is chosen in such a way that the final
theory has $\NN=4$ supersymmetry. 
In keeping with the convention described below \refb{echsq} we shall
take the coordinate radii of $S^1/\ZZZ_N$ and $\wt S^1$ to be 1.
In this convention
the original $S^1$ before orbifolding has coordinate
radius $N$ and the $\ZZZ_N$ action involves $2\pi$ translation along
$S^1$. Since under this translation various modes 
get transformed by $\wt g$ twist instead of remaining invariant,
the momentum along $S^1$ is quantized in multiples of 
$1/N$ instead of being integers.
Following \cite{0505094} we consider in this theory
 a configuration with a single D5-brane wrapped on 
 $\MM\times S^1$, $Q_1$ D1-branes wrapped on $S^1$, a single
 Kaluza-Klein monopole associated with the circle $\wt S^1$ with
 negative magnetic charge,
 momentum $-n/N$ along $S^1$ and momentum $J$
 along $\wt S^1$.\footnote{In short, we have a BMPV black 
hole\cite{9603078} at the 
center of Taub-NUT space.}
Since a D5-brane wrapped on $\MM$
 carries, besides the D5-brane charge, $-\beta$ units of D1-brane charge
 with $\beta$ given by the Euler character of $\MM$ 
  divided by 24 \cite{9511222}, the net 
  D1-brane charge carried by the system
  is $Q_1-\beta$.

As has already been discussed in \S\ref{sphys}, there is a second 
description of the theory obtained by an S-duality transformation of type 
IIB string theory, followed by a T-duality transformation along $\wt S^1$ 
that takes us to type IIA string theory on $(\MM\times \wh S^1\times 
S^1)/\ZZZ_N$, and finally a string-string duality transformation that 
takes us to heterotic (type IIA) string theory on $(T^4\times \wh 
S^1\times S^1)/\ZZZ_N$ for $\MM=K3$ ($\MM=T^4$). 
 By following 
 the duality transformation rules and the sign conventions given in 
appendix \ref{s3.1} we can find the physical
 interpretation of various charges carried by the system in the
 second description.
 We find that it corresponds to a state
 with momentum $-n/N$ 
 along $S^1$, a single Kaluza-Klein monopole associated with 
 $\wh S^1$,
  $(-Q_1+\beta)$ units of NS 5-brane charge along $T^4\times
  S^1$, $-J$ units of NS 5-brane charge along 
  $T^4\times \wh S^1$
  and a single fundamental string wound along 
  $S^1$\cite{0605210}. In particular the Kaluza-Klein monopole
  charge associated with $\wt S^1$ in the first description gets mapped to
  the fundamental string winding number along $S^1$ in the second
  description and the D5-brane wrapped on $\MM\times S^1$ in the
  first description gets mapped to Kaluza-Klein monopole charge
  associated with $\wh S^1$ in the second description.
  Using the convention of \refb{e2dcharge} we see that 
  this corresponds to the charge 
vectors\footnote{Ref.\cite{0605210} actually
considered a more general charge vector where $Q_5$, representing the
number of D5-branes wrapped along $K3\times S^1$, was arbitrary and
found that $d(\vec Q,\vec P)$, expressed as a function of $Q^2$, $P^2$
and $Q\cdot P$, continues to be given by the same function
\refb{egg1int}.
However the analysis of dyon spectrum becomes simpler for
$Q_5=1$. For this reason we have set $Q_5=1$. We shall comment
on the 
more general
case at the end of \S\ref{sadditional} (see paragraph containing
eq.\refb{enewgcd}.).
}
  \be \label{echvec}
  Q=\pmatrix{0 \cr -n/N \cr 0 \cr -1}, \quad P = \pmatrix{Q_1-\beta
 \cr -J\cr 1 \cr 0}\, .
  \ee
 This gives
 \be\label{eqdef}
 Q^2  
 = 2 n/N, \qquad P^2   
 = 2 (Q_1- \beta   ), \qquad Q\cdot P = J\, .
\ee
In the rest of this section we shall summarize the results of
\S\ref{s3.2} - \S\ref{smulti}.

We denote by 
$d(\vec Q,\vec P)$ the number of bosonic minus fermionic
quarter BPS supermultiplets carrying a given set of charges
$(\vec Q, \vec P)$, a supermultiplet being considered bosonic
(fermionic) if it is obtained by tensoring the basic 64 dimensional
quarter BPS supermultiplet, with helicity ranging from $-{3\over 2}$
to ${3\over 2}$,  with a supersymmetry singlet  
bosonic (fermionic) state. For the charge vector given in
\refb{echvec} and in the region of the moduli space where the
type IIB string coupling in the first description of the theory is
small, 
our result for $d(\vec Q,\vec P)$  is\footnote{The overall factor
of $(-1)^{Q\cdot P +1}$ was left out in the analysis of 
\cite{0605210,0607155,0609109}. The $(-1)^{Q\cdot P}$ factor
appeared previously in \cite{0508174,0706.2363} 
and reflects the difference in
statistics between the four and five dimensional viewpoint for
modes carrying odd units of $Q\cdot P$ quantum number.
As we shall see in \refb{ezfin}, the $-1$ 
factor appears from the spectrum
of bound state of a D1-D5 system to the 
Kaluza-Klein monopole.}
\be\label{egg1int}
d(\vec Q,\vec P) = (-1)^{Q\cdot P+1}\,
{1\over N}\, \int _\CC d\wt\rho \, 
d\wt\sigma \,
d\wt v \, e^{-\pi i ( N\wt \rho Q^2
+ \wt \sigma P^2/N +2\wt v Q\cdot P)}\, {1
\over \wt\Phi(\wt \rho,\wt \sigma, \wt v)}\, ,
\ee
where $\CC$ is a three real dimensional subspace of the
three complex dimensional space labelled by $(\wt\rho,\ws,\wv)
\equiv (\wrh_1+i\wrh_2,\ws_1+i\ws_2,\wv_1+i\wv_2)$,
given by
\bea{ep2kk}
\wt \rho_2=M_1, \quad \wt\sigma_2 = M_2, \quad
\wt v_2 = -M_3, \nonumber \\
 0\le \wt\rho_1\le 1, \quad
0\le \wt\sigma_1\le N, \quad 0\le \wt v_1\le 1\, ,
\een
$M_1$, $M_2$ and $M_3$ being large but fixed positive
numbers with $M_3<< M_1, M_2$, and
\bea{edefwtphi}
&& \wt \Phi(\wt \rho,\wt \sigma,\wt v ) =
e^{2\pi i (\wt \alpha\wt\rho + \wt \gamma\ws 
+ \wt v)} \nonumber \\
&& \qquad \times \prod_{b=0}^1\, 
 \prod_{r=0}^{N-1}
\prod_{k'\in \zzz+{r\over N},l\in\zzz,j\in 2\zzz+b
\atop k',l\ge 0, j<0 \, {\rm for}
\, k'=l=0}
\left[ 1 - \exp\left\{2\pi i ( k'\wt \sigma   +  l\wt \rho +  j\wt v)
\right\}\right]^{
\sum_{s=0}^{N-1} e^{-2\pi i sl/N } c^{(r,s)}_b(4k'l - j^2)} \, .
\nonumber \\
\eea
The coefficients $c^{(r,s)}_b(u)$ have been defined through 
eqs. \refb{esi4aint}, \refb{enewint}:
 \ben\label{esi4aintpre}
&& F^{(r,s)}(\tau,z)\equiv
\sum_{b=0}^1\sum_{j\in2\zzz+b, n\in \zzz/N} 
c^{(r,s)}_b(4n -j^2)
e^{2\pi i n\tau + 2\pi i jz}\nonumber \\
&&= {1\over N} Tr_{RR;\wt g^r} \left(\wt g^s
(-1)^{F_L+F_R}
e^{2\pi i \tau L_0} 
e^{-2\pi i \bar\tau \bar L_0}
e^{2\pi i F_L z}\right),  \qquad\qquad
0\le r,s\le N-1\, ,
 \een
where $Tr_{RR;\wt g^r}$ denotes trace over $\wt g^r$ twisted RR-sector 
states in the two dimensional superconformal field theory with target 
space $\MM$, $F_L$, $F_R$ denote the left- and right-handed fermion 
numbers in this world-sheet theory, and $L_n$ and $\bar L_n$ denote the 
left- and right-handed Virasoro generators. The additive constants in 
$L_0$ and $\bar L_0$ are adjusted so that supersymmetric ground states in 
the RR sector have $L_0=\bar L_0=0$, -- this is a convention we shall 
follow throughout the rest of the article. The constants $\wt\alpha$
and $\wt\gamma$ are given in terms of the coefficients $c_b^{(r,s)}(u)$
via eqs.\refb{eqrsrev}, \refb{enn9d}:
\be\label{eqrsrevpre}
Q_{r,s} = N\, 
\left( c^{(r,s)}_0(0)+ 2 \, c^{(r,s)}_1(-1)\right)\, ,
\ee
\be\label{enn9dpre}
\wt \alpha={1\over 24N} \, Q_{0,0} - {1\over 2N}
\, \sum_{s=1}^{N-1} Q_{0,s}\, {e^{-2\pi i s/N}\over
(1-e^{-2\pi i s/N})^2 } \, 
, \qquad 
\wt \gamma= {1\over 24N} \, Q_{0,0} = {1\over 24N}\, 
\chi(\MM) \, .
\ee

As has been shown in \refb{especial}, $\wt\Phi$ satisfies the 
periodicity conditions
\be \label{especialpre}
\wt\Phi(\wt\rho+1,\wt\sigma,\wt v) = \wt\Phi(
\wt\rho,\wt\sigma+N,\wt v)
=\wt\Phi(\wt\rho,\wt\sigma,\wt v+1) = \wt\Phi(\wt\rho,\wt\sigma,
\wt v)\, .
\ee
Using this we can express $d(\vec Q,\vec P)$ as
\be\label{efo1}
d(\vec Q, \vec P) = (-1)^{Q\cdot P+1}\,
g\left({N\over 2} Q^2 , {1\over 2\, N}\, P^2,
Q\cdot P\right)\, ,
\ee
where $g(m,n,p)$ are
the coefficients of Fourier expansion of the function
$1/ \wt\Phi(\wt \rho,\wt \sigma, \wt v)$:
\be\label{efo2}
{1
\over \wt\Phi(\wt \rho,\wt \sigma, \wt v)}
=\sum_{m,n,p} g(m,n,p) \, e^{2\pi i (m\, \wt \rho + n\,
\wt\sigma
+ p\, \wt v)}\, .
\ee

Although the formul\ae\ \refb{egg1int}, \refb{efo1} 
were originally
derived for charge vectors of the form \refb{echvec}, we show 
in \S\ref{sadditional}
that the same formula holds for a more general class of charge
vectors for weak type IIB string coupling in the first
description\cite{0705.1433}. In the subspace $\VV$ introduced in 
\refb{e2dcharge} 
this general charge vector takes the form
\be \label{nne1.6newrp}
Q = \pmatrix{k_3\cr k_4 \cr k_5 \cr -1}, \qquad
P = \pmatrix{l_3\cr l_4 \cr 1 \cr 0}, \qquad k_4\in \ZZZ/N,
\quad k_i, l_i\in \ZZZ\quad \hbox{otherwise}
\, .
\ee 

In writing down the formula \refb{egg1int} we have
made an implicit assumption.
Eqs.\refb{egg1int} and \refb{efo1} are equivalent
only if the sums over $m$, $n$, $p$ in \refb{efo2}
are convergent for large
imaginary $\wrh$, $\ws$ and $-\wv$, -- the
region in which the contour $\CC$ lies. This in particular 
requires that the sum over $m$ and $n$ are bounded from
below, and that for fixed $m$ and $n$
the sum over $p$ is bounded from above.  By examining the
formula \refb{edefwtphi} for $\wt\Phi$ we can see that the
sum over $m$ and $n$ are indeed bounded from below.
Furthermore, using the fact that the
coefficients $c^{(r,s)}_b(u)$ are non-zero only for $4u\ge -b^2$,
we can verify that with the exception of the contribution from the
$k'=l=0$ term in this product, the other terms, when expanded
in a power series expansion in $e^{2\pi i\wrh}$, $e^{2\pi i\ws}$
and $e^{2\pi i\wv}$, 
does have the form of \refb{efo2} with $p$ bounded from above (and
below)
for fixed $m$, $n$.
However for the $k'=l=0$ term, which  gives a contribution
$e^{-2\pi i \wv} / (1 - e^{-2\pi i \wv})^2$, 
there is an ambiguity in
carrying out the series expansion. We could either use the form given
above and expand the denominator in a series 
expansion in $e^{-2\pi i \wv}$, or express it as
$e^{2\pi i \wv} / (1 - e^{2\pi i \wv})^2$ and expand it in 
a series 
expansion in $e^{2\pi i \wv}$. As will be discussed below
\refb{ezfin},
depending on the angle between $S^1$ and $\wt S^1$ in the
first description,
only one of these expansions produce the degeneracy formula
correctly via \refb{efo2}\cite{0605210}. The physical 
spectrum actually changes as this angle
passes through $90^\circ$ since at this point the system is only
marginally stable.
On the other hand our degeneracy formula
\refb{egg1int}, \refb{ep2kk} 
implicitly requires that we expand this factor in powers
of $e^{-2\pi i \wv}$ since only in this case the sum over $p$ in
\refb{efo2} is bounded from above for fixed $m$, $n$.
Thus as it stands the formula is valid for a
specific range of values of the angle between $S^1$ and $\wt S^1$.
In the second
description of the system this 
corresponds to a region in the moduli space where
the axion field $a$, obtained by dualizing the NS sector 2-form
field, has a positive sign.
For the negative sign of the axion 
the correct formula for $d(\vec Q,\vec P)$ is
obtained by expanding $e^{-2\pi i\wt v}/ (1 - e^{-2\pi i\wt v})^2$
in positive powers of $e^{2\pi i \wt v}$. In this case the sum over $p$ in 
\refb{efo2} is bounded from below, and in
\refb{egg1int}
we need to take a different
contour $\wh C$ to get the correct formula for the degeneracy:
\ben \label{ecchat}
 \wt \rho_2=M_1, \quad  \wt\sigma_2 = M_2, \quad
 \wt v_2 = M_3, \nonumber \\
 0\le  \wt\rho_1\le 1, \quad
0\le  \wt\sigma_1\le N, \quad 0\le  \wt v_1\le 1\, ,
 \een
where $M_1$, $M_2$ and $M_3$ are large positive numbers
with $M_3<<M_1,M_2$.

It turns out that walls of marginal stability, -- codimension
one subspaces of the asymptotic moduli space on which
the BPS mass of the 
system becomes equal to the sum of masses of two  or more
other
states carrying the same total charge, -- are quite generic for
quarter BPS states in $\NN=4$ supersymmetric string
theories\cite{9712211}. The shapes of these walls have been
analyzed in detail in \S\ref{smarginal}. 
For fixed values of the other moduli, the
marginal stability walls in the axion-dilaton moduli space
are either circles or straight lines, with the property
that they never intersect in the interior of the upper half plane,
but can intersect either at $i\infty$ or at rational points on the real
axis. Thus a given domain bounded by the walls of marginal 
stability has vertices either at rational points on the real axis or
at $i\infty$. As we vary the other moduli, the shapes of the walls
in the axion-dilaton moduli space changes, but the vertices do not
change. Thus every domain may be given an invariant characterization
by specifying the vertices of the domain. While comparing these domains 
for different states carrying different charges and / or different 
asymptotic values of the other moduli, we shall call them the same if 
their vertices in the axion-dilaton moduli space coincide.

We expect the 
spectrum of quarter BPS states
to change discontinuously as
the asymptotic moduli fields pass through any of these 
walls of marginal 
stability.\footnote{What we refer to as a wall 
is actually a codimension
one subspace of the full moduli space. If a state becomes
marginally stable on a surface of codimension $\geq 2$, then we can
always move around this subspace in going from one point to 
another and hence the spectrum cannot change discontinuously.}
Thus the expression for the degeneracy given above holds 
only in
a finite domain of the moduli space, bounded by the walls of 
marginal stability. More precisely, we have the formula for the 
degeneracy in two different domains separated by the 
domain wall on which the axion field in the second description vanishes. 
We denote by $\RR$ the domain in which the original formula
\refb{egg1int}, \refb{ep2kk} is valid, and by $\LL$ the domain
in which the same formula with the modified integration contour
given in \refb{ecchat} is valid.
An important question is: 
how does the degeneracy formula 
look inside other domains? 
It turns out that invariance of the theory under S- and T-duality
symmetries gives non-trivial information about the degeneracy
formula inside other domains and for other charge
vectors. However before describing the logic
behind this analysis, we need to say a few words about duality
invariance.

First note that that although 
\refb{egg1int} has
been expressed as a function of the T-duality invariant combinations
$P^2$, $Q^2$ and $Q\cdot P$, it was
derived initially for special
charge vectors $\vec Q$, $\vec P$ described in \refb{echvec},  and
extended to more general charge vectors 
of the form
\refb{nne1.6newrp} in \S\ref{sadditional}.
Even then this is not the most general charge
vector of the theory. One can try to extend this to more general
charge vectors using T-duality symmetry of the theory. However
here we encounter two problems. First of all
two charge vectors carrying the same values of $P^2$, $Q^2$
and $Q\cdot P$ may not necessarily
be related by a T-duality
transformation.\footnote{Generically they are related by a
continuous T-duality transformation but only a discrete subgroup
of this is a genuine symmetry of the theory.} In that case the
degeneracy of states for these two charge vectors could be
different. An example of this is that a state carrying
fundamental heterotic string
winding charge $w'$ along
$S^1$ with $w'\ne 0$ mod $N$ can never be related to a state carrying
$w'=0$ mod $N$ even if they have the
same values of $Q^2$, $P^2$ and $Q\cdot P$,
since the former carries twisted sector electric
charge and the latter carries untwisted sector electric charge.
Thus
the degeneracy formula we have derived holds at best for charges which 
are in the same T-duality orbit as the general charge vector 
\refb{nne1.6newrp}.
Second,
even though we expect T-duality to be  a symmetry of the theory,
we should remember that
it acts not
only on the charges but also on the asymptotic moduli. 
Had the spectrum been
independent of the asymptotic moduli, we could have demanded
that the spectrum remains invariant under T-duality transformation
of the charges. However if 
a T-duality
transformation takes the asymptotic moduli fields 
across a wall of marginal stability, then
all we can say is that the spectrum remains unchanged under a
simultaneous T-duality transformation of the moduli fields
and the charges. We show in \S\ref{sdualtwo} 
that a T-duality transformation
on the moduli space preserves the domains bounded by marginal
stability walls in the sense that it preserves the 
vertices of the domain while changing the shapes of the
walls.
Hence we do expect that the degeneracy
formula within a given domain characterized by a fixed set
of vertices will be invariant under a T-duality transformation
acting on the charges only.
In the subpace $\VV$ of electric and magnetic charges 
introduced in \refb{e2dcharge}, -- spanned by the momenta, 
fundamental string winding charge, H- and Kaluza-Klein 
monopole charges 
along the circles $\wh S^1$ and $S^1$ in the second description, 
-- the 
T-duality orbit of \refb{nne1.6newrp} has been analyzed at the end of 
\S\ref{sadditional}.
We find that the orbits contain
four dimensional electric and magnetic charge vectors 
$\vec Q$ and $\vec P$ in this subspace  satisfying charge 
quantization
laws and the following additional restrictions: 
\begin{enumerate}
\item The electric
charge vector must correspond to the charge carried by a $g$
twisted state, \i.e.\ in the second description
the fundamental string winding charge $-k_6$ along
$S^1$ must be 1 mod $N$.
\item  The Kaluza-Klein monopole
charge $l_5$ associated with the circle $\wh S^1$ in the second 
description must 
be 1 mod $N$.
\item The electric and the magnetic charge vectors 
$Q=\pmatrix{k_3\cr k_4\cr k_5\cr k_6}$ and 
$P=\pmatrix{l_3\cr l_4\cr l_5\cr l_6}$ must satisfy the
primitivity conditions:
\ben \label{egcd1}
&&\hbox{g.c.d.}(Nk_3 l_4 -Nk_4 l_3 , k_5 l_6 -k_6  l_5 ,
k_3 l_5 -k_5 l_3 +k_4 l_6  -k_6 l_4 ) = 1\, . 
\een
\end{enumerate}
These are necessary conditions for the charge vectors to be in the
orbit of \refb{nne1.6newrp}, but we have not proven that these
conditions are sufficient.
Thus our formula \refb{egg1int}, \refb{ep2kk} 
holds in the domain $\RR$ for  charge 
vectors satisfying these criteria, and possibly some
additional criteria.\footnote{Possible dependence of the
degeneracy formula on invariants other than the continuous T-duality
invariants have been anticipated in \cite{0401049}}.
For the same charge vectors, the
formula \refb{egg1int} with the contour $\CC$ replaced by a new
contour $\wh \CC$ given in \refb{ecchat}, hold in the domain
$\LL$.

For the choice $\MM=K3$, \i.e.\ theories for which
the second
description is based on orbifolds of heterotic string theory,
we can relax the conditions somewhat
by taking an
initial configuration with multiple D5-branes\cite{0605210}, with
the number of D1 and D5-branes being relatively prime.
This analysis has been described in
appendix \ref{ssym} and, after being combined with the analysis
of \S\ref{sadditional}, shows that our degeneracy formula
\refb{egg1int}, \refb{ep2kk} holds in the domain $\RR$
for a general charge vector of the form
\be \label{enewseed}
Q = \pmatrix{k_3\cr k_4 \cr k_5 \cr -1}, \qquad
P = \pmatrix{l_3\cr l_4 \cr l_5 \cr 0}, \qquad k_4\in \ZZZ/N,
\quad k_i, l_i\in \ZZZ\quad \hbox{otherwise}, \quad
\hbox{g.c.d.}(l_3, l_5)=1
\, .
\ee 
Thus the degeneracy formula will continue to hold for any charge
vector related to \refb{enewseed} by a T-duality transformation.
This allows us to relax the condition 2 on $l_5$  given above.
This also relaxes the condition \refb{egcd1} to
\ben \label{egcd3}
&& \hbox{g.c.d.} \{ k_i l_j - k_j l_i, \, N k_4 l_s - N k_s l_4,
\, k_4 l_6 - k_6 l_4; \quad i,j=3,5,6, \, s=3,5\} = 1\, . 
\een
It may also be possible to relax these conditions in
a similar manner for orbifolds of type IIB string
theory on $T^4\times \wt S^1\times S^1$ by taking multiple
D5-branes as in the case of $K3\times \wt S^1\times S^1$. However
a careful analysis, taking into account the dynamics of Wilson
lines on multiple D5-branes, has not been carried out so far.

Given this, we can now study the consequences of S-duality
invariance. It turns out that unlike T-duality,
typically an S-duality transformation takes us from
one domain to another. Thus invariance under S-duality can be used
to derive the formula for the degeneracy in domains other than
the original domain where it was computed. 
We find that
the degeneracies inside the other domains
formally look the same as \refb{egg1int}
but the
contour $\CC$ over which we need to carry out the integration is
different in different domains.
On the other
hand there are
a few S-duality transformations which preserve a given domain.
These must be symmetries of the degeneracy formula, -- namely
two charge vectors related by such an S-duality transformation must
have the same degeneracy. This is indeed borne out by explicit
computation. 
These issues have been discussed in detail in \S\ref{sdualtwo}.

Finally let us turn to the comparison between statistical entropy
-- defined as the logarithm of the degeneracy of states -- with
the black hole entropy. For this we need to extract the
behaviour of the degeneracy $d(\vec Q,\vec P)$ for large charges.
By performing one of the integrals in \refb{egg1int}
by picking up residues at
the poles of the integrand, and other two integrals 
by saddle point approximation, we can extract this
behaviour. The result is that up
to first non-leading order, the entropy is given by extremizing a
statistical entropy function
\be\label{enn18int}
-\wt\Gamma_B(\vec\tau) = {\pi\over 2 \tau_{2}} \, |Q -\tau P|^2
- \ln g(\tau) -\ln g(-\bar\tau)
- (k+2) \ln (2\tau_{2}) + \hbox{constant} + \OO(Q^{-2})
\ee
with respect to real and imaginary parts of the complex variable
$\tau=\tau_1+i\tau_2$. The functon $g(\tau)$
and the constant $k$ are the same as the ones which appear
in the expression \refb{eh10bint} for the function $\phi(a,S)$
multiplying the coefficient of the Gauss-Bonnet term in the low
energy effective action.

With the identification $\tau=u_a + i u_S$ the statistical entropy
function \refb{enn18int} matches the black hole entropy function
of the same theory given in \refb{es11}. Thus the statistical entropy
and the black hole entropy, given by the values of the corresponding
entropy functions at their extrema, also agree to this order.

Since the expression for $d(\vec Q, \vec P)$ changes across 
the walls of marginal stability, one might wonder how this affects
the large $\vec Q,\vec P$  behaviour of
$d(\vec Q,\vec P)$. One finds that these changes
are exponentially small compared to the leading contribution.
Nevertheless one could ask if the changes in $d(\vec Q,\vec P)$
across walls of marginal stability can be seen on the black hole
side.  
It turns out that
this is indeed possible. 
First of all, due to the attractor mechanism the entropy of a single
centered extremal black hole of the kind analyzed in \S\ref{s2.2}
does not change as the asymptotic values of the moduli vary across
a wall of marginal stability. However
on the black hole side the contribution
to the total entropy comes not only from single centered
black holes but
also from multi-centered black holes carrying the same total
charge. It turns out that as we cross a wall of marginal stability,
typically a two centered black hole solution (dis)appears, \i.e.\
these solutions exist only on one side of the marginal stability
wall\cite{0005049,0010222,0101135,0304094,0702146,0705.2564}. 
As a result there is a change in the total entropy on the black hole
side as well. This change precisely agrees with the
result predicted from   the
exact degeneracy formula\cite{0705.3874,0706.2363}. 
We shall illustrate
this in \S\ref{smulti}.

\subsectiono{The counting} \label{s3.2}

We shall now describe the counting of BPS states of the
configuration described in \refb{echvec}, -- this will eventually lead to
the expressions \refb{egg1int}, \refb{ep2kk} for $d(\vec Q, \vec P)$.
For this configuration
the charges $(\vec Q,\vec P)$ are labelled by the
set of integers $Q_1$, $n$ and $J$. The other two charges, namely
the 
number of D5-branes along $\MM\times S^1$ and the
number of Kaluza-Klein monopoles associated with the circle
$\wt S^1$ in the first description have 
been taken to be 1. 
We shall denote by $h(Q_1,n,J)$ the number of bosonic supermultiplets
minus the number of fermionic supermultiplets carrying quantum
numbers $(Q_1,n,J)$. Computation of $h(Q_1,n,J)$ is best done
in the weak coupling limit of the
first description of the system where 
the quantum numbers  $n$ and $J$ can
arise from three different sources\cite{0605210}: the
excitations of the Kaluza-Klein monopole   carrying
certain amount of momentum $-l_0'/N$ along $S^1$,\footnote{A
Kaluza-Klein monopole associated with the compactification
circle $\wt S^1$ cannot carry any momentum along $\wt S^1$
since the solution is invariant under translation along
$\wt S^1$\cite{9705212}.} 
the  overall 
motion of the D1-D5 system in the background of the
Kaluza-Klein monopole  carrying certain amount of
momentum $-l_0/N$ along $S^1$ and $j_0$ along $\wt S^1$ and
the  motion of the $Q_1$ D1-branes  in the plane of the
D5-brane carrying  total  momentum $-L/N$ along $S^1$ and $J'$
along $\wt S^1$. Thus we have
\be\label{esi1}
  l_0'+l_0+
L = n, \qquad j_0+  J'=J\, .
\ee
Let
\be\label{esi8}
f(\wt \rho,\wt \sigma,\wt v) = \sum_{Q_1, n, J} 
(-1)^{J+1} \, h(Q_1,n, J)
e^{2\pi i ( \wt \rho n + \wt \sigma Q_1/N +\wt v J)} \, ,
\ee
denote the partition function of the system. Then in the
weak coupling limit we can ignore the interaction between the
three different sets of degrees of freedom described above, and
$f(\wt \rho,\wt \sigma,\wt v)$ is obtained as a product of
three separate partition functions:\footnote{Even though the mutual 
interaction between these three systems vanish, the individual systems may 
be interacting. In particular we shall see that the dynamics of the D1-D5 
system in the Kaluza-Klein monopole background has a strongly interacting 
component that is responsible for binding the D1-D5 system to the 
Kaluza-Klein monopole.} 
\bea{esi9} f(\wt \rho,\wt \sigma,\wt v ) &=& 
-{1\over 64} \, \left(
\sum_{Q_1,L,J'} (-1)^{J'} \, d_{D1}(Q_1,L,J') 
e^{2\pi i ( 
\wt \sigma Q_1 /N +\wt \rho L + \wt v J')}\right)
\nonumber \\
&& \, \left(\sum_{l_0, j_0} (-1)^{j_0}
d_{CM}(l_0, j_0) e^{2\pi i l_0\wt\rho + 2\pi i
j_0\wt v}\right) \, 
\left(\sum_{l_0'} d_{KK}(l_0') e^{2\pi i l_0' \wt\rho} \right)\, ,
\eea
where $d_{D1}(Q_1,L,J')$ is the degeneracy of $Q_1$ D1-branes
moving in the plane of the D5-brane carrying momenta $(-L/N,J')$
along $(S^1,\wt S^1)$, $d_{CM}(l_0,j_0)$ is
the  degeneracy associated with the overall
motion of the D1-D5 system
in the background of the Kaluza-Klein monopole carrying
momenta $(-l_0/N, j_0)$ along $(S^1,\wt S^1)$ and
$d_{KK}(l_0')$ denotes the degeneracy associated with
the excitations of a Kaluza-Klein monopole carrying momentum
$-l_0'/N$ along $S^1$. The factor of 1/64
in \refb{esi9} accounts for the fact that a single quarter BPS
supermultiplet has 64 states. In each of these sectors we count
the degeneracy weighted by $(-1)^F$ with $F$ denoting space-time
fermion number of the state, except for the parts obtained by
quantizing the fermion zero-modes associated with the broken
supersymmetry generators. Since a Kaluza-Klein monopole
in type IIB string theory on $\MM\times \wt S^1\times 
S^1/\ZZZ_N$ breaks 8
of the 16 supersymmetries, quantization of the fermion
zero modes associated with the broken supersymmetry generators give
rise to a $2^{8/2}=16$-fold degeneracy with equal number of bosonic and
fermionic states. This appears as a factor
in $d_{KK}(l_0')$. Furthermore since a D1-D5 system in the
background of such 
a Kaluza-Klein monopole 
breaks 4 of the 8 remaining supersymmetry generators,
we get, from the associated fermion zero modes, a 
$2^{4/2}=4$-fold degeneracy 
appearing as a factor in $d_{CM}(l_0, j_0)$,
with equal number of fermionic and bosonic
states. 
This factor of $16\times 4$
cancel the $1/64$ factor in \refb{esi9}. After separating out this
factor, we count the contribution to the degeneracy from the rest
of the degrees of freedom 
weighted by a factor of $(-1)^F$.

We shall now compute each of the three pieces, $d_{KK}(l_0')$,
$d_{CM}(l_0, j_0)$ and $d_{D1}(Q_1,L,J')$ separately. Before we
go on however, we shall fix some conventions. 
The world-volume of the Kaluza-Klein monopole as well as that
of the D5-brane is 5+1 
dimensional with the five spatial directions lying along $\MM
\times S^1$. By taking the size of $\MM$ to be much smaller
than that of $S^1$ we can regard these as 1+1 dimensional
world-sheet theories, obtained by dimensional reduction of the original
5+1 dimensional theory on $\MM$. We shall follow this viewpoint
throughout the rest of this section although we shall often refer to this 
as the world-volume theory. In particular left- and right-moving
modes on the world-volume theory will refer respectively
to the modes which
move to the left- and right along $S^1$.
The D1-brane world volume theory
is of course naturally 1+1 dimensional. We
shall choose a convention in which 
the four unbroken supersymmetry generators of the full
configuration act on the right-moving modes in the 
world-volume theory. This in turn means that the supersymmetry
transformation parameters themselves are represented by left-chiral
spinors since the transformation laws of the scalars, being proportional
to $\bar \epsilon \psi$, is non-zero only if the supersymmetry
transformation parameter $\epsilon$ and the fermion field $\psi$
have opposite chirality. In contrast if a supersymmetry is
spontaneously broken then the associated goldstino fermion field
$\psi$ on the world-volume has the same chirality as the
transformation parameter $\epsilon$ due to the transformation
law $\delta\psi\propto \epsilon$.

\subsubsection{Counting states of the 
Kaluza-Klein monopole} \label{s1xxx}

We consider type IIB string theory 
in the background 
$\MM\times TN\times S^1$
where $TN$ denotes Taub-NUT space described by the
metric\cite{eguchi}
\be\label{tnutgeom}
ds^2 = \left(1+\frac{R_0}{r}\right) 
\left( dr^2 +  r^2 ( d\theta^2 + \sin^2 \theta d\phi^2) \right)
+ R_0^2\left( 1 + \frac{R_0}{r}
\right)^{-1} ( 2\, d\psi + \cos\theta d\phi)^2
\ee
with the identifications:
\be\label{eident}
(\theta,\phi,\psi) \equiv (2\pi -\theta,\phi+\pi, \psi+{\pi\over 2})
\equiv (\theta,\phi+2\pi,\psi+\pi)\equiv (\theta,\phi,\psi+2\pi)\, .
\ee
Here $R_0$ is a constant determining the size of the Taub-NUT space.
This describes
type IIB string theory
compactified on 
$\MM\times \wt S^1\times  S^1$ in the presence of
a Kaluza-Klein monopole, with $\wt S^1$ identified with the
asymptotic circle of the Taub-NUT space labeled by the coordinate
$\psi$ in \refb{tnutgeom}.
The metric \refb{tnutgeom}
admits a normalizable self-dual harmonic form $\omega$, given
by\cite{brill,pope}
\be \label{nneomegadefrp}
\omega \propto
{r\over r +R_0}
d\sigma_3 + {R_0\over (r+R_0)^2}dr
\wedge \sigma_3\, , \qquad \sigma_3 \equiv
\left (d\psi + {1\over 2} \cos\theta d\phi\right)\, .
\ee

We now  
take an
orbifold of the theory by a $\ZZZ_N$ group
generated by the transformation $g$. 
Our goal is to compute the degeneracy
of the half-BPS states of the Kaluza-Klein 
monopole carrying momentum $-l_0'/N$ along $S^1$.
For $\MM=K3$,  the world-volume supersymmetry on the 
Kaluza-Klein monopole will be chiral
since the supersymmetry
generators of type IIB string theory
on K3 are chiral. According to our
convention there will be eight left-chiral 
supersymmetry transformation
parameters,
acting only on the right-moving
degrees of freedom. Thus the
BPS states of the Kaluza-Klein monopole will correspond to
states in this field theory where the right-moving
oscillators are in their ground state. 
For $\MM=T^4$ the world-volume theory of the Kaluza-Klein 
monopole will have eight left-chiral and eight right-chiral
supersymmetry transformation parameters. However right-chiral
supersymmetries will be broken once we take the $\ZZZ_N$
orbifold, and the unbroken supersymmetries of the theory will
again come from the left-chiral spinors. 
Since the latter act on
the right-moving modes, the BPS condition will again require
that the right-moving oscillators are in their ground state.

In order to
count these states we proceed as follows:
\begin{enumerate}
\item First we determine the spectrum of massless fields in the
world-volume theory of the Kaluza-Klein monopole 
solution described above. In particular we show that the world-volume
theory always contains eight right-moving massless scalar fields and
eight right-moving massless fermion fields. In addition for $\MM=K3$
there are twenty four left-moving massless scalar fields whereas for
$\MM=T^4$ there are eight left-moving massless scalar fields and
eight left-moving massless fermion fields.
\item Next we identify the transformation laws of various fields
under the orbifold group generator $\wt g$. We show that all the
right-moving fields are $\wt g$ invariant, whereas the action of $\wt g$
on the left-moving massless bosonic (fermionic) fields is identical to the
action of $\wt g$ on the even (odd) degree harmonic forms of $\MM$.
\item We now use this information to determine all the $g$-invariant
modes on the Kaluza-Klein monopole. This will essentially
require that a field that picks up a $\wt g$ phase $e^{2\pi i k/N}$
must carry momentum $n-k/N$ ($n,k\in \ZZZ$) 
along $S^1$ so that the phase
obtained due to translation along $S^1$ cancels the $\wt g$ phase.
\item Finally we count the number of ways a total momentum 
$-l_0'/N$ along $S^1$ can be partitioned into these various 
$g$-invariant modes,
taking into account that part of this momentum comes from the
momenum of the Kaluza-Klein monopole vacuum without any
excitations. This vacuum momentum is calculated by mapping the
Kaluza-Klein monopole to a fundamental string state in a dual
description of the theory.
\end{enumerate}

We begin by analyzing
the spectrum of massless fields in the world-volume
theory. First of all
there are
three non-chiral massless
scalar fields  associated with oscillations
in the three transverse directions of the Kaluza-Klein
monopole solution. There are 
two additional non-chiral scalar 
fields obtained by reducing the two 2-form fields of type IIB
string theory along
the harmonic
2-form \refb{nneomegadefrp}.  Finally,
the self-dual four form field of type IIB theory, 
reduced along the tensor product of the harmonic
2-form \refb{nneomegadefrp} and a harmonic 2-form on $\MM$, can give
rise to a chiral scalar field on the world-volume. The chirality of
the scalar field is correlated with whether the corresponding
harmonic 2-form on $\MM$ is self-dual or anti-self-dual.
Since $T^4$ has three self-dual and 
three anti-selfdual harmonic 2-forms
and $K3$ has three self-dual and nineteen 
anti-selfdual harmonic 2-forms,
we get  
3 right-moving and $P$ left-moving scalars where $P=3$ for 
$\MM=T^4$ and 19 for $\MM=K3$. Thus we have
altogether 8 right-moving massless scalar fields 
and $P+5$ left-moving massless scalar
fields on the world-volume of the Kaluza-Klein monopole.

Next we turn to the spectrum of massless fermions
in this world-volume theory.
These typically arise from the Goldstino fermions associated
with broken supersymmetry generators.
Since type IIB string theory on K3 has 16 unbroken 
supersymmetries\footnote{In this section we shall refer
to unbroken supersymmetries in various context. Some time it
may refer to the symmetry of the given compactification, 
and some time it
will refer to  the symmetry of a given brane
configuration.  Also some time it may refer to the number of unbroken 
supersymmetries before taking the $\ZZZ_N$ orbifold and at other times 
it may refer to the number of unbroken supersymmetries in the orbifold 
theory. The reader must carefully examine the context in which the 
symmetry is being discussed, since the number of unbroken generators and
their action on various fields depend crucially on this
information.}
of which 8 are broken in the presence of the Taub-NUT space,
we have 8 massless goldstino fermion fields on the world-volume
of the Kaluza-Klein monopole. Since 
according to our convention the eight unbroken supersymmetry
transformation parameters are left-chiral on $S^1$, the
broken supersymmetry transformation parameters must be
right-chiral. As a result
the goldstino fermion fields associated with
broken supersymmetries are also
right-moving on the 
world-volume. 
On the other hand if we take type IIB on $T^4$ 
we have altogether 32
unbroken supercharges of which 16 are broken in the presence of the
Taub-NUT space. 
This produces 16 goldstino fermion fields on the world-volume of the 
Kaluza-Klein
monopole.
Since type IIB on $T^4$ is a non-chiral theory,
eight of these fermion fields are right-moving and eight 
are left-moving. 

To summarize, the world-volume theory describing the dynamics
of the Kaluza-Klein monopole always
contains 8 bosonic and 8  fermionic right-moving massless fields.
For $\MM=K3$ the world-volume theory has 24 left-moving
massles bosonic fields and no left-moving massless
fermionic fields whereas for $\MM=T^4$ the world-volume theory
has 8 left-moving bosonic and 8
left-moving fermionic fields. This is consistent with the fact the under 
the duality transformation that takes us from the first to the second 
description of the theory, the Kaluza-Klein monopole associated with $\wt 
S^1$ is mapped to a fundamental heterotic (type IIA) string wrapped on 
$S^1$ for $\MM=K3$ ($\MM=T^4$). 

We shall now determine the $\wt g$
transformation properties of these modes. For this we note that
irrespective of whether $\MM$ is $T^4$ or $K3$, after taking the
$\ZZZ_N$ orbifold the unbroken supersymmetries of the theory
are in one to one correspondence with those of type IIB string theory
on $K3\times \wt S^1\times S^1$. Thus
$\wt g$ commutes with the supersymmetries of type IIB on
$K3$. Half of these $\wt g$ invariant supersymmetries are broken
in the presence of Kaluza-Klein
monopole and give rise to right-moving goldstino fermions
on the world-volume of the Kaluza-Klein monopole.
Thus these fermions must be neutral under $\wt g$. The other half
of the $\wt g$ invariant supersymmetry generators, which remain unbroken 
in the
presence of the Kaluza-Klein monopole, transform the right-moving
fermions into right-moving world-volume scalars. Thus the eight
right-moving scalars must also be invariant under $\wt g$.
Five of the left-moving
scalars, associated with the  3 transverse  degree of freedom
and the modes of the 2-form fields along the harmonic 2-form
of $TN$ are
also invariant under $\wt g$ since
$\wt g$ acts trivially on the Taub-NUT space. 
The action  of $\wt g$ on the other $P$ left-moving
scalars, associated with the modes of the 4-form field along the
tensor product of the harmonic 2-form of $TN$ and the $P$ left-handed
harmonic
2-forms of $\MM$, is represented
by the action of $\wt g$ on the $P$
left-handed 2-forms on $\MM$.  
This completely determines the action of $\wt g$ on all the
$P+5$ left-moving scalars. Now
it has been shown in appendix 
\ref{s0} that 
$\wt g$ leaves invariant the harmonic 0-form, 4-form
and all the three right-handed 2-forms on $\MM$. 
Associating these five $\wt g$-invariant harmonic forms with the
five $\wt g$ invariant left-moving scalars found above, 
we can represent
the net action
of $\wt g$ on the $(P+5)$ left-handed scalar fields  
by the action of $\wt g$
on the $(P+5)$ even degree harmonic forms of $\MM$.
 
What remains is to determine the action of $\wt g$ on the left-moving
fermions.
We shall now show that this
can be represented by the action of $\wt g$
on the harmonic 1- and 3-forms of $\MM$. For $\MM=K3$ there
are no 1- or 3-forms and no left-moving fermions on
the world-volume of the Kaluza-Klein monopole. Hence the result
holds trivially. 
For $\MM=T^4$ there are eight left-moving fermions and eight
right-moving fermions
associated with the sixteen  supersymmetry
generators which are broken in the presence of
a Kaluza-Klein monopole in type IIB string theory on $T^4
\times \wt S^1\times S^1$. 
Although we have already argued that the right-moving fermions
are $\wt g$ neutral, let us forget this result for a while and analyze
the $\wt g$ transformation properties of the full set of 16 fermions.
Clearly these transform in the spinor
representation of the tangent space $SO(4)_\parallel$ 
group associated
with the $T^4$ direction.
Now $\wt g$ is an element of this group describing $2\pi/N$ rotation
in one plane and $-2\pi/N$ rotation in an orthogonal plane.
Translating this into the spinor representation we see that $\wt g$
must leave half of the sixteen fermions invariant, rotate two
pairs of fermions by $2\pi/N$ and rotate the other two pairs of fermions
by $-2\pi/N$. Now we use the information that the right-moving
fermions are neutral under $\wt g$. Thus the action of $\wt g$ on the
left-moving fermions is to rotate two pairs of fermions by 
$2\pi/N$ and another two pairs of fermions by $-2\pi/N$.
This is identical to the action of
$\wt g$ on the harmonic 1- and 3-forms of $T^4$ given
in \refb{eoneform}
and \refb{ethreeform}:
\ben\label{eoneformpre}
&& dz^1 \to e^{2\pi i/N} dz^1, \quad dz^2\to e^{-2\pi i/N} dz^2\, ,
\nonumber \\
&& dz^1\wedge dz^2\wedge d\bar z^1\to e^{-2\pi i/N}\,
dz^1\wedge dz^2\wedge d\bar z^1, \quad
dz^1\wedge dz^2\wedge d\bar z^2\to e^{2\pi i/N}\,
dz^1\wedge dz^2\wedge d\bar z^2, \nonumber \\
\eea
where $(z^1,z^2)$ denote complex coordinates on $T^4$.

Thus the problem of studying the $\wt g$ transformation properties
of the left-moving bosonic and fermionic degrees of freedom on the 
world-volume
reduces to the problem of finding
the action of $\wt g$ on the even and odd degree
harmonic forms of $\MM$. Let $q_s$ be
the difference between the number of even
degree harmonic forms and odd degree harmonic forms, 
weighted
by $\wt g^s$. It has been shown in appendix \ref{s0} 
(see
the 
discussion below \refb{eqrsrev}) that $q_s$ is equal to $Q_{0,s}$ defined 
in 
\refb{eqrsrevpre}. Combining this with the results of our previous 
analysis we now get
\bea{e3xxx}
Q_{0,s} &=& \hbox{number of left handed 
bosons weighted by $\wt g^s$}
\nonumber \\
&& - \hbox{number of left handed fermions weighted by $\wt g^s$}
\, 
\eea
on the world-volume of the Kaluza-Klein monopole.
Let $n_l$ be the number of left-handed bosons minus fermions
carrying $\wt g$ quantum number $e^{2\pi i l/N}$. Then
we have from \refb{e3xxx}, \refb{eqrsrevpre}
\be\label{e4xxx}
n_l = {1\over N} \, \sum_{s=0}^{N-1} 
e^{-2\pi i l s/N}
\, Q_{0,s}\, =
\sum_{s=0}^{N-1}
e^{-2\pi i l s/N}
\, \left(c^{(0,s)}_0(0) + 2 c^{(0,s)}_1(-1)\right).
\ee
Clearly $n_l$ is invariant under $l\to l+N$.

This finishes our analysis of the spectrum of massless fields in
the world-volume theory of the Kaluza-Klein monopole.
We now turn to the problem of counting the spectrum of BPS
excitations of the Kaluza-Klein monopole. 
First of all
note that
since there are eight right-moving fermions neutral under $\wt g$,
the zero modes of these fermions are $\ZZZ_N$ invariant.
These eight fermionic zero modes may be 
associated with the eight supersymmetry generators of type IIB on
$(\MM\times S^1)/\ZZZ_N$ which are broken in the presence of the 
Kaluza-Klein monopole.
Upon quantization this produces a 16-fold degeneracy of states
with equal number of bosonic and fermionic states.
This is the correct degeneracy of a single irreducible
short multiplet representing half BPS states in type IIB string
theory compactified on $\MM\times S^1/\ZZZ_N$, and will eventually
become part of the 64-fold degeneracy of a 1/4 BPS supermultiplet
once we tensor this state with the state of the D1-D5 system.
Thus in computing the index we are interested in, we must associate
weight one with each of the sixteen states irrespective of whether
the state is bosonic or fermionic.
Since supersymmetry
acts on the right-moving sector of the world-volume theory, 
BPS condition requires that all the non-zero mode
right-moving oscillators are in their ground state.
Thus the 
spectrum of BPS states is obtained by taking the tensor product
of the irreducible 16 dimensional supermultiplet 
with either fermionic or bosonic
excitations involving the left-moving degrees of freedom on the
world-volume of the Kaluza-Klein monopole. We shall denote
by $d_{KK}(l_0')/16$ the degeneracy of states associated with
left-moving oscillator excitations carrying total  
momentum $-l_0'/N$, weighted by $(-1)^{F_L}$. 
Thus $d_{KK}(l_0')$
measures the total degeneracy of half-BPS states weighted by
$(-1)^{F'}$ where for a given half-BPS supermultiplet
$F'$ denotes the fermion number of the middle helicity
state of the supermultiplet. 

In order to calculate $d_{KK}(l_0')$ we need to count the number
of ways the total momentum $-l_0'/N$ can be distributed among
the different $\ZZZ_N$ invariant
left-moving oscillator excitations. 
Since a mode carrying momentum $-l/N$ along
$S^1$ picks up a phase of $e^{-2\pi i l/N}$ under 
$2\pi$ translation along $S^1$, it must pick up a phase of 
$e^{2\pi i l/N}$ under $\wt g$. Thus the number of 
left-handed bosonic
minus fermionic modes carrying momentum
$-l/N$ along $S^1$ is equal to the number $n_l$ 
given in eq.\refb{e4xxx}. The
number $d_{KK}(l_0')/16$ can now be identified as the number
of different ways the total momentum $-l_0'/N$ can be distributed
among different oscillators, there being $n_l$ oscillators 
carrying momentum $-l/N$. This gives
\be\label{e5xxx}
\sum_{l_0'} d_{KK}(l_0') e^{2\pi i \wt\rho l_0'} 
= 16\, e^{2\pi i N\, C\wt\rho}\, \prod_{l=1}^\infty (1 -  
e^{2\pi i l\wt\rho})^{-n_l} \, .
\ee
The constant $C$ represents the $l_0'/N$ quantum number
of the vacuum of the
Kaluza-Klein monopole when all oscillators are in their ground state.
In order to determine $C$ let us consider the 
second description of the
system where the Kaluza-Klein monopole 
gets mapped to an elementary
heterotic or type IIA string wound
along $S^1$, and  $C$ represents the contribution
to the ground state $L_0$ eigenvalue 
from all the left-moving oscillators.
If $\wh g$ denotes the
image of $\wt g$ in this description, then the elementary
string wound once along $S^1$ is in the sector twisted by 
$\wh g$. Since the modes of the
Kaluza-Klein monopole get mapped to the degrees of freedom of the
fundamental heterotic or type IIA string,  there are $n_l$ left moving
bosonic minus fermionic modes which pick up a 
phase of $e^{2\pi i l/N}$ under the action of $\wh g$.
Since
a bosonic and a fermionic mode twisted by a phase of $e^{2\pi i \vp}$
for $0\le \vp\le 1$
gives a contribution of $-{1\over 24}+ {1\over 4} \, \vp\, (1-\vp)$ 
and ${1\over 24}-
{1\over 4} \, \vp\, (1-\vp)$ respectively 
to the ground state $L_0$ 
eigenvalue,\footnote{We are counting the contribution from a mode and
 its complex conjugate separately.}
we
 have
 \ben\label{ecv1}
 C &=& -{1\over 24} \sum_{l=0}^{N-1} \, n_l + {1\over 4}\, 
 \sum_{l=0}^{N-1} \, n_l \, {l\over N} \, \left( 1 -{l\over N}\right)\, 
 \nonumber \\
&=& -{1\over 24N} \sum_{s=0}^{N-1}\, Q_{0,s}
\sum_{l=0}^{N-1} \, 
  e^{-2\pi i ls/N}\, +{1\over 4N} \, \sum_{s=0}^{N-1}\, Q_{0,s}\, 
  \sum_{l=0}^{N-1} \, {l\over N} \, \left( 1 -{l\over N}\right)
  e^{-2\pi i ls/N} \nonumber \\
  \een
where in the last step we have used
 the expression for $n_l$ given in \refb{e4xxx}. 
  The sum over $l$ can be performed separately for $s=0$ and $s\ne 0$,
  and yields the answer
\be\label{ecvalue}
C = -\wt \alpha/N\, ,
\ee
with $\wt \alpha$ defined as in \refb{enn9dpre}.
The left-right level matching condition in the
second description guarantees that $N\, C$ and 
hence $\wt\alpha$ must be an integer.
Using \refb{e4xxx},  \refb{ecvalue} 
we can rewrite \refb{e5xxx} as
\be\label{e8xxx}
\sum_{l_0'} d_{KK}(l_0') e^{2\pi i \wt\rho l_0'} 
=16\, e^{-2\pi i \wt \alpha\wt\rho}\, \prod_{l=1}^\infty (1 -  
e^{2\pi i l\wt\rho})^{-\sum_{s=0}^{N-1} 
e^{-2\pi i l s/N}
\, \left(c^{(0,s)}_0(0) + 2 c^{(0,s)}_1(-1)\right)}\, .
\ee

\subsubsection{Counting states associated with the 
overall motion of the
D1-D5 system} \label{s2xxx}

We shall now analyze the contribution to the partition
function from the overall motion of the D1-D5 system. This has two
components, -- the center of mass motion of the D1-D5 system along
the Taub-NUT space transverse
to the plane of the D5-brane, and the dynamics of the Wilson lines on
the D5-brane along $\MM$. The first component is present irrespective
of the choice of $\MM$ but the second component exits only if $\MM$
has non-contractible one cycles, \i.e.\ for $\MM=T^4$.

\noindent {\bf Dynamics of D1-D5 motion in Taub-NUT space:}
The contribution from this component is
independent of the choice
of $\MM$. We shall take  $\MM=K3$ and analyze the contribution
following  \cite{0605210}.
The D1-D5-system wrapped on $K3$ in flat transverse
space
has four massless scalar fields associated with
the four
transverse coordinates and eight massless goldstino 
fermionic fields
associated with the breaking of 
eight out of sixteen supersymmetries of type IIB
string theory on $K3$ by the D1-D5 system.
Thus when the transverse space
is Taub-NUT, we expect
the low energy  dynamics of this system to be
described by a (1+1) dimensional supersymmetric field theory 
with four scalar and eight fermion
fields, with the scalar fields taking value in the 
Taub-NUT 
target space.  
Since $\wt g$ does not act on the Taub-NUT space, the scalar fields
are $\wt g$ invariant. Furthermore, since $\wt g$ commutes with the
supersymmetry generators of type IIB string theory on K3, all the
massless fermions, being associated with the broken supersymmetry
generators, are also  $\wt g$ invariant.\footnote{This is consistent
with the fact that the unbroken supersymmetry generators
-- also $\wt g$ invariant -- transform the scalar fields to
fermions and vice versa.}

Our goal will be to calculate the spectrum of BPS
states in this theory carrying momentum $-l_0/N$ along $S^1$
and $j_0$ along the asymptotic circle $\wt S^1$ of the Taub-NUT
space after taking the $\ZZZ_N$ orbifold. 
This will be done as follows:
\begin{enumerate}
\item We first find the interpretation of the quantum 
numbers $l_0$ and $j_0$
in this (1+1)
dimensional supersymmetric field theory and then
determine the
values of $l_0$ and $j_0$ quantum numbers carried by various
world-volume fields.
\item Since a $\sigma$-model with Taub-NUT target space is an
interacting theory, we cannot carry out the counting of states by
regarding the world-volume theory as free. We show that the
world-volume theory actually contains two 
mutually non-interacting pieces,
-- a theory of free left-moving fermions and an
interacting theory of scalars and right-moving fermions.
\item The contribution to the partition function from the free left-moving
fermions is easily computed. In computing the contribution from
the scalars and the
right-moving fermions, we split the system into two
parts: the zero mode part and the non-zero mode part. By taking
the size of the Taub-NUT space to be large we argue that the
non-zero mode part can be treated essentially as a free field
theory and we can evaluate the contribution to the partition function
by simple counting.
\item The problem of studying the effect of the zero mode part can be
mapped to counting of bound states in a supersymmetric quantum
mechanics describing the motion of a superparticle in Taub-NUT
space. This problem had 
been studied earlier in
\cite{pope,9912082}. Using the results of these papers we 
compute the contribution to the partition function from the
zero modes.
\item Finally multiplying the contribution to the partition function
from different sources we get the net contribution to the partition
function from the overall motion of the D1-D5 system 
in the Taub-NUT space.
\item Note that since all the world-volume fields involved in this
analysis are neutral under $\wt g$, the orbifold projection will
force the momentum of various modes along $S^1$
to be integers. Other than that it plays no role.
\end{enumerate}

We begin by identifying the quantum numbers $l_0$ and $j_0$ in
the world-volume theory. $-l_0/N$ is, by definition, the
momentum along $S^1$. According to the point 6 above, all the
modes on the world-volume carry integer values of $l_0/N$.
Conversely, for every world-volume field all non-negative
integer values of $l_0/N$ are allowed, -- the positivity constraint
being a consequence of the BPS condition which allows only
left-moving modes carrying negative momentum along $S^1$
to be excited.

To identify the quantum number $j_0$ we need to 
examine closely the
metric of the Taub-NUT space given in \refb{tnutgeom}, \refb{eident}.
Close to the origin $r=0$ the metric reduces to that of flat space 
$\RRR^4$ 
written in terms of Euler angles $\theta, \phi, \psi$ and the radial
coordinate $\rho\equiv \sqrt r$,
while for large $r$ it is that of $\RRR^3\times \wt S^1$, with
$\wt S^1$ parametrized by the angular coordinate $\psi$ and 
$\RRR^3$
parametrized by the spherical polar coordinates $(r,\theta,\phi)$.
In terms of the coordinates
\bea{ecart}
&& x^1 = 2\sqrt r \cos{\theta\over 2}\cos\left(
\psi+{\phi\over 2} \right), \qquad x^2 =  
2\sqrt r \cos{\theta\over 2}\sin\left(
\psi+ {\phi\over 2} \right), \nonumber \\ &&
 x^3 =  2\sqrt r \sin{\theta\over 2}\cos\left(\psi -
{\phi\over 2}\right), 
\qquad x^4 =  2\sqrt r \sin{\theta\over 2}\sin\left(\psi -
{\phi\over 2}\right)\, 
\eea
the metric at the origin $r=0$ takes
the form of the flat Euclidean metric written in Cartesian
coordinates. As a result it has the
usual $SO(4)\equiv SU(2)_L\times SU(2)_R$ rotation symmetry
acting on the $x^i$'s as:
\be \label{etan1}
\pmatrix{x^1+ix^2 & x^3 + i x^4\cr x^3 - i x^4 & x^1 - i x^2}
\to U_L \pmatrix{x^1+ix^2 & x^3 + i x^4\cr x^3 
- i x^4 & x^1 - i x^2} U_R^T, \qquad U_L, U_R\in SU(2)\, .
\ee
It is easy to see that only the $U(1)_L\times SU(2)_R$
subgroup of this is a symmetry of the the full metric
\refb{tnutgeom}. 
The $SU(2)_R$ symmetry generated by the matrix $U_R$ 
acts as the usual rotation
group on the three dimensional space labelled by
$(r,\theta,\phi)$.
The $U(1)_L$ symmetry generated by ${\rm diag}(e^{i\eps/2},
e^{-i\eps/2})$ acts as
\be\label{eu12}
\psi\to\psi+{1\over 2}\epsilon\, ,
\ee
with no action on any of the other coordinates. From the point of view 
of an asymptotic observer 
this is just a translation along the compact circle $\wt S^1$
parametrized by $\psi$, and the
corresponding conserved charge is the quantum number $j_0/2$.
On the other hand using \refb{ecart} 
we see that near the origin
the $\psi$ translation acts as simultaneous rotation
in the 1-2 and 3-4 planes. Thus near the origin
the contribution to the $j_0$ charge can be identified
as the sum of the angular momentum in the 1-2 and 3-4 
planes\cite{0503217}.

Since the metric at the origin is the usual four dimensional
Euclidean metric, we can describe it in terms of a set of
vierbeins proportional to the identity matrix. The 
$SU(2)_L\times SU(2)_R$ transformation described in \refb{etan1}
will leave the vierbeins invariant only if they are accompanied
by a compensating $SU(2)_L^T\times SU(2)_R^T$ rotation in 
the tangent space. In particular the global $U(1)_L$ symmetry
\refb{eu12} will induce a tangent space rotation 
belonging to the $U(1)_L^T\subset SU(2)_L^T$ group, and
the quantum number $j_0$ can be interpreted as 
a $U(1)_L^T$ charge 
$j_0/2$. This
has a subtle effect on the statistics of various degrees of
freedom carrying $j_0$ charge. {}From the point of view of
an asymptotic observer, the angular momentum generators
are those of $SU(2)_R$ and hence the statistics $(-1)^F$
of an excitation is equal to $(-1)^{2J_{3R}}$. On the other
hand the same excitation, viewed from the center of the
Kaluza-Klein monopole will have statistics $(-1)^{2J_{3L}
+ 2J_{3R}} = (-1)^F (-1)^{j_0}$. Since we shall be interested
in identifying the statistics of the states from the point of
view of the asymptotic observer, but  a large part of
the actual counting of states will be done by analyzing
the modes near the center of Taub-NUT, we must take into
account this difference in our analysis.

Since the massless scalar fields describing transverse oscillations of 
the D1-D5 system 
take values in the Taub-NUT target
space, \refb{eu12} automatically 
determines the transformation laws of these scalar fields
under the $U(1)_L$ transformation and hence the $j_0$ charges carried by 
these fields. In particular at the origin of the Taub-NUT space the 
scalar fields are in one to one correspondence with the coordinates 
$x^i$, and belong to the $(2,2)$ representation of the $SU(2)_L\times 
SU(2)_R$ group. Identifying $j_0$ with $2U(1)_L$ we see that
two of 
these fields carry $j_0$ charge 1 and other two fields carry $j_0$ charge 
$-1$. 
On 
the other hand, 
since the fermions transform in the $(1,2)+(2,1)$
representation
of the tangent space $SU(2)_L^T\times SU(2)_R^T$ group,
half of the fermions are neutral under $SU(2)_L^T$
and hence also under the global $U(1)_L$. As a result these do
not carry any $j_0$ quantum number. The other half of the fermions
are neutral under $SU(2)_R$ but transform in the spinor representation of 
the $SU(2)_L$ group. Thus they carry $j_0=\pm 1$.

The world-volume field theory involving these bosons and fermions will in 
general be an 
interacting field theory since the target space metric \refb{tnutgeom} is
non-trivial.
The bosonic part of the theory is just that of a $\sigma$-model
with  Taub-NUT as the target space. The coupling of the fermions
may be determined as follows. As discussed in \S\ref{s1xxx}, type
IIB string theory on $K3\times TN$ has 8 unbroken left-chiral
supersymmetries on the 1+1
dimensional world-volume. These are singlets of the holonomy
group of $K3\times TN$ and must also commute with the generator
of translation along $\wt S^1$. Thus they carry zero $j_0$ charge.
The presence of D1-D5 system breaks 4 of these supersymmetries.
The associated goldstino fermions must be left-moving and carry
zero $j_0$ charge. Furthermore, being goldstino fermions  they must
be
non-interacting in the low energy limit. The four unbroken 
supersymmetry generators, which are also $j_0$ neutral and 
left-moving, would mix the four bosons with four right-moving
fermions carrying the same $j_0$ charge as the bosons. Furthermore
since the bosons are interacting, their superpartner 
right-moving fermions must also be interacting. Thus we have four
interacting right-moving fermions carrying $j_0$ charges $\pm 1$.
This correctly accounts for all the eight fermions and their $j_0$
charges on the D1-D5 world-volume.

To summarize, the (1+1) dimensional world-volume theory
associated with the overall motion of the D1-D5 system in the
Taub-NUT target space 
is described by a set of four free left-moving
$U(1)_L$ invariant fermion fields, together with an
interacting theory of four bosons and four right-moving
$U(1)_L$ non-invariant fermions. For a D1-D5 system
placed at the origin of the Taub-NUT space,
two of the bosons and two of the right-moving fermions carry $j_0=1$, and
the other two bosons and right-moving fermions carry $j_0=-1$.  
The unbroken supersymmetry 
transformations leave the
free left-moving fermions untouched but acts on the scalars and
the right-moving fermions. All the fields carry integer momenta
along $S^1$. The above classification of various fields into fermions
and bosons is from the point of view of a five dimensional
observer sitting at the center of Taub-NUT space, -- this is related
to the statistics measured from the point of view of the
asymptotic four dimensional observer by a multiplicative factor
of $(-1)^{j_0}$.

We now turn to the computation of the partition function.
Let us first
calculate the contribution to the partition function due to the free
left-moving
fermions. Since these fermons do not carry any $j_0$ charge, and
carry $l_0$ quantum numbers in units of $N$, their
contribution is given by:
\be\label{ezfree}
Z_{{\rm free}}(\wt \rho) \equiv {\rm Tr}_{{\rm free \, 
left-moving\,
fermions}}
( (-1)^{F} (-1)^{j_0} e^{2\pi i \wt \rho l_0} e^{2\pi i \wt v j_0}) 
=4\, \prod_{n=1}^\infty( 1 - e^{2\pi i n N \wt \rho})^4 \, ,
\ee
where $F$ denotes the total contribution to the space-time
fermion number, -- except from the fermion zero-modes
associated with the broken supersymmetry generators, --
from the point of view of an asymptotic four 
dimensional observer.
The multiplicative
factor of $4$ comes from the quantization of the free 
fermion
zero modes. The latter in turn can be interpreted as due to the four
broken supersymmetries of the D1-D5-system 
on $K3\times TN\times S^1$.

Now we turn to the interacting part of the theory. Since we are
computing an index we can assume that it does not change under
continuous variation of
the moduli
parameters. Let us take the size $R_0$ of the
Taub-NUT space to be large so that the metric is almost flat and
in a local region of the Taub-NUT space the world-volume
theory of the D1-D5 system is almost
free. In this case we should be able to compute the contribution due
to the
non-zero mode bosonic and fermonic oscillators by placing the
D1-D5 system at the origin of the Taub-NUT space
and treating them as oscillators of free fields. 
The contribution from the 
zero modes however is sensitive
to the global geometry of the Taub-NUT space and should be
computed separately.
 
Since supersymmetry acts on the right-moving bosons and fermions,
in order to get a BPS state the right-moving bosonic and fermionic
oscillators must be in their ground state. Thus as far as the
contribution due to the non-zero mode oscillators are
concerned, we only need to
examine the effect of  left-moving
bosonic oscillators carrying
momentum $-l_0/N$ along $S^1$
and angular momentum $j_0$. We have already argued that two of the four 
bosons carry $j_0=1$ and the other two bosons carry $j_0=-1$, and
that each of these bosons carry arbitrary positive integer values of
$l_0/N$. The 
contribution to the partition function from these oscillators 
can be easily computed\cite{9405117} and yields the answer  
\bea{rusosus}
Z_{{\rm osc}}(\wt \rho, \wt v) &\equiv&  {\rm Tr}_{\rm{oscillators}} 
( (-1)^{F} (-1)^{j_0}
e^{2\pi i \wt \rho l_0 + 2\pi i \wt v j_0} ) \cr
&=& 
\prod_{n=1}^\infty \frac{1}
{(1- e^{2\pi i n N \wt \rho + 2\pi i \wt v})^2
(1- e^{2\pi i n N \wt \rho - 2\pi i \wt v})^2 }\, .
\eea
In arriving at this equation we have used the fact that since these 
oscillators are bosonic from the five dimensional viewpoint, they have 
statistics $(-1)^F=(-1)^{j_0}$ from the four dimensional viewpoint.

Finally we turn to the contribution $Z_{{\rm zero}}$ to the
partition function from the bosonic and
fermionic zero modes of the interacting part of the theory. 
Since the intercting theory has four bosonic and four fermionic
fields, the dynamics of zero modes is that
of a
superparticle with four bosonic and four fermionic coordinates
moving in Taub-NUT space. 
Under the holonomy group
$SU(2)^T_L$ 
of the Taub-NUT space the fermions and hence also their
superpartner bosons
transform in a pair of spinor representations. This system is
described by
an $\NN=4$ supersymmetric quantum mechanics. 
Thus in order to look for
BPS states of the D1-D5 system we need to look for supersymmetric
ground states of this  quantum mechanics.

So far we have been working at a special point in the moduli
space of the CHL string theory where the circles $S^1$ and $\wt S^1$
are orthonormal in the asymptotic geometry. In this case the BPS mass
of the D1-D5-Kaluza-Klein monopole system is
equal to the sum of the BPS masses of the D1-D5 system and
the Kaluza-Klein monopole. As a result there is no potential
term in the D1-D5 world-volume action and analysis of bound 
states is difficult.
But this is not a generic situation. As we shall see,
once we
switch on a component of the metric that mixes $S^1$
and $\wt S^1$ we get a potential that binds
the D1-D5 system to the Kaluza-Klein monopole 
and the 
problem is easier to analyze. On the
other hand by taking the potential to be sufficiently mild we can
ensure that
the analysis of the dynamics of non-zero modes will not
be
affected by this modification. Hence $Z_{\rm free}$ and
$Z_{\rm osc}$ should remain unchanged.

The mixing between $S^1$ and $\wt S^1$ can be achieved
by replacing the $d\psi$ term
in the expression for the metric given in \refb{tnutgeom}
by $d\psi + \lambda d y$ 
where $y$ is the coordinate along $S^1$ and $\lambda$
is a small deformation parameter. This clearly remains a solution
of the equations of motion but gives an $r$ dependent contribution
to the $yy$ component of the metric:
\be\label{egyy}
\Delta g_{yy} = 4\, \lambda^2\,  R_0^2 \left( 1 +{R_0\over r}\right)^{-1}\, .
\ee
As a result the tension of the D1-D5 system, being proportional to
$\sqrt{g_{yy}}$, acquires an $r$-dependent contribution proportional
to
\be\label{egyy1}
\lambda^2 R_0^2 \left( 1 +{R_0\over r}\right)^{-1}
\ee
to first order in $\lambda^2$. Supersymmetrization of this term gives rise to
other fermionic terms.

Thus we have to analyze the dynamics of a superparticle with $\NN=4$
supersymmetry moving in 
Taub-NUT space under a potential proportional to \refb{egyy1}.
This is precisely the problem solved in 
\cite{pope,9912082}. 
The result of this analysis can be summarized as follows.
Depending on the sign of the deformation parameter $\lambda$ we have
supersymmetric bound states for $j_0>0$ or $j_0<0$, where $j_0$
is the momentum conjugate to the coordinate 
$\psi$.\footnote{Since the potential given in \refb{egyy1} does not
depend on the sign of $\lambda$, the reader may wonder why the
spectrum depends on the sign of $\lambda$. It turns out that other
terms in the world-volume action related to \refb{egyy1}
by supersymmetry do
depend on the sign of $\lambda$.}
In the weak
coupling limit the
number of bound states for a given value of $j_0$ is given by
$|j_0|$ and they carry angular momentum 
$(|j_0|-1)/2$.\footnote{Strictly speaking there is 
an upper bound on the
possible value of $|j_0|$ which goes to $\infty$ as the type IIB
coupling goes to zero. Put another way, the degeneracy formula  given
by the partition function  \refb{ezfin} is valid only for
type IIB coupling below a certain value determined
by the magnitude of $j_0$.
This bound is
related to the existence of the walls of marginal 
stability to be discussed
in \S\ref{smarginal}.} Thus for these states $(-1)^F=(-1)^{j_0-1}$.
If for
definiteness we choose the sign of $\lambda$ 
such that we get bound states
for positive $j_0$, then this gives the zero mode partition function
\be\label{ezfin}
Z_{\rm zero}(\wt v) 
\equiv Tr_{\rm zero\, modes} \left((-1)^{F}  (-1)^{j_0}
e^{2\pi i \wt v j_0}\right)
= -\sum_{j_0=1}^\infty j_0\, e^{2\pi i \wt v j_0} 
=-{e^{2\pi i\wt v}\over (1 - e^{2\pi i\wt v})^2}\, .
\ee
Since this is invariant under $\wt v\to -\wt v$ we shall get the
same answer if we had chosen to work with the opposite sign
of $\lambda$ that produces bound states with negative $j_0$.
However in order to extract the degeneracy of the states from the
partition function we have to make a decision as to whether
we should expand the right hand side of \refb{ezfin} in powers of
$e^{2\pi i \wt v}$ or $e^{-2\pi i \wt v}$, and this depends on the
sign of $\lambda$. Since in the heterotic description the 
complex structure modulus of the torus $\wt S^1\times  
S^1/\ZZZ_N$
corresponds to the axion-dilaton field, the sign of $\lambda$ would
correspond to the sign of the asymptotic value of the axion field.
A careful analysis shows that for positive sign of the axion field
the degeneracy is given by expanding \refb{ezfin} in powers of
$e^{-2\pi i\wt v}$, whereas for negative sign of the axion field we
need to expand \refb{ezfin} in powers of $e^{2\pi i \wt v}$.

Finally putting all the ingredients together the partition 
function of states associated with the centre of mass motion 
of the D1-D5 system in Taub-NUT space is given by
\bea{fullparttn}
&& \sum_{l_0, j_0} d_{transverse} (l_0, j_0) (-1)^{j_0}
e^{2\pi i l_0 \wt \rho + 2\pi i j_0 \wt v} 
= Z_{{\rm free}}(\wt\rho) 
Z_{\rm osc}(\wt\rho, \wt v)  
Z_{{\rm zero}} (\wt v)  
\\ \nonumber
&& \qquad = -4 e^{-2\pi i \wt v} ( 1- e^{-2\pi i \wt v}) ^{-2}
\\ \nonumber 
&& \qquad \times \prod_{n=1}^\infty \{
( 1- e^{2\pi i nN \wt \rho})^4  
( 1 - e^{2\pi i nN \wt \rho + 2\pi i \wt v} )^{-2} 
( 1 - e^{2\pi i nN \wt \rho - 2\pi i \wt v} )^{-2} 
\}.
\eea
 
\noindent{\bf The dynamics of Wilson lines along $\MM$:}
Let us now compute the contribution to the 
partition function from the
dynamics of the Wilson lines along $\MM=T^4$. For this
we can ignore the presence of the Kaluza-Klein monopole and 
the D1-branes, and consider the dynamics of a D5-brane wrapped
on $T^4\times S^1$, -- the sole effect of the Kaluza-Klein monopole
will be in the identification between the angular momentum carried by
the system from the point of view of a five dimensional
observer sitting at the center of Taub-NUT and 
the momentum along $\wt S^1$ from the point
of view of an asymptotic four dimensional observer. 
Taking the $T^4$ to have small size we
can regard the world-volume theory of the D5-brane
as (1+1) dimensional.
This contains eight scalars associated with four Wilson lines
and four transverse coordinates.
It also has a total of 16 massless fermions of which eight are left-moving
and eight are right-moving, -- these can be regarded as the
goldstino fermions associated with 16 broken supersymmetry
generators of type IIB string theory on $T^4$ in the presence of the
D5-brane. The eight bosons and the sixteen fermions mix under the
action of the sixteen unbroken supersymmetry
generators on the D5-brane world-volume. However
only eight of these generators commute with the orbifold
group generator $\wt g$, -- these 
coincide with the unbroken
supersymmetries of a D5-brane wrapped on K3
and consist of four left-chiral 
and four right-chiral generators. Under these $\wt g$ invariant
supersymmetry transformations
the scalars associated with the coordinates transverse to the D5-brane
mix with eight of the sixteen fermions on the D5-brane
world-volume.  The scalars associated with the
Wilson lines mix with 
the other eight fermions. Since the contribution to
the partition function from the first set of fermions and scalars, -- 
associated with the transverse coordinates and their superpartners, --
have  already been taken into account in \refb{fullparttn}, we
shall focus on the second set of world-volume fields consisting
of the Wilson lines and their superpartners. Since the 
$\wt g$ invariant
supersymmetry
generators are non-chiral, the superpartners of the Wilson line
must also be non-chiral. Thus this set consists of four left-moving
and four right-moving fermion fields.

Our goal is to compute the contribution to the partition function
of BPS states from this sector. We proceed in the following steps:
\begin{enumerate}
\item First we shall determine the $\wt g$ transformation properties
as well as the $l_0$ and $j_0$ quantum numbers of various
world-volume fields.
\item Since the bosonic world-volume fields represent Wilson lines
along $T^4$, the world-volume theory is free. Thus once we have
determined the quantum numbers carried by various fields,
computation of the partition function may be done by simple
counting.
\end{enumerate}

We begin by determining the various quantum numbers carried by
the world-volume fields.
Since $\wt g$ represents a $2\pi/N$ rotation in one plane of
$T^4$ and $-2\pi/N$ rotation in an orthogonal plane,
$\wt g$ acts as a rotation by
$2\pi/N$ on one pair of Wilson lines 
and as a rotation by $-2\pi/N$
on the other pair. 
Thus it must act in the same way on the left- and right-moving
fermionic fields related to these Wilson lines by $\wt g$ invariant
supersymmetry transformations.
In order to be $\ZZZ_N$
invariant,  the modes which pick a phase of $e^{2\pi i/N}$
under $\wt g$ must carry momentum along $S^1$ of the form
$k-{1\over N}$ for integer $k$, whereas modes which 
pick a phase of $e^{-2\pi i/N}$
under $\wt g$ must carry momentum along $S^1$ of the form
$k+{1\over N}$ for integer $k$. 
As a result, both in the left and 
the right-moving sector, we have a pair of bosons and a pair
of fermions carrying $S^1$ momentum of the form 
$k+{1\over N}$, and
a pair of bosons and a pair of fermions carrying $S^1$
momentum of the form $k-{1\over N}$.

Eventually when we place this in the background of the
Kaluza-Klein monopole, only the left-moving 
supersymmetry acting on
the right-moving modes remain unbroken. 
Thus in order to get a BPS state
of the final supersymmetry algebra we must put all the right-moving
oscillators in their ground state and consider only 
left moving excitations.

In order to calculate the partition
function associated with these modes we also need information about
their $j_0$ quantum numbers. Since the system is eventually placed
at the centre of Taub-NUT space which converts angular
momentum at the centre into momentum along $\wt S^1$ at
infinity, $j_0$ is still given by the sum of angular momenta in the
1-2 and 3-4 planes transverse to the D5-brane world-volume. Since the
scalars represented by the Wilson lines along $T^4$ are
neutral under rotation in the transverse plane, they do not carry any
$j_0$ quantum number. The story however is different for the fermions.
First of all, since type IIB string theory on $T^4$ is a non-chiral
theory, there should be no correlation between the $SU(2)_L$
quantum number $j_0$ and the world-volume chirality of the massless
fermions living on a D5-brane on $T^4$. Now
from our previous analysis of the D1-D5 system in the background of
$K3\times TN$ we know that
the world-volume fermions in this system
have the property that the left-movers have $j_0=0$ 
and the right-movers carry $j_0$ quantum numbers
$\pm 1$. Such fermions also exist on a D5-brane on $T^4$
as partners of the transverse coordinates under $\wt g$ invariant
supersymmetry transformation, but 
the contribution to the partition function from these fermions
have already been taken into account in \refb{fullparttn}. 
The rest of the fermions 
must have opposite correlation between $SU(2)^T_L$
quantum numbers and world-volume chirality, \i.e.\
the right-movers must have $j_0=0$ 
and the left-movers must carry $j_0$ quantum numbers
$\pm 1$.  Furthermore, since $\wt g$ has no action on the
transverse coordinates,  it commutes with $SU(2)^T_L$ and 
there should be 
no correlation between the $\wt g$ quantum numbers
and the sign of the
$U(1)_L\subset SU(2)^T_L$ quantum numbers. 
Thus the two left-moving fermions
carrying $\wt g$ quantum number $e^{2\pi i/N}$ 
must have $j_0=\pm 1$
and the two left-moving fermions
carrying $\wt g$ quantum number $e^{-2\pi i/N}$ 
must have $j_0=\pm 1$.

To summarize, when we choose $\MM=T^4$ instead of $K3$,
the additional left-moving excitations on the D5-brane
world-volume
consist of four bosonic and four fermionic modes. Invariance under
the orbifold group generator $g$ requires that
two of the four
bosonic modes carry momentum along $S^1$ of the form $k+{1
\over N}$
and the other two carry momentum along $S^1$ of the form
$k-{1\over N}$, but neither of them carry any momentum along
$\wt S^1$. On the other hand two of the left-moving
fermionic modes
carry momentum along $S^1$ of the form $k+{1
\over N}$ and $\pm 1$
unit of momentum
along $\wt S^1$, and the other two left-moving
fermionic modes 
carry momentum along $S^1$ of the form $k-{1\over N}$ and $\pm 1$
unit of momentum
along $\wt S^1$. 
As before the statistics of these oscillators are altered by a factor of
$(-1)^{j_0}$ as we come down from five to four dimensions.
Thus if $d_{wilson}(l_0, j_0)$ denotes
the number of bosonic minus fermionic
states associated with these modes carrying  total
momentum
$-l_0/N$  along $S^1$ and total momentum $j_0$ 
along $\wt S^1$, then
\bea{eone2}
&& \sum_{l_0, j_0} d_{wilson}(l_0, j_0) (-1)^{j_0}
e^{2\pi i l_0\wt\rho + 2\pi i
j_0\wt v} \nonumber \\ 
 &=& \prod_{l\in N\zzz+1\atop l>0} 
(1 - e^{2\pi il \wt\rho})^{-2} \, 
\prod_{l\in N\zzz-1\atop l>0} 
(1 - e^{2\pi il \wt\rho})^{-2}
\prod_{l\in N\zzz+1\atop l>0} 
(1 - e^{2\pi il \wt\rho+2\pi i \wv})   \nonumber \\
&&
\prod_{l\in N\zzz+1\atop l>0} 
(1 - e^{2\pi il \wt\rho-2\pi i \wv}) \,
\prod_{l\in N\zzz-1\atop l>0} 
(1 - e^{2\pi il \wt\rho+2\pi i \wv})  \,
\prod_{l\in N\zzz-1\atop l>0} 
(1 - e^{2\pi il \wt\rho-2\pi i \wv})  \, .  
\eea

The partition function
associated with the overall dynamics of the D1-D5 system is
given by the product of the contribution \refb{fullparttn}
from the dynamics of the transverse modes and (in case $\MM=T^4$)
the contribution \refb{eone2} from the dynamics of the Wilson lines
along $T^4$. 
The final result can be written in a compact form using the
coefficients $c_b^{(r,s)}(u)$ introduced in 
\refb{eqrsrevpre}.
In particular if we use the relations \refb{eck3}, \refb{ect4}:
\be \label{erelagain}
c^{(0,s)}_1(-1) =\cases{
{2\over N} \quad \hbox{for $\MM=K3$}\cr 
{1\over N} 
\left(2 - e^{2\pi is/N} - e^{-2\pi is/N}\right) \quad 
\hbox{for $\MM=T^4$}}
\ee
then the product of \refb{fullparttn} and (for $\MM=T^4$)
\refb{eone2}
 can be written as
\bea{ecmfin}
&& \sum_{l_0,j_0} d_{CM}(l_0, j_0) (-1)^{j_0}\,
e^{2\pi i l_0\wt\rho + 2\pi i
j_0\wt v} =
-4\, e^{-2\pi i \wt v}\, \prod_{l=1}^\infty (1 -  
e^{2\pi i l\wt\rho})^{2\sum_{s=0}^{N-1} 
e^{-2\pi i l s/N}
\, c^{(0,s)}_1(-1)} \nonumber \\
&& \quad \prod_{l=1}^\infty (1 -  
e^{2\pi i l\wt\rho+2\pi i \wv})^{-\sum_{s=0}^{N-1} 
e^{-2\pi i l s/N}
\, c^{(0,s)}_1(-1)} \, 
\prod_{l=0}^\infty (1 -
e^{2\pi i l\wt\rho-2\pi i \wv})^{-\sum_{s=0}^{N-1} 
e^{-2\pi i l s/N}
\, c^{(0,s)}_1(-1)} \nonumber \\
\eea
both for $\MM=K3$ and $\MM=T^4$.

\subsubsection{Counting states associated with the 
relative motion of
the D1-D5 system} \label{srel}

Finally we turn to the problem of counting states associated with the
motion of the D1-brane in the plane of the D5-brane. 
This is a well known system that appears in the original analysis
of the five dimensional black holes in \cite{9601029} 
(see \cite{0203048} for a
review of this system). Our analysis will follow the approach
described in \cite{0605210}, which in turn is a generalization
of the analysis given in 
\cite{9608096,0505094} for the case of type 
IIB string theory on $K3\times S^1$. 

At the special point in the moduli space at which we have been
working so far, the D1-D5 system in the absence of the
Kaluza-Klein monopole is marginally bound and hence the
D1-brane can move freely in directions transverse to D5. This makes
it difficult to analyze this system. We shall avoid this problem by
switching on a small amount of NS-NS sector 2-form field along the
D5-brane world-volume. This binds the D1-brane on to the D5-brane,
-- the D1-brane being identified as non-commutative U(1) instanton
of the gauge theory on the D5-brane 
world-volume\cite{9802068,9908142}. Thus the only possible motion
of the D1-brane inside the D5-brane is along the directions
tangential to the D5-brane.

We can now proceed in the following steps:
\begin{enumerate}
\item We first analyze the world-volume theory of a single
D1-brane inside a D5-brane. This is given by a superconformal
field theory with target space $\MM$. $\wt g$ represents a
$\ZZZ_N$ symmetry of this superconformal field theory. We also
identify the
momenta along $S^1$ and $\wt S^1$ as specific
quantum numbers in this superconformal field theory.
\item We then compute of the degeneracy $n(w,l,j)$ 
of a single D1-brane
moving inside the D5-brane,
wound $w$ times along $S^1/\ZZZ_N$,  and carrying
momenta $-l/N$ along $S^1$  and 
$j$ along $\wt S^1$.  The result is
expressed in terms of the set of coefficients $c_b^{(r,s)}(u)$
introduced in \refb{esi4aintpre}.
\item Finally we consider the contribution to the
partition function from multiple D1-branes moving inside the
D5-brane and, using straightforward combinatoric analysis,
express the result in terms of the degeneracies $n(w,l,j)$ of
a single D1-brane.
\end{enumerate}

We begin our analysis with a single D1-brane moving inside a
D5-brane.
Let $\sigma$ denote the coordinate along the length of the
D1-brane, $\sigma$ being normalized so that it coincides with the
target space coordinate in which  $S^1/\ZZZ_N$ has period
$2\pi$. If the
D1-brane winds $w$ times along $S^1/\ZZZ_N$, then
$\sigma$ changes by $2\pi w$
as we traverse the whole length of the string, regarded as a configuration
in the orbifold.
Under $\sigma\to \sigma+ 2\pi w$, the physical coordinate
of the D1-brane shifts by $2\pi r$  along $S^1$ where
\be\label{esi2}
r=w \quad \hbox{mod $N$}\, .
\ee
Identification in $\MM\times S^1$ under
$\ZZZ_N$ then
requires that under $\sigma\to \sigma+ 2\pi w$ the location
of the D1-brane along $\MM$ gets transformed by $\wt g^r=\wt g^w$.

Since in the weak coupling limit the dynamics of the D1-brane
inside a D5-brane is insensitive to the presence of the Kaluza-Klein
monopole, the 2-dimensional theory describing this system has
(4,4) supersymmetry. Thus
we expect the low energy dynamics of this D-brane system to be
described by a superconformal field theory  (SCFT)
with target space $\MM$ subject to the above boundary condition.
In particular the state must be twisted by $\wt g^r$. 
Furthermore, since the supersymmetry generators commute with
$\wt g$, the supercurrents will satisfy periodic boundary condition
under $\sigma\to \sigma+2\pi w$. Thus the state belongs to the RR
sector.
Since the D1-brane has coordinate length 
$2\pi w$, the momentum along $S^1$ can be
identified as the $(\bar L_0 - L_0)/w$ eigenvalue of this
state.  Thus a total momentum $-l/N$ corresponds to 
$\bar L_0-L_0$
eigenvalue $-lw/N$.
On the other hand the BPS condition on the
D1-brane requires $\bar
L_0$ to vanish.\footnote{Even though the D1-D5 system has 
supersymmetry acting on both the right- and the left-moving fields,
only the supersymmetries acting on the right-moving fields survive
when we place the system in a Kaluza-Klein monopole background.
Thus we only require invariance under the supersymmetries acting
on the right-moving fields.}
Thus we are looking for a state in the $\wt g^r$ twisted
RR sector of this SCFT with
\be\label{esi3}
L_0=lw/N, \qquad \bar L_0 = 0\, .
\ee

Eqs.\refb{esi2}, \refb{esi3} give interpretation of the quantum
numbers $w$ and $l$ in the D1-brane world-volume theory.
What about the quantum number $j$?
This superconformal field theory has an $R$-symmetry group
$SO(4)^T=SU(2)_L^T\times SU(2)_R^T$ associated with tangent
space rotation along directions transverse to the D1-brane and the
D5-brane. All the bosonic fields in the world-volume theory are
neutral under this group but the fermions  transform non-trivially.
As discussed in the paragraphs below \refb{eu12},  
in the presence of the Kaluza-Klein
monopole background a translation $\epsilon$ along $\wt S^1$
must be accompanied by a rotation $2\epsilon$ in $U(1)_L
\subset SU(2)_L$. Thus if $F_L$ denotes twice the $U(1)_L$
generator, then the quantum number $j$ can be identified
as the $F_L$ eigenvalue of the state\cite{9602065}. 
$F_L$ is also referred to as the world-sheet fermion number
associated with the left-moving sector of the (4,4) 
superconformal field theory.

Let $F_R$ denote twice the $U(1)_R\subset SU(2)_R$ generator,
or equivalently, the world-sheet fermion number associated with
the right-moving modes of the (4,4) superconformal field theory.
Since in our system the world-sheet supersymmetry originates from
space-time supersymmetry, the total world-sheet fermion number
$F_L+F_R$ can be interpreted as the space-time fermion number
from the point of view of a five dimensional observer at the
center of Taub-NUT space. Taking into account the fact that the four
and five dimensional statistics differ by a factor of $(-1)^{j}$
we see that in four dimensions,
in counting the total number of bosonic minus fermionic
states weighted by $(-1)^{j}$
with a given set of charges, we must calculate the number
of states weighted by $(-1)^{F_L+F_R}$. 

Finally we must remember that not all the states of the
superconformal field theory are allowed states of the D-brane, --
we must pick $\ZZZ_N$ invariant states. 
Since the total momentum along
$S^1$ is $-l/N$, under translation by $2\pi$
along $S^1$ this state picks up a phase $e^{-2\pi i l/N}$. Thus the
projection operator onto $\ZZZ_N$ invariant states is
given by
\be\label{esi6}
{1\over N}\, \sum_{s=0}^{N-1} e^{-2\pi i sl/N} \wt g^s \, .
\ee
Putting these results together we see that the total number of
$\ZZZ_N$ invariant bosonic minus fermionic
states weighted by $(-1)^{j}$
of the single D1-brane carrying quantum
numbers $w,l,j$ is given by
\be\label{esi7}
n(w,l,j) \equiv {1\over N}
\, \sum_{s=0}^{N-1} e^{-2\pi i sl/N}
Tr_{RR;\wt g^r} \left(\wt g^s(-1)^{F_L+F_R} \delta_{NL_0,lw}
\delta_{{F_L}, j}\right), \quad \hbox{$r=w$ mod $N$} \, .
\ee
Here $Tr_{RR;\wt g^r}$ denotes trace, in the
superconformal $\sigma$-model with target space $\MM$, 
over RR sector states
twisted by $\wt g^r$. 
Insertion of
 $(-1)^{F_R}$ in the trace automatically projects 
 onto $\bar L_0=0$
 states, -- so we do not need to insert a $\delta_{\bar L_0,0}$
 factor.

Let us define\cite{0602254}
\be\label{esi4a}
F^{(r,s)}(\tau,z) \equiv {1\over N} Tr_{RR;\wt g^r} \left(\wt g^s
(-1)^{F_L+F_R}
e^{2\pi i \tau L_0} e^{2\pi i {F_L} z}\right) \, .
\ee
As shown in \refb{enewint},
$F^{(r,s)}(\tau, z)$ has power series expansion of the 
form
\be\label{esi4b}
F^{(r,s)}(\tau,z)= \sum_{b=0}^1\sum_{j\in2\zzz+b,
n\in \zzz
/N} c_b^{(r,s)}(4n - j^2) e^{2\pi in\tau} 
e^{2\pi i  j z}\, ,
\ee
for appropriate coefficients $c_b^{(r,s)}(4n - j^2)$.
{}From  
\refb{esi4a}, \refb{esi4b} it now follows that
 \be\label{esi5}
{1\over N}\,
tr_{RR,\wt g^r} \left(\wt g^s(-1)^{F_L+F_R}
\delta_{NL_0,lw}
\delta_{{F_L}, j}
\right) = c_b^{(r,s)}(4lw/N - j^2)\, , \qquad \hbox{$b=j$ mod 2}\, .
\ee
Hence \refb{esi7} gives
\be\label{esi7.1}
n(w,l,j)  
= 
\sum_{s=0}^{N-1} e^{-2\pi i sl/N} c_b^{(r,s)}(4lw/N - j^2)\, ,
\qquad \hbox{$r=w$ mod $N$}, \quad \hbox{$b=j$ mod 2}\, .
\ee

Using this result for single D1-brane spectrum, we can now evaluate
the degeneracy of multiple D1-branes moving inside the D5-brane. Let
$d_{D1}(W,L,J')$ denote the total number of bosonic minus
fermionic states of this system, weighted by $(-1)^{J'}$
and carrying total
D1-brane charge $W$, total momentum $-L/N$
along $S^1$ and total  momentum $J'$ along $\wt S^1$. 
This represents the number of ways we can distribute the
quantum numbers $W$, $L$ and $J'$ into individual D1-branes
carrying quantum number $(w_i,l_i,j_i)$ subject to the constraint
\be \label{ecccon}
W=\sum_i w_i, \qquad L
=\sum_i l_i, \qquad J'=\sum_i j_i, \qquad
w_i,l_i, j_i\in \ZZZ, \qquad w_i\ge 1, \qquad 
l_i\ge 0\, .
\ee
A
straightforward combinatoric analysis shows that
\be\label{emulti}
\sum_{W,L,J'} d_{D1}(W,L,J') (-1)^{J'}\, e^{2\pi i ( 
\wt \sigma W /N +\wt \rho L + \wt v J')}
= 
\prod_{w,l,j\in \zzz\atop w>0, l\ge 0}  
\left( 1 - e^{2\pi i (\wt \sigma w / N + \wt \rho l + 
\wt v j)}\right)^{-n(w,l,j)}\, .
\ee
Using \refb{esi7.1} and defining $k'=w/N$, we can express
\refb{emulti} as
\ben \label{enewexp}
&& \sum_{W,L,J'} d_{D1}(W,L,J') (-1)^{J'}\, e^{2\pi i ( 
\wt \sigma W /N +\wt \rho L + \wt v J')} \nonumber \\
&=& \prod_{r=0}^{N-1}\prod_{b=0}^1
\prod_{k'\in \zzz+{r\over N},l\in \zzz, j\in 2\zzz+b
\atop k'>0, l\ge 0}
\left( 1 - e^{2\pi i (\wt \sigma k'   + \wt \rho l + \wt v j)}\right)^{
-\sum_{s=0}^{N-1} e^{-2\pi i sl/N } c_b^{(r,s)}(4lk' - j^2)}\, .
 \een

\subsubsection{The full partition function} \label{sfull}

Using \refb{esi9}, \refb{e8xxx}, \refb{ecmfin} 
and \refb{enewexp} we now get the full partition
function:
\be\label{esi9ex}
f(\wt \rho,\wt \sigma,\wt v )= e^{-2\pi i (\wt \alpha\wt\rho 
+ \wt v)}
 \prod_{b=0}^1\, \prod_{r=0}^{N-1}
\prod_{k'\in \zzz+{r\over N},l\in \zzz, j\in 2\zzz+b
\atop k',l\ge 0, j<0 \, {\rm for}
\, k'=l=0}
\left( 1 - e^{2\pi i (\wt \sigma k'   + \wt \rho l + \wt v j)}\right)^{
-\sum_{s=0}^{N-1} e^{-2\pi i sl/N } c_b^{(r,s)}(4lk' - j^2)}\, .
 \ee
The multiplicative factor $e^{-2\pi i (\wt \alpha\wt\rho
+ \wt v)}$ as well as the $k'=0$ term in 
this expression comes from the terms involving
$d_{CM}(l_0, j_0)$ and $d_{KK}(l_0')$.  
 Comparing the right hand side of this equation with the expression
 for the function $\wt\Phi(\wt\rho,\wt\sigma,\wt v)$ 
 given in \refb{edefwtphi}
we can rewrite \refb{esi9ex} as
\be\label{esi9a}
f(\wt \rho,\wt \sigma,\wt v )
= {e^{2\pi i  \wt \gamma \wt \sigma }
\over \wt\Phi(\wt \rho,\wt \sigma, \wt v)}\, .
\ee
According to \refb{esi8} we can identify $(-1)^{J+1}
h(Q_1, n, J)$, --
the number of bosonic minus fermionic
quarter BPS supermultiplets weighted by $(-1)^{J+1}$,
carrying $Q_1$
units of D1-brane winding charge along $S^1$, $-n/N$ units
of momentum along $S^1$ and $J$ units of momentum 
along $\wt S^1$, --
 as the
coefficients of the expansion of $f(\wrh,\ws,\wv)$ in powers of
$e^{2\pi i \wrh}$, $e^{2\pi i\ws}$ and $e^{2\pi i \wv}$. Except
for the overall multiplicative factor of $e^{-2\pi i \wt\alpha\wrh}$
in \refb{esi9ex},
this expansion involves positive powers of $e^{2\pi i\wrh}$ and
$e^{2\pi i \ws}$. Furthermore, except for the $k'=l=0$ term,
the power of $e^{2\pi i \wv}$ for any given power of
$e^{2\pi i\wrh}$ and $e^{2\pi i \ws}$ is bounded both from
above and below. For the $k'=l=0$ term we need to carry out the
expansion in positive or negative powers of $e^{2\pi i \wv}$
depending on the sign of the angle between $S^1$ and $\wt S^1$.
If the expansion is in positive powers of $e^{-2\pi i\wv}$, 
-- as in the
case of positive value of the axion field, -- we
can extract the Fourier coefficient $h(Q_1,n,J)$ from the equation:
\be\label{ehexp}
h(Q_1,n,J) =  (-1)^{J+1}\, 
{1\over N}\, \int _\CC d\wt\rho \, d\wt\sigma \,
d\wt v \, e^{-2\pi i ( \wt \rho n 
+ \wt \sigma (Q_1-\wt \gamma\, N)/N +\wt v J)}\, {1
\over \wt\Phi(\wt \rho,\wt \sigma, \wt v)}\, ,
\ee
where $\CC$ is a three real dimensional subspace of the
three complex dimensional space labelled by $(\wt\rho,\ws,\wv)$,
given by
\bea{ep2int}
\wt \rho_2=M_1, \quad \wt\sigma_2 = M_2, \quad
\wt v_2 = -M_3, \nonumber \\
 0\le \wt\rho_1\le 1, \quad
0\le \wt\sigma_1\le N, \quad 0\le \wt v_1\le 1\, .
\een
$M_1$, $M_2$ and $M_3$ are large but fixed 
positive numbers with $M_3<< M_1, M_2$.
The choice of the $M_i$'s is determined from the requirement that
the Fourier expansion is convergent in the region of integration.
On the other hand if the $k'=l=0$ term in $f(\wrh,\ws,\wv)$ is to be
expanded in positive powers of $e^{2\pi i\wv}$, 
-- as in the case of negative value of the axion field,
-- then $h(Q_1,n,J)$
is given by an expression similar to \refb{ehexp}, except that $\wv_2$
is now set equal to a positive number $M_3$ instead of a negative
number $-M_3$.

Identifying $h(Q_1,n,J)$ with the degeneracy $d(\vec Q,\vec P)$,
using \refb{eqdef}, and noting that
$\beta$, being equal to $\chi(\MM)/24$,  
is equal to $\wt \gamma N$ given in \refb{enn9dpre},
we can rewrite \refb{ehexp} as
\be\label{egg1}
d(\vec Q,\vec P) = (-1)^{Q\cdot P+1}\,
{1\over N}\, \int _\CC d\wt\rho \, 
d\wt\sigma \,
d\wt v \, e^{-\pi i ( N\wt \rho Q^2
+ \wt \sigma P^2/N +2\wt v Q\cdot P)}\, {1
\over \wt\Phi(\wt \rho,\wt \sigma, \wt v)}\, .
\ee
Various useful properties of the function 
$\wt\Phi(\wt\rho,\wt\sigma,\wt v)$ and a related function
$\wh\Phi(\rho,\sigma,v)$ have been discussed in appendices
\ref{ssiegel} and \ref{szero}.

\subsectiono{Additional charges from collective modes}
\label{sadditional}

The analysis so far has been carried out for a restricted class of charge
vectors. We shall now extend our result to a more general class of
charge vectors by considering collective excitations of the system
analyzed above.  Our analysis will follow \cite{0705.1433}.
For simplicity we shall restrict our analysis to 
type II string theory compactified on  $K3\times \wt S^1\times S^1$
or equivalently heterotic string theory on $T^4\times \wh S^1\times
S^1$.
Generalizing this to the case of $\NN=4$ supersymmetric orbifolds
of this theory will require setting some of the charges, which
are not invariant under the orbifold group, to zero. The analysis
for $\NN=4$ supersymmetric $\ZZZ_N$ orbifolds of type II string
theory compactified on $T^6$ can also be done in an identical manner.

The compactified theory
has 28 U(1) gauge fields and hence a given state
is characterized by 28 dimensional electric and
magnetic charge vectors $\vec Q$ and 
$\vec P$ as defined in \S\ref{sugra}.
We shall choose a basis in which the matrix $L$  has the form
\be \label{nne1}
L =  \pmatrix{\wh L & & & & \cr & 0 & 1 & & \cr & 1 & 0 &&\cr
&&& 0_2 & I_2\cr &&& I_2 & 0_2}\, ,
\ee
where $\wh L$ is a matrix with 3 eigenvalues $+1$ and 19
eigenvalues $-1$. The charge vectors
will be labelled as
\be \label{nne1.1}
Q = \pmatrix{\wh Q\cr \qq_1 \cr \qq_2 \cr \qq_3 \cr \qq_4
\cr
\qq_5\cr \qq_6}\, , \qquad P = \pmatrix{\wh P\cr \pp_1\cr \pp_2 \cr \pp_3
\cr \pp_4\cr \pp_5 \cr \pp_6}
\, .
\ee
The last four elements of $\vec Q$ and $\vec
P$ are to be identified with the
four dimensional electric and magnetic charge vectors introduced
in \refb{e2dcharge}:
\be \label{echargeiden}
\pmatrix{\qq_3 \cr \qq_4
\cr
\qq_5\cr \qq_6}=\pmatrix{\wh n\cr n'\cr \wh w\cr w'}
\, , \qquad  \pmatrix{\pp_3
\cr \pp_4\cr \pp_5 \cr \pp_6}= 
\pmatrix{\wh W\cr W'\cr
\wh N\cr N'}
\, .
\ee
Thus
in the second description of the theory $\qq_3$, $\qq_4$, $-\qq_5$ and
$-\qq_6$ label respectively the momenta along $\wh S^1$, $S^1$
and fundamental string winding along $\wh S^1$ and $S^1$. 
On
the other hand $-\pp_3$, $\pp_4$, $\pp_5$ and $\pp_6$ label 
respectively the number of
NS 5-branes wrapped along $S^1\times T^4$ and $\wh S^1\times T^4$, 
and Kaluza-Klein
monopole charges associated with $\wh S^1$ and $S^1$.
The rest of the charges label the momentum/winding
or monopole charges associated with the other internal directions. 
By following the duality chain that relates the first and second
description of the theory and the sign conventions described in
appendix \ref{s3.1},
the different components of $\vec P$ and $\vec Q$
can be given the following interpretation in
the first description of the theory.
$\qq_3$ represents 
the D-string winding charge along $\wt S^1$, $\qq_4$
is the momentum along $S^1$, $\qq_5$ is the D5-brane charge
along $K3\times \wt S^1$, $\qq_6$ is the number of Kaluza-Klein
monopoles associated with the compact circle $\wt S^1$,
$\pp_3$ is the D-string winding charge along $ S^1$, $-\pp_4$
is the momentum along $\wt S^1$, $\pp_5$ is the D5-brane charge
along $K3\times   S^1$ and $\pp_6$ is the number of Kaluza-Klein
monopoles associated with the compact circle $ S^1$.
Other components of $\vec Q$ ($\vec P$) 
represent various other branes of
type IIB string theory wrapped on $\wt S^1$ ($S^1$) times various
cycles of $K3$. We shall choose a convention in which
the 22-dimensional 
charge vector $\wh Q$ represents 
3-branes wrapped along the 22 2-cycles of K3 times $\tilde S^1$,
$\qq_1$ represents fundamental type IIB string winding charge
along $\wt S^1$, $\qq_2$ represents the number of
NS 5-branes of type IIB wrapped along $K3\times \wt S^1$, 
 the 
22-dimensional 
charge vector $\wh P$ represents 
3-branes wrapped along the 22 2-cycles of K3 times $ S^1$,
$\pp_1$ represents fundamental type IIB string winding charge
along $S^1$ and $\pp_2$ represents the number of
NS 5-branes of type IIB wrapped along $K3\times  S^1$.
In this convention
$\wh L$ represents the intersection matrix of 2-cycles of $K3$.

In our new notation
the original configuration described in \refb{echvec}
has charge vectors of the form:
\be \label{nne1.2}
Q = \pmatrix{\wh 0\cr 0 \cr 0 \cr 0 \cr -n
\cr
0\cr -1}\, , \qquad P = \pmatrix{\wh 0\cr 0\cr 0 \cr Q_1
-Q_5=Q_1-1 \cr -J  \cr Q_5=1 \cr 0}
\, ,
\ee
with
\be \label{nne1.3}
Q^2 = 2n, \qquad P^2 = 2(Q_1-1), \qquad Q\cdot P = J\, .
\ee
Note that we have set $N=1$.
The degeneracy of this system calculated in \S\ref{sfull} may be
written as:
\be \label{nne1.4}
d(\vec Q,\vec
P) = h(Q_1, n, J) = h\left( {1\over 2} P^2+1, {1\over 2}Q^2, 
Q\cdot P\right)\, ,
\ee
where the function $h$ is given in \refb{ehexp} with the 
choice $\MM=K3$
and $N=1$.
Our goal will be to consider charge vectors more general than
the ones given in \refb{nne1.2} and check if the degeneracy is
still given by \refb{nne1.4}. We shall do this by adding charges to the
existing system by exciting appropriate collective modes of the
system. These collective modes come from three sources:
\begin{enumerate}
\item The original configuraion in the type IIB theory contains
a Kaluza-Klein monopole associated with the circle $\wt S^1$.
This solution has been given in \refb{tnutgeom}, with
the coordinate $\psi$ labelling the coordinate of $\wt S^1$,
and $(r,\theta,\phi)$ representing spherical polar coordinates of the
non-compact space.
These coordinates label the geometry of the space `transverse'
to the Kaluza-Klein monopole. 
The world-volume of the Kaluza-Klein
monopole spans the $K3$ surface, the 
circle $S^1$ which we shall label 
by $y$, and time $t$. As in \S\ref{s3.2} 
we shall take the size of $K3$ to be
small compared to that of $S^1$ and use dimensional reduction on
K3 to regard the world-volume as two dimensional, spanned by $y$
and $t$.

Type IIB string theory compactified on K3 has various 2-form fields,
-- the original NSNS and RR 2-form fields $B$ and $C^{(2)}$
of the ten dimensional
type IIB string theory 
 as well as the components of the 4-form field $C^{(4)}$ along
various 2-cycles of $K3$.
Given any such 2-form field $C_{MN}$, 
we can introduce a scalar
mode $\phi$ by considering deformations of the form\cite{9705212}:
\be \label{nne1.5}
C = \phi(y,t) \, \omega\, ,
\ee
where $\omega$ is the harmonic 2-form of Taub-NUT space
introduced in \refb{nneomegadefrp}:
\be \label{nneomegadef}
\omega \propto 
{r\over r +R_0} 
d\sigma_3 + {R_0\over (r+R_0)^2}dr
\wedge \sigma_3\, , \qquad \sigma_3 \equiv
\left (d\psi + {1\over 2} \cos\theta d\phi\right)\, .
\ee

If the field strength
$dC$ associated with $C$ is
self-dual or anti-selfdual in six dimensions transverse to
$K3$ then the corresponding
scalar field $\phi$ is chiral in the $y-t$ space; otherwise it represents
a non-chiral scalar field. The non-zero mode oscillators associated
with the left-moving components of these scalar fields were used
in \S\ref{s1xxx} for counting 
the number of BPS states of the Kaluza-Klein
monopole. Our focus in this section will be on the zero modes
of these scalar fields. In particular
we shall consider configurations which carry momentum conjugate
to $\phi$ or
winding number along $y$
of  $\phi$,  represented
by a solution where $\phi$ is linear in $t$ or $y$. In the six
dimensional language this corresponds to $dC \propto dt\wedge \omega$
or
$dy\wedge\omega$. {}From \refb{nneomegadef} we see that
$dC\propto  dt\wedge \omega$
will have a component proportional to $r^{-2}\, 
dt\wedge dr\wedge d\psi$ for large $r$, and hence 
the coefficient of this term represents the charge associated with
a string, electrically charged under $C$, wrapped 
along $\wt S^1$. On the
other hand $dC\propto  dy\wedge \omega$
will have a component proportional to $\sin\theta\,
dy\wedge d\theta\wedge d\phi$
and the coefficient of this term will represent 
the charge associated with
a string, magnetically charged under $C$, wrapped 
along $\wt S^1$. If the 2-form field $C$ represents the original
RR or
NSNS 2-form field of type IIB string theory in ten dimensions,
then the electrically charged string would correspond to a D-string or
a fundamental type IIB string and the magnetically charged string
would correspond to a D5-brane or NS 5-brane wrapped on $K3$.
On the other hand if the 2-form $C$ represents the
component of the 4-form field along a 2-cycle of K3, then the
corresponding string represents a D3-brane wrapped on a
2-cycle times $\wt S^1$. Recalling the interpretation
of the charges $\wh Q$ and $\qq_i$ appearing in \refb{nne1.1} we now
see that
the momentum and winding modes of
$\phi$ correspond to the charges $\wh Q$, 
$\qq_1$, $\qq_2$, $\qq_3$ and
$\qq_5$. More specifically, after taking into account the sign
conventions described in appendix \ref{s3.1}, 
these charges correspond to switching on
deformations of the form:
\ben \label{nnep1}
&& dB \propto -\qq_1 dt\wedge \omega,  \quad
dB\propto \qq_2 dy\wedge \omega, \quad 
d C^{(2)} \propto -\qq_3 dt\wedge \omega,  
\quad dC^{(2)}\propto \qq_5 dy\wedge \omega,
\nonumber \\ 
&& dC^{(4)} \propto \sum_\alpha \wh 
Q_\alpha (1 + *) \, \Omega_\alpha
\wedge dy\wedge \omega \, ,
\een
where $\{\Omega_\alpha\}$ denote a basis of harmonic 2-forms
on $K3$ ($1\le\alpha\le 22$) satisfying
$\int_{K3}\Omega_\alpha\wedge\Omega_\beta=\wh L_{\alpha\beta}$.
Thus in the presence of these deformations 
we have a more general electric charge vector of the form
\be \label{nne1.6}
Q_0 = \pmatrix{\wh Q\cr \qq_1 \cr \qq_2 \cr \qq_3 \cr -n
\cr
\qq_5\cr -1}\, .
\ee

As can be easily seen from \refb{nnep1}, 
$\qq_2$ represents NS 5-brane charge wrapped along 
$K3\times \wt S^1$. However for weakly coupled type IIB string theory,
the presence of this charge could have large backreaction on the
geometry. We can avoid this by choosing
\be \label{nne1.7}
\qq_2 = 0\, .
\ee
Alternatively by taking the radius of $S^1$ to be large we could make 
the
Kaluza-Klein monopole much heavier than the NS 5-brane
wrapped on $K3\times\wt S^1$. This will keep the backreaction of the
NS 5-brane under control. We shall continue to take $\qq_2=0$
for simplicity.

\item The original configuration considered in \S\ref{s3.s} also
contains a D5-brane wrapped around $K3\times S^1$. We can
switch on flux of gauge field strengths $\FF$
on the D5-brane world-volume
along the various 2-cycles of K3 that it wraps. 
The net coupling of the RR gauge fields to the D-brane world-volume in
the presence of the gauge fields may be expressed 
as\cite{9910053}
\be \label{nnetot}
\int \left[ C^{(6)} + C^{(4)}\wedge \FF + {1\over 2} C^{(2)}\wedge \FF
\wedge \FF + \cdots\right]\, ,
\ee
up to a constant of proportionality. The integral is over the D5-brane
world-volume spanned by $y$, $t$ and the coordinates of $K3$.
Via the coupling
\be \label{nne1.8}
\int C^{(4)} \wedge \FF \, ,
\ee
the gauge field
configuration will produce the charges of a D3-brane wrapped on
a 2-cycle of $K3$ times $S^1$, -- \i.e.\ the 22 dimensional 
magneic charge vector $\wh P$ appearing in \refb{nne1.1}.

In order to be compatible with our convention 
of appendix \ref{s3.1} that the D5-brane
wrapped on $K3\times S^1$ carries negative 
$(dC^{(6)})_{(K3)yrt}$ asymptotically, we need to take the
integration measure in the $yt$ plane in \refb{nnetot}
as $dy\wedge dt$, \i.e.\ $\epsilon^{yt}>0$.
Using this information one can show that 
the gauge field flux required to produce a specific
magnetic charge vector $\wh P$ is
\be \label{nneflux1}
\FF\propto -\sum_\alpha \wh P_\alpha \, \Omega_\alpha\, .
\ee

\item The D1-D5 system can also carry
electric flux along $S^1$. This will induce the charge of a fundamental
type IIB string wrapped along $S^1$. According to the physical
interpretation of various charges given earlier,
this gives the component $\pp_1$ of the magnetic charge vector $P$.

The net result of switching on both the electric and magnetic flux
along the D5-brane world-volume is to generate a magnetic charge
vector of the form:
\be \label{nnemag}
P_0 = \pmatrix{\wh P\cr \pp_1\cr 0 \cr Q_1
-1 \cr -J  \cr 1 \cr 0}
\, .
\ee

\end{enumerate}

This however is not the end of the story. So far we have
discussed the effect of the
various collective mode excitations on the charge vector to first
order in the charges, without taking into account the effect of the
interaction of the deformations produced by the collective
modes with the background fields already present in the system,
or the background fields produced by other collective modes.
Taking into account these effects produces further shifts in the
charge vector as described below.
\begin{enumerate}
\item As seen from \refb{nnetot}, 
the D5-brane world-volume theory has a coupling
proportional to
$\int C^{(2)}\wedge \FF\wedge \FF$.
Thus in the presence of magnetic flux
$\FF$ the D5-brane wrapped on $K3\times S^1$ acts as a source of the
D1-brane charge wrapped on $S^1$. The effect is a shift in the magnetic
charge quantum number $\pp_3$ that is quadratic 
in $\FF$ and hence 
quadratic in $\wh P$ due to
\refb{nneflux1}. A careful calculation, taking into
various signs and normalization factors,  shows that the
net effect of this term is to give an additional contribution to $\pp_3$ 
of the form:
\be \label{nnemag2}
\Delta_1 \pp_3 =  -\wh P^2 /2  
\, .
\ee
\item Let $C$ be a 2-form in the six dimensional theory obtained by
compactifying type IIB string theory on $K3$ and $F=dC$ be
its field strength. As summarized in \refb{nnep1}, switching on various
components of the electric charge vector $\vec Q$ 
requires us to switch
on $F$ proportional to $
dt\wedge\omega$ or $ dy\wedge\omega$.
The presence  of such background induces a coupling 
proportional to
\be \label{nneq1}
-\int \sqrt{-\det g} g^{yt} F_{ymn} F_{t}^{~mn} 
\ee
with the indices $m,n$ running over the coordinates 
of the Taub-NUT space. This
produces a source for $g^{yt}$, \i.e.\ momentum along $S^1$. 
The effect of such terms is to shift the component $\qq_4$ of the charge
vector $\vec Q$. A careful calculation shows that the net change in
$\qq_4$ induced due to this coupling is given by
\be \label{nneq2.1}
\Delta_2 \qq_4 =  \qq_3 \qq_5 + \wh Q^2/2
\, ,
\ee
where we have used the fact that $\qq_2$ has been set to zero.
The $\qq_3\qq_5$ term comes from taking $F$ in
\refb{nneq1} to be the
field strength of the RR 2-form field, and $\wh Q^2/2$ term
comes from taking $F$ to be the
field strength of the components of the RR 4-form field along various
2-cycles of $K3$. For non-zero $k_2$ there will also be an additive
contribution of $k_1k_2$ to the right hand side of eq.\refb{nneq2.1}.
\item The D5-brane wrapped on $K3\times S^1$
or the magnetic flux on this brane along any of the 2-cycles of $K3$
produces a magnetic type 2-form field configuration of the form:
\be \label{nneq2}
F\equiv dC \propto \sin\theta\, d\psi \wedge d\theta\wedge d\phi\, ,
\ee
where $C$ is any of the RR 2-form fields in six dimensional
theory obtained by compactifying type IIB string theory on $K3$.
One can verify that the 3-form appearing on the right hand side
of \refb{nneq2} is both closed and co-closed in the Taub-NUT
background and hence $F$ given in \refb{nneq2} satisfies both Bianchi
identity and the linearized equations of motion.
The coefficients of the term given in \refb{nneq2} for various
2-form fields $C$ are determined in terms of $\wh P$ and the
D5-brane charge along $K3\times S^1$ which has been set equal
to 1.
This together with the term in $F$ proportional to 
$dt\wedge \omega$ coming from the collective coordinate
excitation of the Kaluza-Klein monopole  generates a source
of the component $g^{\psi t}$ of the metric via the coupling
proportional to
\be \label{nneq3}
-\int \sqrt{-\det g} g^{\psi t} F_{\psi mn} F_{t}^{~mn} 
\ee
This induces a net momentum along $\wt S^1$ and gives a contribution
to the component $\pp_4$ of the magnetic charge vector $P$. A careful
calculation shows that the net additional contribution to $\pp_4$ 
due to this coupling is given by
\be \label{nneq4}
\Delta_3 \pp_4 =
\qq_3  + \wh Q\cdot \wh P\, .
\ee
In this expression the contribution proportional to $\qq_3$ comes
from taking $F$ in \refb{nneq3} to be the field strength associated
with the RR 2-form field of IIB, whereas the term proportional
to $\wh Q\cdot \wh P$ arises 
from taking $F$ to be the field strength associated
with the components of the RR 4-form field along various 2-cycles
of $K3$.
\item Eqs.\refb{nnemag} and \refb{nnemag2} show that we have a net
D1-brane charge along $S^1$ equal to
\be \label{nneq5}
\pp_3 = Q_1 - 1 -\wh P^2 / 2\, .
\ee
If we denote by $C^{(2)}$ the 2-form field of the original ten
dimensional type IIB string theory, then the effect of this charge is
to produce a background of the form:
\be \label{nneq6}
dC^{(2)} \propto (Q_1 - 1 - \wh P^2/2) \, r^{-2}\, dr \wedge dt\wedge
dy\, .
\ee
Again one can verify explicitly that the right hand side of \refb{nneq6}
is both closed and co-closed in the Taub-NUT background. If we
superimpose this on the background 
\be \label{nneq7}
dC^{(2)} \propto \qq_5 \, dy\wedge \omega \, ,
\ee 
coming from the excitation of the collective coordinate of the
Kaluza-Klein monopole, then we get a source term for
$g^{\psi t}$ via the coupling proportional to
\be \label{nneq8}
-\int \sqrt{-\det g} g^{\psi t} F_{\psi ry} F_{t}^{~ry} 
\ee
This  gives an additional contribution to the charge $\pp_4$
of the form
\be \label{nneq9}
\Delta_4 \pp_4 = 
\qq_5(Q_1 - 1 - \wh P^2/2) \, .
\ee
\end{enumerate}
For non-zero $k_2$ there will also be an additional 
contribution of $k_2 l_1$ to $l_4$ from the $-\int g^{\psi t}
(dB)_{\psi r y} (dB)_t^{~ry}$ term in the action.

This finishes our analysis of the possible additional
sources produced by the quadratic terms in the fields. What about
higher order terms? It is straightforward to show that the possible
effect of the
higher order terms on the shift in the charges 
will involve one or more powers
of the type IIB string coupling. Since the shift in the charges must
be quantized, they cannot depend on continuous moduli. Thus at least
in the weakly coupled type IIB string theory there are
no additional corrections
to the charges.

Combining all the results we see that we have a net electric charge
vector $\vec Q$ and a magnetic charge vector $\vec P$ of the form:
\be \label{nne1.6new}
Q = \pmatrix{\wh Q\cr \qq_1 \cr 0 \cr \qq_3 \cr -n + \qq_3 \qq_5 
+ \wh Q^2/2
\cr
\qq_5\cr -1}, \qquad P = \pmatrix{\wh P\cr \pp_1\cr 0 \cr Q_1
-1 -\wh P^2/2 \cr -J +\qq_3  + \wh Q\cdot \wh P
+  \qq_5(Q_1 - 1 - \wh P^2/2)\cr 1 \cr 0}
\, .
\ee
This has
\be \label{nneq2new}
Q^2=2n, \qquad P^2 = 2(Q_1-1), \qquad Q\cdot P = J\, .
\ee
Thus the additional charges do not affect the relationship
between the invariants $Q^2$, $P^2$, $Q\cdot P$ and the original
quantum numbers $n$, $Q_1$ and $J$. 

Let us now turn to the analysis of the dyon spectrum in the presence
of these charges. For this we recall that in \S\ref{s3.2}
the dyon
spectrum was computed from three mutually non-interacting
parts, -- the dynamics of
the Kaluza-Klein monopole, the overall motion of the D1-D5 system
in the background of the Kaluza-Klein monopole and the motion of
the D1-branes relative to the D5-brane. 
The precise dynamics of the D1-branes relative to the D5-brane is affected
by the presence of the gauge field flux on the D5-brane since it
changes the non-commutativity parameter describing the dynamics
of the gauge field on the D5-brane world-volume\cite{9908142}. 
As a result the
moduli space of D1-branes, described as non-commutative
instantons in this gauge theory\cite{9802068}, 
gets deformed. However we do not
expect this to change the elliptic genus of the corresponding conformal
field theory\cite{9608096} that enters the degeneracy formula.
With the exception of the
zero mode associated with the D1-D5 center of mass motion in the
Kaluza-Klein monopole background, the rest of the contribution
to the degeneracy came from the excitations involving non-zero
mode oscillators and this is not affected either by switching on
gauge field fluxes on the D5-brane world-volume or the momenta or
winding number of the collective coordinates of the Kaluza-Klein 
monopole. On the other hand the dynamics of the D1-D5-brane center
of mass motion in the background geometry is also not expected to
be modified in the weakly coupled type IIB string theory since in this
limit the additional background fields due to the additional  charges
are small compared to the one due to the Kaluza-Klein monopole.
(For this it is important that the additional charges do not involve any
other Kaluza-Klein monopole or NS 5-brane charge.) Thus we expect
the degeneracy to be given the same function $h(Q_1,n,J)$ that
appeared in the absence of the additional charges. Using \refb{nneq2new}
we can now write
\be \label{nnefin}
d(\vec Q,\vec P) = h\left( {1\over 2}P^2+1, 
{1\over 2} Q^2, Q\cdot P\right)\, .
\ee
This is a generalization of \refb{nne1.4} and shows that for the charge
vectors given in \refb{nne1.6new}, 
the degeneracy $d(\vec Q,\vec P)$ depend on the charges
only through the combination $Q^2$, $P^2$ and $Q\cdot P$.

This analysis
easily generalizes to $\ZZZ_N$ orbifolds of type IIB string theory on
$K3\times \wt S^1\times S^1$,
with the only change that the quantum
number $n$, instead of being an integer, will be an integer
multiple of $1/N$ and the charge vectors $\vec Q$, $\vec P$ are
restricted to the $\ZZZ_N$ invariant subspace. For $\ZZZ_N$ orbifolds
of type IIB on $T^4\times \wt S^1\times S^1$ there is an additional
change, -- the $(Q_1-1)$ factors in 
\refb{nne1.6new}, \refb{nneq2new} are replaced by $Q_1$.

It is instructive to focus on the four dimensional subspace of the
charge lattice spanned by the last four elements of 
$\vec Q$ and $\vec P$. 
In the second description of the theory
these correspond to momenta and 
fundamental string winding charges
and H- and Kaluza-Klein monopole charges along the circles
$\wh S^1$ and $S^1$.
It
follows from \refb{nne1.6new}, generalized to the
$\ZZZ_N$ orbifold cases so as to allow $n$ to be multiples
of $1/N$,  that our formula for the degeneracy
is valid for a 
general charge vector of the form
\be \label{egk1}
Q = \pmatrix{k_3\cr k_4 \cr k_5 \cr -1}, \qquad
P = \pmatrix{l_3\cr l_4 \cr 1 \cr 0}, \qquad k_4\in \ZZZ/N,
\quad k_i, l_i\in \ZZZ\quad \hbox{otherwise}
\, ,
\ee
in this subspace.

For use in \S\ref{sdualtwo} we shall now analyze
the T-duality orbit of these charge vectors. This is equivalent
to the question:
What is the most
general charge vector that can be reached from \refb{egk1} by
a T-duality transformation acting within this four dimensional
subspace? For this recall that 
T-duality transformation is the 
$\Gamma_1(N)\times \Gamma_1(N)$ subgroup of 
$SO(2,2;\ZZZ)$ matrices 
described in \refb{efulltduality}.
Such matrices, acting on the set \refb{egk1},  produce
charge vectors of the form:\footnote{While we do not know of
any further constraint on the charges, we have not proven that
\refb{efgg1} is the complete set of conditions on the T-duality
orbit of \refb{egk1}. Thus it is possible that the actual T-duality
orbit has additional conditions on $k_i$, $l_i$. \label{fsp}}
\ben \label{efgg1}
&& \qquad \qquad Q = \pmatrix{k_3\cr k_4 \cr k_5 \cr k_6}, \quad
P = \pmatrix{l_3\cr l_4 \cr l_5 \cr l_6}, \nonumber \\
&&
k_4\in \ZZZ/N,
\quad k_6\in N\ZZZ -1, \quad l_5\in N\ZZZ+1,
\quad l_6 \in N\ZZZ, \quad
k_i, l_i\in \ZZZ \, \, \hbox{otherwise}
\, , \nonumber \\
&& \hbox{g.c.d.}(Nk_3 l_4 -Nk_4 l_3 , k_5 l_6 -k_6  l_5 ,
k_3 l_5 -k_5 l_3 +k_4 l_6  -k_6 l_4 ) = 1
\, .
\een
The condition involving the g.c.d. comes from the 
observation that the left 
hand side
is preserved under the T-duality 
transformation\footnote{This in turn
follows from the fact that under the transformation
\refb{efulltduality} the arguments of g.c.d. transform into
linear combinations of each other with integer coefficients.
Thus the final g.c.d. must be an integer multiple of the initial
g.c.d.. Applying the inverse of the transformation 
\refb{efulltduality} on the final variables we can prove that the
initial g.c.d. is an integer multiple of the final g.c.d. Thus the
initial and final g.c.d.'s must be equal.
}
generated by the
matrices \refb{efulltduality}
and for the initial charge vector \refb{egk1} the left hand side
is 1 since $k_5 l_6 - k_6 l_5=1$. 
Note that
quantization law allows $k_6$ to be an arbitrary integer but a
T-duality transformation on the original charge vector can only
produce those $k_6$ which are $-1$ modulo $N$. 
Since in the
second description $-k_6$ measures the fundamental string
winding charge $w'$
along $S^1$ and the orbifold group generator
involves translation by $2\pi$ along $S^1$, 
requiring $k_6$ to be $-1$ mod
$N$ corresponds to restriction to states whose electric charge
vector lies in the sector twisted by a single power of the
orbifold group. 
Similarly quantization law allows $l_5$ to be an arbitrary
integer but a T-duality transformation of \refb{egk1} can
only produce charge vectors for which $l_5=1$ modulo 
$N$. In the second description this corresponds to 
requiring the total Kaluza-Klein monopole charge associated with the
$\wh S^1$ direction to be 1 modulo $N$.
We shall use the T-duality invariance of the theory
to argue in \S\ref{sdualtwo} that our results 
for degeneracy are valid for the
class of charge vectors given in \refb{efgg1}
in the same domain of the moduli space where
the original calculation was performed.

For orbifolds 
of type IIB string theory on
$K3\times \wt S^1\times S^1$ the condition
$l_5\in N\ZZZ+1$ 
can be relaxed by considering a more general configuration with 
arbitrary number $Q_5$ of D5-branes 
instead of a single D5-brane, subject to the condition
g.c.d.$(Q_5, Q_1)$=1. The 
counting of quarter BPS states for this more general configuration has 
been carried out in \cite{0605210} and reproduced in appendix
\ref{ssym}.
Also in this case the g.c.d. appearing in \refb{efgg1} is no longer 1
since for the initial charge vector $l_5=Q_5$ and
hence $k_5 l_6 - k_6 l_5=Q_5$.
Nevertheless the condition $\hbox{g.c.d.}
(Q_5, Q_1-Q_5)=1$, -- which translates to g.c.d.$(l_3,l_5)=1$, --
can be used
to argue that the charge vectors on the T-duality orbit of the initial
charge vector still satisfy the condition\cite{0702150}
\ben \label{enewgcd}
&& 
\hbox{g.c.d.} \{ k_i l_j - k_j l_i, \, N k_4 l_s - N k_s l_4,
\, k_4 l_6 - k_6 l_4; \quad i,j=3,5,6, \, s=3,5\} = 1\, .
\een
This follows from the fact that the left hand side is invariant
under T-duality and that it takes
the value 1 for the initial charge vector.
These relaxed constraints are also
consistent with the existence of 
the extra duality transformation \refb{esemidirect} in
these
theories. This duality transformation does not preserve the $l_5=1$ 
modulo $N$ condition, nor does it preserve the g.c.d. apperaing
in \refb{efgg1}. 

It may also be possible to relax the $l_5\in N\ZZZ+1$ condition and
the last condition in \refb{efgg1} for orbifolds of type IIB string
theory on $T^4\times \wt S^1\times S^1$ by taking multiple
D5-branes as in the case of $K3\times \wt S^1\times S^1$. This
would require
a careful analysis taking into account the dynamics of Wilson
lines on multiple D5-branes.

Let us now consider the effect of S-duality transformation on these
charge vectors. The action of this transformation on the charges
has been described in \refb{tdual3}, \refb{tdual4}. 
It follows from this that an S-duality
transformation acting on a charge vector of the form given in
\refb{efgg1} gives us back another charge vector of the same form. 
Thus the 
subset of the
charge lattice consisting of elements of the form \refb{efgg1} is
invariant under both the $\Gamma_1(N)$ S-duality and 
$\Gamma_1(N)\times \Gamma_1(N)$ T-duality transformations.
This is also the case if we relax the $l_5\in N\ZZZ+1$ condition and
replace the last condition on \refb{efgg1} by the 
condition \refb{enewgcd}, -- in this case the conditions are also
invariant under the additional $\ZZZ_2$ T-duality transformation
given in \refb{esemidirect}.

Since we have not shown that T-duality orbits of  \refb{egk1}
generate all charge vectors of the form \refb{efgg1} (see footnote
\ref{fsp}), it will
be useful to prove a slightly different result, -- namely that
the set of charge vectors in the 
T-duality orbits of the charges of the form
\refb{egk1} is closed under S-duality transformation. For this
we need to prove that an arbitrary
S-duality transform of any charge  vector
in the orbit can be brought to the form \refb{egk1} by a
T-duality transformation. To see that let us consider a charge
vector obtained from \refb{egk1} by left multiplication by
a T-duality transformation
matrix $\Omega_0$, followed by an 
arbitrary S-duality transformation
$\pmatrix{\alpha &\beta\cr \gamma & \delta}$. This produces a
charge vector:
\be \label{enewchvec}
Q= \Omega_0\pmatrix{\alpha k_3 + \beta l_3\cr \alpha k_4 + \beta l_4\cr
\alpha k_5 + \beta \cr -\alpha}, \qquad
P = \Omega_0\pmatrix{\gamma k_3 + \delta l_3\cr 
\gamma k_4 + \delta l_4\cr
\gamma k_5 + \delta \cr -\gamma}\, .
\ee
It is easy to see that left multiplication by the T-duality transformation
matrix
\be \label{efintd}
\pmatrix{ \gamma k_5 +\delta&-\gamma  & 0 & 0\cr 
-\alpha k_5 - \beta &  \alpha & 0 & 0\cr
0 & 0 & \alpha & \alpha k_5 
+ \beta\cr
0 & 0 & \gamma & \gamma k_5 +\delta} \Omega_0^{-1}\, ,
\ee
brings \refb{enewchvec} to the form \refb{egk1} with
$k_5=0$. Thus the
set of charge vectors which can be obtained from charge vectors
of the form \refb{egk1} by T-duality transformations
is closed under an S-duality transformation.

\subsectiono{Walls of marginal stability}
\label{smarginal}

As has been briefly mentioned in \S\ref{s3.s}, the degeneracy formula
derived here is expected to be valid within a certain region of the
moduli space bounded by codimension one subspaces on which the
BPS state under consideration becomes marginally stable. As we
cross this subspace of the moduli space, the spectrum can change
discontinuously. In this section we 
shall study in some detail the locations
of these walls of marginal stability so that we can identify the region
within which our degeneracy formula will remain valid. Our
analysis will follow the one given in \cite{0702141}. Some related
work can be found 
in \cite{0702150,0706.2363,0707.1563,0707.3035}.

Let us 
consider a state carrying electric charge $\vec Q$ and magnetic charge
$\vec P$ and examine under what condition it can decay into a
pair of half-BPS states.\footnote{There are also 
subspaces of the asymptotic moduli space where the mass of
a quarter BPS state becomes equal to the sum of the masses of a pair
of {\it quarter} BPS states, or a quarter BPS state and
a half-BPS state or more than two half or quarter BPS states. 
However it has been shown 
in \cite{0707.1563,0707.3035} that
such subspaces are
of codimension larger than one. Hence in going from one
generic point in the moduli space to another one can avoid them
by going around them. As a result we do not expect them to affect
the dyon spectrum.}
This happens
when its mass is equal to the sum of the masses of a pair of half BPS
states whose electric and magnetic charges add up to $\vec Q$ and
$\vec P$ respectively.
Since for half BPS states the electric and magnetic
charges must be parallel, these pair of states must
have charge vectors of
the form $(a\vec M, c \vec M)$ and $(b \vec N, d \vec N)$ for some
constants $a$, $b$, $c$, $d$ and a pair of $r$-dimensional vectors
$\vec M$, $\vec N$. We shall normalize $\vec M$, 
$\vec N$ such that 
\be \label{enorm}
ad-bc=1\, .
\ee
Then the requirement that the charges add up to $(\vec Q, \vec P)$ gives
\be \label{esm2}
\vec M = d\vec Q - b\vec P, \qquad \vec N = -c\vec Q + a\vec P\, .
\ee
Thus the charges of the decay products are given by
\be \label{escale0}
(ad\vec Q-ab\vec P, cd\vec Q-cb\vec P) \quad \hbox{and} \quad
(-bc\vec Q + ab\vec P, -cd\vec Q+ad\vec P)\, .
\ee
Under the scale
transformation
\be \label{escale1}
\pmatrix{a & b\cr c & d}\to 
\pmatrix{a & b\cr c & d} \, \pmatrix{\lambda & 0\cr 0 
& \lambda^{-1}}
\ee
eqs.\refb{enorm} and \refb{escale0} remain unchanged.
There is also a discrete transformation
\be \label{esc12}
\pmatrix{a & b\cr c & d}\to 
\pmatrix{a & b\cr c & d} \, \pmatrix{0 & 1\cr -1 & 0}\, ,
\ee
which  leaves \refb{enorm} unchanged and exchanges the
two decay products in \refb{escale0}.
Thus a pair of matrices $\pmatrix{a & b\cr c & d}$ related by
\refb{escale1} or \refb{esc12} describe identical decay channels.

In order that the charge vectors of the decay products
given in \refb{escale0} satisfy the
charge quantization rules we must ensure that $a\vec M
=ad\vec Q-ab\vec P$
and $b\vec N=-bc\vec Q + ab\vec P$ belong to the lattice of electric
charges and that $c\vec M=cd\vec Q-cb\vec P$ and 
$d\vec N=-cd\vec Q+ad\vec P$ belong to
the lattice of magnetic charges. 
For the charge vectors $\vec Q$, $\vec P$ given in
\refb{echvec},  or more generally \refb{nne1.6new} 
or \refb{efgg1},
this would require 
\be \label{egen}
ad,ab,bc\in \ZZZ, \qquad 
cd\in N\ZZZ\, .
\ee
The condition $cd\in N\ZZZ$ comes from the requirement
that $cd\vec Q-cb\vec P$ is an allowed magnetic charge.
In particular for a $\vec Q$ of the form \refb{echvec}, a magnetic
charge $cd\vec Q$ represents, in either description, 
a state with
Kaluza-Klein monopole charge $-cd$
associated with $S^1$. Since this charge is quantized in units
of $N$, $cd$ must be a multiple of $N$.
We shall denote by $\AAA$ the set of matrices $\pmatrix{a & b\cr c
& d}$ subject to the equivalence relations \refb{escale1}, \refb{esc12}
and satisfying \refb{enorm}, \refb{egen}. One can show that
using the scaling freedom \refb{escale1} one can always choose
$a,b,c,d$ to be integers satisfying \refb{enorm}\cite{0702141}. 
Eq.\refb{egen}
then gives further restriction on the integers $c$ and $d$. 

We shall now determine the wall of marginal stability corresponding
to the decay channel given in \refb{escale0}.
Our starting point will be the formula for the mass $m(\vec Q,
\vec P)$
of a BPS state carrying electric charge $\vec Q$ and magnetic charge
$\vec P$\cite{9507090,9508094}
\ben \label{esm1}
m(\vec Q, \vec P)^2 &=& {1\over S_\infty} 
(Q - \bar\tau_\infty P)^T
(M_\infty + L) (Q -\tau_\infty P) \nonumber \\
&&
+ 2 \left[ (Q^T (M_\infty + L) Q) (P^T (M_\infty + L) P)
- (P^T (M_\infty + L) Q)^2\right]^{1/2}\, , \nonumber \\
\een
where $\tau=a+iS$ and the subscript $\infty$ denotes asymptotic
values of various fields. This expression is manifestly invariant under
the T- and S-duality transformations described in 
eqs.\refb{tdual1}-\refb{tdual4} in \S\ref{sdual}.
In order that the state $(\vec Q,\vec P)$ is marginally stable against decay
into $(ad\vec Q-ab\vec P, cd\vec Q-cb\vec P)$ and $(-bc\vec Q
+ab\vec P, -cd\vec Q+ad\vec P)$, we need
\be \label{esm3}
m(\vec Q,\vec P) = m(ad\vec Q-ab\vec P, cd\vec Q-cb\vec P)
+ m(-bc\vec Q
+ab\vec P, -cd\vec Q+ad\vec P)\, .
\ee
Using \refb{esm1}, \refb{esm3} 
and some tedious algebra, we arrive at the
condition
\be \label{esm4}
\left(a_\infty - {ad+bc\over 2cd}\right)^2 
+ \left( S_\infty +{E\over 2 cd}
\right)^2 = {1\over 4 c^2 d^2} (1 + E^2)\, ,
\ee
where
\be \label{esm5}
E \equiv { cd (Q^T (M_\infty + L) Q)
+ ab (P^T (M_\infty + L) P) -(ad + bc) (P^T (M_\infty + L) Q) \over 
\left[ (Q^T (M_\infty + L) Q) (P^T (M_\infty + L) P)
- (P^T (M_\infty + L) Q)^2\right]^{1/2}
}\, .
\ee
Note that $E$ depends on $M_\infty$, the constants $a,b,c,d$ and
the charges $\vec Q$, $\vec P$, but is 
independent of $\tau_\infty$.
Thus for fixed $\vec P$, $\vec Q$ and 
$M_\infty$, the wall of marginal
stability describes a circle in the $(a_\infty, S_\infty)$
plane with radius 
\be \label{esm6}
R = \sqrt{1 + E^2}/2|cd|\, ,
\ee 
and center at
\be \label{esm7}
C = \left( {ad+bc\over 2cd}, -{E\over 2 cd}\right)\, .
\ee
This circle 
intersects the real $\tau_\infty$
axis at
\be \label{eireal}
{a/c} \quad \hbox{and} \quad {b/d}\, .
\ee

The cases where either $c$ or $d$ vanish require special attention.
First consider the case $c=0$. In this case the condition
$ad-bc=1$ implies that $ad=1$, \i.e.\ either $a=d=1$
or $a=d=-1$. Using the scaling freedom
\refb{escale1} we can choose $a=d=1$. By taking the $c\to 0$
limit of \refb{esm4}, \refb{esm5} 
we see that  the wall of marginal
stability becomes a straight line in the $(a_\infty, S_\infty)$ plane
for a fixed $M_\infty$:
\be\label{ecurve1}
a_\infty - {b (P^T (M_\infty + L) P)-(P^T (M_\infty + L) Q)  \over 
\left[ (Q^T (M_\infty + L) Q) (P^T (M_\infty + L) P)
- (P^T (M_\infty + L) Q)^2\right]^{1/2}
}\, S_\infty -b =0\, .
\ee
The $d=0$ case is related to the $c=0$ case 
by the equivalence relation \refb{esc12}, and hence do not give
rise to new walls of marginal stability.

In order to get some insight into the geometric structure of the domain
bounded by these marginal stability walls it will be useful to
study the possible intersection points of these walls in the upper
half $\tau_\infty$ plane\cite{0702141}. 
By a careful analysis one finds that these walls
never intersect in the interior of the upper half plane. The only
possible intersections are on the real axis, at $i\infty$ 
or in the lower half
plane. Furthermore due to \refb{eireal} the intersection points on
the real axis are independent of the charges or the moduli $M_\infty$.
Thus a domain bounded by the marginal stability walls has
universal vertices although the precise shape of the walls do 
depend on the moduli $M_\infty$ as well as the charges $(\vec Q,
\vec P)$. This allows us to give a universal classification of domains
in terms of their vertices. 
In fact since the integers $a,b,c,d$ are related to the charges of the
decay products on the corresponding wall via \refb{escale0}, this
universal characterization of a domain corresponds to specifying
how the charges of a decay product are related to the charges
of the original system on the different walls bordering a
particular domain.
Various other
geometric properties of these
walls have been discussed in \cite{0707.3035}.

This finishes our general analysis of marginal stability walls and
domains bounded by them. An important question that we need to
address now is: in which of the many domains in the $\tau_\infty$
plane is our degeneracy formula given in \refb{egg1int}, \refb{ep2kk}
valid? This question can be answered by recalling that in the
first description we work in the weak
coupling limit of type IIB string theory
at finite values of the other moduli. Following the
chain of dualities one can translate this to information about
$\tau_\infty$  and the matrix $M_\infty$ 
in the second description, and work out how this region is situated
with respect to various marginal stability lines in the $\tau_\infty$
plane.
It turns out\cite{0702141} in this limit $a_\infty$, $S_{\infty}$
are finite and $P^T(M_\infty+L)P\sim |Q^T(M_\infty+L)P|
<< Q^T (M_\infty+L)Q$. As a result for $cd\ne 0$, $E$ defined
in \refb{esm5} is large and in the 
upper half plane
the circles \refb{esm5} 
lie close to the real axis. On the other hand the straight lines
\refb{ecurve1} become almost vertical lines passing through the integers
$b$. Thus for $-1<a_\infty<1$ the weak coupling
region in the first description gets mapped to one
of two domains, -- the
right domain $\RR$ bounded by the lines corresponding to
$b=0$ and $b=1$ 
in \refb{ecurve1} together with a set of circle segments at the
bottom, and the left domain $\LL$ 
bounded by the lines corresponding to $b=-1$ and $b=0$
in \refb{ecurve1} together with a set of circle segments at the
bottom. The marginal stability wall corresponding to $b=0$
corresponds to the wall of marginal stability encountered in
the analysis of the supersymmetric quantum mechanics describing
the D1-D5 center of mass motion in the Kaluza-Klein monopole
background. For the right domain $\RR$ 
our original formula
given in \refb{egg1int}, \refb{ep2kk} is valid. For the left domain
$\LL$ we need to change the integration contour to the one given
in \refb{ecchat}.
We shall denote by $\BB_R$ the set
of  matrices $\pmatrix{a & b\cr c & d}$
corresponding to the marginal stability
walls which form the boundary of the region $\RR$ and by $\BB_L$
the set of matrices describing the marginal stability
walls which form the boundary of the region $\LL$. These sets
have been determined explicitly in \cite{0702141} for various
values of $N$. 
For example for $N=1$ the set $\BB_R$ is given by
\be \label{esetbr}
\BB_R: \quad \left\{
\pmatrix{1 & 0\cr 0 & 1}, \quad \pmatrix{1 & 1\cr 0 & 1}, 
\quad \pmatrix{1 & 0\cr 1 & 1}\right\}\, ,
\ee
representing respectively  a straight line passing through
0, a straight line passing through 1 and a circle 
passing through 0 and 1. Thus the vertices of $\RR$ are at 0, 1 and 
$\infty$. One can show that for $N=2$ the vertices of $\RR$ are at 0, 
$1/2$, 1 and $\infty$ and for $N=3$ they are at 0, $1/3$, $1/2$, $2/3$ and 
$\infty$\cite{0702141}. For $N\ge 4$ each domain bounded by walls of 
marginal stability has infinite number of vertices.

An interesting question is: how does the degeneracy formula change as
we cross other marginal stability walls? We shall argue in 
\S\ref{sdualtwo} that the changes are such that the expression
for the degeneracy in the other domains can also be expressed as
an integral of the type \refb{egg1int}, but with an integration
contour that is different from \refb{ep2kk}.

\subsectiono{Duality transformation of the degeneracy formula} 
\label{sdualtwo}

As noted in \S\ref{s3.s}, 
the degeneracy formula \refb{egg1int},
\refb{ep2kk} has been 
written in terms
of T-duality invariant combinations $Q^2$, $P^2$ and $Q\cdot P$
although we have derived the formula only for a special class of
charge vectors \refb{echvec} or more generally 
\refb{nne1.6new}. 
In this section we shall discuss what information about
the degeneracy formula can be extracted using the 
T- and S-duality symmetries of the theory.

We begin by studying the consequences of 
the T-duality symmetries of the theory. It follows from
\refb{tdual1}, \refb{tdual2} that if a T-duality transformation takes
a charge vector $(\vec Q, \vec P)$ to $(\vec Q',\vec P')$ then
\be \label{sadd2}
Q^{\prime 2}=Q^2, \quad P^{\prime 2}=P^2, \quad Q'\cdot P'
= Q\cdot P\, .
\ee
However there may be
pairs of charge vectors with the same $Q^2$, $P^2$ and
$Q\cdot P$ which are not related by a T-duality transformation.
Clearly T-duality invariance of the theory cannot give us
any relation between the degeneracies associated with
such a pair of charge vectors.
In what follows we shall focus on charge vectors $(\vec Q',\vec P')$
which are in the same T-duality orbit of a charge vector $(\vec Q,
\vec P)$ for which we have derived \refb{egg1int}. 

In \S\ref{smarginal}
we have denoted by $\RR$ the  domain of the
region of the moduli space  in which
the original formula \refb{egg1int}, \refb{ep2kk}
for $d(\vec Q, \vec P)$ is valid. It is bounded
by a set of marginal stability walls labelled by 
$\pmatrix{a & b\cr c & d} \in \BB_R$.  
Let
$\RR'$ denote the image of
$\RR$ under some particular T-duality map. 
In this case we
expect $d(\vec Q', \vec P')$ in the region
$\RR'$ to be equal to $d(\vec Q, \vec P)$ given
in \refb{egg1int}:
\ben\label{egg1intrp}
d(\vec Q',\vec P') &=& (-1)^{Q\cdot P+1}\,
{1\over N}\, \int _\CC d\wt\rho \, 
d\wt\sigma \,
d\wt v \, e^{-\pi i ( N\wt \rho Q^2
+ \wt \sigma P^2/N +2\wt v Q\cdot P)}\, {1
\over \wt\Phi(\wt \rho,\wt \sigma, \wt v)}\, , \nonumber \\
&=& (-1)^{Q'\cdot P'+1}\,
{1\over N}\, \int _\CC d\wt\rho \, 
d\wt\sigma \,
d\wt v \, e^{-\pi i ( N\wt \rho Q^{\prime 2}
+ \wt \sigma P^{\prime 2}2/N +2\wt v Q'\cdot P')}\, {1
\over \wt\Phi(\wt \rho,\wt \sigma, \wt v)}\, ,
\een
where $\CC$ has been defined in \refb{ep2kk}.
In going from the first
to the second line of \refb{egg1intrp} we have used \refb{sadd2}.

Let us now determine the region $\RR'$.
Since under a T-duality
transformation $M\to \Omega M \Omega^T$, and since $\RR'$
is the image of $\RR$ under this map, $\RR'$
is bounded by walls of marginal stability described in \refb{esm4},
\refb{esm5} with $\pmatrix{a & b\cr c & d}\in \BB_R$
and $M_\infty$ in \refb{esm5}
replaced by $\Omega^{-1} M_\infty (\Omega^T)^{-1}$.
Using \refb{tdual1}
we see that this effectively replaces $(\vec Q, \vec P)$ by $(\vec Q',
\vec P')$ in \refb{esm5}. Thus
$\RR'$ is the region of the upper half plane bounded by the
circles:
\be \label{esm4rp}
\left(a_\infty - {ad+bc\over 2cd}\right)^2 
+ \left( S_\infty +{E'\over 2 cd}
\right)^2 = {1\over 4 c^2 d^2} (1 + E^{\prime 2})\, , 
\quad \pmatrix{a & b\cr c & d}\in \BB_R\, ,
\ee
where
\be \label{esm5rp}
E' \equiv {   cd (Q^{\prime T} (M_\infty + L) Q')
+ ab (P^{\prime T} (M_\infty + L) P') 
-(ad + bc) (P^{\prime
T} (M_\infty + L) Q') \over 
\left[ (Q^{\prime T} (M_\infty + L) Q') (P^{\prime T} (M_\infty + L) 
P')
- (P^{\prime T} (M_\infty + L) Q')^2\right]^{1/2}
}\, .
\ee
In particular $\RR'$ has the same vertices $a/c$ and $b/d$ as $\RR$.
Thus in the universal classification scheme described in 
\S\ref{smarginal}, $\RR'$ and $\RR$ correspond to the same
domains although the precise shape of the
domain walls differ for $\RR$ and $\RR'$ due to the replacement of
$(\vec Q, \vec P)$ by the new charge vector $(\vec Q',\vec P')$.
Since eqs.\refb{egg1intrp}-\refb{esm5rp} are valid for any charge vector
$(\vec Q',\vec P')$ which can be related to the charge vectors
given in  
\refb{nne1.6new} via a T-duality transformation,
we conclude that any $(\vec Q, \vec P)$ which is in the T-duality
orbit of the charge vectors \refb{nne1.6new}, the degeneracy
$d(\vec Q, \vec P)$ is given by \refb{egg1int}, \refb{ep2kk}
in the domain $\RR$.
In the four dimensional subspace of charge vectors, consisting
of momenta, fundamental string winding charge, H-monopole
charge and Kaluza-Klein
monopole charge along the circles $\wh S^1$ and $S^1$ in the
second description, the T-duality orbit consists of  charge
vectors of the type given in \refb{efgg1}. Thus for these charge
vectors, 
eqs.\refb{egg1int},
\refb{ep2kk} give us the correct expression for $d(\vec Q,
\vec P)$ in the domain $\RR$.

Next we shall analyze the consequences of S-duality symmetry.
We begin with a charge vector $(\vec Q, \vec P)$ for which
\refb{egg1int}, \refb{ep2kk} hold in the domain $\RR$, {\it e.g.} 
a charge
vector in the T-duality orbit of \refb{egk1}.
An S-duality transformation changes the vector $(\vec Q, \vec P)$
to another vector $(\vec Q'', \vec P'')$ 
and $\tau$ to $\tau''$ 
via the formul\ae\ \refb{tdual3}, \refb{tdual4}.
Thus if $\RR''$ denotes
the image of the region $\RR$ under the  map \refb{tdual4}, then
S-duality invariance implies that
inside $\RR''$ the degeneracy $d(\vec Q'', \vec P'')$ is given by
the same expression \refb{egg1int} for $d(\vec Q, \vec P)$:
\be\label{sadd5}
d(\vec Q'',\vec P'') = (-1)^{Q\cdot P+1}\,
{1\over N}\, \int _\CC d\wt\rho \, 
d\wt\sigma \,
d\wt v \, e^{-\pi i ( N\wt \rho Q^2
+ \wt \sigma P^2/N +2\wt v Q\cdot P)}\, {1
\over \wt\Phi(\wt \rho,\wt \sigma, \wt v)}\, .
\ee
We would like to express the right hand side of \refb{sadd5} in
terms of the vectors $\vec Q''$ and $\vec P''$. For this we define
\be\label{epm}  
\pmatrix{\tilde \alpha &\tilde \beta\cr\tilde \gamma &\tilde \delta} = 
\pmatrix{\delta  & \gamma /N \cr \beta  N & \alpha }\in
\Gamma_1(N)\, ,
\ee
and
\be\label{e3.3}
  \pmatrix{\wrh''\cr \ws''\cr \wv''}  \equiv 
  \pmatrix{\wrh''_1+i\wrh''_2\cr \ws''_1+i\ws''_2\cr \wv''_1
  +i\wv''_2}=
  \pmatrix{\tilde \alpha^2& \tilde \beta^2&- 2\tilde 
   \alpha
   \tilde \beta\cr
   \tilde \gamma^2&\tilde \delta^2 & - 
   2\tilde
   \gamma\tilde \delta\cr
    -\tilde \alpha\tilde \gamma& - \tilde \beta\tilde 
   \delta & (\tilde \alpha\tilde \delta + \tilde 
   \beta\tilde \gamma)} \pmatrix{\wrh\cr \ws\cr \wv}\, .
   \ee
Using \refb{tdual3}, \refb{tdual4}, \refb{epm},
\refb{e3.3} one can easily verify that
\be\label{e3.4}
e^{-\pi i ( N\wt \rho Q^2
+ \wt \sigma P^2/N +2\wt v Q\cdot P)} = 
e^{-\pi i ( N\wt \rho'' Q^{\prime\prime 2}
+ \wt \sigma'' P^{\prime \prime 2}/N +2\wt v'' 
Q''\cdot P'')}   \, ,
\ee
and
\be\label{e3.5}
d\wt\rho \, 
d\wt\sigma \,
d\wt v = d\wt\rho'' \, 
d\wt\sigma'' \,
d\wt v'' \, .
\ee
Furthermore, with the help of eq.\refb{ewtphitrs}
one can show that\cite{0609109}
\be\label{e3.6}
   \wt\Phi(\wt \rho'',\wt\sigma'',\wt v'') = \wt\Phi(
   \wt \rho,\wt\sigma,\wt v)\, .
\ee
If $\CC''$ denotes the image of $\CC$ under the map \refb{e3.3}
then 
eqs.\refb{e3.4}-\refb{e3.6} allow us to express
\refb{sadd5} as\footnote{In arriving at \refb{e3.7} we have used
that $(-1)^{Q\cdot P}=(-1)^{Q''.P''}$. This follows from the S-duality
transformation laws of the charges and the observation that
$NQ^2$ and $P^2$ are even integers.}
\be\label{e3.7}
d(\vec Q'',\vec P'') = (-1)^{Q''\cdot P''+1}\,
{1\over N}\, \int _{\CC''} d\wt\rho'' \, 
d\wt\sigma'' \,
d\wt v'' \, e^{-\pi i ( N\wt \rho'' Q^{\prime\prime 2}
+ \wt \sigma'' P^{\prime \prime 2}/N +2\wt v Q''\cdot P'')}\, {1
\over \wt\Phi(\wt \rho'',\wt \sigma'', \wt v'')}  \, .
\ee
To find the location of $\CC''$ we note that under the map
\refb{e3.3} the real parts of $\wrh$, $\ws$ and $\wv$ mix
among themselves and the imaginary parts of 
$\wrh$, $\ws$ and $\wv$ mix among themselves.
The initial contour $\CC$ corresponded to a unit cell of
the cubic lattice in the
($\wrh_1$,$\ws_1$,$\wv_1$) space spanned by the basis vectors
$(1,0,0)$, $(0,N,0)$ and $(0,0,1)$. The unimodular
map \refb{e3.3} transforms this into a different unit cell
of the same lattice. We can now use
the 
shift symmetries
\be\label{eshift}
\wt\Phi(\wt\rho,\wt\sigma,\wt v) =
\wt\Phi(\wt\rho+1,\wt\sigma,\wt v) =
\wt\Phi(\wt\rho,\wt\sigma+N,\wt v) =
\wt\Phi(\wt\rho,\wt\sigma,\wt v+1)\, ,
\ee
which are manifest from \refb{edefwtphi}, to
bring the integration region back to the original unit cell.
Thus $\CC''$ and $\CC$ differ only in the values of the imaginary
parts of $\wrh$, $\ws$ and $\wv$. Using \refb{ep2kk},
\refb{e3.3} we see that for the contour $\CC''$,
\ben \label{sadd7}
    \wt \rho''_2 =\tilde \alpha^2\, M_1 +\tilde \beta^2\, M_2 + 2\tilde 
   \alpha
   \tilde \beta \, M_3 \, ,
   \nonumber \\
    \wt\sigma''_2 =\tilde \gamma^2\, M_1 +\tilde \delta^2 \, M_2 + 
   2\tilde
   \gamma\tilde \delta\, M_3 \, ,
   \nonumber \\
    v''_2 = -\tilde \alpha\tilde \gamma\, M_1 - \tilde \beta\tilde 
   \delta \, M_2 - (\tilde \alpha\tilde \delta + \tilde 
   \beta\tilde \gamma)\, M_3 \, .
\een
Thus $\CC''$ is not identical to $\CC$.
We could try to deform $\CC''$ back to $\CC$, but in that process
we might pick up contribution from the residues at the poles of
$1/\wt\Phi(\wt \rho'',\wt \sigma'', \wt v'')$. Thus we see that the
degeneracy formula \refb{e3.7} for $d(\vec Q'', \vec P'')$ is not
obtained by simply replacing $(\vec Q, \vec P)$ by
$(\vec Q'', \vec P'')$ in the
expression for $d(\vec Q, \vec P)$. The integration contour $\CC$
also gets deformed to a new contour $\CC''$.

Let us now analyze the region $\RR''$
of the asymptotic moduli space in
which \refb{e3.7} is valid. This is obtained by taking the image
of the region $\RR$ under the transformation \refb{tdual4}.
To determine this region we need to first study the images of the
curves described in \refb{esm4} in the
$a_\infty-S_\infty$ plane. A straightforward analysis shows that
the image of \refb{esm4} is described by the curve
\be \label{esm4rprp}
\left(a_\infty - {a''d''+b''c''\over 2c''d''}\right)^2 
+ \left( S_\infty +{E''\over 2 c''d''}
\right)^2 = {1\over 4 c^{\prime\prime 2} d^{\prime\prime 2}} (1 + 
E^{\prime\prime 2})\, ,
\ee
where
\be \label{sadd8}
\pmatrix{a'' & b'' \cr c'' & d''} = \pmatrix{\alpha & \beta\cr \gamma
& \delta} \pmatrix{a & b\cr c & d}\, ,
\ee
and
\be \label{esm5rprp}
E'' \equiv { c''d'' (Q^{\prime\prime 
T} (M_\infty + L) Q'')
+ a''b'' (P^{\prime\prime  T} (M_\infty + L) P'') 
 -(a''d'' + b''c'') (P^{\prime\prime
T} (M_\infty + L) Q'') \over 
\left[ (Q^{\prime \prime T} (M_\infty + L) Q'') 
(P^{\prime \prime T} (M_\infty + L) 
P'')
- (P^{\prime \prime T} (M_\infty + L) Q'')^2\right]^{1/2}
}\, .
\ee
This is identical in form to \refb{esm4} with 
$\pmatrix{a & b \cr c & d}$ replaced by
$\pmatrix{a'' & b'' \cr c'' & d''}$ and $(\vec Q,\vec P)$
replaced by $(\vec Q'',\vec P'')$.
Since the original domain $\RR$ was bounded by a
set of marginal stability walls $\left\{\pmatrix{a & b\cr
c & d}\right\}\in\BB_R$,
the domain $\RR''$ 
is bounded by the collection of walls described by  
\refb{esm4rprp}, \refb{esm5rprp} with
$\pmatrix{a'' & b''\cr c'' & d''}\in \pmatrix{\alpha & \beta\cr
\gamma & \delta}\BB_R$. Since these walls end on vertices 
$a''/c''\ne a/c$ and $b''/d''\ne b/d$, under the universal
classification scheme $\RR''$ and $\RR$ describe different
domains. Thus the result of this analysis may be summarized
in the statement that $d(\vec Q'',\vec P'')$ inside the domain
$\RR''$ is given by the same integral formula \refb{egg1int} with
$(\vec Q,\vec P)$ replaced by $(\vec Q'', \vec P'')$ and the
contour $\CC$ replaced by the new contour $\CC''$ given 
in \refb{sadd7}.

Now, as has been argued at the end of \S\ref{sadditional}, an
S-duality transformation acting on a charge vector in the T-duality
orbit of \refb{egk1} gives us back another charge vector
in the T-duality orbit of \refb{egk1}. 
Thus $(\vec Q'', \vec P'')$ is in the T-duality orbit
of \refb{egk1}, and $d(\vec Q'',\vec P'')$ inside the domain $\RR$
would have been given by eqs.\refb{egg1int}, \refb{ep2kk}
with $(\vec Q, \vec P)$ replaced by $(\vec Q'', \vec P'')$.
Thus we now have expressions for $d(\vec Q'', \vec P'')$ in two
different domains,  -- $\RR$ and $\RR''$. In both domains the
degeneracy is given by an integral. The integrand in both cases are
same, but in one case the integration contour is $\CC$ while in
the other case it is $\CC''$.
This shows that
as we cross the walls of marginal stability to move from the
domain $\RR$ to $\RR''$ the expression
for $d(\vec Q'', \vec P'')$ changes by a modification in the location
of the contour of integration. The precise modification of the contour
for a given wall crossing will be
given by \refb{sadd7}.  Following this line of
argument one finds that 
for charge vectors
in the T-duality orbit of \refb{egk1}
the degeneracy formula in different
domains bounded by walls of marginal stability are given by the
three dimensional contour integral of the same integrand as
in \refb{egg1int} but with different choices 
of the integration contour.

The jump in the degeneracy across a marginal stability wall
can be calculated in terms of residues of the integrand at the poles
we encounter as we deform the relevant contour on one side of
the wall to the relevant contour on the other side of the wall.
One finds that if the wall corresponds to the decay of the original
dyon into a pair of half BPS dyons with charges 
$(\vec Q_1,\vec P_1)$
and $(\vec Q_2,\vec P_2)$, and if it is related to the wall
corresponding to the decay into $(\vec Q, 0)+(0,\vec P)$
by an S-duality transformation,
then up to a sign that depends on the
direction in which
we cross the wall, the jump is given by\cite{0702141}
\be \label{ejump}
(-1)^{\vec Q_1\cdot \vec P_2-\vec Q_2\cdot \vec P_1 +1}
(\vec Q_1\cdot 
\vec P_2-\vec Q_2\cdot \vec P_1) \,
d_{half}(\vec Q_1,\vec P_1)\,
d_{half}(\vec Q_2,\vec P_2)\, ,
\ee
where $d_{half}(\vec Q_i, \vec P_i)$ denotes the number of 
bosonic minus fermionic half BPS supermultiplets carrying
charges $(\vec Q_i,\vec P_i)$.
For $(\vec Q_1,\vec P_1)=(\vec Q, 0)$
and $(\vec Q_2,\vec P_2)=(0,\vec P)$ this result will be proved in
eq.\refb{mes2}. 
For the other cases the formula
can be obtained by duality transformation of this result 
as long as the  corresponding marginal stability wall is related 
to the wall associated with the decay into
$(\vec Q,0)$ and $(0,\vec P)$ by an S-duality
transformation. For $N=1,2,3$, \i.e.\ for heterotic string theory
on $T^6$ and asymmetric
$\ZZZ_2$ and $\ZZZ_3$ orbifolds of heterotic
or type II string theory on $T^6$, this includes all the 
walls\cite{0702141}.
The
result \refb{ejump} is identical to the wall crossing
formula proposed in \cite{0702146} for $\NN=2$ supersymmetric
string theories and will be relevant for the analysis in \S\ref{smulti}.

So far we have used S-duality invariance to determine the
locations of the integration contour in the degeneracy formula in
different domains in the moduli space, but have not carried out
any test of S-duality. We shall now
describe some tests of
S-duality that one could perform.
\begin{enumerate}
\item If there is an S-duality transformation that leaves the set
$\BB_R$ invariant, then under such a transformation the contour
$\CC$ either should not transform, or should transform to another
contour that is continuously deformable to $\CC$ without passing
through any poles.
\item Analysis of \S\ref{s3.2} has shown that inside the 
left domain
$\LL$ corresponding to the set of matrices $\BB_L$, the degeneracy
is obtained by performing integration over the
contour $\wh\CC$ described in \refb{ecchat}.
Thus if there is an S-duality transformation that maps the set
$\BB_R$ to the set $\BB_L$ then such a transformation must
map the contour $\CC$ to the contour $\wh\CC$ or another
contour deformable to $\wh\CC$ without passing through
any pole.
\end{enumerate}
In fact for all values of $N$ 
one can identify 
a pair of S-duality transformations which map
$\BB_R$ to $\BB_L$\cite{0702141}. They are given by
\be \label{egiven}
\pmatrix{\alpha & \beta \cr \gamma & \delta}
= \pmatrix{1 & -1\cr 0 & 1}\equiv g_1, \qquad 
\pmatrix{\alpha & \beta \cr \gamma & \delta}=\pmatrix{1 & 0\cr
-N & 1}\equiv g_2\, .
\ee
This in turn implies that
$g_1 g_2^{-1}$ maps the set $\BB_R$ to itself.
One can show that under both transformations given in
\refb{egiven} the
contour $\CC$ gets mapped to another contour that is deformable
to $\wh\CC$\cite{0702141}.  This in turn implies
that $g_1g_2^{-1}$ takes the contour $\CC$ to another contour
deformable to $\CC$. This 
provides non-trivial test of S-duality symmetry of the
degeneracy formula. For $N\le 6$ all transformations which map
$\BB_R$ to $\BB_R$ may be obtained by
taking positive and negative powers of $g_1 g_2^{-1}$, and those
which takes $\BB_R$ to $\BB_L$ are obtained by taking positive
and negative powers of $g_1 g_2^{-1}$ followed by a single
power of $g_1$ (or $g_2$)\cite{0702141}. Thus for these cases
our test of S-duality is complete.

\subsectiono{The statistical entropy function} \label{s3.3}

Although \refb{egg1int} gives an exact formula for the degeneracy
of dyons, it is hard to compare this directly with the results for
the black hole entropy derived earlier in \S\ref{s2.2}.
In this section we shall describe a systematic procedure for
extracting the
behaviour of $d(\vec Q,\vec P)$
for large charges:
\be \label{elargecharge}
Q^2>>0, \qquad P^2>>0, \qquad Q^2 P^2 - (Q\cdot P)^2 >> 0,
\ee
and also explicitly compute
the first order corrections to the
leading asymptotic formula. Our analysis will follow the one given
in \cite{0605210}; this in turn is based on the earlier analysis
of \cite{9607026,0412287,0510147,0601108}.

Our starting point will be the general expression for
$d(\vec Q,\vec P)$
\be\label{egg1rp}
d(\vec Q,\vec P) = (-1)^{Q\cdot P+1}\,
{1\over N}\,  \int\, d\wt\rho \, 
d\wt\sigma \,
d\wt v \, e^{-\pi i ( N\wt \rho Q^2
+ \wt \sigma P^2/N +2\wt v Q\cdot P)}\, {1
\over \wt\Phi(\wt \rho,\wt \sigma, \wt v)}\, ,
\ee
with the integration contour chosen according to the domain in which
we want to compute the degeneracy.
As we shall discuss at the
end of this section, the change in the degeneracy across a wall
of marginal stability is exponentially suppressed compared to the
leading term in the limit of large charges. Thus the asymptotic
expansion of the statistical entropy in inverse powers of charges will
be independent of the domain in which we calculate the
entropy.
For definiteness 
we shall choose our
asymptotic moduli to lie inside the domain $\RR$ introduced
at the end of \S\ref{smarginal}.
In this case the contour of integration in \refb{egg1rp} 
will be taken
along fixed but large positive values $M_1$ and $M_2$ of
$\wrh_2$
and $\ws_2$, and fixed but large  negative value $-M_3$ of
$\wv_2$, with $|M_3|<< M_1, M_2$. At a typical point on the
contour the exponent in the integrand is quadratic in $Q$ and $P$
with large coefficients governed by the $M_i$'s. However the
phase of the integrand oscillates rapidly and so we cannot estimate
the integral from the magnitude of the integrand. We remedy this
problem by deforming the integration contour so as to make the
factor in the exponent as small as possible. In particular if we bring
the 
integration contour to a new position $\wt \CC$,
defined by
\be \label{enewcontour1}
\wrh_2= \eta_1, \quad \ws_2 = \eta_2, \quad \wv_2 = \eta_3\, ,
\ee
where $\eta_1$, $\eta_2$ are small but fixed positive numbers and 
$\eta_3$ is a small positive or negative number, then  the
exponential factor in the new integrand is quadratic in the charges
with small coefficients. Since the expected
entropy should grow quadratically with charges with finite
coefficients, we conclude that the integral over the contour $\wt \CC$
is exponentially suppressed compared to the leading
contribution. Thus
the dominant contribution must come from the residue at the poles
through which the contour passes as we deform it from $\CC$ to
$\wt \CC$. For this reason we shall tentatively neglect the
contribution from $\wt\CC$ and focus on the contribution from
the poles, -- as long as this contribution to the statistical entropy
grows quadratically with the charges with finite coefficients, our
ansatz is self-consistent. 

Let $\DD$ denote the four dimensional region
bounded by $\CC$ and $\wt\CC$ along which we deform the
contour. For evaluating the contribution to the integral
from the residues at the poles 
we need to locate the poles of the integrand in \refb{egg1rp}
inside the region $\DD$.
These poles in turn must come from the zeroes of 
$\wt\Phi(\wrh,\ws,\wv)$. 
According to the analysis given in appendix \ref{szero} 
(eq.\refb{eonly}), 
$\wt\Phi(\wrh,\ws,\wv)$ has second order zeroes at
\bea{ep4}
&&  n_2 ( \ws \wrh  -\wv ^2) + j\wv  
+ n_1 \ws  -m_1 \wrh + m_2
 = 0, \nonumber \\
&& \hbox{for} \quad  \hbox{$m_1\in N\ZZZ$, 
$n_1\in\ZZZ$,
$j\in 2\ZZZ+1$, $m_2, n_2\in \ZZZ$}, \quad
m_1 n_1 + m_2 n_2 +\frac{j^2}{4} = {1\over 4}\, .
%\nonumber \\
\eea
For fixed integers $m_1$, $m_2$, $n_1$, $n_2$ and $j$ this describes a
surface of real codimension two, \i.e.\ of real dimension four. Typically
this
will intersect $\DD$ in a two dimensional subspace $\BB$
bounded by
the intersection of $\wt \CC - \CC$ with \refb{ep4}, and
the contribution to the integral from the pole at \refb{ep4}
is obtained by  integrating the residue over
$\BB$. In fact since we
expect the integrand to be non-singular over the original 
contour $\CC$,
$\CC$ should not have any intersection with \refb{ep4} representing
locations of the poles of the integrand. 
Thus the boundary of the two
dimensional subspace $\BB$
is a one dimensional curve $\p\BB$ given by the intersection
of $\wt \CC$ with \refb{ep4}.

Since $\BB$ is an open two dimensional
space we cannot again use the residue theorem to carry out
the integral over the region $\BB$. 
Neither can we estimate the integral by examining the maximum
value of the integrand inside $\BB$ since typically the integrand
will still have rapid oscillations inside $\BB$ and there are large
cancellations. We remedy this by
using
the saddle point method. Keeping the
boundary $\p\BB$ fixed we deform the integration region $\BB$
to a new region $\BB'$
inside the subspace \refb{ep4} 
such that the maximum value of the integrand inside $\BB'$ takes
the minimum possible value.\footnote{In terms 
of the original construction such deformations of $\BB$
amounts to deforming the  region $\DD$ bounded by $\CC$
and $\wt\CC$ so that it
contains the saddle point of the integrand.} The location of this
maximum is a saddle point, -- the integrand decreases as we move
away from the saddle point along $\BB'$ and increases as we 
move away from the saddle point in directions transverse to $\BB'$.
We can now
evaluate the leading contribution to the integral as well
as power law corrections to it
by systematically expanding the integrand
around the saddle point and carrying out the integration.

In order to carry out this procedure we shall proceed in three steps.\
\begin{enumerate}
\item First we shall determine which of the zeroes of $\wt\Phi$
described in \refb{ep4} would give dominant contribution to the
integrand at the saddle point.

\item Then we shall verify that this particular zero has a non-trivial
intersection $\BB$ with the region $\DD$, and that this two dimensional
surface $\BB$ can be deformed to another surface $\BB'$ passing through
the saddle point so that the integrand on $\BB'$
has maximum magnitude at the saddle point.

\item Finally we shall evaluate the contribution from the integral
using the saddle point method.
\end{enumerate}

We begin with the first step.
Let us define
\be\label{ep3}
A = n_2, \quad B = (n_1, -m_1, {1\over 2} j), 
\quad y = (\wt\rho, \wt\sigma, 
-\wt v), \qquad
C = m_2\, , \quad q = (P^2/N, NQ^2,  Q\cdot P)\, ,
\ee
and denote by $\cdot$ the $SO(2,1)$ invariant inner product
\be\label{ep5}
(x^1, x^2, x^3)\cdot (y^1, y^2, y^3) = x^1 y^2 + x^2 y^1 - 2 x^3 y^3\, .
\ee
Then we have
\be\label{eex}
y^2\equiv y\cdot y = 2(\wt\rho\wt\sigma - \wt v^2), \qquad
B\cdot y = j\wt v + n_1\wt\sigma - m_1\wt\rho\, ,
\ee
and the first equation of \refb{ep4} may be rewritten as
\be\label{efourtwo}
{1\over 2} A y^2 + B\cdot y + C 
=0\, .
\ee
 Picking up residue at the pole forces us to evaluate the exponent
 in \refb{egg1rp}
 \be\label{ep6}
 -i\pi\left(\wt \rho NQ^2 + \wt \sigma P^2/N +
2\wt v Q\cdot P\right) = -i\, \pi \, q \cdot y\, ,
\ee
at \refb{efourtwo}. For a given zero of $\wt\Phi$ labelled
by $(\vec m, \vec n, j)$ 
the location of the saddle point 
to leading approximation is now determined by
 extremizing \refb{ep6} with respect to $y$
subject to the condition \refb{efourtwo}. This gives, for
$n_2\ne 0$,
\be\label{ep7}
q + \lambda (A y + B) = 0\, ,
\ee
where $\lambda$ is a lagrange multiplier. 
\refb{efourtwo} and \refb{ep7}
now give:
\be\label{ep8}
\lambda = \pm
\sqrt{q^2 \over B^2 - 2 A C}, \qquad y = -{1\over A} \left(
{q\over \lambda} + B\right)\, .
\ee 
Since
\be\label{eq1}
B^2 - 2 A C = - 2 (m_1 n_1 + m_2 n_2 + {j^2\over 4}) 
= -{1\over 2}
\ee
due to the last equation in \refb{ep4}, we get
\be\label{eq3pre}
 \lambda = \pm\,  i\, \sqrt{2q^2}
 \ee
 for $q^2\equiv 2(Q^2P^2 - (Q\cdot P)^2)>0$.
 The correct sign in \refb{eq3pre} is determined as follows. First of all,
 note that \refb{ep4} describes the same surface if we change the signs
 of $m_i$, $n_i$ and $j$. Using this freedom we can choose $n_2$,
 \i.e.\ $A$ to be positive. Since $Q^2$ and $P^2$ are
 positive, \refb{ep3}, \refb{ep8}
 shows that in order for $\wrh_2$ and $\ws_2$
 to be positive we must have
 $Im~\lambda>0$. Thus we have 
 \be\label{eq3}
 \lambda =    i\, \sqrt{2q^2}\, ,
 \ee
and at the saddle point the exponential $e^{-i\pi q\cdot y}$
takes the form:
\be\label{eq2}
E \equiv  e^{-i\pi\, q\cdot y} =
e^{i\pi (q^2 / \lambda + q\cdot B)/A}= e^{\left(
\pi \sqrt{q^2 / 2}
+ i\pi q\cdot B\right)/A}\, .
\ee
Since $q\cdot B/A$ is a rational number, 
 the second term only gives a phase.
Hence
 \be\label{eq4}
 |E| = e^{{\pi\over A} \sqrt{q^2 / 2}  } = e^{\pi \sqrt{Q^2 P^2 
 - (Q\cdot P)^2} / n_2}\, .
 \ee
  
 Eq.\refb{eq4} shows that the leading contribution to the integral
 comes from the saddle point corresponding to $n_2=1$. In this case
 a $\wrh \to \wrh +1$ transformation in \refb{ep4}, which is a
 symmetry of $\wt\Phi(\wrh,\ws,\wv)$ due to \refb{especialpre},
 induces $n_1\to n_1+1$, $m_2\to 
 m_2-m_1$. Since $n_1\in \ZZZ$,
 we can use this symmetry to bring the saddle point
 to $n_1=0$. On the other hand a $\ws \to \ws +N$
 transformation in eq.\refb{ep4}
 induces $m_1\to m_1-N$, $m_2\to m_2+n_1N$. Since $m_1\in
 N\ZZZ$, we can use this transformation to bring $m_1$ to 0.
 Finally the $\wv \to \wv +1$ transformation in \refb{ep4} induces
 $j\to j-2$, $m_2\to m_2 +j- 1$. Since $j\in 2\ZZZ+1$, we can use this
 transformation to set $j=1$. $m_2$ is now determined to be zero
 from the last equation in \refb{ep4}. Thus we have 
 \be\label{eq5}
 m_1=m_2=n_1=0, \quad n_2 = 1, \quad j=1\, .
 \ee
 The corresponding zero of $\wt\Phi$ is at
\be\label{epolepre}
\wt\rho\wt\sigma -\wt v^2 +\wt v=0 \, .
\ee 
Also eqs.\refb{ep3}, \refb{ep8} give the location of the saddle point
to be at
\be \label{esa11}
\wrh = i{P^2\over 2 N\sqrt{Q^2 P^2 - (Q\cdot P)^2}}\, ,
\qquad
\ws = i{N Q^2 \over 2\sqrt{Q^2 P^2 - (Q\cdot P)^2}}, \qquad
\wv = {1\over 2} - i {Q\cdot P\over 2 \sqrt{Q^2 P^2 - (Q\cdot P)^2}}
\, .
\ee
Since we have used the freedom of $\wrh$, $\ws$ and $\wv$ 
 translations to pick this particular pole, we can no longer choose the
 range of $\wrh_1$, $\ws_1$ and $\wv_1$ to be (0,1), $(0,N)$
 and $(0,1)$ respectively. Instead we should allow them to run all the
 way from $-\infty$ to $\infty$ and intersection with 
 \refb{epolepre} will
 pick up the appropriate subspace over which the integration needs to
 be performed.

Let us now verify that the region $\DD$ has a non-trivial
intersection with \refb{epolepre}.
This will be done by showing that 
$\wt \CC$ and hence $\p\DD$
has a non-trivial intersection with \refb{epolepre}.
The equation determining the
intersection of \refb{enewcontour1} 
with \refb{epolepre} is given by\footnote{I wish to thank Justin
David for pointing out an error in eq.\refb{einsect1}
in an earlier version of the
manuscript.}
\ben \label{einsect1}
&& \wrh_2=\eta_1, \quad \ws_2=\eta_2, \quad \wv_2=\eta_3,
\nonumber \\
&& \wrh_1= -{\eta_1\ws_1-(2\wv_1-1)\eta_3\over \eta_2}\, , \quad
{\eta_1\over \eta_2} (\ws_1)^2 + \left( \wv_1 -{1\over 2}\right)^2
- 2 \, {\eta_3\over \eta_2}\, \ws_1\,
\left(\wv_1 -{1\over2}\right)
= {1\over 4} -( \eta_1 \eta_2 -\eta_3^2)\, .\nonumber \\
\een
As long as $0<4(\eta_1\eta_2 -\eta_3^2)<1$, 
this describes an ellipse in the
$\ws_1$-$\wv_1$ plane. Thus the two dimensional surface $\BB$
will be a surface inside
the 4-dimensional subspace \refb{epolepre}, bounded by the
curve \refb{einsect1}.

Finally we need to show that there exists a
surface $\BB'$ inside \refb{epolepre} 
bounded by the same curve \refb{einsect1} such that it
passes through the saddle point
\refb{esa11} and the integrand has a global maximum at
the saddle point. In that case we can deform the surface $\BB$
to $\BB'$ and compute the integral over $\BB'$ by expanding the
integrand around the saddle point. 
We can explicitly construct such a surface $\BB'$ by considering
the family of curves $\CC(\lambda)$ defined as
\ben \label{esa12}
&& \wrh_2=\eta_1(\lambda), \quad 
\ws_2=\eta_2(\lambda) , \quad \wv_2=\eta_3(\lambda),
\qquad \wrh_1= -{\eta_1(\lambda)\ws_1 - (2 \wv_1-1)
\eta_3(\lambda)
\over \eta_2(\lambda)}\, ,
\nonumber \\
&& 
{\eta_1(\lambda)\over \eta_2(\lambda)} (\ws_1)^2 
+ \left( \wv_1 -{1\over 2}\right)^2
- 2 \, {\eta_3(\lambda)\over \eta_2(\lambda)}\, \ws_1\,
\left(\wv_1 -{1\over2}\right)
= {1\over 4} -( \eta_1(\lambda) \eta_2(\lambda) -\eta_3(\lambda)^2)\, ,
\nonumber \\
\een
where
\ben \label{esa13}
&& \eta_1(\lambda) =
\lambda {P^2\over 2 N\sqrt{Q^2 P^2 - (Q\cdot P)^2}}\, ,
\qquad
\eta_2(\lambda) 
= \lambda 
{N Q^2 \over 2\sqrt{Q^2 P^2 - (Q\cdot P)^2}}, \nonumber \\ &&
\eta_3(\lambda) =  -  \lambda {Q\cdot P\over 2 
\sqrt{Q^2 P^2 - (Q\cdot P)^2}}
\, .
\een
Since \refb{esa12} is the same as eq.\refb{einsect1} with $\eta_i$
replaced by $\eta_i(\lambda)$, the curve \refb{esa12} lies
on the surface \refb{epolepre}. Identifying $\eta_i(\epsilon)$
with $\eta_i$ for some small positive number $\epsilon$, we see that
at $\lambda=\epsilon$ the curve coincides with $\wt C$. On the
other hand at $\lambda=1$ the curve \refb{esa12} shrinks to the
saddle point \refb{esa11}. Finally, the magnitude of the exponential
factor in the integrand, -- which depends only on the imaginary parts
of $\wrh$, $\ws$ and $\wv$, --
 takes value $\exp(\pi\lambda \sqrt{Q^2 P^2
-(Q\cdot P)^2})$ on the curve \refb{esa12}. Thus in the range
$\eps\le\lambda\le 1$ it reaches a maximum at $\lambda=1$,
\i.e.\ at the saddle point \refb{esa11}. 
{}From this we see that the surface foliated by the
family of curves $\CC(\lambda)$ for $\eps\le\lambda\le 1$ has all 
 the right properties to be
identified as the surface $\BB'$.

We now turn to the evaluation of the contribution to the integral
from the residue at \refb{epolepre}.
For this we introduce a new set of variables $(\rho,\sigma,v)$ related
to $(\wt\rho,\wt\sigma, \wt v)$ via the relations
\be\label{e6narep}
   \wc\rho={1\over N}\, 
   {1\over 2v-\rho-\sigma}, \qquad
   \wc\sigma = N\,
   {v^2-\rho\sigma \over 2v-\rho-\sigma}, \qquad
    \wc v =
   {v-\rho \over 2v-\rho-\sigma}\, ,
\ee
or equivalently,
\be\label{e5nrep}
\rho 
   = {\wc \rho \wc\sigma - \wc v^2\over N\wc\rho}, 
   \qquad \sigma = {\wc\rho \wc \sigma - (\wc v - 1)^2\over  
   N\wc\rho}, \qquad
   v 
=   {\wc\rho \wc\sigma - \wc v^2 + \wc v\over N\wc\rho}\, .
\ee
Under this map \refb{epolepre} takes the form
\be \label{epole}
v=0\, .
\ee
Now it has been shown in eq.\refb{enn11} that
\be\label{enn11rep}
\wt\Phi(\wc\rho,\wc\sigma,\wc v)=-(i)^k\, C_1\, 
(2v -\rho-\sigma)^k\, \cp(\rho,\sigma,v)
\ee
where $C_1$ is a real positive
constant and $\wh\Phi(\rho,\sigma,v)$ is a new
function defined in \refb{enn9c}:
\bea{enn9cpre}
\cp(\rho,\sigma,v) &=&  e^{2\pi i \left(
\wh \alpha\rho+\wh \gamma\sigma+v\right) } \nonumber \\
&& \prod_{b=0}^1\,
\prod_{r=0}^{N-1}\,  \prod_{(k',l)\in \zzz,j\in 2\zzz+b\atop
k',l\ge 0, j<0 \, {\rm for}
\, k'=l=0}
\Big\{ 1 - e^{2\pi i r / N} \, e^{ 2\pi i ( k' \sigma + l \rho + j v) 
}
\Big\}^{ \sum_{s=0}^{N-1} e^{-2\pi i sr/N } c^{(0,s)}_b(4k'l - j^2)
}  \nonumber \\
&&    \wh \alpha= \wh \gamma 
= {1\over 24}\, \chi(\MM)\, .
\eea
Using \refb{enn11rep} and the identity
\be\label{ejac}
d\wrh \wedge d\ws \wedge d\wv = -(2v -\rho-\sigma)^{-3} 
d\rho \wedge d\sigma \wedge d v\, ,
\ee
we can express the contribution to \refb{egg1rp} from the residue
at the pole at
$v=0$ as
\bea{enn17a}
d(\vec Q, \vec P)&\simeq&  (-1)^{Q\cdot P+1}\,
{(i)^{-k}\over N\, C_1}
\int_{\CC'}  d\rho \wedge   d\sigma \wedge  
dv \, (2v -\rho-
\sigma)^{-k-3} \, {1\over \wh\Phi(\rho, \sigma,  v)} \nonumber \\
&& \exp\left[ -i\pi \left\{ {v^2 -\rho\sigma\over 2v 
-\rho-\sigma}
P^2 +{1\over 2v -\rho-\sigma} Q^2 +{2(v-\rho)\over 2v -\rho-
\sigma} Q\cdot P\right\}\right]  \nonumber \\
\eea
where $\CC'$ denotes a contour around $v=0$. 
Note that we have used the wedge product notation to keep track of
the orientation of the integration region and have implicitly used
the convention that the original integral over $\wt\rho_1$, 
$\wt\sigma_1$ and
$\wt v_1$ corresponds to the measure 
$d\wrh\wedge d\ws\wedge d\wv$.

We shall evaluate this integral by first 
performing the $v$ integral using
Cauchy's formula and then carrying out the $\rho$ and $\sigma$
integrals by saddle point approximation. 
According to 
\refb{enn12} near $v=0$
\be\label{enn12rep}
\wh\Phi(\rho, \sigma,  v)=   -4\pi^2 \, v^2 \, g( \rho)\, 
g( \sigma) + \OO(  v^4)\, ,
\ee
where the function
$g(\rho)$ has been introduced in \refb{enn13pre}.
Thus the 
 contribution to \refb{enn17a} from the pole at $v=0$ is
given by
\bea{enn17}
d(\vec Q, \vec P)&\simeq& 
-(-1)^{Q\cdot P+1}\,
{(i)^{-k}\over 4\pi^2 N C_1}\, \int_{\CC'}  
d\rho \wedge  d\sigma  
\wedge dv \, v^{-2} \, (2v -\rho-
\sigma)^{-k-3} \, \left(g(\rho) g(\sigma)
\right)^{-1} \nonumber \\
&& \exp\left[ -i\pi \left\{ {v^2 -\rho\sigma\over 2v 
-\rho-\sigma}
P^2 +{1\over 2v -\rho-\sigma} Q^2 +{2(v-\rho)\over 2v -\rho-
\sigma} Q\cdot P\right\}\right]\, . \nonumber \\
\een
Evaluating the $v$ integral in \refb{enn17}
by Cauchy's formula, we get
\bea{er2}
d(\vec Q, \vec P) &\simeq& {(i)^{-k+1}
\gamma\over 2\pi N C_1} \, 
(-1)^{k} \, \int 
{d\rho \wedge  d\sigma
\over (\rho+\sigma)^2} \nonumber \\ && 
\left[ -2 (k+3) + 2 \pi i \, \left\{ {\rho\sigma\over \rho+\sigma} P^2
-{1\over \rho+\sigma} Q^2 + {\rho-\sigma\over \rho+\sigma}
Q\cdot P\right\} \right]\nonumber \\
&&\exp\Bigg[ -i\pi \left\{ {\rho\sigma\over \rho+\sigma} P^2
-{1\over \rho+\sigma} Q^2 + {\rho-\sigma\over \rho+\sigma}
Q\cdot P\right\}  \nonumber \\
&& -\ln g(\rho)-\ln g(\sigma)
- (k+2) \ln (\rho+\sigma)\Bigg]\, , 
\een
where $\gamma=1$ or $-1$ depending on whether the $\wv$ (or
$v$) contour encloses the pole anti-clockwise or clockwise, or
equivalently whether the pole from which the dominant contribution
comes was above or below the $\wv$ integration contour for the
original contour $\CC$. It
may be determined as follows.  First note that the zeroes of
$\wt\Phi$ at $\wrh\ws-\wv^2+\wv=0$ 
correspond to $\wt v = {1\over 2} \pm
\sqrt{{1\over 4} +\wrh\ws}$. For the original contour $\CC$, $\wrh_2$
and $\ws_2$ are large and hence the poles are located at large positive
or negative imaginary values of $\wt v$. As we reduce $\wrh_2$ 
and $\ws_2$ in the process of deforming the $\wrh$ and
$\ws$ contours, these poles
approach the $\wt v$ integration contour. We can however avoid
them by deforming the $\wt v$ integration contour with the
ultimate goal that we try to minimize the maximum value of the
integrand over the integration contour. For $Q\cdot P>0$ this
can be done by deforming the $\wt v$ contour into the lower half
plane since the exponential factor decreases as we reduce $\wt v_2$.
Thus we can always avoid the pole coming down from above; but at
some point we encounter the pole coming up from below and the
major contribution to the integral would come from the residue at
this pole. Since this pole was below the original
$\wt v$ integration contour, the residue from this is calculated
by enclosing it in the clockwise direction. This gives $\gamma=-1$.
An exactly similar argument shows that for $Q\cdot P<0$, the main
obstruction to our ability to reduce the integrand by deforming the
$\wt v$ contour comes from the pole  above the original
$\wt v$ contour. This
gives $\gamma=1$. Thus we have
\be \label{esigngamma}
\gamma = -Sign(Q\cdot P)\, .
\ee

The correction to \refb{er2} involves contribution from other
poles for which $n_2\ne 1$, and are
suppressed by fractional powers of $e^{-\pi\sqrt{Q^2P^2 -
(Q\cdot P)^2}}$. Thus \refb{er2} contains information not only
about the leading contribution to the statistical entropy but also about
all the subleading corrections involving inverse powers of charges.

Let us now introduce new complex variables $\tau_1$ and 
$\tau_2$ through the
relations:
\be\label{er3}
\rho = \tau_1+i\tau_2, \qquad \sigma = -\tau_1+i\tau_2\, .
\ee
Then \refb{er2} may be rewritten as
\bea{er4}
d(\vec Q, \vec P) &\simeq& {\gamma\over 4\pi N C_1} \, 
 \int {d\tau_1  \wedge
 d\tau_2 \over \tau_2^2} \nonumber \\
&&
\left[ 2(k+3) + {\pi\over \tau_2} \left\{ (\tau_1^2 + \tau_2^2)  P^2 
+  Q^2
- 2\tau_1 Q\cdot P\right\} \right]\nonumber \\
&& \exp\Bigg[ {\pi\over 2 \tau_2} \left\{ (\tau_1^2 + \tau_2^2)  
P^2 +  Q^2
- 2\tau_1 Q\cdot P\right\} \nonumber \\ &&
-\ln g(\tau_1 + i\tau_2) -\ln g(-\tau_1 + i\tau_2)
- (k+2) \ln (2 \tau_2)\Bigg]\, .
\een
The integral in \refb{er4} is carried out over the image
of the surface \refb{esa12} in the complex
$\tau_1$-$\tau_2$ coordinate system.
In the leading
approximation the saddle point, obtained by extremizing
\be\label{er5}
{\pi\over 2 \tau_2} \left\{ (\tau_1^2 + \tau_2^2)  P^2 +  Q^2
- 2\tau_1 Q\cdot P\right\}
\ee
occurs at
\be \label{esaddle}
\tau_1 = Q\cdot P/ P^2, \qquad \tau_2 =\sqrt{Q^2 P^2 -
(Q\cdot P)^2} / P^2
\, .
\ee
On the other hand the steepest descent of the integrand occurs as we
move away from the saddle point along the imaginary directions
in the $\tau_1$ and $\tau_2$ plane. Hence the integration contour
is such that it passes through the saddle
point \refb{esaddle}
and lies along imaginary $\tau_1$ and $\tau_2$ direction at
the saddle point.
If we now regard $\tau_1$ and $\tau_2$ as real variables and define
\be\label{er6}
\tau =  \tau_1 + i \tau_2\, ,
\ee
then \refb{er4} may be  formally reexpressed as
\bea{er7pre}
d(\vec Q, \vec P) &\simeq& {\gamma\over 4\pi N C_1} \, 
 \, 
\int {d\tau_1 \wedge d\tau_2\over \tau_2^2}\, 
\left[ 2(k+3) + {\pi\over \tau_2} |Q -\tau P|^2\right] \nonumber \\
&& \exp\Bigg[ {\pi\over 2 \tau_2} \, |Q -\tau P|^2
-\ln g(\tau) -\ln g(-\bar\tau) - (k+2) \ln (2\tau_2)
\bigg]\, . \nonumber \\
\een
In evaluating this integral 
we must first express the integrand in terms of $\tau_1$ and $\tau_2$
treating them as real variables, and then carry out the integral by
analytically rotating the integration contours along the
imaginary axis both in the complex $\tau_1$ and complex $\tau_2$
plane.

In order to determine the overall sign, we need to determine 
whether $d\tau_1\wedge d\tau_2$ defines a positive or
negative integration measure along the deformed contour at the
saddle point. This can be determined by expressing the original
$(\wrh,\ws,\wv)$ coordinates in terms of the new coordinates
$(\tau_1,\tau_2)$. Setting $v=0$ in \refb{e6narep} and using
\refb{er3} we get
\be \label{eorsad}
\wrh = {i\over 2 N\tau_2}, \qquad \ws = iN {\tau_1^2 +\tau_2^2
\over 2\tau_2}, \qquad \wt v = {1\over 2} - i{\tau_1\over 2\tau_2}\, .
\ee
{}From this we see that
\be \label{esimes}
d\wrh \wedge d\ws = -{\tau_1 
\over 2\tau_2^3}
d\tau_1\wedge d\tau_2 \, .
\ee
Since $d\wrh\wedge d\ws$ represents positive integration measure
$d\wrh_1d\ws_1$ according to the convention introduced below
\refb{enn17a}, we see that 
around the saddle point $d\tau_1\wedge
d\tau_2$ describes positive (negative) integration measure
if $\tau_1$ is negative (positive), \i.e, $Q\cdot P$ 
is negative (positive). Combining
this with \refb{esigngamma} we see that \refb{er7pre} may be
expressed as
\bea{er7}
d(\vec Q, \vec P) &\simeq& {1\over 4\pi N C_1} \, 
\int {d\tau_1  d\tau_2\over \tau_2^2}\, 
\left[ 2(k+3) + {\pi\over \tau_2} |Q -\tau P|^2\right] \nonumber \\
&& \exp\Bigg[ {\pi\over 2 \tau_2} \, |Q -\tau P|^2
-\ln g(\tau) -\ln g(-\bar\tau) - (k+2) \ln (2\tau_2)
\bigg]\, , \nonumber \\
\een
where $d\tau_1 d\tau_2$ is defined such that it describes a positive
integration measure when the integration contour lies along
imaginary $\tau_1$ and $\tau_2$ direction at the saddle point.

Identifying $d(\vec Q, \vec P)$ with $e^{S_{stat}(\vec Q, \vec P)}$ where
$S_{stat}$ denotes the statistical entropy, we can rewrite
\refb{er7} as
\be\label{ek1}
e^{S_{stat}(\vec Q, \vec P)} \simeq
\int{d^2\tau\over \tau_2^2} \, e^{-F(\vec \tau)}\, ,
\ee
where $\vec\tau = (\tau_1, \tau_2)$ or $(\tau, \bar\tau)$ depending on
the basis we choose to use, and
\bea{ek2}
F(\vec\tau) &=& -\Bigg[ {\pi\over 2 \tau_2} \, |Q -\tau P|^2
-\ln g(\tau) -\ln g(-\bar\tau) - (k+2) \ln (2\tau_2)
\nonumber \\
&& +\ln\bigg\{K_0 \, \left(
2(k+3) + {\pi\over \tau_2} |Q -\tau P|^2\right)
\bigg\}\Bigg] \, , \nonumber \\
K_0 &\equiv& {1\over 4\pi N C_1}\, .
\een
 Note that $F(\vec\tau)$ also depends on 
 the charge vectors $\vec Q$,
$\vec P$, but we have not explicitly
displayed these in its argument. 

As described in eq.\refb{egrhotrspre} (and proved in 
appendix \ref{ssiegel}, eq.\refb{egrhotrs}),  
$g(\tau)$ transforms as
a modular form of weight $(k+2)$ under the $\Gamma_1(N)$ group.
Using this fact, and \refb{ek2},
one can show that 
in \refb{ek1}
the integration measure $d^2\tau / (\tau_2)^2$ as well as the integrand
$e^{-F(\vtau)}$ 
are manifestly invariant under the $\Gamma_1(N)$
transformation:
\bea{egamman}
&& \vec Q \to \alpha \vec Q +\beta \vec 
P, \quad \vec P \to \gamma \vec Q + \delta \vec P, \quad
\tau \to {\alpha\tau + \beta\over \gamma\tau + \delta}\, , \nonumber \\
&& \quad \alpha,\beta,\gamma,\delta
\in \ZZZ, \quad \alpha\delta-\beta\gamma = 1, \quad \alpha,\delta=
\hbox{1 mod $N$},
\quad \gamma = \hbox{0 mod $N$}\, .
\een
Thus $S_{stat}(\vec Q, \vec P)$ computed from \refb{ek1} is 
invariant under $\Gamma_1(N)$. From \refb{tdual4} 
one can see that this
corresponds to the S-duality symmetry of the theory in the second
description.

We shall now describe a systematic procedure for
evaluating $S_{stat}$ given in \refb{ek1}
as an expansion in inverse powers of
the charges. For this we
introduce the generating function:
\be\label{ek3}
e^{W(\vec J)} = \int{d^2\tau\over \tau_2^2} 
\, e^{-F(\vec \tau) + \vec J\cdot
\vec \tau}\, ,
\ee
for a two dimensional vector $\vec J$, and define $\Gamma(\vec u)$ as
the Legendre transform of $W(\vec J)$:
\be\label{ek4}
\Gamma(\vec u) =  \vec J\cdot \vec u - W(\vec J) \, ,
\qquad u_i = {\p W(\vec J)\over \p J_i}\, .
\ee
It follows from \refb{ek4} that 
\be\label{ek5}
J_i = {\p \Gamma(\vec u)\over \p u_i}\, .
\ee
As a result if
\be\label{ek6}
{\p \Gamma(\vec u)\over \p
\vec u_i} = 0 \quad \hbox{at $\vec u 
=\vec u_0$}\, ,
\ee
then it follows from \refb{ek3}-\refb{ek5},
\refb{ek1} that
\be\label{ek7}
\Gamma(\vec u_0) = -W(\vj=0)= -S_{stat}  \, .
\ee
Thus the computation of $S_{stat}$ can be done by first calculating
$\Gamma(\vec u)$ and then evaluating it at its extremum. 
$\Gamma(\vec u)$ in turn can be calculated by regarding this as a
sum of one particle irreducible (1PI) Feynman diagrams in the zero
dimensional field theory with action $F(\vec\tau)+2\ln\tau_2$.  
Since $S_{stat}$ is given
by the value of the function $-\Gamma(\vu)$ at its extremum, we
can identify $-\Gamma(\vu)$ as the entropy function for the
statistical entropy in analogy with the corresponding result for
black hole entropy\cite{0506177,0508042}.

A convenient method of
calculating $\Gamma(\vu)$ is the so called background field method.
For this we choose some arbitrary base point $\vtau_B$ and define
\be\label{ek3a}
e^{W_B(\vtau_B,\vec J)} = \int{d^2\eta\over (\tau_{B2}+\eta_2)^2} 
\, e^{-F(\vtau_B+\htau) + \vec J\cdot
\htau}\, ,
\ee
\be\label{ek4a}
\Gamma_B(\vtau_B, \vc) =  \vec J\cdot \vc - W_B(\vtau_B,\vec J) \, ,
\qquad \chi_i = {\p W_B(\vtau_B,\vec J)\over \p J_i}\, .
\ee
By shifting the integration variable in \refb{ek3a} to $\vtau=\vtau_B
+\htau$ it follows easily that
\be\label{ek5a}
W_B(\vtau_B,\vec J)=W(\vec J) - \vtau_B\cdot \vj\, ,
\ee
and hence
\be\label{ek6a}
\Gamma_B(\vtau_B, \vc) = \Gamma(\vtau_B+\vc)\, .
\ee
Thus the computation of $\Gamma(\vu)$ reduces to the computation
of $\Gamma_B(\vu, \vc=0)$. The latter in turn can be computed
as the sum of 1PI vacuum diagrams in the 0-dimensional field theory
with action 
$F(\vu +\htau)+2\ln (u_{2}+\eta_2)$, 
with $\htau$ regarded as fundamental fields, and 
$\vu$ regarded as some fixed background.

While this gives a definition of the statistical
entropy function whose
extremization leads to the statistical entropy, the entropy function
constructed this way is not manifestly invariant under the S-duality
transformation given in \refb{tdual4}. This is due
to the fact that since the S-duality transformation has a non-linear action
on $(\tau_1, \tau_2)$, the generating function $W(\vj)$ defined in 
\refb{ek3} and hence also the effective action $\Gamma(\vu)$ 
defined in \refb{ek4} does not have manifest S-duality symmetries. Of
course the statistical entropy obtained by extremizing $\Gamma(\vu)$
will be duality invariant since this is given in terms
of the manifestly duality
invariant integral \refb{ek2}. In appendix \ref{s3xyz} 
we have described a
slightly different construction based on Riemann
normal coordinates which yields a manifestly duality invariant 
statistical entropy function. The result of this analysis is that instead 
of using the function $-\Gamma_B(\vec\tau)$ as the statistical entropy 
function we can use a different manifestly duality invariant function 
$-\wt 
\Gamma_B(\vec\tau)$ as the 
statistical entropy function. $\wt\Gamma_B(\vec\tau_B)$ is defined as
the sum of
1PI vacuum diagrams computed from the action 
\be\label{einfo1}
-\ln\left( {1\over |\vec\xi|} \sinh{|\vec\xi|} \right) 
- \sum_{n=0}^\infty {1\over n!} (\tau_{B2})^n
\xi_{i_1}\ldots
\xi_{i_n}\, D_{i_1}
\cdots D_{i_n} F(\vtau)\bigg|_{\vtau=\vtau_B}\, ,
\ee
where $\xi=\xi_1+i\xi_2$ is to be regarded as a zero dimensional
complex quantum field, $\tau_B$ is a fixed zero dimensional
complex background field and
$D_\tau$, $D_{\bar\tau}$ are duality invariant covariant derivatives
defined recursively through the relation:
\bea{einfo3}
D_\tau (D_\tau^m D_{\bar\tau}^n F(\vec\tau))
&=& (\p_\tau - im/\tau_2) (D_\tau^m D_{\bar\tau}^n F(\vec\tau)),
\nonumber \\
D_{\bar\tau} (D_\tau^m D_{\bar\tau}^n F(\vec\tau))
&=& (\p_{\bar\tau} + in/\tau_2)
(D_\tau^m D_{\bar\tau}^n F(\vec\tau))\, ,
\een
for any arbitrary ordering of $D_\tau$ and $D_{\bar\tau}$
in $D_\tau^m D_{\bar\tau}^n F(\vec\tau)$. 
The result of this
computation expresses $\wt\Gamma_B(\vec\tau_B)$ in terms
of manifestly duality invariant quantity $F(\vec\tau)$ and its
duality covariant derivatives.
It has been shown in appendix \ref{s3xyz} that explicit evaluation of 
$\wt\Gamma_B(\vec\tau)$ gives
\be\label{einfo4}
-\wt\Gamma_B(\vec\tau) = {\pi\over 2 \tau_{2}} \, |Q -\tau P|^2
- \ln g(\tau) -\ln g(-\bar\tau)
- (k+2) \ln (2\tau_{2}) + \ln (4\pi K_0) + \OO(Q^{-2})\, .
\ee
In order to see how good this approximation is, we give below the
exact results for $d(\vec Q, \vec P)$ calculated from
\refb{egg1int} and $S_{stat}
\equiv \ln\, d(\vec Q,\vec P)$ for certain values of
$Q^2$, $P^2$ and $Q\cdot P$ in the case of heterotic string theory
on $T^6$, \i.e.\ the $N=1$, $\MM=K3$ model.\footnote{In this case
$K_0=1/4\pi$ and the additive constant in \refb{einfo4} vanishes.}
We also give the
approximate statistical entropies $S^{(0)}_{stat}$ calculated with
the `tree level' statistical entropy function, and $S^{(1)}_{stat}$
calculated with the `tree level' plus `one loop' statistical entropy
function.\footnote{In this table the sign of $Q\cdot P$ has
been chosen in such a way that in deforming the original contour
lying at $\wt v_2<0$ to the final contour corresponding to
$\wrh_2$, $\wv_2$, $\ws_2$ given by $\eta_1(\eps)$, 
$\eta_2(\eps)$, $\eta_3(\eps)$ given in \refb{esa13} we do not
pass through the pole at $\wv=0$. For the other choice of the sign
of $Q\cdot P$ we need to pass through this pole. This will give
an additional contribution to $S_{stat}$. Although this contribution
is exponentially suppressed for large charges, it may not be negligible
for the charges used in this table.}
\begin{center}\def\st{\vrule height 3ex width 0ex}
\begin{tabular}{|l|l|l|l|l|l|l|} \hline 
$Q^2$ & $P^2$ & $Q\cdot P$ & $d(Q,P)$ & $S_{stat}$
& $S^{(0)}_{stat}$ & $S^{(1)}_{stat}$ \st\\[1ex] \hline \hline
2 & 2 & 0 &  $50064$ & 10.82 &  6.28
&  10.62 \st\\[1ex] \hline
4 & 4 & 0 &  $32861184$ & 17.31 &  12.57
&  16.90 \st\\[1ex] \hline
6 & 6 & 0 &  $16193130552$ & 23.51 &  18.85
&  23.19 \st\\[1ex] \hline
6 & 6 & 1 &  $11232685725$ & 23.14 &  18.59
&  22.88 \st\\[1ex] \hline
6 & 6 & 2 &  $4173501828$ & 22.15 &  17.77
&  21.94 \st\\[1ex] \hline
6 & 6 & 3 &  $920577636$ & 20.64 &  16.32
&  20.41 \st\\[1ex] \hline
6 & 6 & 4 &  $110910300$ & 18.52 &  14.05
&  18.40 \st\\[1ex] \hline
 \hline 
\end{tabular} 
\end{center}

Up to an additive constant \refb{einfo4} 
agrees with the black hole entropy
function for these models
as given in \refb{es11} if we identify $\tau$ as $u_a+i u_S$.
This in turn implies agreement between black hole entropy and
statistical entropy to this order.
We should however keep in mind that the computation of the
black hole entropy function has been done by including in the
effective action only a certain subset of four derivative terms
proportional to the Gauss-Bonnet term. As discussed at the
end of \S\ref{sgauss}, for large $u_S$
or equivalently $\sqrt{Q^2P^2-(Q\cdot P)^2}>>P^2$, the argument
of \cite{0506176,0508218,0609074} tell us that the result
computed using the Gauss-Bonnet term is the correct answer
to this order. But there is no such concrete 
result for finite $u_S$.
Nevertheless the agreement between the black hole and the
statistical entropy indicates that there might exist such a
non-renormalization theorem even for finite $u_S$.

Finally we would like to note that even though the complete
spectrum changes discontinuously as we cross a wall of
marginal stability, the large charge expansion
is not affected by this change. 
We can illustrate this by considering the example of the marginal
stability wall separating the domains $\RR$ and $\LL$ defined at
the end of \S\ref{smarginal}. As we cross this wall, the prescription
for the integration contour changes from $\wv_2=-M_3$ to 
$\wv_2=M_3$, keeping $\wrh_2$ and $\ws_2$ at fixed values
$M_1,M_2>>M_3$. 
It is easy to see that
the only pole through which the contour passes during this
deformation is the one at $\wv=0$.
Thus the difference between the two cases is the residue of the
integrand at $\wv = 0$. 
Since this pole is not in the class given in 
\refb{ep4} with $n_2=1$, the contribution from 
this pole is exponentially
suppressed compared to the leading contribution.
A careful analysis shows that
the same story is repeated as we cross other marginal stability
walls\cite{0702141}.

This result is consistent with the fact that on the
black hole side the attractor point described in \S\ref{sugra} is a
stable supersymmetric attractor for $P^2>0$, $Q^2>0$,
$P^2 Q^2 > (Q\cdot P)^2$. Thus the near horizon geometry of these
black holes is always given by this attractor point and is independent
of the asymptotic moduli even if the asymptotic moduli 
cross one or more walls of marginal stability.
We shall see in \S\ref{smulti} that the exponentially suppressed
changes in the degeneracy across the walls of marginal stability
can be related to the (dis)appearance of two centered black hole
solutions as we cross these walls.

\subsectiono{Jump in degeneracy and two centered black holes}
\label{smulti}

Given the success of the analysis of the previous section relating
the black hole entropy to the statistical entropy, we can now ask:
can we understand the
jump in the degeneracy across walls of marginal stability on the
black hole side?
The issue is somewhat tricky since these 
jumps in the degeneracy are
exponentially small compared to the leading contribution
to the entropy. 
Nevertheless since the change is 
discontinuous, one might hope that there is a clear mechanism on the
black hole side which produces these discontinuous changes
across  the walls of marginal stability
and if we can
identify this mechanism then we may be able to reproduce these jumps
on the black hole side. In this section 
we shall show that there is
indeed a clear mechanism on the black hole side that describes these
jumps, -- this is the phenomenon of (dis)appearance of multicentered
black hole solutions for a given total charge as we cross various
walls of marginal stability in the space of asymptotic values of the
moduli fields\cite{0005049,0101135,0206072,
0304094,0702146,0705.2564}.
In particular the exponential of the entropy 
associated with these multi-centered black holes will reproduce
the jump in the degeneracy computed from the exact dyon spectrum.
Our analysis will follow \cite{0705.3874}. Related observations have
been made in \cite{0702150,0706.2363}.

In order to keep the analysis simple
we shall restrict our analysis to the four dimensional charge 
vectors of the type given in \refb{echvec}
\be \label{me5}
\vec 
Q =\pmatrix{0\cr -n/N \cr 0 \cr -1}, \qquad 
\vec P = \pmatrix{Q_1-1\cr
-J\cr Q_5 \cr 0}\, , \qquad n, J, Q_1, Q_5\in \ZZZ, \quad
n, Q_1\ge 0, \quad Q_5 > 0\, ,
\ee
and
consider $M_\infty$ 
of the form:
\be \label{me4}
M_\infty = \pmatrix{\wh R^{-2} &&&\cr & R^{-2} &&\cr && 
\wh R^2 &\cr
&&& R^2}\, .
\ee
In this case   eq.\refb{ecurve1} describing the marginal
stability wall takes the form
\be \label{me6}
a_\infty = a_c, \quad
a_c \equiv -{J \, \wh R\over R \{Q_1 - 1 + \wh R^2 Q_5\}}\,
S_\infty\, 
\ee
for $b=0$.
By following carefully the duality chain relating the first
and the second description one finds that
the weak coupling region of the first description
corresponds to large values of $R$ parametrizing the matrix
$M_\infty$ in \refb{me4}. 
In this region the 
degeneracy formul\ae\ for $a_\infty>a_c$ and $a_\infty<a_c$
are
given respectively
by \refb{egg1int}, \refb{ep2kk} and a similar formula
with the contour $\CC$ replaced by the contour $\wh\CC$
given in \refb{ecchat}. If we denote them by $d_>(\vec Q,
\vec P)$ and $d_<(\vec Q,
\vec P)$ respectively, then the difference between them can be
computed by evaluating the contribution from the pole of
the integrand at $\wt v=0$ since this is the pole we encounter while 
deforming $\CC$ to $\wh\CC$\cite{0702141}. Now from \refb{ephilimit}   
we know that
for $\wt v\simeq 0$
$\wt\Phi$ takes the form:
\be \label{mes1}
\wt\Phi(\wrh,\ws,\wv) = -4\pi^2\, \wt v^2 \, f_1(N\wrh) f_2(\ws / N)
+ \OO(\wt v^4)\, ,
\ee
where $(f_1(\tau))^{-1}$ and 
$(f_2(\tau))^{-1}$ have the interpretation of the generating
function for the degeneracies of
purely electric half-BPS states and purely magnetic
half-BPS states respectively. 
Using \refb{mes1} and
evaluating the residue at the pole of the integrand
in \refb{egg1int} at $\wt v=0$ we get
\be \label{mes2}
d_>(Q,P)- d_<(Q,P)
= - (-1)^{Q\cdot P+1}\, 
Q\cdot P \, d_{el}(Q) \, d_{mag}(P)\, ,
\ee
where
\be \label{mes3}
d_{el}(Q) = \int_0^1d\wrh \, e^{-i\pi N\wrh Q^2} \left(f_1(N\wrh)
\right)^{-1}\, ,
\qquad
d_{mag}(P) ={1\over N}\int_0^N \, d\ws\,
e^{-i\pi \ws P^2 / N} 
\left( f_2(\ws/N)\right)^{-1}\, 
\ee
are the degeneracies of purely 
electric and purely magnetic half-BPS states carrying
charges $Q$ and $P$ respectively. 
Thus $\ln d_{el}(Q)$ and $\ln d_{mag}(P)$ 
can be identified as the entropies of small
black holes of electric charge $Q$ and magnetic charge $P$.
Since $\ln|Q\cdot P|$ is subleading compared to these
entropies for large $Q^2$ and $P^2$ 
\i.e.\ for
\be \label{menewrange}
n, Q_1, Q_5 >> 1\, ,
\ee
we see that $\ln |d_>(Q,P)- d_<(Q,P)|$ 
can be identified as the sum of the entropies of a small electric black
hole of charge $Q$ and a small magnetic black hole of charge $P$.
In carrying out the analysis on the black hole side we shall choose
charge vectors satisfying \refb{menewrange}.

It is known from the general analysis of \cite{0206072,0702146,0706.3193} 
that a two centered black hole carrying charges $(Q,0)$ and $(0,P)$ 
contributes to the index with a sign $(-1)^{Q\cdot P+1}$. Indeed,
according to these papers the contribution to the index due to a 
two centered 
black hole is given precisely by $(-1)^{Q\cdot P+1}\,
|Q\cdot P| \, d_{el}(Q) \, d_{mag}(P)$.\footnote{This formula was derived 
for black holes in $\NN=2$ supersymmetric string theories, -- we shall 
assume that it holds also in the $\NN=4$ theories. This is not 
unreasonable since these black holes can be embedded in the $\NN=2$ 
supersymmetric S-T-U models.} 
Taking into account the sign of the right 
hand side of 
\refb{mes2}, and assuming that this phenomenon has a description in terms
of (dis)appearance of a two centered black hole carrying charges 
$(Q,0)$ and $(0,P)$, 
we can draw the following conclusion:\footnote{Note that
when a new configuration with same
charge appears in the black hole system, its degeneracy (or
more precisely the index), \i.e.\
exponential of the entropy, will add to the degeneracy of the
other configurations of the same charge.}

\noindent {\it For $J(=Q\cdot P)>0$, as we cross the wall of 
marginal stability
\refb{me6} from $a_\infty>a_c$ to $a_\infty<a_c$, 
a new two centered small black hole solution should
appear with an electric center 
of charge $Q$ and a  magnetic center of charge $P$.
On the other hand for $J(=Q\cdot P)<0$, 
as we cross the wall of 
marginal stability
\refb{me6} from $a_\infty<a_c$ to $a_\infty>a_c$, 
a new two centered small black hole solution should
appear with an electric center 
of charge $Q$ and a  magnetic center of charge $P$.}

We shall now verify this explicitly.
For describing the two centered black hole we shall use the
$\NN=2$ supersymmetric description of the same system given
in \S\ref{stu} in terms of the metric $g_{\mu\nu}$,
four scalar fields $X^I$, four gauge fields
$\AAA_\mu^I$ and some additional auxiliary fields.
We shall work in the supergravity approximation in which case
the prepotential \refb{e13ns} takes the form:
\be \label{me2.1}
F = - {X^1 X^2 X^3\over X^0}\, .
\ee
Using the relations \refb{e15ns} between the charges $\wt
q_I$, $\wt p^I$
in the STU model and the charge vector $Q$, $P$ in the $\NN=4$
description we see that 
for the configuration \refb{me5} we have
\be \label{me2.5}
(\wt q_0,\wt q_1,\wt q_2,\wt q_3) = (0, Q_1-1, -1, -n/N),  \qquad
(\wt p^0, \wt p^1, \wt p^2, \wt p^3) = (Q_5, 0,-J,  0)\, .
\ee
As discussed in \S\ref{stu}, 
the theory has an underlying `gauge invariance' that allows for a
scaling of all the $X^I$'s by a complex function. We shall fix this gauge
using the gauge condition:
\be \label{me2.6}
i (\bar X^I F_I - X^I \bar F_I) = {1}\, , \qquad F_I \equiv \p F
/ \p X^I\, .
\ee
This fixes the normalization but not the overall phase 
of the $X^I$'s.
While studying a black hole solution carrying a given set
of charges, it will be convenient to fix the overall phase of
the $X^I$'s such that
\be \label{me2.7}
Arg(\wt q_I X^I - \wt p^I F_I) = 
\pi \quad \hbox{at $\vec r=\infty$}\, .
\ee
In this gauge one can construct 
a general multi-centered black hole solution with
charges $(\wt q^{(s)}, \wt p^{(s)})$ located at $\vec r_s$
following the procedure described in
\cite{0005049,0010222,0101135,0304094}. 
The locations $\vec r_s$ are constrained by the 
equations\cite{0005049,0010222,0101135,0304094}
\be \label{me2.8a}
h_I \wt p^{(s)I} - h^I \wt q^{(s)}_I + \sum_{t\ne s} {
\wt p^{(s)I} \wt q^{(t)}_I - \wt q^{(s)}_I 
\wt p^{(t) I} \over |\vec r_s - \vec r_t|} = 0
\ee
where $h^I$ and $h_I$ are constants defined through the
equations
\be \label{me2.8}
X^I_\infty -\bar X^I_\infty = i  h^I \, , \qquad
F_{I\infty} - \bar F_{I\infty} = i   h_I \, .
\ee
We shall not give the complete solution but will give the
solution for the scalars for illustration. They are obtained by
solving the equations:
\be \label{me2.8full}
X^I  -\bar X^I  = i  \left( h^I  + \sum_s {\wt 
p^{(s)I}\over |\vec r - 
\vec r_s|} 
\right)\, , \qquad
F_{I} - \bar F_{I} = i   \left(h_I + \sum_s {\wt q^{(s)}_I\over 
|\vec r - \vec r_s|} 
\right)\, .
\ee
If we
define $\alpha$ and $\beta$ via the relations
\be \label{me2.9}
X^0_\infty = \alpha+i\beta\, ,
\ee
then using 
\refb{edilrel}, \refb{expar}, \refb{me2.1} and \refb{me2.8} we get
\ben \label{me2.10}
&& h^0 = 2\beta, \quad h^1 = 2 (\beta a_\infty + \alpha S_\infty), \quad
h^2 = 2\wh R R \alpha, \quad h^3 = 2\wh R \alpha / R, 
\nonumber \\
&&
h_0 = -2\wh R^2 (\alpha S_\infty +\beta a_\infty),  
\quad h_1 = 2\beta \wh R^2,
\quad h_2 = 2 \wh R (\beta S_\infty -\alpha a_\infty) / R, 
\nonumber \\ &&
h_3 = 2\wh R R (\beta S_\infty - \alpha a_\infty)\, . 
\een
The gauge condition \refb{me2.6} gives
\be \label{me2.11}
\alpha^2 + \beta^2 = (8 \wh R^2 S_\infty)^{-1}\, .
\ee

To proceed further we need to focus on a specific 
two centered solution with electric charge $Q$ at one center
and a magnetic charge $P$ at the other 
center.\footnote{In the supergravity approximation the
solution is singular at each center, 
but once higher derivative corrections are
taken into account each center is transformed into the near horizon
geometry of a 
non-singular extremal black
hole with finite entropy equal to the statistical entropy of the
corresponding microstates as described in \S\ref{s2.3},
\S\ref{smallre}. As we have seen, this phenomenon can be
demonstrated explicitly for purely electrically
charged small black holes representing 
fundamental heterotic string\cite{0409148,0410076,
0411255,0411272,0501014,0502126,0502157,
0507014,0506176} and hence also 
their S-dual purely magnetic states. In this case 
the modifications of the two centered solution
due to higher derivative
corrections can be found using the method developed in
\cite{0009234}. This approach fails for small black holes
describing fundamental type II string compactification and hence
also their S-dual purely magnetic states. However it is expected that
once the effect of full set of higher derivative terms are taken into
account the entropy of a small black hole in type II string theory
will also reproduce the statistical entropy of the corresponding
microstates\cite{9504147,9712150}, -- 
see \cite{0707.3818} for some recent 
progress on this issue.}
Using \refb{e15ns}, \refb{me5} 
we see that the charges at the two centers
are given by:
\be \label{me2.12}
\wt q^{(1)} = (0,0,-1,-n/N), \quad \wt p^{(1)} = (0,0,0,0),
\quad \wt q^{(2)} =(0, Q_1-1, 0,0), \quad 
\wt p^{(2)} =(Q_5, 0, -J, 0)\, .
\ee
Eqs.\refb{me2.8a} for $s=1$ and 2 now give:
\be \label{me2.13a}
h^2 + {n\over N} h^3 = {J\over L}, 
\ee
\be \label{me2.13b}
h_0 Q_5 - h_2 J - h^1 (Q_1-1)
+ {J\over L} = 0\, ,
\ee
where $L=|\vec r_1 -\vec r_2|$ is the separation between the two
centers.\footnote{Note that this is the coordinate separation. In order to 
express this in physical units {\it e.g.} string length, we need to 
examine the asymptotic metric associated with this solution and also the 
relation between the metric appearing in the $\NN=2$ supersymmetric S-T-U 
model and the string metric $G_{\mu\nu}$ that appears naturally in the 
$\NN=4$ supergravity action \refb{eag3}.} Using \refb{me2.10} and 
\refb{me2.13a} we get \be \label{me2.14}
\alpha = {J\over 2L} {1\over R\wh R + {n\over N}{\wh R\over R}}\, .
\ee
Using \refb{me2.10} and \refb{me2.14} 
we may now express
\refb{me2.13b} as
\be \label{me2.15}
\beta \left(a_\infty (Q_1 - 1 + \wh R^2 Q_5) + {\wh R J 
S_\infty \over R} 
\right) 
+ \alpha \left( (Q_1 - 1 + \wh R^2 Q_5) S_\infty -\wh R R -{n\over N}
{\wh R\over R} - {\wh R J a_\infty \over R}
\right) = 0
\, .
\ee
Substituting $\alpha$ and $\beta$ computed from
\refb{me2.14}, \refb{me2.15} 
into \refb{me2.11} we can
determine $L$. The ambiguity in determining the sign of $L$
can be fixed using \refb{me2.7}.

We are interested in determining under what conditions the two
centered black hole solution described above exists. For this we note
that a sensible solution should have positive value of $L$. Typically
as we change the values of the asymptotic moduli keeping the
charges fixed, the value of $L$ changes. On some subspace of 
codimension
1 the value of $L$ becomes infinite and beyond that the solution gives
negative values of $L$ which means that a
physical solution does not exist.
To determine this codimension 1 subspace we simply need to determine
the conditions on the asymptotic moduli for which $L=\infty$. 
{}From \refb{me2.14} we see that in this case $\alpha=0$. 
Since eq.\refb{me2.11} now requires $\beta$ to be non-zero, we see from
\refb{me2.15} that
\be \label{me2.16}
a_\infty (Q_1 - 1 + \wh R^2 Q_5) + \wh R J S_\infty / R =0\, .
\ee
This is identical to the condition \refb{me6} for marginal 
stability\cite{0005049}. 
Thus we conclude that as $a_\infty$ 
passes through
$a_c$, the two centered black hole solution carrying an entropy
equal to the sum of the entropies of a small electric black hole of
charge $Q$ and a small magnetic black hole of charge $P$, (dis)appears
from the spectrum.
This is precisely what was predicted earlier
by analyzing the exact
formula for the degeneracy of dyons.

In order to complete the verification of the 
predictions made earlier we need to determine on which side of the 
$a_\infty=a_c$ line
the two centered solution exists. 
For this we use eq.\refb{me2.7}. For the solution under consideration
this gives, using \refb{me2.15},
\be \label{mesign}
\alpha \left(a_\infty (Q_1 - 1 + \wh R^2 Q_5) + {\wh R J S_\infty 
\over R} 
\right) \left\{ 1 + 
{\left( (Q_1 - 1 + \wh R^2 Q_5) S_\infty -\wh R R -{n\over N}
{\wh R\over R} - {\wh R J a_\infty \over R}
\right)^2 \over  \left(a_\infty (Q_1 - 1 + \wh R^2 Q_5) + {\wh R J S_\infty 
\over R} 
\right)^2} \right\} < 0\, .
\ee
First consider the case $J>0$. 
Since $L$ must be positive for the two centered solution to
exist, we see from \refb{me2.14} that $\alpha>0$. 
In this case the term on the left hand side of \refb{mesign} is
negative for $a_\infty<a_c$ and positive for 
$a_\infty>a_c$. Thus the inequality
is satisfied only for $a_\infty<a_c$, leading to the conclusion that the
two centered black hole exists only for $a_\infty<a_c$. A similar analysis
shows that for $J<0$, the two centered black hole exists only for
$a_\infty>a_c$. This is exactly what has been predicted earlier 
from the analysis of the exact dyon spectrum of the theory.

\renewcommand{\theequation}{\thesection.\arabic{equation}}

\sectiono{Open Questions and Speculation on $\NN=2$} 
\label{sopen}

We end by reviewing some of the questions left open in our
analysis.
Some of these issues are technical in nature while
some others are conceptual.
We also speculate on the degeneracy of dyons in $\NN=2$
supersymmetric string theories. 

\begin{itemize}

\item {\bf Non-locality of the 1PI action:}
The entropy function formalism described in this article
gives us a way to calculate the entropy of an extremal black hole
with a given set of charges in a given theory assuming that
such a black hole solution exists.
The main ingredient in this computation is the assumption that the
underlying theory is described by a local generally covariant and
gauge invariant action and also that the near horizon geometry of
an extremal black hole has $SO(2,1)$ isometry associated with
an $AdS_2$ factor.

It is natural to expect that the formalism also works for studying
quantum corrected entropy provided we replace the classical
Lagrangian density in the expression for the entropy function 
by the quantum Lagrangian density 
associated
with one particle irreducible effective action. 
Indeed this was an underlying
assumption in studying the effect of Gauss-Bonnet term on the
entropy of $\NN=4$ supersymmetric black holes, since some 
part
of the Gauss-Bonnet term comes from quantum corrections. This
is also natural from the point of view of duality symmetries since
classical higher derivative corrections in one description of the theory
may appear as quantum effects in a different description and the
entropy of the black hole must be given by a duality invariant
expression. However this assumption suffers from an
intrinsic problem that sooner or later we shall encounter 
non-local terms in the quantum effective action, and the current
formulation of the entropy function does not allow us to deal with
non-local terms in the action. It seems natural that the way to deal
with such terms is to generalize the notion of entropy being an
extremum of entropy function to the entropy being the result of
an appropriate functional integral in the background of an
$AdS_2$ geometry. It will be interesting to make this into a more
precise conjecture.

\item {\bf $\NN=2$ dyon spectrum:}
Given the success in finding the dyon spectrum in $\NN=4$
supersymmetric string compactifications, one could hope that
the dyon degeneracy in $\NN=2$
supersymmetric string theories will also be given by a similar
formula:
\be\label{en2pred}
d(\vec Q, \vec P) = \int_{\CC} dM \, f(\vec Q, \vec P,M)\, ,
\ee
where $M$ denotes  a set of complex variables, $\CC$ is
a contour in the complex manifold labelled by the variables
$M$, and $f(\vec Q, \vec P,M)$ is an appropriate function of
the charges and the complex variables $M$.  One could further
speculate that $f(\vec Q, \vec P,M)$ has the structure
\be \label{efurther}
f(\vec Q, \vec P,M) = \exp\left(f^{(1)}_{ij}(M) Q_i Q_j
+ f^{(2)}_{ij}(M) P_i P_j + f^{(3)}_{ij}(M) Q_i P_j\right)
\, g(M)\, ,
\ee
where $f^{(s)}_{ij}(M)$ are simple functions of $M$ and $g(M)$
encodes all the non-trivial properties of the theory. 
The degeneracies in different domains 
in the moduli space bounded by walls
of marginal stability will correspond to different choices of 
contour $\CC$,
and as we cross the walls of marginal stability the choice of the
contour will change. The jump in the
degeneracy can then
be computed by evaluating the residues of the integrand
at the poles that we encounter as we deform the contour. 
On the
other hand from the analysis of \cite{0206072,0702146,0706.3193} 
we know that for a
decay $(\vec Q, \vec P)\to (\vec Q_1, \vec P_1) + (\vec Q-\vec Q_1,
\vec P - \vec P_1)$ this jump is given by
\be \label{en2jump}
\Delta d(\vec Q, \vec P) = (-1)^{Q_1\cdot P-Q\cdot P_1+1} 
\, |Q_1\cdot P- Q\cdot P_1|\, 
d(\vec Q_1, \vec P_1) d(\vec Q-\vec Q_1, \vec P - \vec P_1)\, .
\ee
Since for $\NN=2$ supersymmetric string theories all BPS
states are half-BPS, the above equation would
relate the residues
of the function $f(\vec Q,\vec P, M)$ at various poles to
contour integrals of the functions 
$f(\vec Q_1, \vec P_1, M)$ and $f(\vec Q-\vec Q_1, 
\vec P-\vec P_1, M)$.
This in effect would give a set of 
bootstrap relations involving
the integrands $f(\vec Q, \vec P, M)$ for different $\vec Q$, $\vec P$,
$M$. Under favourble conditions we may even be able to solve
these bootstrap equations to extract the form of the functions
$f(\vec Q, \vec P, M)$.

\item {\bf Non-renormalization of the Gauss-Bonnet 
contribution:} One of the small miracles in our analysis has been that in a class
of string theories, by taking
into account either the tree level Gauss-Bonnet term or
the $\NN=2$ supersymmetric generalization of the curvature squared
terms, we recover the exact results for the entropy of a dyonic
black hole in the large electric charge approximation.
This is surprising because 
there are additional corrections to the tree level
effective action which
could affect the calculation of the entropy. 
Indeed for a similar non-supersymmetric black hole,  related to
the supersymmetric black hole by the reversal of sign of one of the
charges, neither the Gauss-Bonnet term nor the supersymmetric
generalization of the curvature squared term gives the correct answer
for the entropy.
It will be important to
understand the origin of the underlying non-renormalization theorem
in more detail. It is especially important in view of the fact that for the
case where the electric and the magnetic charges are comparable the
complete answer for the black hole entropy
is not known. Thus if there is an underlying
non-renormalization theorem that tells us that including just the
effect of Gauss-Bonnet term is sufficient even in this case, then it
would give us a way to calculate the entropy of the dyonic black holes
for comparable electric and magnetic charges. The agreement between
the statistical entropy and the answer for the entropy computed from the
Gauss-Bonnet action seems to point towards the existence of such a
non-renormalization theorem.

\item {\bf Small black holes:}
The statistical entropy of a small black hole in heterotic string
theory, whose microscopic description involves excitations of a
fundamental heterotic string, agree with the black hole entropy
after taking into account the effect of
higher derivative corrections in the effective action. 
Although initial analysis took into account only a small
part of the higher derivative corrections and the agreement between the
two answers was puzzling, we now have a good understanding of this
agreement under the assumption that the near horizon geometry
of these black holes have an underlying locally $AdS_3$ space.
Unfortunately the same analysis does not produce a similar agreement
for small black holes describing excited states of the fundamental
type II string. The most naive explanation seems to be that these
black holes do not have an underlying $AdS_3$ factor in their
near horizon geometry; however a different explanation
has been suggested in 
\cite{0611062,0707.3818,0707.4303,0708.0016}. It will
be useful to resolve this issue one way or the other and also to
extend the analysis to small black holes in higher dimensions.

\item {\bf OSV conjecture:}
One question that naturally comes to mind is: 
is there a possible connection
between the results on black hole entropy discussed here with the
OSV conjecture\cite{0405146}? 
There are two aspects of the OSV conjecture. The first one gives a 
prescription
for relating the microscopic
degeneracy of states to the topological string 
partition
function. 
Most of the precision tests of the OSV conjecture has been carried
out in this context.
Since we have an exact expression
for the degeneracy of states in a class of $\NN=4$ supersymmetric
string theories, one can in principle try to see if this can be reproduced
using the OSV conjecture. This would be a test of the first aspect
of the conjecture. 
Some attempts in this direction have been 
made in \cite{0508174}. 

The second aspect of OSV conjecture -- which is much less
studied\footnote{Even the precision tests of
\cite{0502157,0507014} involving small black holes 
really tests the relation between
the topological string partition function and the microscopic
degeneracy since the `Wald entropy' that was used in this test was
computed using only the F-term corrections, and in this
approximation it is directly related to the topological string partition
function.} --
relates Wald's entropy to the
statistical entropy. However the statistical entropy, instead of being 
computed in the microcanonical ensemble, is
computed in a mixed ensemble where
magnetic charges and the chemical potentials dual to the electric
charges are fixed. 
If both aspects of the OSV conjecture are correct then this would
amount to the statement that Wald's entropy of a black hole
in $\NN\ge 2$ supersymmetric string theory receives
contribution only from the F-terms modulo some corrections discussed in 
\cite{0508174,0702146}. This seems strange since the
effective action is known to receive other corrections. It is
conceivable (although by no means proven)
that if one computes Wald's entropy using the
Wilsonian effective action then only the F-terms contribute to the
entropy.
We have not followed this approach.
Instead we calculate Wald's entropy using 
the one particle irreducible (1PI) 
effective action\footnote{This is reflected for 
example in the $\ln S$ term
in the coefficient of the Gauss-Bonnet term as given in 
eqs.\refb{ecorg1},
\refb{eh10bint}.} 
and relate it directly to the statistical entropy
computed in the microcanonical ensemble. This approach has
the advantage that 
both the Wald's
entropy computed using the one particle irreducible (1PI) 
effective action and the microcanonical entropy are manifestly
duality invariant in the sense described at the end of
\S\ref{s2.1}.  Thus we compare two duality invariant
quantities.
The disadvantage of this approach is the intrinsic non-local
nature of the 1PI effective action discussed earlier.

\end{itemize}

Before concluding I would like to emphasize again that this article
reviews only some special aspects of extremal black hole entropy.
Various other recent references on entropy function and attractor
mechanism can be found in 
\cite{0508110,0510024,0511117,0511306,0512048,0512138,0601016,
0601183,0602005,0602022,0602292,0603003,0604021,0604028,
0604106,0605279,0606026,0606084,0606098,0606218,
0606263,0607132,0607202,0607227,0608091,0608222,
0611140,0611240,0611345,0612181,0612225,0612308,
0701004,0701090,0701158,0701216,0701221,0702019,0702040,
0702088,0702142,0702170,0703035,0703036,0703037,
0703178,
0703214,0703223,0703260,0704.0955,0704.1239,0704.1405,0704.1452,
0704.1819,0704.3295,0705.1847,0705.2149,0705.2892,
0705.3866,0705.4554,
0706.1167,0706.1667,0706.1847,0706.2046,0707.0105,0707.0964,
0707.1464,0707.2730,0707.4554,0708.0240}.

\bigskip

\noindent {\bf Acknowledgement:} I would like to thank my 
collaborators Dumitru Astefanesei, Nabamita Banerjee,
Atish Dabholkar, Justin David, 
Kevin Goldstein, Norihiro Iizuka, Ashik Iqubal, Dileep Jatkar, 
Rudra Jena, 
Bindusar Sahoo, Masaki Shigemori and
Sandip Trivedi for many useful discussions. Much of this
review is based on the work done in various collaborations
with them. I would also like to thank Nabamita Banerjee
and Rajesh Gupta for their
comments on the manuscript. 
These notes arose out of lectures given at the IPM string school at
Tehran (April 2006), IOPB string school at Bhubaneswar
(September 2006), the first Asian string school at Seoul
(January 2007) and Prestrings meeting at Granada (June 2007).
I would like to thank the organisers of these meetings for
support.
Finally I would like to thank the
people of India for their generous support to string theory.

\appendix

\sectiono{The Sign Conventions} \label{s3.1}

  In this appendix we shall
 fix the sign conventions for
 the charges carried by various branes, as well as the sign convention
 for the duality transformations relating the two descriptions
of the $\NN=4$ supersymmetric string 
theories introduced in \S\ref{sphys}:
as a $\ZZZ_N$ orbifold of type IIB string theory on $\MM\times \wt 
S^1\times S^1$ for $\MM=K3$ or $T^4$ and as an asymmetric $\ZZZ_N$ 
orbifold of heterotic or type 
IIA string theory on $T^4\times \wh S^1\times S^1$.
The series of duality transformations taking us from the first to the 
second description are: an S-duality transformation on type IIB string 
theory, a T-duality on $\wt S^1$ mapping the theory to a $\ZZZ_N$ orbifold 
of type IIA string 
theory on $\MM\times \wh S^1\times S^1$, and finally a string-string 
duality 
transformation taking this to an asymmetric $\ZZZ_N$ orbifold of heterotic 
/ type IIA string theory on $T^4\times \wh S^1\times S^1$ for $\MM=K3$ / 
$\MM=T^4$.

 We denote by $y$ the coordinate along $S^1$, by $\psi$ the coordinate
along $\wt S^1$, by $\chi$ the coordinate along the circle
$\wh S^1$ dual to $\wt S^1$, by $t$ the time coordinate and by
$(r,\theta,\phi)$ the spherical polar coordinates of the non-compact
space.
Let 
$B_{MN}$ denote the NSNS 
2-form fields, $g_{MN}$ denote the metric, and
$C^{(k)}_{MN}$ denote the RR $k$-form field subject to the
relations
\be \label{eckc8}
* dC^{(k)} = (-1)^{k(k-1)/2} \, dC^{(8-k)} + \cdots\, ,
\ee
where $*$ denotes Hodge dual and $\cdots$ represent non-linear
terms. In defining the Hodge dual in the first description we choose
the $\epsilon$ tensor such that after dimensional reduction on K3,
$\epsilon^{ty\psi r\theta \phi}>0$. 
The chain of duality transformations taking us from the first to
the second description are chosen so that 
at the linearized
level the
first S-duality transformation of IIB acts as $C^{(2)}\to B$,
$B\to -C^{(2)}$,  the next $R\to 1/R$ 
duality transformations of $\wt S^1$
acts as $g_{\psi\mu}\to - B_{\chi\mu}$, $B_{\psi\mu}\to
- g_{\chi\mu}$ together with appropriate transformations on the
various RR gauge fields, and
the final string string duality transformation
acts via a Hodge duality transformation in six dimensions 
on the NS sector 3-form field
strength with $\epsilon^{t\chi y r\theta\phi}>0$, and maps various
four dimensional gauge fields arising from various components
of the RR sector fields to the 24 (8) NS sector gauge fields in heterotic 
(type IIA) string
theory on $T^4$. 
Finally, we
use the following
convention for the signs of
the charges carried by various branes.
If $F^{(3)}\equiv d C^{(2)}$ denotes the RR 3-form field strength, then 
asymptotically a
D1-brane along $S^1$  carries positive
$F^{(3)}_{yrt}$,  a D5-brane along $\wt S^1\times K3$ carries
positive $F^{(3)}_{\theta y\phi}$,
a D1-brane along $\wt S^1$ carries positive
$F^{(3)}_{\psi rt}$ and a 
D5-brane along $S^1\times K3$ carries
negative $F^{(3)}_{\theta \psi\phi}$. 
The same convention is followed
for fundamental string and NS 5-brane with $F^{(3)}$ replaced by the
NSNS 3-form field strength $H=dB$. 
A
state carrying positive momentum along $S^1$ or $\wt S^1$ is
defined to 
be the one which
produces positive $\p_r g_{yt}$ or $\p_r g_{\psi t}$, 
and a 
positively charged Kaluza-Klein monopole 
associated with the circle $S^1$ or $\wt S^1$
is defined to be the one that carries 
positive $\p_\theta g_{y\phi}$ or $\p_\theta g_{\psi\phi}$.
Using the relation between $C^{(2)}$ and $C^{(6)}$ given
in \refb{eckc8} one can verify that a D5-brane wrapped on
$K3\times S^1$ carries negative $(dC^{(6)})_{K3,yrt}$
asymptotically.

\sectiono{A Class of (4,4) Superconformal Field Theories} \label{s0}

In this appendix we shall introduce a class of (4,4) superconformal
field theories and study their properties.
These have been used in \S\ref{s3} for precision
counting of dyon states in  $\NN=4$ supersymmetric
string theories based on $\ZZZ_N$ orbifolds. These will also be
used in appendix \ref{ssgauss} for computing the coefficient of
the Gauss-Bonnet term in the effective action of these string
theories.

Let $\MM$ be either a $K3$ or a $T^4$ manifold, and let
$\wt g$ be an order $N$ discrete symmetry transformation 
acting on $\MM$.
We shall choose $\wt g$ in such a way that it 
satisfies the following properties (not all of which are independent):
\begin{enumerate}

\item We require that in an appropriate complex coordinate system
of $\MM$, $\wt g$  preserves the (0,2) and (2,0)
harmonic forms of $\MM$.

\item Let $\wt\ZZZ_N$ denote the group generated by $\wt g$.
We shall require that the orbifold   
$\wh\MM=\MM/\wt\ZZZ_N$ has
SU(2) holonomy.  

\item Let $\omega_i$ denote the harmonic
2-forms of $\MM$ and
\be\label{einter}
I_{ij}=\int_\MM  \omega_i \wedge \omega_j 
\ee
denote 
the intersection matrix of these
2-forms in $\MM$. When we 
diagonalize $I$ we get 3 eigenvalues $-1$ and a certain number
(say $P$) of the 
eigenvalues $+1$ ($P=19$ for $K3$ and 3 for $T^4$). We call
the 2-forms carrying eigenvalue $-1$  right-handed or self-dual
2-forms and the 2-forms carrying eigenvalues $+1$ left-handed or
anti-selfdual
2-forms. We shall choose $\wt g$ such that it
leaves invariant all the right-handed
2-forms.\footnote{We should note that there is no correlation
between left- and right-handed 2-forms and left- and right-moving
degrees of freedom on the world-sheet of the sigma model
with target space $\MM$.}

\item The $(4,4)$ superconformal field theory
with target space
$\MM$ has $SU(2)_L\times SU(2)_R$ R-symmetry group. 
We shall require that the transformation $\wt g$ 
commutes
with the (4,4) superconformal symmetry and the
$SU(2)_L\times SU(2)_R$ R-symmetry group of
the theory.
(For
$\MM=T^4$ the supersymmetry and the
R-symmetry groups are bigger, but $\wt g$ must be such that
only the (4,4) superconformal
symmetry and the 
$SU(2)_L\times SU(2)_R$ part of the R-symmetry group
commute with $\wt g$.)

\end{enumerate}

Let us now take an orbifold of this
(4,4) superconformal field theory by the
group $\wt \ZZZ_N$ generated by the transformation $\wt g$, and
define\cite{9306096}
 \be\label{esi4aint}
F^{(r,s)}(\tau,z) \equiv {1\over N} Tr_{RR;\wt g^r} \left(\wt g^s
(-1)^{F_L+F_R}
e^{2\pi i \tau L_0} 
e^{-2\pi i \bar\tau \bar L_0}
e^{2\pi i F_L z}\right), \qquad 0\le r,s\le N-1\, ,
 \ee
where
$Tr_{RR;\wt g^r}$ denotes trace 
over all the RR sector 
states twisted by $\wt g^r$ in the SCFT described above
{\it before we project on to $\wt g$ invariant
states}, $L_n$, $\bar L_n$ denote the left- and right-moving
Virasoro generators and $F_L$ and $F_R$ 
denote the world-sheet fermion 
numbers
associated with left and right-moving sectors in this 
SCFT.\footnote{At this stage we are describing an abstract
conformal field theory without connecting it to string theory.
In all cases where we use this conformal field theory
to describe a fundamental string world-sheet theory
or world-volume theory
of some soliton, we 
shall use the Green-Schwarz formulation. Thus the world-sheet
fermion number of this SCFT
will represent the space-time fermion number in string theory.}
As mentioned earlier we include in the definition
of $L_0$, $\bar L_0$ additive factors of $-c_L/24$ and $-c_R/24$
respectively, so that RR sector ground state has $L_0=\bar L_0=0$.
Due to the insertion of $(-1)^{F_R}$ factor in the trace the
contribution to $F^{(r,s)}$ comes only from the $\bar L_0=0$
states. As a result $F^{(r,s)}$ does not depend on $\bar\tau$.

The quantities $F_L$, $F_R$ can be identified as twice the third
generators of the $SU(2)_L$ and $SU(2)_R$ R-symmetry algebras
respectively. As a result the $z$-dependence of $F^{(r,s)}(\tau,z)$
is determined by the characters of the $SU(2)_L$ current algebra.
Since for the SCFT under consideration the $SU(2)_L$, $SU(2)_R$
current algebras have level 1, the only $SU(2)_L$ primaries which can
appear in the spectrum are those corresponding to isospins 0 
and ${1\over 2}$. The associated characters are given by
$\vt_3(2\tau, 2z)$ and $\vt_2(2\tau,2 z)$ respectively. Thus
the functions $F^{(r,s)}(\tau,z)$ have the form
\be\label{efhrel}
F^{(r,s)}(\tau,z) = h_0^{(r,s)}(\tau) \, \vt_3(2\tau,2z) +
h_1^{(r,s)}(\tau) \, \vt_2(2\tau,2z)\, 
\ee
for some functions $h_0^{(r,s)}(\tau)$ and $h_1^{(r,s)}(\tau)$.
The functions 
$h_b^{(r,s)}(\tau)$ in turn have
expansions of the form
\be\label{ehbexp}
h_b^{(r,s)}(\tau) = \sum_{k\in{1\over N}\zzz -{b^2\over 4}}
c_b^{(r,s)}(4k) e^{2\pi i k\tau}\, .
\ee
This defines the coefficients $c^{(r,s)}_b(u)$. We shall justify
the restriction on the allowed values of $k$ shortly.
Using the known expansion of $\vt_3$ and $\vt_2$:
\be\label{eg1theta}
\vartheta_3(2\tau, 2 z) = \sum_{j\in 2\zzz} e^{2\pi i j z} e^{\pi i
\tau j^2/2}, 
\qquad \vartheta_2(2\tau, 2 z) = \sum_{j\in 2\zzz+{1}} e^{2\pi i j z} 
e^{\pi i \tau j^2/2},
\ee
 we get
\be\label{enewint}
F^{(r,s)}(\tau,z) =\sum_{b=0}^1\sum_{j\in2\zzz+b, n\in \zzz/N} 
c^{(r,s)}_b(4n -j^2)
e^{2\pi i n\tau + 2\pi i jz}\, .
\ee

$F^{(r,s)}(\tau,0)$ defined from \refb{esi4aint} 
can be regarded as the 
partition function of the
superconformal field theory on a torus with appropriate twisted
boundary condition along the $a$ and the $b$ cycles. {}From
this it follows that
\be \label{efollows}
F^{(r,s)}\left( {a\tau + b\over c\tau +d}, 0\right)
= 
F^{(cs + ar,ds+br)}(\tau,0)
\, .
\ee
Using \refb{efhrel} and the known modular transformation properties
of Jacobi $\vt$-functions it then follows that
\be \label{efrsmod}
F^{(r,s)}\left( {a\tau + b\over c\tau +d}, {z\over c\tau+d}\right)
= \exp\left( 2\pi i {cz^2 \over c\tau+d}\right) 
F^{(cs + ar,ds+br)}(\tau,z)
\, .
\ee

The functions $F^{(r,s)}(\tau, z)$ or equivalently the coefficients
$c_b^{(r,s)}(u)$ contain information about the spectrum of
$\wt g^r$ twisted $\bar L_0=0$ states of the superconformal field
theory, carrying definite $\wt g$, $L_0$ 
and $F_L$ quantum numbers.
In the rest of this appendix we shall study various properties of
these coefficients. First of all, 
since in the RR sector the  $L_0$ eigenvalue is $\ge 0$ for any
state, 
it follows from \refb{efhrel}-\refb{eg1theta} that
\be\label{evanish}
c^{(r,s)}_0(u)=0 \quad \hbox{for $u<0$},
\qquad c^{(r,s)}_1(u)=0 \quad \hbox{for $u<-1$}\, .
\ee
Using $\pmatrix{a & b\cr c & d}=\pmatrix{-1 & 0\cr 0 & -1}$
in \refb{efrsmod} we get
$F^{(r,s)}(\tau,z)=F^{(-r,-s)}(\tau, -z)$.
It then follows
from \refb{efhrel}, \refb{enewint},  that
\be\label{ecrev}
h^{(r,s)}_b(\tau)
= h^{(-r,-s)}_b(\tau)\, , \qquad c^{(r,s)}_b(u) = c^{(-r,-s)}_b(u)\, ,
\qquad \hbox{for $b=0,1$}\, .
\ee
Furthermore, taking $\pmatrix{a & b\cr c & d}=
\pmatrix{1 & 1\cr 0 & 1}$ in \refb{efrsmod} we get
$F^{(r,s)}(\tau+1,z)=F^{(r,s+r)}(\tau, z)$. Since
$(r,s)$ are defined modulo $N$ we get $F^{(r,s)}(\tau, z)
=F^{(r,s)}(\tau+N, z)$. This is the physical origin of the 
restriction $n\in \ZZZ/N$ in \refb{enewint} and $k\in \ZZZ/N-b^2/4$
in \refb{ehbexp}.

The $n=0$ terms in the expansion \refb{enewint}
is given by the contribution to \refb{esi4aint} 
from the RR sector
states with $L_0=\bar L_0=0$.
For $r=0$, \i.e.\ in the untwisted
sector, these states
are in one to one correspondence with 
harmonic $(p,q)$ forms on $\MM$, with
$(p-1)$ and $(q-1)$ measuring the quantum numbers
$F_L$ and $F_R$\cite{wittop,vafatop}. 
Comparing \refb{esi4aint} with \refb{enewint} we now see that
$N\, c^{(0,s)}_0(0)$, being $N\times$
the coefficient of the $n=0$, $j=0$ term in \refb{enewint},
measures the number
of harmonic $(1,q)$ forms weighted by $(-1)^{q-1} \wt g^s$
and summed over $q$,
and $N\, c^{(0,s)}_1(-1)$, being $N\times$
the coefficient of the $n=0$,
$j=-1$ (or $j=1$) term in \refb{enewint}, 
measures the number of harmonic
$(0,q)$ (or $(2,q))$ forms weighted by $(-1)^{q}\wt g^s$
and summed over $q$.
If $\MM=K3$ then the only $(0,q)$ forms are $(0,0)$ and $(0,2)$
forms both of which are invariant under $\wt g$. Thus we have
\be\label{eck3}
c^{(0,s)}_1(-1) ={2\over N} \quad \hbox{for $\MM=K3$}\, .
\ee
On the other hand for $\MM=T^4$
one can represent the explicit action of $\wt g$
in an appropriate complex coordinate system $(z^1,z^2)$ as
\be\label{eoneform}
dz^1 \to e^{2\pi i/N} dz^1, \quad dz^2\to e^{-2\pi i/N} dz^2\, ,
\quad d\bar z^1 \to e^{-2\pi i/N} d\bar
z^1, \quad d\bar z^2\to e^{2\pi i/N} d\bar z^2\, .
\ee
Using this one can  work out 
its action on all the 2-, 3- and 4-forms:
\bea{etwoform}
&&dz^1\wedge dz^2\to dz^1\wedge dz^2, \quad
dz^1\wedge d\bar z^1\to dz^1\wedge d\bar z^1, \quad
dz^1\wedge d\bar z^2\to e^{4\pi i/N} \, dz^1\wedge d\bar z^2,
\nonumber \\
&&d\bar z^1\wedge d\bar z^2\to d\bar z^1\wedge d\bar z^2, \quad
dz^2\wedge d\bar z^2\to dz^2\wedge d\bar z^2, \quad
d\bar z^1\wedge d z^2\to e^{-4\pi i/N} \, d\bar z^1\wedge d z^2\, ,
\nonumber \\
\eea
\bea{ethreeform}
&& dz^1\wedge dz^2\wedge d\bar z^1\to e^{-2\pi i/N}\,
dz^1\wedge dz^2\wedge d\bar z^1, \quad
dz^1\wedge dz^2\wedge d\bar z^2\to e^{2\pi i/N}\,
dz^1\wedge dz^2\wedge d\bar z^2, \nonumber \\
&& d\bar z^1\wedge d\bar z^2\wedge d  z^1\to e^{2\pi i/N}\,
d\bar z^1\wedge d\bar z^2\wedge d  z^1 , \quad
d\bar z^1\wedge d\bar z^2\wedge d  z^2\to e^{-2\pi i/N}\,
d\bar z^1\wedge d\bar z^2\wedge d  z^2, \nonumber \\
\eea
\be\label{efourform}
dz^1\wedge dz^2\wedge d\bar z^1\wedge d\bar z^2 \to
dz^1\wedge dz^2\wedge d\bar z^1\wedge d\bar z^2\, .
\ee
This shows that
the $(0,0)$ and $(0,2)$ forms are
invariant under $\wt g$ but the two $(0,1)$ forms carry 
$\wt g$ eigenvalues $e^{\pm 2\pi i/N}$. Thus we have
\be\label{ect4}
c^{(0,s)}_1(-1) ={1\over N} 
\left(2 - e^{2\pi is/N} - e^{-2\pi is/N}\right) \quad 
\hbox{for $\MM=T^4$}\, .
\ee
\refb{etwoform} also shows that  
$\wt g$ acts trivially on four of the 2-forms, and acts as a
rotation by $4\pi/N$ in the two dimensional subspace spanned by
the other two 2-forms. By writing the
2-forms in the real basis one can easily verify that the 2-forms
which transform non-trivially under $\wt g$ correspond to 
left-handed 2-forms. 
This is one of the requirements on $\wt g$ listed at the
beginning of this appendix.

Another useful set of results emerges by taking the $z\to 0$
limit of eqs.\refb{esi4aint} and \refb{enewint}. This gives
\be\label{efqrs}
\sum_{b=0}^1\sum_{j\in2\zzz+b, n\in \zzz/N} 
c^{(r,s)}_b(4n -j^2)
e^{2\pi i n\tau} = {1\over N}\, Q_{r,s}\, ,
\ee
where
\be\label{eqrs}
Q_{r,s}= Tr_{RR;\wt g^r} \left(\wt g^s
(-1)^{F_L+F_R}
e^{2\pi i \tau L_0} e^{-2\pi i \bar\tau \bar L_0}
\right), \qquad 0\le r,s\le N-1\, .
 \ee
$Q_{r,s}$ is independent of $\tau$ and $\bar\tau$ since the 
$(-1)^{F_L+F_R}$
insertion in the trace makes the contribution from the $(L_0,\bar L_0)
\ne 
(0,0)$
states cancel. Thus \refb{efqrs} gives
\be\label{ecqrel}
\sum_{b=0}^1\sum_{j\in2\zzz+b} 
c^{(r,s)}_b(4n -j^2) = {1\over N} \, Q_{r,s}\, \delta_{n,0}\, .
\ee
Setting $n=0$ in the above equation and using eq.\refb{evanish} we get
\be\label{eqrsrev}
Q_{r,s} = N\, 
\left( c^{(r,s)}_0(0)+ 2 \, c^{(r,s)}_1(-1)\right)\, .
\ee
For $r=0$, \i.e.\ in the untwisted sector, the trace in \refb{eqrs}
reduces to a sum over the harmonic  
forms of $\MM$. Since $F_L+F_R+2$ is mapped to the degree of
the harmonic form, $Q_{0,s}$ has the interpretation of the trace of $\wt 
g^s$ over the even 
degree harmonic forms of $\MM$ minus the trace of $\wt g^s$ over the 
odd degree
harmonic forms of $\MM$. In 
particular we have
\be\label{eeuler}
Q_{0,0}= \chi(\MM)\, ,
\ee
where $\chi(\MM)$ denotes the Euler number of $\MM$.

So far our definitions of various quantities have been somewhat
abstract. 
For $\MM=K3$ and prime values of $N$ ($N=$ 1, 2, 3, 5, 7)
the functions 
$F^{(r,s)}(\tau)$ are known explicitly and
are given by\cite{0602254}
\bea{fifthn}
F^{(0,0)}(\tau, z) &=& {8\over N} A(\tau, z)\, ,
\nonumber \\
F^{(0,s)}(\tau, z) &=& {8\over N(N+1)} \, A(\tau, z) -{2\over N+1}
\, B(\tau, z) \, E_N(\tau) \qquad \hbox{for $1\le s\le (N-1)$}
\, , \nonumber \\
F^{(r,rk)}(\tau, z) &=& {8\over N(N+1)} \, A(\tau, z)
+ {2\over N(N+1)} \, E_N\left({\tau+k\over N}\right)\, B(\tau, z)\, ,
\nonumber \\
&& \qquad \qquad \qquad \qquad 
\hbox{for $1\le r \le (N-1)$, \quad $0\le k\le (N-1)$}
\, ,\nonumber \\
\eea
where
\be\label{efirstn}
 A(\tau, z) =  \left[ {\vartheta_2(\tau,z)^2
\over \vartheta_2(\tau,0)^2} +
{\vartheta_3(\tau,z)^2\over \vartheta_3(\tau,0)^2}
+ {\vartheta_4(\tau,z)^2\over \vartheta_4(\tau,0)^2}\right]\, ,
\ee
\be\label{secondn}
B(\tau, z) = \eta(\tau)^{-6} \vartheta_1(\tau, z)^2\, ,
\ee
and
\be\label{thirdn}
E_N(\tau) = {12 i\over \pi(N-1)} \, \p_\tau \left[ \ln\eta(\tau)
-\ln\eta(N\tau)\right]= 1 + {24\over N-1} \, \sum_{n_1,n_2\ge 1\atop
n_1 \ne 0 \,  mod \, N} n_1 e^{2\pi i n_1 n_2 \tau}\, .
\ee
On the other hand
for $\MM=T^4$ the function $F^{(r,s)}(\tau,z)$ are known for
$N=2,3$ and are given by\cite{0607155}
\bea{valindalt}
F^{(0,s)}(\tau, z) &=& {16\over N}\, \sin^4\left(
{\pi s\over N}\right)\,  {\vt_1\left(\tau, z+{s\over N}
\right)\vt_1\left(\tau, -z+{s\over N}
\right)\over \vt_1\left(\tau,{s\over N}\right)^2} \nonumber \\
F^{(r,s)}(\tau,z) &=& {4\, N\over(N-1)^2}
\, {\vt_1\left(\tau, z+{s\over N}
+{r\over N}\tau
\right)\vt_1\left(\tau, -z+{s\over N}+{r\over N}\tau
\right)\over \vt_1\left(\tau,
{s\over N}+{r\over N}\tau\right)^2} \, ,
\nonumber \\
&& \qquad \hbox{for $1\le r\le N-1$,  $0\le s\le N-1$}
\, . 
\eea
Using these results we can compute the coefficients $c_b^{(r,s)}(u)$
and $Q_{r,s}$ for these theories.

\sectiono{Siegel Modular Forms from Threshold Integrals} 
\label{ssiegel}

In this section we shall prove various properties of the
function $\wt\Phi(\wrh,\ws,\wv)$ defined
in \refb{edefwtphi} by
relating it to a `threshold integral'. These techniques
were first developed in 
\cite{dixon,9510182,9512046,9607029} 
and generalized to the cases under study
in \cite{0602254}.

We begin 
by defining:
\be\label{edefomega}
\Omega=\pmatrix{\rho  & v \cr v  & \sigma}\, ,
\ee
and
\bea{e7n}
{1\over 2} p_R^2 &=& {1\over 4 \det Im  \Omega} |-m_1 \rho  +
m_2 + n_1 \sigma + n_2 (\sigma\rho -v^2) + j v |^2, \nonumber \\
{1\over 2} p_L^2
&=& {1\over 2}  p_R^2 + m_1 n_1 + m_2 n_2 + {1\over 4} j^2\, ,
\eea
where $\rho$, $\sigma$ and $v$ are three
complex variables. We now consider the `threshold integrals'
\be\label{rwthrint}
\tI(\rho , \sigma, v ) = \sum_ {r, s =0}^{N-1}\sum_{b=0}^1
\tI_{r, s, b}\, , \qquad 
\hI(\rho , \sigma, v)
= \sum_ {r, s =0}^{N-1}\sum_{b=0}^1\, \hI_{r,s,b}\, ,
\ee
where
\be\label{defirsl}
\tI_{r,s,b} = \int_{\FF} \frac{d^2\tau}{\tau_2} 
\left[\sum_{\stackrel{m_1, m_2, n_2 \in 
\zzz}{n_1\in \zzz+ \frac{r}{N}, 
j\in 2\zzz + b}}
q^{p_L^2/2} \bar q^{ p_R^2/2} e^{2\pi i  m_1 s/N}
h_b^{(r,s)}(\tau) - \delta_{b,0} \delta_{r,0} c^{(0,s)}_0(0)
\right]\, ,
\ee
and
\be\label{enn6}
\hI_{r,s,b} = \int_{\FF} \frac{d^2\tau}{\tau_2} 
\left[\sum_{\stackrel{m_1, n_1\in \zzz, m_2\in 
\zzz/N}{n_2\in N\zzz+ r, 
j\in 2\zzz + b}}
q^{p_L^2/2} \bar q^{ p_R^2/2} e^{2\pi i m_2 s} 
h_b^{(r,s)}(\tau)- \delta_{b,0} \delta_{r,0} c^{(0,s)}_0(0)
\right]\, ,
\ee
with
\be\label{edefq}
q\equiv e^{2\pi i\tau}\, .
\ee
$\FF$ denotes the fundamental region of $SL(2,\ZZZ)$ in the
upper half plane. The subtraction terms 
proportional to $c_0^{(0,s)}(0)$ have been chosen so that the
integrand vanishes faster than $1/\tau_2$
in the $\tau\to i\infty$ limit, rendering the
integral finite.
Let us now introduce another
set of variables 
$(\wc\rho,\wc\sigma,\wc v)$ related
to $(\rho,\sigma, v)$ via the relations
\be\label{e6na}
   \wc\rho={1\over N}\, 
   {1\over 2v-\rho-\sigma}, \qquad
   \wc\sigma = N\,
   {v^2-\rho\sigma \over 2v-\rho-\sigma}, \qquad
    \wc v =
   {v-\rho \over 2v-\rho-\sigma}\, ,
\ee
or equivalently,
\be\label{e5n}
\rho 
   = {\wc \rho \wc\sigma - \wc v^2\over N\wc\rho}, 
   \qquad \sigma = {\wc\rho \wc \sigma - (\wc v - 1)^2\over  
   N\wc\rho}, \qquad
   v 
=   {\wc\rho \wc\sigma - \wc v^2 + \wc v\over N\wc\rho}\, .
\ee
We also define
\be\label{edefomega2}
\wc\Omega=\pmatrix{\wc\rho  & \wc v \cr \wc v  & \wc \sigma}\, .
\ee
Then we have the relations:
\ben \label{efirst}
&& (\det Im  \Omega)^{-1}\, |-m_1 \rho  +
m_2 + n_1 \sigma + n_2 (\sigma\rho -v^2) + j v |^2 \cr
&=& ( \det Im  \wt\Omega)^{-1}\,  |-m_1' \wt\rho  +
m_2' + n_1' \wt\sigma + n_2' (\wt\sigma\wt\rho -\wt v^2) + j' 
\wt v |^2 \nonumber \\  
&& m_1 n_1 + m_2 n_2 +{1\over 4} j^2 = 
m_1' n_1' + m_2' n_2' +{1\over 4} j^{\prime 2} 
\een
where
\be \label{esecond}
m_1'=m_2 N, \quad m_2'=n_1, \quad n_1'=n_2/N,
\quad n_2'=m_1 - n_1 - j, \quad
j'=-j - 2 n_1 \, .
\ee
Using these relations and
relabeling the indices $m_1$, $m_2$, $n_1$, $n_2$ in
eqs.\refb{defirsl}-\refb{enn6} one can easily prove the relations
\be\label{enn4}
\hI(\rho,\sigma,v)=\tI(\wc\rho,
\wc\sigma,\wc v)\, .
\ee
In the same way one can show that under a transformation of the
form
\be\label{etr1}
 \Omega\to (A \Omega+B)(C \Omega+D)^{-1}\, ,
\ee
$\hI(\rho,\sigma, v)$ remains invariant for the following
choices of the matrices $A$, $B$, $C$, $D$:
\ben \label{egroup}
   \pmatrix{A&B\cr C&D}  &=& \pmatrix{ a & 0 & b & 0 \cr
     0 & 1 & 0 & 0\cr c & 0 & d & 0\cr 0 & 0 & 0 & 1}\, ,
   \qquad ad-bc=1, \quad \hbox{$c=0$ mod $N$, \quad $a,d=1$
     mod $N$}
   \nonumber \\
   \pmatrix{A&B\cr C&D}  &=& 
   \pmatrix{0 & 1 & 0 & 0 \cr -1 & 0 & 0 & 0\cr
     0 & 0 & 0 & 1\cr 0 & 0 & -1 & 0}\, , \nonumber \\
   \pmatrix{A&B\cr C&D}  &=& 
   \pmatrix{ 1 & 0 & 0 & \mu \cr
     \lambda & 1 & \mu & 0\cr 0 & 0 & 1 & -\lambda\cr
     0 & 0 & 0 & 1}\, , \qquad \lambda, \mu \in \ZZZ.
   \nonumber \\
\een
The group of transformations generated by these matrices is a subgroup
of the Siegel modular group $Sp(2,\ZZZ)$; 
we shall denote this subgroup by
$\wh G$. Via 
eq.\refb{enn4} this also induces a
group of symmetry transformations of $\tI(\wrh,\ws,\wv)$;
we shall denote this group by
$\wt G$.

We can now use the modular property of $F^{(r,s)}(\tau,z)$
given in \refb{efrsmod} to
evaluate the
integrals $\tI$ and $\hI$. We refer the reader to
\cite{0602254} for detailed calculations in a specific
example and quote here the final results:
\be\label{enn7}
\tI(\rho ,\sigma,v ) = -2 \ln \left[
 ( \det\,{\rm Im}\Omega)^k \right] - 2\ln \wt\Phi (\rho ,\sigma, v )
- 2\ln \bar{\wt\Phi}(\rho ,\sigma, v ) +\hbox{constant}
\ee
and
\be\label{enn9}
\hI(\rho ,\sigma, v ) = -2 \ln \left[
 ( \det\,{\rm Im}\Omega)^k \right] - 2\ln \cp (\rho ,\sigma,v )
- 2\ln \bar{\cp}(\rho ,\sigma,v ) +\hbox{constant}
\ee
where
\be\label{ekvalue}
k={1\over 2}\, \sum_{s=0}^{N-1} \, c_0^{(0,s)}(0)\, ,
\ee
\bea{enn9a}
&& \wt \Phi(\rho ,\sigma, v ) =  
e^{2\pi i (\wt \alpha \rho + \wt \gamma\sigma 
+   v)} \nonumber \\
&& \qquad \times \prod_{b=0}^1\, 
 \prod_{r=0}^{N-1}
\prod_{k'\in \zzz+{r\over N},l\in\zzz,j\in 2\zzz+b
\atop k',l\ge 0, j<0 \, {\rm for}
\, k'=l=0}
\left( 1 - e^{2\pi i ( k'  \sigma   +  l  \rho +  j  v)
}\right)^{
\sum_{s=0}^{N-1} e^{-2\pi i sl/N } c^{(r,s)}_b(4k'l - j^2)} 
\nonumber \\
\eea
and
\bea{enn9c}
\cp(\rho,\sigma,v) &=&  e^{2\pi i \left(
\wh \alpha\rho+\wh \gamma\sigma+\wh\beta v\right) } \nonumber \\
&& \prod_{b=0}^1\,
\prod_{r=0}^{N-1}\,  \prod_{(k',l)\in \zzz,j\in 2\zzz+b\atop
k',l\ge 0, j<0 \, {\rm for}
\, k'=l=0}
\Big\{ 1 - e^{2\pi i r / N} \, e^{ 2\pi i ( k' \sigma + l \rho + j v) 
}
\Big\}^{ \sum_{s=0}^{N-1}
e^{-2\pi i rs/N}  c^{(0,s)}_b(4k'l - j^2)
}  \nonumber \\
\eea
\bea{enn9d}
&& \wt \alpha={1\over 24N} \, Q_{0,0} - {1\over 2N}
\, \sum_{s=1}^{N-1} Q_{0,s}\, {e^{-2\pi i s/N}\over
(1-e^{-2\pi i s/N})^2 } \, 
, \qquad 
\wt \gamma= {1\over 24N} \, Q_{0,0} = {1\over 24N}\, 
\chi(\MM), \nonumber \\
&&    \wh \alpha= \wh\beta = \wh \gamma 
= {1\over 24} 
\, Q_{0,0}= {1\over 24}\, \chi(\MM)\, .
\eea
The quantities $Q_{r,s}$ have been given in terms of the coefficients
$c^{(r,s)}(u)$ in \refb{eqrsrev}. In arriving at
\refb{enn9a},\refb{enn9c} one needs to use the relations
\refb{ecrev}, \refb{ecqrel}, \refb{eeuler} and also \refb{eck3}, \refb{ect4}.
Since $Nc_0^{(0,s)}(0)$ measures the number of harmonic
$(1,q)$-forms weighted by $(-1)^q \wt g^s$ and summed over $q$, 
the constant $k$ defined in \refb{ekvalue} 
has the interpretation of being
half the number of $\wt g$ invariant $(1,q)$ forms weighted by
$(-1)^{q+1}$ and summer over $q$. Since both for $K3$ and $T^4$
the only $\wt g$ invariant harmonic $(1,q)$ forms are (1,1) forms
(see eqs.\refb{eoneform}-\refb{efourform}), 
$k$ can be regarded as half the number of
$\wt g$ invariant harmonic (1,1) forms on $\MM$.

It  follows from  \refb{enn4},
\refb{enn7} and \refb{enn9} that
\be\label{enn11}
\wt\Phi(\wc\rho,\wc\sigma,\wc v)=-(i)^k\, C_1\, 
(2v -\rho-\sigma)^k\, \cp(\rho,\sigma,v)
\ee
where $C_1$ is a constant. The factor of $-(i)^k$
has been included to ensure that $C_1$ is real and positive.
To see this we can consider the case where $\rho$, $\sigma$
and $v$ are all purely imaginary, with $|v_2|<<
\rho_2,\sigma_2$. In this case $\wh\Phi(\rho,\sigma,v)$ 
defined in \refb{enn9c} is real
and positive. On the other hand from \refb{e6na} 
we get $\wrh$ and $\ws$
purely imaginary and $\wv$ real. Eq.\refb{enn9a}, together with the
relation $\sum_{s=0}^{N-1} c_1^{(0,s)}(-1)=2$ then tells us that
$\wt\Phi(\wrh,\ws,\wv)$ is real and negative. Hence in order to
satisfy \refb{enn11} we must have $C_1$ real and positive.
The magnitude of $C_1$ can be calculated by carefully evaluating
the constants in eqs.\refb{enn7}, \refb{enn9} 
but we shall not do it here.

Furthermore given the
invariance of  $\tI$ and $\hI$ under the groups  $\wt G$
and $\wh G$, it follows 
from
eqs.\refb{enn7}, \refb{enn9} that
\be \label{ewtphitrs}
\wt\Phi((A\Omega+B)(C\Omega+D)^{-1}) 
= \det(C\Omega+D)^k \, \wt\Phi(\Omega) \quad
\hbox{ for $\pmatrix{A & B\cr C & D}\in\wt G$}\, ,
\ee
and
\be \label{ewhphitrs}
\wh\Phi((A\Omega+B)(C\Omega+D)^{-1}) 
= \det(C\Omega+D)^k \, \wh\Phi(\Omega) \quad
\hbox{ for $\pmatrix{A & B\cr C & D}\in\wh G$}\, .
\ee
In principle these transformation laws could have arbitrary
phases but one can show that the phases are trivial.
Thus $\wt\Phi$ and $\cp$
transform as modular forms of weight $k$ under the groups
$\wt G$ and $\wh G$ respectively. As special cases of 
\refb{ewtphitrs}, we have
\be \label{especial}
\wt\Phi(\wt\rho+1,\wt\sigma,\wt v) = \wt\Phi(
\wt\rho,\wt\sigma+N,\wt v)
=\wt\Phi(\wt\rho,\wt\sigma,\wt v+1) = \wt\Phi(\wt\rho,\wt\sigma,
\wt v)\, .
\ee
This also follows from \refb{enn9a} 
and the integrality of $\wt\alpha$
and $N\wt\gamma$. Integrality of $N\wt\gamma$ is
manifest from \refb{enn9d} and that of $\wt\alpha$ has been argued
below \refb{ecvalue}.

{}From \refb{enn9a}, \refb{enn9c},  
\refb{ecqrel}, \refb{eqrsrev},
\refb{eck3} and \refb{ect4}
 it is easy to see that for small $v$
\be\label{enn12}
\wh\Phi(\rho, \sigma,  v)=   -4\pi^2 \, v^2 \, g( \rho)\, 
g( \sigma) + \OO(  v^4)
\ee
\be \label{ephilimit}
\wt\Phi(\rho,\sigma,v) = -4\pi^2\, 
v^2 \, f_1(N\rho) f_2(\sigma/N)  + \OO(  v^4)\, ,
\ee
where
\be\label{enn13}
g(\rho) = e^{2\pi i \wh\alpha\rho}\,
\prod_{n=1}^\infty \prod_{r=0}^{N-1}
 \left( 1 - e^{2\pi i r/N}
e^{2\pi i n\rho}\right)^{s_{r}}\, ,
\ee
\be \label{edeff1}
f_1(N\rho) = e^{2\pi i \wt\alpha \rho} \prod_{l=1}^\infty
(1 - e^{2\pi i l\rho})^{s_l}\, ,
\ee
\be \label{edeff2}
f_2(\sigma/N) = e^{2\pi i \wt\gamma \sigma} \, \prod_{r=0}^{N-1}
\prod_{k'\in\zzz +r/N\atop k'>0} (1 - e^{2\pi i k'\sigma})^{t_r}\, ,
\ee
\be\label{enn14}
s_{r} = {1\over N} \sum_{s'=0}^{N-1}
e^{-2\pi i r s'/N} \, Q_{0,s'}=
\sum_{s'=0}^{N-1}
e^{-2\pi i r s'/N} \, \left( c_0^{(0,s')}(0)+ 2 c_1^{(0,s')}(-1)
\right)\,,
\ee
\be \label{edeftr}
t_r = {1\over N} \sum_{s=0}^{N-1} Q_{r,s}
= \sum_{s=0}^{N-1} \left( c_0^{(r,s)}(0) + 2 c_1^{(r,s)}(-1)\right)\, .
\ee
Eq.\refb{enn11} then gives, for small $v$, \i.e.\ small 
$\wc\rho\wc\sigma-\wc v^2+\wc v$,
\be\label{enn15}
\wt\Phi(\wc\rho,\wc\sigma,\wc v)=-4\pi^2\,  C_1\, 
(2v -\rho-\sigma)^k\, v^2 g(\rho) \, g(\sigma) + \OO(v^4)\, .
\ee
Since $Q_{0,s}$ is the trace of $(-1)^p \wt g^s$ over the harmonic
$p$-forms of $\MM$, 
$s_r$ has the interpretation of being the number of harmonic
$p$-forms in $\MM$ with $\wt g$ eigenvalue $e^{2\pi i r/N}$
weighted by $(-1)^p$. Thus it is an integer.
On the other hand it follows from the definition
\refb{eqrs} of $Q_{r,s}$ that
$b_r$ is the number of $\wt g$ invariant, $\wt g^r$ twisted
state. Thus it is also an integer. 

$(f_1(N\wrh))^{-1}$
computed from  \refb{edeff1} coincides with the partition function
\refb{e8xxx} of a single Kaluza-Klein monopole in the first
description. Since this corresponds to an elementary twisted
sector string in the second description, we see that $(f_1(\tau))^{-1}$
can be interpreted as the partition function of purely
electrically charged twisted sector states in the second description.
On the other hand $(f_2(\ws/N))^{-1}$ coincides with the
$l=0$ term in \refb{enewexp} with $\wt v=0$. 
This leads to the conclusion
that $(f_2(\ws/N))^{-1}$ can be interpreted as the partition
function of the D1-D5 system in the absence of Kaluza-Klein
monopole, with arbitrary angular momentum,
zero momentum along $\wt S^1$
and zero
momentum along $S^1$.\footnote{Note that in the absence of
the Kaluza-Klein monopole the momentum along $\wt S^1$
can no longer be identified with angular momentum.} 
Since in the second description this
gets mapped to a purely magnetically charged half-BPS
state, we conclude that $(f_2(\tau))^{-1}$ describes the
partition function of purely magnetically charged half-BPS
states in the second description.

Using 
\ben \label{especsdual}
&& \pmatrix{A & B\cr C & D}=\pmatrix{a & 0 & b & 0\cr
0 & 1 & 0 & 0\cr c & 0 & d & 0\cr 0 & 0 & 0 & 1}\subset \wh G,
\nonumber \\
&&
\quad ad-bc=1, \quad a,b,c,d\in \ZZZ, \quad a,d = \hbox{1 mod $N$},
\quad c = \hbox{0 mod $N$},
\een
in \refb{ewhphitrs}, and eq.\refb{enn12}, one can show that 
\be \label{egrhotrs}
g((a\rho+b)(c\rho +d)^{-1}) = (c\rho + d)^{k+2} g(\rho)\, .
\ee
Thus $g(\rho)$ transforms as a modular form of weight $(k+2)$
under $\Gamma_1(N)$. The behaviour of $g(\rho)$ for large
$\rho_2$ is governed by the
constant $\wh\alpha$ defined in
\refb{enn9d}. 

For $\MM=K3$ and prime values of $N$ ($N=$ 1, 2, 3, 5, 7) we can
use \refb{fifthn} to find the explicit expressions for
$g(\tau)$ and $k$\cite{0602254}:
\be \label{egtauh}
g(\tau) = \eta(\tau)^{k+2}\, \eta(N\tau)^{k+2}\, ,
\ee
and
\be \label{ekssh}
k = {24\over N+1}-2\, .
\ee
On the other hand for $\MM=T^4$ and $N=2,3$ we get, using
\refb{valindalt}\cite{0607155},
\be \label{egtauii}
g(\tau) = \eta(\tau)^{2N(k+2)/(N-1)}
\eta(N\tau)^{-2(k+2)/(N-1)}\, ,
\ee
and
\be \label{ekssii}
k = {12\over N+1}-2\, .
\ee

For $\MM=K3$ and $N=1$, the function $\wt\Phi$ constructed
in this appendix is the well known weight 10 cusp form of the
genus two Siegel modular 
group\cite{igusa1,igusa2,borcherds,9504006}.
For $\MM=K3$ and $N=2,3$ the function $\wt\Phi$ was
found in \cite{ibu1,ibu2,ibu3}. A general discussion on
construction of Siegel modular forms can be found in
\cite{eichler}. Different ways of constructing the same functions
$\wt\Phi$ can be found in \cite{skor,rama,0510147,0603066}.

\sectiono{Zeroes and Poles of $\wt\Phi$} \label{szero}

In this appendix we shalll determine the zeroes 
and poles of the function
$\wt\Phi(\wrh,\ws,\wv)$ introduced in appendix \ref{ssiegel}.
Via eq.\refb{enn11} this also determines the zeroes and poles of
the function $\wh\Phi(\rho,\sigma, v)$. The zeroes of $\wt\Phi$
found in \refb{ephilimit} and \refb{enn15} will be special cases
of the general set of zeroes we shall find.

Using \refb{rwthrint}, \refb{defirsl}, \refb{enn7} we see that 
the Siegel modular form $\wt\Phi(\wrh ,\ws,\wv )$ 
satisfies the relation: 
\bea{eu1}
&& -2\ln \wt\Phi(\wt\rho,\wt\sigma,\wt v) 
- 2\ln \bar{\wt\Phi}(\wt\rho,\wt\sigma,\wt v) -2 \, k\, \ln \det Im
\wt\Omega + \hbox{constant}
\nonumber \\
 &=& 
 \int_\FF
\frac{d^2\tau}{\tau_2} \Bigg[
\sum_{r,s=0}^{N-1}\sum_{b=0}^1\,
\sum_{{m_1, m_2, n_2\in \zzz, \atop
n_1\in \zzz+{r\over N}, j\in 2\zzz + b}}
\exp\left[ 2\pi i \tau( m_1 n_1 + m_2 n_2 +\frac{j^2}{4} ) \right]
\times \cr
&\;& \;\;\; \exp \left(\frac{-\pi \tau_2}{\wt Y} \left|
n_2 ( \ws \wrh  -\wv ^2) + j\wv  + n_1 \ws  -m_1  \wrh 
+ m_2 \right|^2 \right)\,
e^{2\pi i m_1 s / N} \, h^{(r,s)}_b(\tau) \nonumber \\
&& \qquad \qquad - 
\sum_{s=0}^{N-1}\, c^{(0,s)}_0(0)
\Bigg]\, , \nonumber \\
\een
where
\be\label{eu1a}
\wt\Omega=\pmatrix{ \ws  & \wv \cr \wv  & \wrh }\, , \qquad 
\wt Y = \det
Im\wt\Omega\, .
\ee
Eq.\refb{eu1} shows that the zeroes and poles of $\wt\Phi$ appear only
when the $\tau$ integral on the right hand side of this equation diverges 
from the region near $\tau=i\infty$. 
Now, if we consider a term proportional to $e^{2\pi i n\tau}$
in the expansion of $h_b^{(r,s)}$, then the $\tau_1$ dependent
term in the integrand is of the form
$\exp\left(2\pi i \tau_1 (n + m_1 n_1 + m_2 n_2 
+{1\over 4}j^2)\right)$. Since
for large $\tau_2$ 
the $\tau_1$ integral runs from $-{1\over 2}$ to 
${1\over 2}$, it gives a
non-vanishing answer only if
\be\label{eu8}
n + m_1 n_1 + m_2 n_2 +\frac{j^2}{4} = 0\, .
\ee
Thus after performing the $\tau_1$ integral, 
the only $\tau_2$ dependence of the integrand
in the large $\tau_2$ region
comes from the
\be\label{eu8a}
\exp\left[- \frac{\pi \tau_2}{\wt Y} \left|
n_2 ( \ws \wrh  -\wv ^2) + j\wv  + n_1 \ws  
-\wrh m_1 + m_2 \right|^2  \right]
\ee
factor.  As long as the coefficient of $\tau_2$ in the exponent is
non-zero the integrand is exponentially suppressed for large
$\tau_2$ and as a result the integral is convergent.
Thus the only way the
integral can diverge from the large $\tau_2$ region is if this vanishes:
 \be\label{eu7}
n_2 ( \ws \wrh  -\wv ^2) + j\wv  + n_1 \ws  -\wrh m_1 + m_2 =0\, , 
\ee
for some $m_1$, $m_2$, $n_1$, $n_2$, $j$ appearing in the sum in 
\refb{eu1}.  

Now  we have the identity
\be\label{eu7a}
 m_1 n_1 + m_2 n_2 +\frac{j^2}{4} ={1\over 2} (p_L^2 - p_R^2)\, ,
 \ee
 where
 \bea{eu7aa}
 p_R^2 &=& {1\over 2 \wt Y} \left|
n_2 ( \ws \wrh  -\wv ^2) + j\wv  + n_1 \ws  -\wrh m_1 + m_2 \right|^2, \nonumber \\
p_L^2 &=& {1\over 2\wt Y} \left\{ m_2 
+ n_2(\ws_2\wrh_2+\ws_1\wrh_1-\wv_1^2-\wv_2^2) - m_1 \wrh_1 + n_1 \ws_1 
+ j \wv_1\right\}^2 \nonumber \\
&& + {1\over 2 \wt Y} \left\{ n_2 \left(\ws_1\wrh_1 - \ws_1 \wrh_2 
+ 2 \wv_1\wv_2 - {2\wv_2^2\wrh_1\over
\wrh_2}\right) + m_1\wrh_2
+ n_1 \left(\ws_2 - 
{2 \wv_2^2\over \wrh_2}\right) - j\wv_2\right\}^2\nonumber \\
&& + 2 \left\{ {j\over 2} + n_1 {\wv_2\over \wrh_2} - n_2\wv_1
+ n_2 {\wv_2 \wrh_1\over \wrh_2}\right\}^2\, .
\een
 Since $p_L^2$ is positive semi-definite,  and
since $p_R^2$ vanishes when \refb{eu7} holds,
\refb{eu7a}
 shows that
we must have
 \be\label{eu7b}
 m_1 n_1 + m_2 n_2 +\frac{j^2}{4}  \ge 0\, .
 \ee
 Furthermore the equality sign holds only when $p_L^2$ also
 vanishes. This requires $m_1=m_2=n_1
 =n_2=j=0$. The corresponding divergence is present for all
 $\ws $, $\wrh $, $\wv $ and is removed by the subtraction 
 term proportional to $\sum_s c_0^{(0,s)}(0)$ in \refb{eu1}. 
 Thus the
 divergences which depend on $\ws $, $\wrh $, $\wv $ come from those
 values of $m_i$, $n_i$, $j$ which satisfy \refb{eu7}
 and for which
 \be\label{eu7ba}
 m_1 n_1 + m_2 n_2 +\frac{j^2}{4}  > 0\, .
 \ee
 This, together with
 eq.\refb{eu8}, now show that we must have
 \be\label{eu7c}
 n < 0\, .
 \ee
 In other words the only terms in the expansion of $h_b^{(r,s)}$ 
 responsible for a divergent contribution to the integral 
 \refb{eu1} are the ones involving negative powers of 
 $e^{2\pi i\tau}$. 
 Eqs.\refb{ehbexp}, \refb{evanish}
 now imply that the divergent contribution
 comes from terms proportional to 
$c_1^{(r,s)}(-1)$ for all $N$,
and $c_1^{(r,s)}(-1+{4p\over N})$ ($p\in\ZZZ$, ${N\over 4}
> p\ge 1$)
for
$N\ge 5$. The corresponding values of $n$ are $-{1\over 4}$ and
$-{1\over 4}+{p\over N}$ respectively.

First consider the contribution from the $c_1^{(r,s)}(-1)$ term.
Putting $n=-1/4$ in \refb{eu8} we get
\be\label{ev2}
m_1 n_1 + m_2 n_2 +\frac{j^2}{4} = {1\over 4}\, .
\ee
By estimating the $\tau_2$ integral in the right hand side
of \refb{eu1}
for $n_2 ( \ws \wrh  -\wv ^2) + 
j\wv  + n_1 \ws  -\wrh m_1 + m_2\simeq 0$, one easily finds that
the 
divergent contribution is given by
\bea{ev2a}
-2\sum_{s=0}^{N-1}
e^{2\pi i m_1 s / N}\, c_1^{(r,s)}(-1)\, 
\ln \left|
n_2 ( \ws \wrh  -\wv ^2) + j\wv  + n_1 \ws  -\wrh m_1 + m_2 \right|^2\, ,
\nonumber \\
\quad \hbox{$r=n_1N$ mod $N$, $j=1$ mod 2}\, , \quad
\een
where we have included a factor of 2 due to the fact that  
the lattice vectors
$(\vec m, \vec n, j)$ and $(-\vec m, -\vec n, -j)$
give identical divergent
contribution.
Comparing this with the left-hand side of \refb{eu1} we see that
near this region $\wt\Phi$ behaves as
\ben\label{ev3}
&& \wt\Phi \sim \left( n_2 ( \ws \wrh  -\wv ^2) + j\wv  + 
n_1 \ws  -\wrh m_1 + m_2
\right)^{\sum_{s=0}^{N-1}
e^{2\pi i m_1 s / N}\, c_1^{(r,s)}(-1)}\, , \cr
&& m_1, m_2, n_2 \in \ZZZ, \quad j\in 2\ZZZ+1, \quad
n_1\in \ZZZ+{r\over N}, \quad 
m_1 n_1 + m_2 n_2 +\frac{j^2}{4} = {1\over 4}\, . \nonumber \\
\een

For $N\ge 5$ we also have divergent contribution to \refb{eu1}
from the
$c_1^{(r,s)}(-1+{4p\over N})$ term. 
In this case \refb{eu8}   gives
\be\label{ev2x}
m_1 n_1 + m_2 n_2 +\frac{j^2}{4} = {1\over 4}-{p\over N}\, .
\ee
The divergent contribution takes the form
\bea{ev2aa}
-2\sum_{s=0}^{N-1}
e^{2\pi i m_1 s / N}\, c_1^{(r,s)}(-1+{4p\over N})\, 
\ln \left|
n_2 ( \ws \wrh  -\wv ^2) + j\wv  + n_1 \ws  -\wrh m_1 + m_2 \right|^2\, ,
\nonumber \\
\quad \hbox{$r=n_1N$ mod $N$, $j=1$ mod 2}\, . 
\een
Thus 
$\wt\Phi$ behaves as
\ben\label{ev3aa}
&& \wt\Phi \sim \left( n_2 ( \ws \wrh  -\wv ^2) + j\wv  + n_1 \ws  
-\wrh m_1 + m_2
\right)^{\sum_{s=0}^{N-1}
e^{2\pi i m_1 s / N}\, c_1^{(r,s)}(-1+{4p\over N})}\, , \cr
&& m_1, m_2, n_2 \in \ZZZ, \quad j\in 2\ZZZ+1, \quad
n_1\in \ZZZ+{r\over N}, \quad 
m_1 n_1 + m_2 n_2 +\frac{j^2}{4} = {1\over 4}-{p\over N}\, .
\nonumber \\
\een

{}From \refb{esi4aint}, \refb{enewint} it follows that
the exponent in
\refb{ev3} has the interpretation as the number of $\wt g^r$ twisted
states with $\wt g$ eigenvalue $e^{-2\pi i m_1/N}$, $F_L=1$ (or $F_L
=-1$) and $L_0=\bar L_0=0$, weighted by $(-1)^{F_L+F_R}$.
On the other hand the exponent in \refb{ev3aa} 
has the interpretation as the number of $\wt g^r$ twisted
states with $\wt g$ eigenvalue $e^{-2\pi i m_1/N}$, $F_L=1$ (or $F_L
=-1$), $L_0=p/N$ and $\bar L_0=0$, weighted by $(-1)^{F_L+F_R}$.
Thus both numbers are integers.

For the analysis in \S\ref{s3.3} we need to know which exponents are
positive, corresponding to the zeroes of $\wt\Phi$, and which
exponents are negative, corresponding to the poles of $\wt\Phi$.
First consider the case $r=0$, \i.e.\ $n_1\in \ZZZ$. In this case using
eqs.\refb{eck3} we see that the exponent in \refb{ev3} for $\MM=K3$
is given by
\be \label{eexplicit1}
\sum_{s=0}^{N-1}
e^{2\pi i m_1 s / N}\, c_1^{(0,s)}(-1) = \cases{
\hbox{2 for $m_1\in N\ZZZ$} \cr \hbox{0 otherwise}}\, .
\ee
On the other hand using eq.\refb{ect4} we see that for $\MM=T^4$
the exponent in \refb{ev3} for $\MM=T^4$ is given by
\be \label{eexplicit2}
\sum_{s=0}^{N-1}
e^{2\pi i m_1 s / N}\, c_1^{(0,s)}(-1) = \cases{
\hbox{2 for $m_1\in N\ZZZ$} \cr \hbox{$-1$ for
$m_1\in N\ZZZ\pm 1$} \cr \hbox{0 otherwise}}\, .
\ee
In the special case of $N=2$, the sets $N\ZZZ\pm 1$ coincide, and
the exponent becomes equal to $-2$ instead of $-1$ for
$m_1\in N\ZZZ\pm 1$.

Since for $r=0$, $m_1,n_1,m_2,n_2\in \ZZZ$, and $j\in 2\ZZZ+1$,
the only way to satisfy \refb{ev2x} is to take $p=0$. Thus there are
no zeroes or poles of the type given in \refb{ev3aa}
with $p\ne 0$.

The zeroes and poles originating in the
$r\ne 0$ mod $N$ sector are more difficult
to evaluate since these require counting twisted sector states with
specific $\wt g$ quantum numbers. 
However one can extract some general information by noting that
since the coefficients $c_b^{(r,s)}(4n)$ do not depend on the shape
and size of $\MM$, we can compute them by taking the size of
$\MM$ to be large so that near any fixed point of $\wt g$ the orbifold
can be regarded as that of ${\bf R}^4$. 
Thus the contribution from a given
twisted sector associated with a given fixed point can be computed
in a free super-conformal field theory. Locally the action of $\wt g$
may be represented as rotation by $2\pi/N$ in one two dimensional
plane and rotation by $-2\pi/N$ in an orthogonal two dimensional plane.
Thus in a twisted RR sector, 
all the bosons and fermions will be twisted
and there are no zero modes. As a result the ground state with
$L_0=\bar L_0=0$ is unique,
carrying $F_L=F_R=0$. Even after we apply left-moving oscillators
to create excited BPS states, these states
will continue to have $F_R=0$. Now
since the computation of the exponents in \refb{ev3}, \refb{ev3aa}
involves counting BPS
states with $F_L=1$ (or $-1$), the weight factor
$(-1)^{F_L+F_R}$ is given by $-(-1)^{F_R}=-1$ for each of these
states. Thus the
exponents in \refb{ev3} or \refb{ev3aa} are always negative for
$r\ne 0$ mod $N$. These correspond to poles of $\wt \Phi$ rather
than zeroes.

The net conclusion of this analysis is that both for $\MM=K3$ and
$\MM=T^4$, the only zeroes of $\wt\Phi(\wrh,\ws,\wv)$ are of the
form:
\ben\label{eonly}
&& \wt\Phi \sim \left( n_2 ( \ws \wrh  -\wv ^2) + j\wv  + 
n_1 \ws  -\wrh m_1 + m_2
\right)^2\, , \cr
&& m_1\in N\ZZZ, \quad 
n_1, m_2, n_2 \in \ZZZ, \quad j\in 2\ZZZ+1, \quad
m_1 n_1 + m_2 n_2 +\frac{j^2}{4} = {1\over 4}\, . \nonumber \\
\een
The rest are poles.

For $\MM=K3$ and prime values of $N$ ($N=1,2,3,5,7$) we can
use eq.\refb{fifthn} to explicitly compute the exponents appearing in
\refb{ev3} and \refb{ev3aa}. We get\cite{0605210}
\be\label{efirstexp}
\sum_{s=0}^{N-1}
e^{2\pi i m_1 s / N}\, c_1^{(r,s)}(-1) = 
\cases{2 \quad \hbox{for $r=0$ mod $N$, 
$m_1=0$ mod $N$} \cr 
0 \quad \hbox{otherwise}}\, ,
\ee
\be\label{eseconsexp}
 \sum_{s=0}^{N-1}
e^{2\pi i m_1 s / N}\, c_1^{(r,s)}(-1+{4\over N}) 
= \cases{-48/(N^2-1) \quad \hbox{for $m_1 r= -1$ 
mod $N$} \cr 0 \quad \hbox{otherwise} }\, .  
\ee
For $\MM=T^4$ and $N=2,3$ we can 
use eq.\refb{valindalt} to explicitly compute 
the exponents appearing in
\refb{ev3} and \refb{ev3aa}. The result is\cite{0607155}
\be\label{ethirdexp}
\sum_{s=0}^{N-1}
e^{2\pi i m_1 s / N}\, c_1^{(r,s)}(-1) = 
\cases{2 \quad \hbox{for $r=0$ mod $N$, $m_1=0$ mod $N$} \cr 
-{2\over N-1} \quad \hbox{for $r=0$ mod $N$, 
$m_1=\pm 1$ mod $N$} \cr
0 \quad \hbox{otherwise}}\, .
\ee

\sectiono{The 
 Case of Multiple D5 branes} 
\label{ssym}

In this appendix we extend the counting of states associated with 
the relative motion of the D1-D5 system to the case when the
number of D5-branes is $Q_5 \geq 1$. We shall restrict our analysis
to the case when $\MM=K3$ and follow the analysis given in
\cite{0605210}.

We shall choose the world-volume space coordinate $\sigma$
along each of the D1-branes to coincide with the coordinate
along $S^1$. Due to the $\ZZZ_N$ orbifolding each D1-brane
satisfies a twisted boundary condition, -- under 
$\sigma\to\sigma+2\pi$ its location along $K3$ must get
transformed by $\wt g$. We shall first ignore the effect of this
twist and pretend that the D1-brane
satisfies periodic boundary condition under $\sigma\to
\sigma+2\pi$, and later take into account the effect of the
twist.

The  dynamics of the relative motion
of
$Q_5$ D5-branes wrapped  on $K3\times S^1$ and $Q_1$ 
D1-branes wrapped on $S^1$ is captured by the ${\cal N} =(4,4)$
superconformal $\sigma$-model with the symmetric product of 
$W=Q_5(Q_1 -Q_5) +1$ copies of $K3$ as the target 
space as long as $Q_5$ and $Q_1$ do not have a
common factor\cite{9512078}.   We shall denote this target space by 
$S^W K3\equiv
(K3)^W/S_W$, where $S_W$ refers to the permutation group of $W$
elements. The world-sheet coordinate $\sigma$ of this conformal
field theory is identified with the coordinate along $S^1$.
We shall first review various aspects of the superconformal field
theory with target space $ (K3)^W/S_W$\cite{9608096}, 
and then discuss the effect
of the $\ZZZ_N$ twist that is required to describe a
D1-D5-brane configuration on the CHL orbifold.

 Let $g$ be an element of  $S_W$ and $[g]$ denote the conjugacy class
 of $g$. 
Then the Hilbert space of the SCFT with target space $(K3)^W/S_W$
decomposes into a direct sum of twisted sectors labelled 
by the conjugacy classes of $S_W$:
\be\label{dechilb}
{\cal H}= \oplus_{[g]} {\cal H}_{g}^{({\cal C}_g)} 
\ee
where $\CC_g$ denotes the centralizer of $[g]$ and
${\cal H}_{g}^{({\cal C}_{g}) }$ refers to the
Hilbert space in the $g$ twisted sector projected by $\CC_g$.
The conjugacy classes of $S_W$ may be labelled as
\be\label{decompg}
[g] = (1)^{P_1} (2)^{P_2} \cdots (s)^{P_s} 
\ee
where $(w)$ denotes cyclic permutation of $w$ elements and
$P_w$ is the number of copies of $(w)$ in $g$. Thus these
conjugacy classes 
are characterized by partitions $P_w$ of $W$ such that
\be\label{consconj}
\sum_w wP_w = W\, .
\ee
The 
centralizer $\CC_g$ of the conjugacy class $[g]$ given in \refb{decompg}
is given by
\be\label{central}
{\cal C}_{g} =
S_{P_1} \times
 (S_{P_2} \times \ZZZ_2^{P_2})   \times
\cdots\times
(S_{P_s} \times \ZZZ_s^{P_s})  \, .
\ee
Let us denote by $\HH_w$ 
the Hilbert space of states twisted by the
generator $\omega$ of the $\ZZZ_w$ group of
cyclic permutation of $w$ elements, and projected 
by the same $\ZZZ_w$ group. Then \refb{central}
shows that for the conjugacy class $[g]$
given in \refb{decompg}
\be\label{symmtenpro}
{\cal H}_{g}^{({\cal C}_{g}) } 
=
\otimes_{w>0}
S^{P_w}{\cal H}_{w} 
\ee

Consider first the Hilbert space ${\cal H}_{w}$.  
This twisted 
sector is represented by the Hilbert space of the sigma model
of $w$ coordinate fields $X_i(\sigma) \in K3$ with the 
cyclic boundary condition
\be\label{cycbc}
X_i(\sigma + 2\pi) = \omega X_i(\sigma) = X_{i+1}(\sigma), \quad
i\in (1, \ldots , w)\, ,
\ee
where $\omega$ acts by $\omega: X_i \rightarrow X_{i+1}$. 
Therefore the $w$ coordinate fields can be glued together as a single
field  but in the interval $0\leq \sigma\leq 2\pi w$, moving
in the target space K3. Thus we now have a string of length
$2\pi w$, -- commonly known as the long string, -- moving in K3.
Whereas for $Q_5=1$ the
quantum number $w$ can be identified with the winding charge of
the D-string, this is not so for $Q_5>1$. Thus we should not regard
the long string as a D-string, -- rather it provides 
some effective description
of the dynamics.  

Once we know the spectrum of $\HH_w$, -- which can be found
from the spectrum of an SCFT with target space $K3$ after a
rescaling of the $L_0$ and $\bar L_0$ eigenvalues by $1/w$ to take
into account the effect of the length of the string, --
the full spectrum of the CFT of the D1-D5 system is obtained by
taking the direct product of the spectrum of $\HH_w$'s
and then carrying out appropriate symmetrization described in
\refb{symmtenpro}.

We now turn to the effect of the $\ZZZ_N$ twisted boundary
condition that is required
in order to get a state of the D1-D5 system 
in the $\ZZZ_N$ CHL model.
For this we need to change \refb{cycbc} to $X_i(\sigma+2\pi)
= \wt g X_{i+1}(\sigma)$. 
Thus effectively we modify the generator of $\ZZZ_w$ by an additional
$\wt g$ transformation leaving unchanged the rest of the analysis.
Since the long string has length $2\pi w$, 
as we
go once around the long string the boundary condition is twisted
by $\wt g^w=\wt g^r$ where $r=w$ mod $N$. 
Let us denote
by $\HH'_w$ the Hilbert space of states of the
long string with $\wt g^r$ twisted boundary condition, and
projected by the new $\ZZZ_w$ group. Then the
full Hilbert space of the D1-D5 system will be obtained simply
by replacing $\HH_w$ by $\HH_w'$ in \refb{symmtenpro}:
\be\label{etar10}
\otimes_{w>0}
S^{P_w}{\cal H}'_{w} \, .
\ee
Clearly $\HH'_w$ can be identified with the Hilbert space of
$\wt g^r$ twisted states in the SCFT described in 
appendix \ref{s0}. 
Since the string has length $2\pi w$, a physical momentum $-l/N$ 
along $S^1$
would correspond to $L_0-\bar L_0$ 
eigenvalue of $lw/N$ in this SCFT.  Since supersymmetry requires
$\bar L_0$ to vanish, we have $L_0=lw/N$.
Let $F_L$ denote the left-moving world-sheet fermion number
of this SCFT.
By the standard argument, the presence of Kaluza-Klein
monopole background converts $F_L$ eigenvalues into
momenta along $\wt S^1$.
Since the projection operator for $\ZZZ_N$ invariant states with
physical momentum $-l/N$ along $S^1$ is
\be\label{etar2}
{1\over N} \sum_s e^{-2\pi i l s/N} \wt g^s\, ,
\ee
the total number of bosonic minus fermionic states
in the single
long string Hilbert space, carrying
momentum $-l/N$ along $S^1$  
and  momentum $j$ along $\wt S^1$ is given by:
\be\label{etar3pre}
n(w,l,j)={1\over N} \sum_s e^{-2\pi i l s/N} 
Tr_{RR;\wt g^r} (\wt g^s (-1)^{F_L+F_R}
\delta_{NL_0,lw} \delta_{F_L, j}) \, .
\ee
Using \refb{esi5} this may be written as
\be \label{etar3}
n(w,l,j)= \sum_s e^{-2\pi i l s/N} 
c_b^{(r,s)}(4lw/N - j^2), \quad \hbox{$b=j$ mod 2}\, .
\ee

According to
\eq{etar10}  the next step is the evaluation
of the partition function for the symmetrized 
tensor products of 
the Hilbert spaces 
${\cal H}_w'$. For this we use the
following formula from \cite{9608096}. 
If $d_{sym}(P_w,w,L,J')$ denotes the number of bosonic minus
fermionic states in $S^{P_w} {\cal H}_w'$ carrying total momentum
$-L/N$ along $S^1$ and total  momentum $J'$ along $\wt S^1$, then
\be\label{etar4}
\sum_{P_w=0}^\infty \sum_{L,J'}
d_{sym}(P_w,
w,L,J') e^{2\pi i L\wt \rho
+2\pi i J'\wt v + 2\pi i \ws P_w/N}
= \prod_{l, j\in \zzz\atop l\ge 0} \left( 1 - e^{2\pi i \ws/N 
+ 2\pi i l\wt\rho
+ 2\pi i j \wt v}\right)^{-n(w,l,j)}\, .
\ee

Using the  identity in \eq{etar4} we can evaluate the generating function
for the bosonic minus fermionic states for the relative
dynamics of the D1-D5 system. 
Eq.\refb{etar10} shows that all we
need to do is to take the product over $w$ of the right hand side of
\refb{etar4}. More specifically, if $d_{D1D5}(W,L,J')$ 
denotes the total number
of bosonic minus fermionic states carrying total string length $2\pi
W =2\pi \sum_{w>0} w P_w$
(counting a single long 
string with quantum number $w$ to have length
$2\pi w$), total momentum $-L/N$ along $S^1$ and total
momentum $J'$ along $\wt S^1$, then we have
\bea{d1d5genfn1}
&\,&\sum_{W,L, J'} d_{D1D5}( W, L,J') 
e^{2\pi i (\wt\rho L + \wt \sigma W/N +
\wt v J' )}  \nonumber \\
&\,& = \prod_{w\in \zzz\atop w>0}
\prod_{l,j\in \zzz\atop l\ge 0} 
\left( 1 - e^{2\pi i \wt\sigma w /N
+ 2\pi i l\wt\rho
+ 2\pi i j \wt v}\right)^{-n(w,l,j)} \nonumber \\
&\,& =  
\prod_{r=0}^{N-1}\prod_{b=0}^1
\prod_{ k'\in\zzz+ \frac{r}{N}, l\in\zzz, j\in2\zzz+b\atop
k'>0, l\ge 0} 
( 1- e^{2\pi i (k'\wt\sigma + l \wt \rho + j\wt v ) })
^{- \sum_{s=0}^{N-1} 
e^{-2\pi i s l/N} c_b^{(r,s)}( 4 k' l - j^2) } \, .
\nonumber \\
\eea
In arriving at the last expression in
\refb{d1d5genfn1} we have defined $k'=w/N$.
Physically $d_{D1D5}(W,L,J')$ counts the number of states
with fixed number $Q_1$ and $Q_5$ of D1 and D5-branes,
and fixed momenta $-L/N$ and $J'$ along $S^1$ and $\wt S^1$,
with $W$ identified with the number
$Q_5(Q_1-Q_5) +1$. Eq.\refb{d1d5genfn1} replaces
\refb{enewexp} for general $Q_5$, and reduces to \refb{enewexp}
for $Q_5=1$. Subsequent analysis leading to the full partition function
of the system proceeds in a manner identical to the one described
in \S\ref{sfull} and the final result for $d(\vec Q, \vec P)$ has the
form of \refb{egg1} with $P^2$ given by $2(W-1)=2Q_5(Q_1-Q_5)$.

\sectiono{Riemann Normal Coordinates and Duality Invariant Statistical
Entropy Function} \label{s3xyz}

In section \ref{s3.3} we considered 
$\htau=\vtau -\vtau_B$ for some fixed base point $\vtau_B$
as the fundamental field
in defining $W_B(\vtau_B,\vec J)$ and $\Gamma_B(\vtau_B, \vc)$. In this
appendix we shall try to generalize this by treating 
\be\label{ef1}
\vxi = \vec g(\htau)
\ee
as a fundamental field. Here $\vec g(\htau)$ is an arbitrary function
of $\htau$ with a Taylor series expansion starting with the linear terms
(\i.e. $\vec g(\vec 0)=\vec 0$). In this case the generating function
of $\vxi$ correlation functions will be given by
\be\label{ef2}
e^{\wt W_B(\vtau_B,\vec J)} = \int{d^2\eta\over (\tau_{B2}+\eta_2)^2} 
\, e^{-F(\vtau_B+\htau) + \vec J\cdot
\vec g(\htau)}\, .
\ee
As before $\wt W_B(\vtau_B,\vec 0)=S_{stat}$.
The corresponding effective action is defined via the equations
\be\label{ef3}
\psi_i = {\p \wt W_B(\vtau_B,\vec J)\over \p J_i}\, ,
\qquad \wt \Gamma_B(\vtau_B, \vpsi) =  \vec J\cdot \vpsi - 
\wt W_B(\vtau_B,\vec J)  \, .
\ee
{}From this it follows that
\be\label{ef3a}
J_i = {\p \wt \Gamma_B(\vtau_B, \vpsi)\over \p \psi_i}\, .
\ee
Now suppose $\vtau^{(0)}$ is a specific value of $\vtau_B$ for which
\be\label{ef4}
{\p \wt
\Gamma_B(\vtau^{(0)}, \vpsi)\over \p\psi_i}\bigg|_{\vpsi =0}
= 0 \quad  \hbox{\i.e.} \quad {\p \wt W_B(\vtau^{(0)},\vec J)
\over \p J_i}\bigg|_{\vj=0}=0\, .
\ee
In this case we have $\vec J=0$ for $\vpsi=0$, and hence
\be\label{ef5}
\wt\Gamma_B(\vtau^{(0)}, \vec 0) = - \wt W_B(\vtau^{(0)},\vec 0)
= - S_{stat}\, .
\ee

We shall now show that $\wt\Gamma_B(\vec \tau_B, \vec 0)$, regarded
as a function of $\vec\tau_B$, has an 
extremum at $\vtau_B=\vtau^{(0)}$. From \refb{ef3}, \refb{ef3a}
we see
that
\be\label{ef6}
\wt\Gamma_B(\vtau^{(0)}+\vec\epsilon, \vec 0)
= - 
\wt W_B\left.\left(\vtau^{(0)}+\vec\epsilon,
\vec J=\p \wt\Gamma_B(\vtau^{(0)}+
\vec\epsilon, \vec\psi) / \p\vec \psi\right)
\right|_{\vpsi=\vec 0} \, .
\ee
Now
\be\label{ef7}
e^{\wt W_B(\vtau_B +\vec\epsilon,
\vec J)} = \int{d^2\eta\over (\tau_{B2}+\epsilon_2+
\eta_2)^2} 
\, e^{-F(\vtau_B+\vec\epsilon+\htau) + \vec J\cdot
\vec g(\htau)}
= \int{d^2\eta\over (\tau_{B2}+\eta_2)^2} 
\, e^{-F(\vtau_B+\htau) + \vec J\cdot
\vec g(\htau-\vec\epsilon)}\, ,
\ee
where in the second step we have made a change of variables
$\vec\eta\to \vec\eta-\vec\epsilon$. Since $g(\htau-\vec\epsilon)
= g(\htau) + O(\vec\epsilon)$, this shows that
\be\label{ef8}
 \wt W_B(\vtau_B +\vec\epsilon,
\vec J) = \wt W_B(\vtau_B ,
\vec J) + O(\eps_i J_k)\, .
\ee
Using this information in \refb{ef6} we get
\be\label{ef9}
\wt\Gamma_B(\vtau^{(0)}+\vec\epsilon, \vec 0)
= - 
\wt W_B(\vtau^{(0)},
\vec J=\p \wt\Gamma_B(\vtau^{(0)}+
\vec\epsilon, \vec\psi) / \p\vec \psi)|_{\vpsi=\vec 0}
+ O\left(\epsilon_i {\p \wt\Gamma_B(\vtau^{(0)}+
\vec\epsilon, \vec\psi)\over \p \psi_j}\bigg|_{\vpsi=0}\right)
\, .
\ee
Eq.\refb{ef4} now
shows that the second term on the right hand side
of this equation is of order
$\eps^2$, and $\vec J$ appearing in the argument of the first term
is of order $\eps$. Expanding the first term in a Taylor series 
expansion in $\vj$, and noting that the linear term vanishes
due to \refb{ef4}, we get
\be\label{ef10}
\wt\Gamma_B(\vtau^{(0)}+\vec\epsilon, \vec 0)
= - 
\wt W_B(\vtau^{(0)},
\vec J=\vec 0)  
+O(\epsilon^2) = \wt\Gamma_B(\vtau^{(0)}, \vec 0) 
+O(\epsilon^2)\, .
\ee
Thus
\be\label{ef11}
{\p \wt\Gamma_B(\vtau, \vec 0)\over \p \tau_i}=0 \quad \hbox{at}
\quad \vec \tau = \vtau^{(0)}\, .
\ee
Using \refb{ef5} and \refb{ef11} we see that the statistical entropy is
given by the value of
$-\wt\Gamma_B(\vtau, \vec 0)$ at its extremum $\vtau = \vtau^{(0)}$.
Thus we can identify $-\wt\Gamma_B(\vtau, \vec 0)$ as the 
statistical entropy
function. This is computed as the sum of
1PI vacuum amplitudes in the theory
with $\xi_i$ regarded as the fundamental fields.

We shall now show that for a suitable choice of the coordinates
$\vxi$, the statistical
entropy function $-\wt\Gamma_B(\vec\tau, \vec 0)$
defined this way can be made manifestly duality invariant. This is done
by choosing $\vec\xi$ as Riemann normal coordinates. For a given
base point $\vtau_B$ the coordinate $\vxi$ for a given
point $\vtau$ in the upper half plane is
defined as follows. We introduce the duality
invariant metric on the upper half plane
\be\label{eg1}
ds^2 = (\tau_2)^{-2} (d\tau_1^2 + d\tau_2^2)\, ,
\ee
and draw a geodesic connecting $\vtau_B$ and $\vtau$. The coordinate
$\vxi$ corresponding to the point $\vtau$ is then defined by identifying
$|\vxi|$ as the distance between $\vec\tau_B$ and $\vtau$ along the
geodesic and the direction of $\vxi$ is taken to be the direction of
the tangent vector to the geodesic at $\vtau_B$.\footnote{Often one
uses the convention that the distance along the geodesic is
$\sqrt{g_{ij}(\vec\tau_B) \xi^i\xi^j}$. This definition differs 
from the one used here by a multiplicative factor of $\tau_{2B}$.
Since this transforms covariantly under a duality transformation,
both choices of $\vec\xi$ would give manifestly covariant Feynman
rules.} 
Since the metric 
\refb{eg1} is
invariant under a duality transformation, it is clear that if a duality
transformation maps $\vtau_B$ to $\vtau_B'$ and $\vtau$ to $\vtau'$,
then the Riemann normal coordinate $\vxi'$ of $\vtau'$ with respect
to $\vtau'_B$ will have the property that $|\vxi'|=|\vxi|$. Thus $\vxi$
and $\vxi'$ are related by a rotation. In order to determine the angle
of rotation, we note that under a duality transformation
\refb{egamman},
\be\label{eg2tau}
d\tau' = (\gamma\tau + \delta)^{-2} d\tau\, .
\ee
Thus
\be\label{eg3tau}
{d\tau'\over |d\tau'|} = {|\gamma\tau + \delta
|^2\over (\gamma\tau+\delta)^2}\, 
{d\tau\over |d\tau|}\, .
\ee
This shows that a geodesic passing through $\tau_B$ gets rotated by
a phase $|\gamma\tau_B + \delta|^2/(\gamma\tau_B+\delta)^2$ under 
a duality transformation. Hence
\be\label{eg4tau}
\xi' = {|\gamma\tau_B + \delta|^2\over (\gamma\tau_B+\delta)^2}\, 
\xi = 
{\gamma\bar\tau_B + \delta \over \gamma\tau_B+\delta }\, \xi \, ,
\ee
where
\be\label{eg5tau}
\xi = \xi_1 + i \xi_2, \qquad \xi' = \xi'_1 + i \xi'_2\, .
\ee

Since for given $\tau_B$ the duality transformation acts linearly
on $\vxi$, the corresponding generating function $\wt W_B(\vtau_B,
\vj)$ and the effective action $\wt\Gamma_B(\vtau_B, \vpsi)$ will
be manifestly duality invariant under simultaneous transformation
of $\vtau_B$, $\vj$ or $\vpsi$ and of course the charges $\vec Q$
and $\vec P$. In particular the 1PI vacuum amplitude 
$\wt\Gamma_B(\vtau, \vec 0)$ will be invariant under
the duality transformation \refb{egamman}.

We shall now give an algorithm for explicitly generating duality
covariant vertices in this 0-dimensional field theory. For this we
need to expand the duality invariant function $F(\vtau)$ in a Taylor
series expansion in $\vxi$. This is given by:
\be\label{eh1}
F(\vtau) = \sum_{n=0}^\infty {1\over n!} (\tau_{B2})^n
\xi_{i_1}\ldots
\xi_{i_n}\, D_{i_1}
\cdots D_{i_n} F(\vtau)\bigg|_{\vtau=\vtau_B}\, ,
\ee
where $D_i$ denotes  covariant derivative with respect to $\tau_i$,
computed using  the
affine connection $\Gamma^i_{jk}$ constructed from the 
metric \refb{eg1}. We arrive at \refb{eh1} by using the result that
in the $\vec\xi$ coordinate system symmetrized covariant derivatives
are equal to ordinary derivatives. Using this we can replace ordinary
derivatives in the Taylor series expansion by covariant derivatives with
respect to $\xi_i$. In the second step we use the tensorial transformation
properties of covariant derivatives to convert covariant derivative with
respect to $\xi_i$ to covariant derivative with respect to $\tau_i$. The
$(\tau_{B2})^n$ factor in \refb{eh1} arises due to the fact that near
$\vtau=\vtau_B$,
\be\label{eh1.9}
d\tau_i = \tau_{B2} d\xi_i\, .
\ee

In the $(\tau,\bar\tau)$ coordinate system
the non-zero components of the connection
are
\be\label{eh2}
\Gamma^\tau_{\tau\tau} = {i\over \tau_2}, \qquad
\Gamma^{\bar\tau}_{\bar\tau\bar\tau} = -{i\over \tau_2}\, .
\ee
The  curvature tensor computed from this connection has the form
\be\label{eh3}
R^i_{~jkl} = - (\delta^i_k g_{jl} - \delta^i_l g_{jk})\, ,
\ee
which shows that the metric \refb{eg1} describes a constant negative
curvature metric. From \refb{eh2} it follows that
\bea{eh4}
D_\tau (D_\tau^m D_{\bar\tau}^n F(\vec\tau))
&=& (\p_\tau - im/\tau_2) (D_\tau^m D_{\bar\tau}^n F(\vec\tau)),
\nonumber \\
D_{\bar\tau} (D_\tau^m D_{\bar\tau}^n F(\vec\tau))
&=& (\p_{\bar\tau} + in/\tau_2)
(D_\tau^m D_{\bar\tau}^n F(\vec\tau))\, ,
\een
for any arbitrary ordering of $D_\tau$ and $D_{\bar\tau}$ 
in $D_\tau^m D_{\bar\tau}^n F(\vec\tau)$. 
\refb{eh4} provides us with explicit expressions for the
covariant derivatives of $F$ appearing in \refb{eh1}. Also
using
\refb{eh4} one can prove iteratively that under a duality
transformation
\be\label{eh5}
(\tau_2)^{m+n} \, D_\tau^m D_{\bar\tau}^n F(\vec\tau)
\to \left({\gamma\tau+\delta\over \gamma\bar\tau+\delta}\right)^{m-n}\,
 (\tau_2)^{m+n} \, D_\tau^m D_{\bar\tau}^n F(\vec\tau)\, .
 \ee
This establishes manifest
 duality covariance of the vertices constructed from
 \refb{eh1}.
 
 We also need to worry about the contribution from the integration
 measure. The original measure $d^2\tau/(\tau_2)^2$ was duality
 invariant. Since duality transformation acts on $\vec\xi$ as a rotation,
 $d^2\xi$ is also a duality invariant measure. Thus we must have
 \be\label{eh5x}
 {d^2\tau\over (\tau_2)^2} = \JJ(\vtau_B, \vxi) \, d^2\xi\, ,
 \ee
 for some duality invariant function $\JJ(\vtau_B,\vxi)$. It has been
 shown in appendix \ref{sa} that
 \be\label{eh6}
 \JJ(\vtau_B,\vxi) = {1\over |\vec\xi|} \sinh{|\vec\xi|}\, .
 \ee
 We can now regard $-\ln \JJ(\vtau_B,\vxi)$ as an additional contribution
 to the action and expand this in a power series expansion in $\vxi$
 to generate additional vertices. Using the expression for $F(\vec\tau)$
 given in \refb{ek2} we now see that the full `action' is given by
 \bea{eh7}
 F(\vtau) -\ln \JJ(\vtau_B,\vxi)
 &=&
  -\Bigg[ {\pi\over 2 \tau_2} \, |Q -\tau P|^2
-\ln g(\tau) -\ln g(-\bar\tau) - (k+2) \ln (2\tau_2)
\nonumber \\
&& +\ln\bigg\{ K_0 {\pi\over \tau_2} |Q -\tau P|^2 
\bigg\} + \ln \JJ(\vtau_B,\vxi) \nonumber \\
&& + \ln\left( 1 + { 
2(k+3) \tau_2 \over \pi  |Q -\tau P|^2 }\right)\Bigg] \, .
 \nonumber \\
\een
In this expression the first term inside the square
bracket is quadratic in the charges, the last term contains terms
of order $Q^{-2n}$ for $n\ge 1$, and
the other terms are of order $Q^0$. Thus in order to carry out a
systematic expansion in powers of inverse charges we need to regard
the first term as the tree level contribution, the last term as two and
higher loop contributions and the other terms as
the 1-loop contribution. 

We can now evaluate the effective action 
$\wt\Gamma_B(\vec\tau_B)$ in a systematic loop expansion.
The leading term in the effective action is then just the first term in 
\refb{eh7} evaluated at $\vtau=\vtau_B$:
\be\label{eh8}
\wt\Gamma_0(\vtau_B) = -{\pi\over 2 \tau_{2B}} \, |Q -
\tau_B P|^2
\, .
\ee
At the next order we need to include the tree level contribution from the
rest of the terms in the action (except the last term which is higher
order) and one loop
contribution from the first term. 
The former corresponds to these
terms being evaluated at $\vtau=\vtau_B$, \i.e. $\vec\xi=0$.
Since $\JJ(\vtau_B, \vxi=0)=1$, we get this contribution to be
\be\label{ei0}
\ln g(\tau_B) +\ln g(-\bar\tau_B) + (k+2) \ln (2\tau_{2B})
- \ln\bigg\{K_0\, {\pi\over \tau_{2B}} |Q -\tau_B P|^2
\bigg\}\, .
\ee
For the one loop contribution from the first term in the action
we need to expand this term
to quadratic order in $\vxi$ using eqs.\refb{eh1}, \refb{eh4}.
The
order $\vxi$ and $\xi^2$ terms are given by
\be\label{ei1}
-{i\pi\over 4\tau_{2B}} \left\{ \bar\xi (Q-\tau_BP)^2
+ \xi (Q-\bar\tau_BP)^2 \right\}
-{\pi\over 4 \tau_{2B}} \, |Q -\tau_B P|^2 \,
\bar\xi \xi\, .
\ee
The linear term in $\vxi$ do not give any contribution to the 1PI
amplitudes. The contribution from the quadratic term gives
\be\label{ei2}
\ln \left({1\over 4 \tau_{2B}} \, |Q -\tau_B P|^2\right)\, .
\ee
Thus the complete one loop contribution to the effective action
is given by
\be\label{ei4}
\wt \Gamma_1(\vtau_B) = \ln g(\tau_B) +\ln g(-\bar\tau_B) 
+ (k+2) \ln (2\tau_{2B}) -\ln (4 \pi K_0) \, .
\ee
Up to an additive constant $-\wt\Gamma_0(\vtau_B)
-\wt\Gamma_1(\vtau_B)$ agrees with the black hole entropy 
function
given in \refb{eh10bint},
\refb{es11} if we identify $\tau_B$ as $u_a + i u_S$.

 \sectiono{The Integration Measure $\JJ(\vec\tau_B, \vec\xi)$} 
 \label{sa}
 
 In this appendix we shall compute the integration measure 
 $\JJ(\vec\tau_B,\vec\xi)$ which arises from a change of variables
 from $\tau_1,\tau_2$ to the Riemann normal coordinates:
 \be\label{eh5xapp}
 {d^2\tau\over (\tau_2)^2} = \JJ(\vec\tau_B, \vxi) \, d^2\xi\, .
 \ee
 We first note that the duality invariant metric
 \be\label{esa1}
 {1\over \tau_2^2} ( d\tau_1^2 + d\tau_2^2)
 \ee
 describes a metric of constant negative curvature $-1$. Since this is
 a homogeneous space, $\JJ(\vtau_B,\vxi)$ cannot depend on
 $\vec\tau_B$. Now, by defining new coordinate $\theta$, $\phi$ via
 the relations
 \be\label{enewcoord}
 \tanh{\theta\over 2} \, \, e^{i\phi} = {1 + i\tau\over 1 - i\tau}\, ,
 \ee
 we can express the metric \refb{esa1} and the measure
 \refb{eh5xapp} as
 \be\label{esa2}
 ds^2 = d\theta^2 + \sinh^2\theta d\phi^2\, , \qquad
 {d^2\tau\over \tau_2^2} = \sinh\theta \, d\theta\, d\phi\, .
 \ee
 Since $\JJ$ is independent of the base point $\vtau_B$, 
 we can calculate it by taking
 the base point to be at $\theta=0$. The geodesics passing through
 this point are constant $\phi$ lines, and the length measured along
 such a geodesic is given by $\theta$. Thus we have
 \be\label{esa3}
 \vxi = (\theta\cos\phi, \theta\sin\phi)\, .
 \ee
 This gives 
 \be\label{esa4}
 d^2\xi = \theta d\theta d\phi\, .
 \ee
 Comparing this with \refb{esa2} we get
 \be\label{esa4a}
 {d^2\tau\over \tau_2^2} = {\sinh\theta\over \theta} d^2\xi
 = {1\over |\vxi|} \sinh|\vxi| \, d^2\xi\, .
 \ee
 Thus
 \be\label{esa5}
 \JJ(\vtau_B,\vxi) = {1\over |\vxi|} \sinh|\vxi|\, .
 \ee

\sectiono{The Coefficient of the Gauss-Bonnet Term
in Type IIB on $(\MM\times \wt S^1\times
 S^1)/\ZZZ_N$}
\label{ssgauss}

An important four derivative correction to the effective
action in the $\NN=4$ supersymmetric string theories
analyzed in this review is  the Gauss-Bonnet term.
In this appendix we shall describe the
computation of this term.

The calculation is best carried out in the original
description of the theory
as type IIB string theory
compactified on $(\MM\times \wt S^1\times S^1)/\ZZZ_N$. 
We shall denote by $t=t_1+it_2$ 
and $u=u_1+iu_2$ the
Kahler and complex structure moduli of the torus 
$\wt S^1\times S^1$, and use the normalization convention that 
is appropriate for the orbifold theory as described below
\refb{echsq}. Thus for example if
$\wt R$ and $R_0$ denote the radii of $\wt S^1$ and $S^1$
measured in the string metric, and if the off-diagonal
components of the metric and the anti-symmetric tensor field
are zero, then we shall take $t_2=\wt R R_0/N$ and 
$u_2=R_0/(\wt R N)$,
taking into account the fact that in the orbifold theory the 
various fields have $\wt g$-twisted boundary condition under
a $2\pi R_0/N$ translation along $S^1$ and
$2\pi \wt R$ translation along $\wt S^1$. In the same spirit
we shall choose the
units of momentum along $S^1$ and $\wt S^1$ to be $N/R_0$ and
$1/\wt R$ respectively, and unit of winding charge along
$S^1$ and $\wt S^1$ to be $2\pi R_0/N$ and $2\pi \wt R$
respectively. As a result one unit of winding charge along
$S^1$ actually represents a twisted sector state, with twist $g$.

It is known that one loop quantum corrections in this theory
give rise to a Gauss-Bonnet contribution to the effective 
Lagrangian density
of the form\cite{9708062}:
\be\label{es2}
\Delta\LL =  \phi(u_1,u_2)\,
\left\{ R_{\mu\nu\rho\sigma} R^{\mu\nu\rho\sigma}
- 4 R_{\mu\nu} R^{\mu\nu}
+ R^2
\right\} \, ,
\ee
where $\phi(u_1,u_2)$ is a function to be determined, and
$R_{\mu\nu\rho\sigma}$, $R_{\mu\nu}$ and $R$ denote respectively
the Riemann tensor, Ricci tensor and scalar curvature computed from
the canonical Einstein metric.
Note in particular
that $\phi$ is independent of the Kahler modulus $t$ of
$\wt S^1\times S^1$. The analysis of \cite{9708062} shows that
$\phi(u_1,u_2)$ is given by the relation:
\be\label{ehg1}
\p_u \phi(u_1,u_2) =  \int_\FF \, {d^2\tau\over \tau_2} \, \p_u B_4\, ,
\ee
where
$B_4$ is defined as follows. Let us consider type IIB string theory
compactified on $(\MM\times \wt S^1\times 
S^1)/\ZZZ_N$ in the
light-cone gauge Green-Schwarz formulation, denote by 
$Tr^f$ the trace over all states in this world-sheet 
theory carrying some fixed
momentum along the non-compact directions
and denote by $L_0^f$,
$\bar L_0^f$ the Virasoro generators associated with the left
and the right-moving modes, excluding the contribution from the
momenta along the non-compact directions.
We also define $F_L^f$, $F_R^f$ to be the contribution to the
space-time fermion numbers from the left and the right-moving
modes on the world-sheet.
In this case
\be\label{ehg2}
B_4 = K \,
Tr^f\left( q^{L_0^f}\bar q^{\bar L_0^f} (-1)^{F_L^f
+F_R^f} h^4\right), \qquad q\equiv e^{2\pi i \tau}\, ,
\ee
where $K$ is a constant to be determined later and $h$ denotes
the total helicity of the state. In defining $L_0^f$, $\bar L_0^f$
we subtract the constants $c_L/24$ and $c_R/24$, so that the RR
vacuum has $L_0^f=\bar L_0^f=0$.

The evaluation of the right hand side of \refb{ehg2} proceeds as
follows. We first note that since $(\MM\times \wt S^1
\times S^1)/\ZZZ_N$ has $SU(2)$ holonomy, and since
a spinor representation of $SO(8)$ transforms as a pair of doublets
and four singlets under this $SU(2)$ group, we have four free
left-moving fermions and four free right-moving fermions
associated with the singlets of $SU(2)$. These give rise to
altogether eight fermion zero modes.  
Since quantization
of a conjugate pair
of fermion zero modes $(\psi_0, \psi_0^\dagger)$
gives rise to a pair of 
states with opposite 
$(-1)^{F_L^f
+F_R^f}$, without the $h^4$ term the trace in \refb{ehg2}
will vanish. This can be avoided if we insert a factor of $h$
in the trace and pick the contribution to $h$ from this
particular conjugate pair of fermions, -- in this case the two
states have the same $(-1)^{F_L^f
+F_R^f}\, h$ quantum numbers. This can be repeated for every
pair of conjugate fermions.  
Altogether we need four factors of $h$ to soak up all the 
eight fermion
zero modes. Thus in effect we can simplify \refb{ehg2} by
expressing it as
\be\label{ehg3}
B_4 = K' Tr^{f\prime}
\left( q^{L_0^f}\bar q^{\bar L_0^f} (-1)^{F_L^f
+F_R^f} \right)
\ee
where $K'$ is a  different normalization constant and the prime in
the trace denotes that we should ignore the effect of fermion zero
modes in evaluating the trace.

Since we are using the Green-Schwarz formulation, the 4 left-moving
and 4 right-moving fermions which are neutral under the 
holonomy group 
satisfy periodic boundary condition. Thus the effect of the
non-zero mode oscillators associated with these fermions cancel
against the contribution from the non-zero mode bosonic oscillators
associated with the circles $S^1$ and $\wt S^1$ and the two
non-compact directions. This leads to a further simplification in
which the trace can be taken over only the degrees of freedom
associated with the compact space $\MM$ and the bosonic zero modes
associated with the circles $S^1$ and $\wt S^1$. The latter includes
the quantum numbers $m_1$ and $m_2$ denoting the number of units of
momentum along $\wt S^1$ and $S^1$, and the quantum numbers $n_1$ and
$n_2$ denoting the number of units of winding along $\wt S^1$ and
$S^1$. The units of momentum and winding along the two circles are
chosen according to the convention described earlier. Thus for example
$m_2$ unit of momentum along $S^1$ will correspond to a physical
momentum of $Nm_2/R_0$ in string units. This shows that $m_2$ can be
fractional, being quantized in units of $1/N$.  On the other hand a
sector with $n_2$ unit of winding along $S^1$ describes a fundamental
string of length $2\pi n_2 R_0/N$, and hence this state belongs to a
sector twisted by $g^{n_2}$.

In this convention the contributions to $\bar L_0^f$ and $L_0^f$ from
the bosonic zero modes associated with $\wt S^1\times 
S^1$ are given
by, respectively,
\be\label{ehg4}
{1\over 2} k_R^2 = {1\over 4 t_2 u_2} |-m_1 u + m_2
+ n_1 t + n_2 tu|^2\, ,
\qquad 
{1\over 2} k_L^2 = {1\over 2} k_R^2 + m_1 n_1 + m_2 n_2\, .
\ee
Furthermore, since under the $SU(2)$ holonomy group
a vector in the tangent space of $\MM$ also transform as a pair
of doublets, the fermions in our system which transform as doublets
of $SU(2)$ may be regarded as tangent space vectors.
As a result, these fermions, together with the scalars associated with
the coordinates of $\MM$, describe a superconformal field theory
with target space $\MM$.
Thus \refb{ehg3} may now be rewritten as
\be\label{ehg5}
B_4 = {K' \over N} \, \sum_{r=0}^{N-1} \sum_{s=0}^{N-1}
\sum_{\stackrel{m_1, n_1\in \zzz, m_2\in \zzz/N}{n_2\in N\zzz+r
}}
q^{k_L^2/2} \bar q^{ k_R^2/2} e^{2\pi i m_2 s} 
Tr_{RR, \wt g^r} \left ((-1)^{F_L+F_R} \wt g^s 
q^{L_0} \bar q^{\bar L_0}\right)
\, ,
\ee
where $Tr_{RR;\wt g^r}$ denotes trace over the $\wt g^r$-twisted
sector RR states of the (4,4) superconformal field theory with target
space $\MM$ and
$L_0$ and $\bar L_0$ denote contribution to the Virasoro
generators in this superconformal field theory with
$c_L/24$ and $c_R/24$ subtracted.
The sum over $s$ in \refb{ehg5} arises from the insertion of the
projection operator ${1\over N}\sum_{s=0}^{N-1}g^s$ in the trace,
while the sum over $r$ represents the sum over various twisted sector
states.    As required, the quantum number $n_2$ that determines
the part of $g$-twist along $S^1$ is correlated with the integer $r$
that determines the amount of $g$-twist along $\MM$. The $e^{2\pi i
  m_2 s}$ factor represents part of $g^s$ that acts as translation
along $S^1$ while the action of $g^s$ on $\MM$ is represented by the
operator $\wt g^s$ inserted into the trace.

We now note that the trace part in \refb{ehg5} is precisely the
quantity $N\, F^{(r,s)}(\tau, z=0)$ defined in \refb{esi4aint} 
for $q=e^{2\pi i\tau}$.  Thus we can rewrite
\refb{ehg5} as
\be\label{ehg6}
B_4 = {K' } \, \sum_{r=0}^{N-1} \sum_{s=0}^{N-1}
\sum_{\stackrel{m_1, n_1\in \zzz, m_2\in \zzz/N}{n_2\in N\zzz+r
}}
q^{k_L^2/2} \bar q^{ k_R^2/2} e^{2\pi i m_2 s} 
F^{(r,s)}(\tau,0)
\, .
\ee

We shall now compare \refb{ehg6} with the expression for 
$\hI(\wt\rho,\wt\sigma,
\wt v)$ given in \refb{rwthrint}, \refb{enn6} at $\wt\rho=u$,
$\wt\sigma=t$ and $\wt v=0$. 
In this case $p_R^2$, $p_L^2$ defined in \refb{e7n} reduces
to $k_R^2$ and $k_L^2 + {1\over 2}j^2$ respectively, with
$k_R^2$, $k_L^2$ given in \refb{ehg4}. As a result we have
\bea{ehg8new}
&& \hI(u,t,0) = \int_{\FF} \frac{d^2\tau}{\tau_2}\, \bigg[
\sum_ {r, s =0}^{N-1}\sum_{b=0}^1\,
\sum_{\stackrel{m_1, n_1\in \zzz, m_2\in \zzz/N}{n_2\in N\zzz+ r, 
j\in 2\zzz + b}}
q^{k_L^2/2} \bar q^{ k_R^2/2} q^{j^2/4}
e^{2\pi i m_2 s} 
h_b^{(r,s)}(\tau) - \sum_{s=0}^{N-1} c_0^{(0,s)}(0)\bigg]
\nonumber \\
&=& \int_{\FF} \frac{d^2\tau}{\tau_2}\, \bigg[
\sum_ {r, s =0}^{N-1}\,
\sum_{\stackrel{m_1, n_1\in \zzz, m_2\in \zzz/N}{n_2\in N\zzz+ r}}
q^{k_L^2/2} \bar q^{ k_R^2/2}  
e^{2\pi i m_2 s} \big(\vt_3(2\tau,0) h_0^{(r,s)}(\tau) \nonumber \\
&& \qquad \qquad 
+\vt_2(2\tau,0) h_1^{(r,s)}(\tau)\big)- \sum_{s=0}^{N-1} 
c_0^{(0,s)}(0)\bigg]
\nonumber \\
&=& \int_{\FF} \frac{d^2\tau}{\tau_2}\, \bigg[
\sum_ {r, s =0}^{N-1}\,
\sum_{\stackrel{m_1, n_1\in \zzz, m_2\in \zzz/N}{n_2\in N\zzz+ r}}
q^{k_L^2/2} \bar q^{ k_R^2/2}  
e^{2\pi i m_2 s} F^{(r,s)}(\tau, 0)- \sum_{s=0}^{N-1} 
c_0^{(0,s)}(0)\bigg]\, ,
\eea
where in the second step we have expressed the result of summing
over $j$ in terms of Jacobi $\vartheta$-functions, and
in the last step we have used eq.\refb{efhrel}.
Comparing \refb{ehg6} with \refb{ehg8new} we see that
\be\label{ehg7}
\int_\FF {d^2\tau\over \tau_2} \left(B_4- K'
\sum_{s=0}^{N-1} 
c_0^{(0,s)}(0)\right)
= K'\, \hI(u,t,0)\, .
\ee
Using \refb{enn9} and \refb{enn12} we get
\bea{ehg9a}
\int_\FF {d^2\tau\over \tau_2} \left(B_4- K'\sum_{s=0}^{N-1} 
c_0^{(0,s)}(0)\right)
&=& -2 \, K' \lim_{v\to 0} \bigg[
\left(   k \ln t_2 + k \ln u_2 + 2\ln v + 2\ln \bar v
\right.\nonumber \\
&&
\left. + \ln g(t) + \ln \overline{g( t)} + \ln g(u) + \ln 
\overline{g(u)} \right)\bigg] 
+ \hbox{constant}\, . \nonumber \\
\eea
Naively the right hand side diverges in the $v\to 0$
limit. The origin of this
infinity lies in the fact that 
$\int d^2\tau B_4/\tau_2$ has divergences from integration
over the large $\tau_2$
region for $\vec m=\vec n =0$, and this divergence
is not completely removed by the subtraction term
proportional to $K'\sum_s c_0^{(0,s)}(0)$ in the integrand. 
The correct subtraction term in the integrand must be equal to
$K'\lim_{\tau\to i\infty} F^{(0,s)}(\tau, 0)$, -- from
\refb{enewint} 
we see that
this is given by $K'\sum_s \left( c_0^{(0,s)}(0) 
+ 2 c_1^{(0,s)}(-1)\right)$. The extra counteterm proportional to
$c_1^{(0,s)}(-1)$ is not needed for regulating $\wh\II$
since the correponding potential divergence
from the term $m_1=n_1=m_2=n_2=0$, $j=\pm 1$ in \refb{enn6}.
takes the form:
\be \label{egiven1}
\int {d^2\tau\over \tau_2^2} \, \exp\left( -{2\pi\over t_2 u_2
-v_2^2} \,  |v|^2\,
\sum_{s=0}^{N-1} c_1^{(0,s)}(-1)\right)
\simeq -2 \sum_{s=0}^{N-1} c_1^{(0,s)}(-1) \ln {|v|^2\over t_2 u_2}
+\hbox{constant} \simeq -4 \ln {|v|^2\over t_2 u_2}
\ee
for small $v$. 
This is divergent in the $v\to 0$ limit but finite for small but
finite $v$. Thus
in order to recover 
\be \label{erecover}
\int\,{d^2\tau\over \tau_2}
\left[ B_4 - K'\sum_s \left( c_0^{(0,s)}(0) 
+ 2 c_1^{(0,s)}(-1)\right)\right]\, 
\ee
 from $\wh \II$
we need to first remove the contribution
$-4 \ln {|v|^2\over t_2 u_2}$ from $\wh\II$ and then take the
$v\to 0$ limit.
This gives
\bea{ehg9}
&& \int_\FF {d^2\tau\over \tau_2} \left(B_4 - K'
\sum_{s=0}^{N-1} 
\left(c_0^{(0,s)}(0) + 2 c_1^{(0,s)}(-1)\right)\right)
 \nonumber \\
&=& -2 \, K' \left(   (k+2) \ln t_2 + (k+2) \ln u_2 + \ln g(t) 
+ \ln g(-\bar t) + \ln g(u) + \ln g(-\bar u) \right) + \hbox{constant}\, .
\nonumber \\
\eea
In writing down \refb{ehg9} we have used $\overline{g(t)}
=g(-\bar t)$, -- this follows from the definition
\refb{enn13} of $g(\rho)$ and the identity 
$s_r=s_{-r}$.
Comparing \refb{ehg1} with \refb{ehg9} we now get
\be\label{ehg10}
\phi(u_1,u_2) = - 2 \, K' \, \left( (k+2) \ln u_2 
+ \ln g(u) + \ln g(-\bar u)\right)
+\hbox{constant}\, .
\ee

We now turn to the determination of $K'$. This constant is universal
independent of the specific theory we are analysing. Thus we can find it
by working with the  type IIB string theory compactified on 
$K3\times \wt S^1\times S^1$. In
this case $k=10$ and $g(\tau)=\eta(\tau)^{24}$. 
This matches with the known
answer\cite{9906094,0007195} 
for $\phi(u_1,u_2)$ if we choose $K'=1/(128\pi^2)$. 
Thus we have
\be\label{ehg10a}
\phi(u_1,u_2) = - {1\over 64\pi^2} \, \left( (k+2) \ln u_2 
+ \ln g(u) + \ln g(-\bar u)\right)
+\hbox{constant}\, .
\ee

Under the duality map that relates type IIB string theory on the 
$\ZZZ_N$ orbifold of $\MM\times \wt
S^1\times S^1$ to an
asymmetric $\ZZZ_N$
orbifold of heterotic or type IIA string theory on $T^6$, the modulus
$u$ of the original 
type IIB string theory gets related to the axion-dilaton
modulus $a+iS$ of the final asymmetric orbifold theory. Thus in
this description the Gauss-Bonnet term in the effective 
Lagrangian density
takes the form
\be\label{ehg10-}
\Delta\LL =  \phi(a,S)\,
\left\{ R_{\mu\nu\rho\sigma} R^{\mu\nu\rho\sigma}
- 4 R_{\mu\nu} R^{\mu\nu}
+ R^2
\right\} \, .
\ee

\end{document}